\documentclass{aa}

\newcommand{\hii}{H{\sc II~}}

\newcommand{\sio}{SiO}
\newcommand{\hcop}{HCO$^{+}$}
\newcommand{\nthp}{N$_{2}$H$^{+}$}
\newcommand{\nht}{NH$_{3}$}

\newcommand{\hnc}{HNC}

\newcommand{\hcn}{HCN}

\newcommand{\hntc}{HN$^{13}$C}
\newcommand{\htcop}{H$^{13}$CO$^{+}$}

\newcommand{\nodata}{-}

\usepackage{amsmath}
\usepackage[utf8]{inputenc}
\usepackage{graphicx}
\usepackage{txfonts}
\usepackage{footnote}
\usepackage{latexsym}
\usepackage{amssymb}
\usepackage{hyperref}
\usepackage{natbib}
\usepackage{color}
\usepackage{threeparttable}
\usepackage{float}
\usepackage{subcaption}
\usepackage{wrapfig}
\usepackage{appendix}

\def\kms{$\displaystyle {\rm{km~s^{-1}}}$}
\def\Kkms{$\displaystyle {\rm{K~km~ s^{-1}}}$}

\def\cm2{cm$^{-2}$}
\def\msun{\ifmmode M_{\odot} \else M$_{\odot}$\fi}

\begin{document}
\title{MALT90 molecular content on high-mass IR-dark clumps}
\titlerunning{Astronomy \& Astrophysics manuscript no. output}
\authorrunning {Gozde Saral et al.}
\author{Gozde Saral\inst{1}, Marc Audard\inst{1}, Yuan Wang\inst{2}}
\institute{Department of Astronomy, University of Geneva, Chemin d`Ecogia 16, CH-1290 Versoix, Switzerland\\
\email{gozde.saral@unige.ch}
\and 
Max Planck Institute for Astronomy, K{\"o}nigstuhl 17, 69117 Heidelberg, Germany} 

\date{Received 7 March 2018 /
Accepted 15 October 2018 }

\abstract{High mass stars form in groups or clusters in dense molecular clumps with sizes of 1~pc and masses of 200~M$_\odot$. Infrared-dark clumps and the individual cores within them with sizes $<$0.1~pc and masses $<$100 M$_\odot$ are important laboratories for high-mass star formation in order to study the initial conditions.
}
{We investigate the physical and chemical properties of high-mass clumps in order to better understand the early evolutionary stages and find targets that show star formation signs such as infall motions or outflows.}
{We selected the high-mass clumps from ATLASGAL survey that were identified as dark at 8/24 $\mu$m wavelengths. We used MALT90 Survey data which provides a molecular line set (\hcop, \hnc, \hcn, \nthp, \htcop, \hntc, \sio) to investigate the physical and chemical conditions in early stages of star formation.}
{(1) Eleven sources have significant \sio~detection (over 3$\sigma$) which usually indicates outflow activity.
(2) Thirteen sources are found with blue profiles in both or either HCO$^+$ and/or HNC lines and clump mass infall rates are estimated to be in the range of 0.2 $\times$ 10$^{-3}$ M$_\odot$yr$^{-1}$ $-$ 1.8 $\times$10$^{-2}$ M$_\odot$yr$^{-1}$.
(3) The excitation temperature is obtained as < 24~K for all sources. 
(4) The column densities for optically thin lines of \htcop~and \hntc~are in the range of 0.4-8.8($\times$10$^{12}$) cm$^{-2}$, and 0.9-11.9($\times$10$^{12}$) cm$^{-2}$, respectively, while it is in the range of 0.1-7.5 ($\times$10$^{14}$) cm$^{-2}$ for \hcop~and \hnc~lines. The column densities for \nthp~were ranging between 4.4-275.7($\times$10$^{12}$) cm$^{-2}$ as expected from cold dense regions.
(5) Large line widths of \nthp~might indicate turbulence and large line widths of \hcop, \hnc, and \sio~indicate outflow activities.
(6) Mean optical depths are 20.32, and 23.19 for optically thick \hcop~and \hcn~lines, and 0.39 and 0.45 for their optically thin isotopologues \htcop~and \hntc, respectively.}
{This study reveals the physical and chemical properties of 30 high-mass IR-dark clumps and the interesting targets among them based on their emission line morphology and kinematics.}

\keywords{Stars: formation -- Stars: massive -- ISM: lines and bands -- ISM: molecules -- ISM: abundance -- Submillimeter: ISM} \authorrunning{Saral et al.}

\maketitle
\section{Introduction}
High-mass stars (L$>$ 10$^{3}$ L$_\odot$) play a vital role in the star formation process from their birth to death, yet their formation is less understood than that of low-mass stars \citep[e.g.,][]{beu07,zin07}. Since they form and evolve quickly and ionize and disrupt their natal molecular cloud by strong winds and intense UV radiation, observations of their earliest evolutionary phases are very difficult. In addition, the large distances and dense material in the molecular clouds require high angular resolution and high sensitivity observations which make systematic surveys difficult. Therefore, there is no well established evolutionary sequence for high-mass stars. Based on observational and theoretical studies the following sequence has been proposed: massive starless cores, gravitationally bound prestellar cores or infrared-dark clouds as the first phase; high-mass protostellar objects or hot molecular cores as the second phase; and \hii regions as the final phase \citep[see the recent review:][]{mot17}. In order to better understand the initial conditions of star formation, recent studies focused on mainly nearby prestellar cores and low-mass infrared dark clouds (IRDCs) to study the morphology of the cores, as well as their temperature structures and chemical properties. While these observations had sensitivity and resolution limitations, new instruments like the IRAM30 and APEX telescopes and interferometers like the SMA, CARMA, and ALMA have access to high sensitivity and resolution observations. 

Until recently it has been thought that most high-mass stars form in clustered environments within dense molecular clouds along with lower-mass stars \citep{mot98,lad03,dif07} and recently interferometric observations of IRDCs revealed that these clumps have sizes of $\sim$1~pc and masses of $\sim$200 M$_\odot$ and they host very young star-forming cores \citep[e.g.,][]{beu05}. On the other hand, it is the dense cores (with sizes $<$0.1 pc and masses $<$100 M$_\odot$) within these clumps which are forming protostars or binaries. Also, high angular resolution observations reveal that the individual prestellar cores in clusters are more compact than isolated cores with scales of 0.02-0.03 pc and they do not show further subfragmentation \citep{beu15}. Although several surveys were performed to study the early evolutionary stages of high-mass stars, they targeted mainly the later stages such as \hii regions \citep[e.g., CORNISH,][]{hoa12} and massive young stellar objects (MYSOs) \citep[e.g., RMS,][]{urq08}. Within this context, to study the earlier stages and have a more complete picture of high-mass star formation, the APEX Telescope Large Area Survey\footnote[1]{APEX is a collaboration between the Max-Planck-Institut für Radioastronomie, the European Southern Observatory, and the Onsala Space Observatory} \citep[ATLASGAL,][]{sch09} mapped the dust continuum emission at 870~$\mu$m systematically, in the inner Galactic plane over several 100 $deg^{2}$ with a uniform sensitivity. The ATLASGAL Survey identified and classified over 5000 compact sources \citep{con13,cse14} at early evolutionary stages from quiescent clumps and protostellar cores to \hii regions and PDRs in a series of papers \citep{sch09,gia14,kon17,cse16}. Later, a subsample of ATLASGAL sources were observed with the 22~m single-dish Australia Telescope National Facility (ATNF) MOPRA Telescope\footnote[2]{The Mopra radio telescope is part of the Australia Telescope National Facility which is funded by the Commonwealth of Australia for operation as a National Facility managed by CSIRO.}, MALT90 Survey to examine their chemical and physical properties as well as evolutionary phases \citep{jac13,hoq13,fos13,fos11,mie14,rat16}.

In this study, we began an investigation of high-mass dense clumps in the ATLASGAL survey which were also observed in the MALT90 Survey, have high masses ($>$100~M$_\odot$), but do not have any infrared counterparts in the IRAC 8~$\mu$m or MIPS 24~$\mu$m band, indicating that they are at very early stages of star formation \citep{gia14}. Based on these criteria, 30 infrared dark clouds were selected and their physical and chemical properties were studied in order to better understand the initial conditions during the early evolutionary phases of high-mass stars. Spectral line profiles are examined in order to determine the infall signatures within the clumps which is one of the crucial signs of early stages of star formation \citep{wu07,bel06,ryg13} and especially for massive star formation \citep{mck03}. The line profiles of \hcop, and \htcop~are generally used to study infall and outflow activities in star-forming regions \citep{wu07,che10,cse11,ryg13,smi13}. Double-peaked lines with stronger blue or stronger red profiles, or single-peaked lines with blue- or red-skewed profiles give clues about outflows, infall motions, or rotation \citep{zho92,mye96,eva99,nar02,sun09}. Infalls are usually accompanied by outflows due to accretion during the early stages of protostar formation \citep{li14} and can be well traced by enhanced abundance of \sio~\citep{sch97}. Strong red profiles are found as outflow signs in a survey of extended green objects (EGOs) in massive star-forming regions \citep{che10}. In order to find possible infall or outflow candidate sources, generally optically thick lines are used based on their double-peaked and strong blue or red peak profiles. On the other hand, optically thin lines, for example of \htcop, will not give a double-peaked profile and can be used to determine the systemic velocities of the clumps. In cases with double or multiple line profiles in optically thin lines, we assumed multiple cores along the line of sight. We picked these sources as outflow candidates in cases of a strong red profile and line wings.

In Section 2, we briefly introduce the selected sample and the ATLASGAL, MALT90, and Spitzer data used in this study. In Section 3, we present the spectral line analysis, and derived physical parameters. In Section 4, we discuss the kinematics, abundance ratios of different molecular species, and their relations in order to understand the evolutionary trends in chemical compositions. Finally, in Section 5 we summarize our results. 

\section{Data}

\subsection{Archival data}

In this study, we have used the ATLASGAL data by downloading the fits files from the online archive\footnote[3]{http://atlasgal.mpifr-bonn.mpg.de/cgi-bin/ATLASGAL$\_$DATABASE.cgi} in order to visually inspect the dust emission maps and compare them to the MALT90 molecular line emission maps. The resolution with the Large APEX Bolometer Camera (LABOCA) at 870~$\mu$m wavelength is 19$''$.2 $FWHM$ which makes the ATLASGAL Survey observations sensitive to $\sim$0.5~pc molecular clumps at a distance of 5~kpc. As a starting point, we selected all the clumps identified by ATLASGAL as dark in the 8/24~$\mu$m bands and retrieved their information such as $V_{LSR}$, mass, distance, and radius from \citet{gia14} and \citet{wie15}. Since the dust emission is generally optically thin in the submm band, the masses of the clumps are proportional to the total flux densities. However, there might be large uncertainty in the mass values due to uncertainties in the dust temperature as stated in \citet{mot07}. Distances of the ATLASGAL clumps are taken from \citet{wie15} where they used HI self-absorption and HI absorption methods to resolve the kinematic distance ambiguity.

We downloaded the MALT90 data cubes of selected targets from the online archive\footnote[4]{http://malt90.bu.edu and http://atoa.atnf.csiro.au/}. The MALT90 Survey mapped the 90 GHz line emission with an angular resolution of 38$''$ and spectral resolution of 0.11 {km s$^{-1}$}. It provides simultaneously obtained 16 molecular emission lines that are typically observed in dense molecular clumps and cores. We used molecular lines of \hcop, \htcop, \hnc, \hcn, and {\nthp} since they are the strongest species with their high critical densities. We also analyzed  isotopologues of these lines, which are used to calculate the optical depths and thus column densities. {\sio} data is used to investigate possible outflows. The \textit{Spitzer}/IRAC 3.6~$\mu$m, 4.5~$\mu$m, 5.8~$\mu$m, and 8~$\mu$m mid-IR images and MIPS 24~$\mu$m images from MIPSGAL were also used to create three-color images in order to visually inspect the infrared appearance of the environment of the clumps.

\subsection{Sample selection}

The first stage was to select the sources which were previously identified in the ATLASGAL catalog as dark clumps at 8/24~$\mu$m wavelengths \citep[from Table A.1 in][]{gia14} which were 45 in total. \citet{gia14} identified a source as 8~$\mu$m- or 24~$\mu$m-dark if its flux density is smaller than the average flux at the same wavelength in the vicinity of the source. Following that, we 
downloaded the molecular line data of 36 high-mass clumps which were observed with MALT90 survey. Such observations of thirty of these sources (with high signal-to-noise ratios) were analyzed in this study. Table \ref{malt90sourcetable} summarizes the physical properties of these clumps taken from the ATLASGAL catalogs \citep{urq18,gia14,wie15}.

\begin{table*}
\centering
\begin{threeparttable}
\caption{\label{malt90sourcetable}Source table.}
\tiny
\begin{tabular}{llllllllll}
\hline\hline
\noalign{\smallskip}
No & Source\tnote{a} & R.A. (J2000) & Dec. (J2000) & $V_{LSR}$ & Distance & Mass & Radius & Comments\tnote{c} & Comments\tnote{d} \\
& & & & (km s$^{-1}$) & (kpc) & (10$^{3}$ M$_\odot$) & (pc) & \\
\hline
1 & G008.684-00.367 & 18:06:23.35  & -21:37:05.2 & 37.6 & 4.4           & 2.2 & 1.7 & D8 & MSF\\
2 & G008.706-00.414 & 18:06:36.72  & -21:37:18.6 & 38.5 & 11.8                   & 5.3 & 2.9 & D24 & YSO \\
3 & G010.444-00.017 & 18:08:44.59  & -19:54:36.4 & 72.4 & 10.2                   & 4.3 & 2.3 & D24 & MSF \\
4 & G013.178+00.059 & 18:14:01.32  & -17:28:38.6 & 48.5 & 4.5           & 1.0 & 1.6 & D8 & MSF \\
5 & G014.114-00.574  & 18:18:13.14 & -16:57:20.0 & 20.1 & 1.9                    & 0.8 & 1.3 & D8 & YSO \\
6 & G014.194-00.194  & 18:16:58.77 & -16:42:17.5 & 39.4 & 3.1                    & 2.7 & 1.4 & D8 & YSO \\
7 & G014.492-00.139 & 18:17:22.09 & -16:24:59.4 & 39.7  & 3.2                   & 2.0 & 2.7 & D24 & YSO \\
8 & G014.632-00.577 & 18:19:15.22 & -16:30:02.3 & 18.2  & 1.9                   & 0.5 & 1.0 & D8 & MSF \\
9 & G018.876-00.489 & 18:27:07.69 & -12:41:36.4 & 65.5  & 5.0                   & 1.2 & 1.5 & D8 & YSO \\
10 & G309.382-00.134 & 13:47:24.03 & -62:18:09.5 & -49.5 & 6.7                  & 2.4 & 3.2 & D8 & MSF \\
11 & G317.867-00.151 & 14:53:16.60 & -59:26:33.9 & -40.2 & 2.5                  & 0.3 & 0.7 & D8 & YSO \\
12 & G318.779-00.137 & 14:59:33.06 & -59:00:33.5 & -37.5 & 10.2                 & 13.5 & 5.2 & D8 & YSO \\
13 & G320.881-00.397 & 15:14:33.18 & -58:11:28.5 & -44.8 & 2.8                  & 0.4 & 0.7 & D24 & YSO \\
14 & G326.987-00.031 & 15:49:08.16 & -54:23:06.9 & -58.2 & 3.5                  & 1.0 & 1.1 & D8 & Protostellar\\
15 & G329.029-00.206 & 16:00:31.96 & -53:12:54.0 & -43.5 & 2.8                  & 2.8 & 1.9 & D8 & MSF \\
16 & G331.708+00.583 & 16:10:06.79 & -50:50:29.4 & -66.8 & 10.8                 & 12.1 & 4.7 & D8 & YSO \\
17 & G333.656+00.059 & 16:21:11.61 & -49:52:16.3 & -84.2 & 5.0                  & 2.7 & 3.1 & D24 & Protostellar \\
18 & G335.789+00.174 & 16:29:47.34 & -48:15:52.1 & -49.5 & 3.3                  & 1.9 & 2.2 & D8 & MSF \\
19 & G336.958-00.224 & 16:36:17.12 & -47:40:44.3 & -71.1 & 10.9                 & 1.5 & 1.8 & D24 & MSF \\
20 & G337.176-00.032 & 16:36:18.78 & -47:23:18.6 & -65.6 & 11.2                 & 23.9 & 8.7 & D8 & MSF \\
21 & G337.258-00.101 & 16:36:56.42 & -47:22:27.0 & -67.6 & 11.1                 & 10.5 & 5.4 & D8 & MSF \\
22 & G338.786+00.476 & 16:40:22.55 & -45:51:04.1 & -64.6 & 11.3                 & 2.2 & 2.6 & D24 & Protostellar\\
23 & G340.784-00.097 & 16:50:14.89 &  -44:42:31.6 & -101.4 & 9.6                & 3.7 & 3.5 & D8 & MSF \\
24 & G342.484+00.182 & 16:55:02.53 & -43:13:03.1 & -41.3 & 12.6                 & 11.0 & 6.1 & D8 & MSF \\
25 & G343.756-00.164 & 17:00:50.08 & -42:26:11.9 & -26.9 & 2.5                  & 1.3 & 1.5 & D8 & MSF \\
26 & G351.444+00.659 & 17:20:54.60 & -35:45:12.8 & -3.9 & 1.4                    & 3.1 & 1.0 & D8 & MSF \\
27 & G351.571+00.762 & 17:20:51.00 & -35:35:25.8 & -2.7 & 1.3                   & 2.0 & 1.3 & D24 & YSO \\
28 & G353.066+00.452 & 17:26:13.50 & -34:31:54.0 & 1.9 & 1.3                    & 0.2 & 0.7 & D24 & YSO \\
29 & G353.417-00.079 & 17:29:18.94  & -34:32:05.9 & -54.2 & 10.2                 & 6.8 & 3.1 & D24 & Protostellar\\
30 & G354.944-00.537 & 17:35:11.69  & -33:30:23.8 & -5.5 & 1.4                   & 0.1 & 0.6 & D24 & YSO \\
\hline
\end{tabular}
\begin{tablenotes}
\item[a]{Massive clumps that were detected by the ATLASGAL survey \citep{gia14} and classified as objects dark at $8.0$~$\mu$m or $24.0$~$\mu$m.}
\item[b]{Kinematic distances, $V_{LSR}$, mass and radius are taken from \citet{urq18} which provides the most updated values based on previous ATLASGAL studies and follow-up studies \citep[e.g.,][]{wie15,gia14}. \citet{urq18} combined the solutions from HI self-absorption and HI absorption methods using molecular line observations, maser parallax and spectroscopic distance measurements when available. Masses are based on 870~$\mu$m dust continuum and \nht~lines. There might be large uncertainty in the mass due to uncertain dust emissivity as stated in \citet{mot07}. Also, note that, the clumps are extended compared to the beam size, thus the mass values are above the sensitivity limit of ATLASGAL.}
\item[c]{\citet{gia14} identified the sources as dark at 8.0~$\mu$m or 24.0~$\mu$m if the flux density is smaller than the average flux at the same wavelength in the vicinity of the source. These sources are marked as D8 or D24.}
\item[d]{\citet{urq18} provides associated sources such as massive star forming (MSF) regions and young stellar objects (YSO) from the follow-up observations of these clumps.}
\end{tablenotes}
\end{threeparttable}
\end{table*}
\subsection{Molecular line data}\label{sec:molecularline}

The MALT90 Survey observed 16 molecular lines at 90~GHz toward $\sim$2000 dense molecular clumps that were identified by the ATLASGAL Survey. Among these lines, {\hcop}(1-0), {\hnc}(1-0), {\hcn}(1-0), and {\nthp}(1-0) are the brightest ones, and are density tracers. They are detected toward all clumps and analyzed in this study for the selected targets. In addition to \hcop, \nthp, \hcn, \hnc, and \sio, the 1-0 lines of the isotopologues \htcop~and \hntc, as density tracers, are also analyzed. Besides the density tracers, another important molecular line observed in star-forming regions where shocks/outflows are present is {\sio}(2-1). There are multiple processes that can produce SiO emission in molecular clumps which can be a good tracer for both medium-velocity shocks ($\sim$10-20 {km s$^{-1}$}) or high-velocity shocks ($\sim$20-50 {km s$^{-1}$}) as discussed in \citet{mie14} and references therein. The line data used in this study are shown in Table~\ref{tab_lines}.

\begin{table}
\centering
\footnotesize
\begin{threeparttable}
\caption{\label{tab_lines} MALT90 Molecular line observation parameters.}
\centering
\begin{tabular}{lcc}
\hline\hline
Lines &  Frequency\tnote{a} & Tracer\\
        & (MHz) & \\ 
\hline
\noalign{\smallskip}
{\hcop}(1-0)\tnote{b}   & 89188.526 & density, kinematics \\
{\nthp}(1-0)\tnote{c}   & 93173.770 & density, chemically robust \\
{\hnc}(1-0)\tnote{b}    & 90663.572 & density, cold chemistry \\
{\hcn}(1-0)\tnote{d}    & 88631.847 & density \\
{\htcop}(1-0)  & 86754.330 & optical depth, column density, V$_{LSR}$ \\
{\hntc}(1-0)   & 87090.859 & optical depth, column density, V$_{LSR}$ \\
{\sio}(2-1)    & 86847.010 & shock, outflow \\
\hline
\end{tabular}
\begin{tablenotes}
\item[a]{The frequencies for the lines were obtained from the Cologne Database for Molecular Spectroscopy (CDMS, \citealt{mul01, mul05}) and the Leiden atomic and molecular database \citep[LAMDA;][]{sch05}.}
\item[b]{Some of the \hcop and \hnc~lines show double peaks due to infall, outflow, rotation, etc.}
\item[c]{\nthp~line has 15 hyperfine components.}
\item[d]{\hcn~line has three hyperfine components.}
\end{tablenotes}
\end{threeparttable}
\end{table}

\section{Analysis}\label{sec:specanalysis}

\subsection{Spectral line fitting}\label{sec:specfit}

\begin{figure}
\centering
\includegraphics[width=0.5\textwidth]{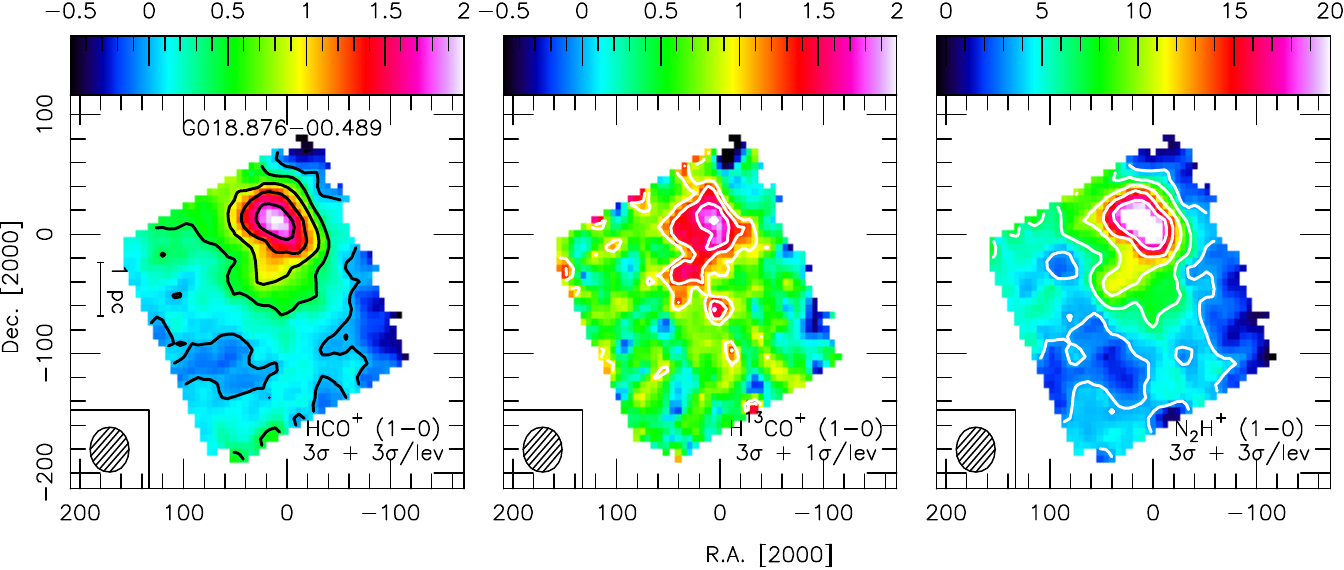}
\includegraphics[width=0.46\textwidth]{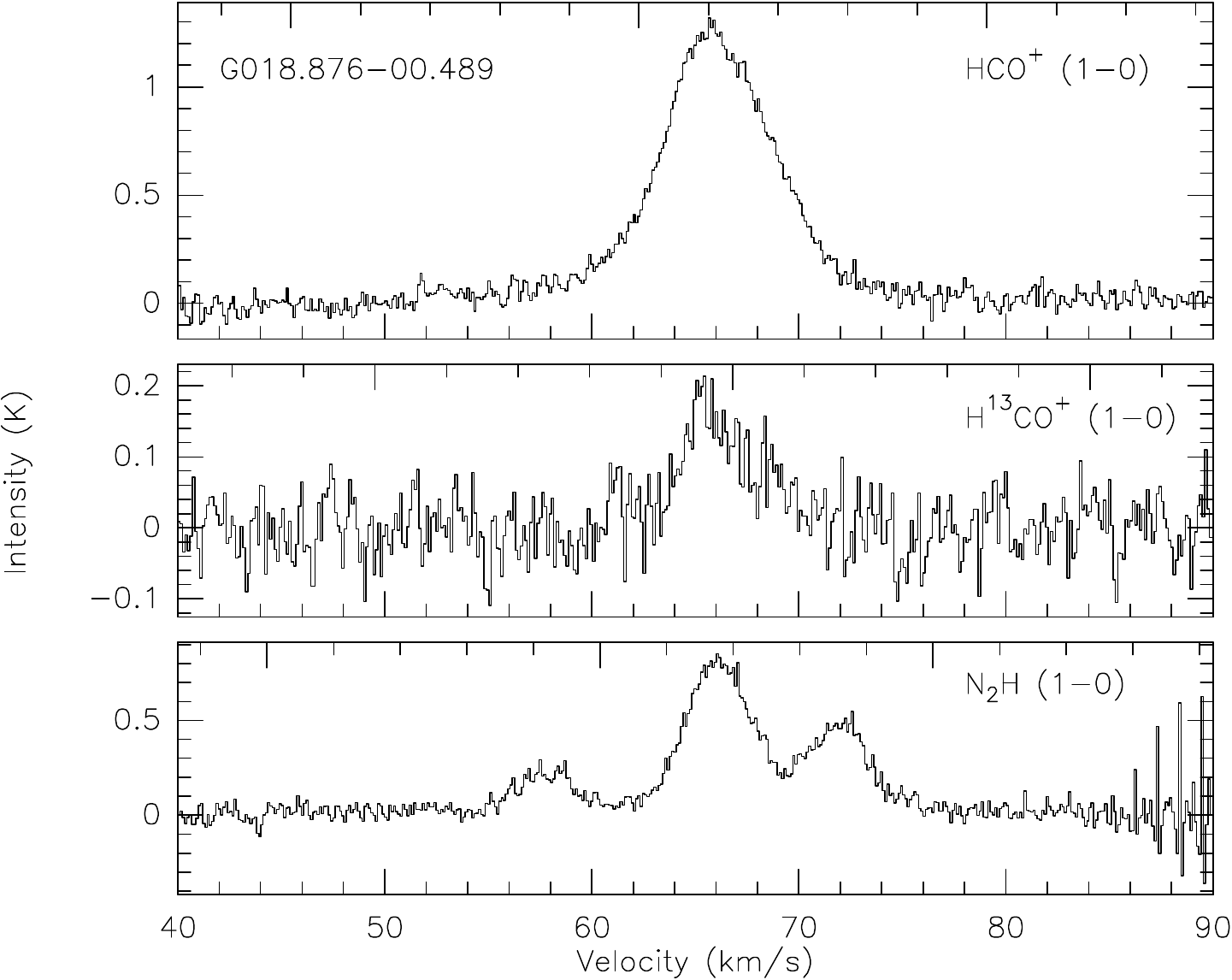}
\caption{\emph{Top}: Integrated intensity map of the \hcop, \htcop, and \nthp~emission for the G018.876-00.489 clump. \emph{Bottom}: Observed averaged spectra over the whole map for \hcop, \htcop,~and \nthp.}\label{fig:averagespectra}
\end{figure}

The CLASS program of the GILDAS software package\footnote[5]{Grenoble Image and Line Data Analysis Software is provided and actively developed by IRAM and is available at http://www.iram.fr/IRAMFR/GILDAS} was used to analyze the spectra. First, the spectra for each line were averaged and checked for single- or double-peaked profiles (e.g., Figure~\ref{fig:averagespectra}) and the first fit was applied. Spectral lines along with their isotopologues were averaged from a 38$''$ x 38$''$ square region (the size of the beam) centered on the peak emission from \nthp(1-0) and the initial guesses from the first fit were used to fit the spectra in the boxed region. The \hcop, \htcop, \sio, \hnc, and \hntc~lines were fit with a single or double Gaussian profile since they do not have hyperfine line structures. Most of the sources have multiple line profiles which might be due to line-of-sight confusion, multiple cores, infall, or outflow. If there were double peaked profiles in the \hcop~or \hnc~lines, they were fitted with two velocity components and then the profiles of optically thin lines (\htcop~and/or \hntc) were checked in order to examine the origin of the observed line profile. If optically thin lines showed double peaked profiles, it was considered that multiple cores were being sampled along the line of sight, otherwise infall, or outflow possibilities were considered. Within the target group only one source (G335.790+00.174) had double peaked shapes in its optically thin lines.

Integrated intensities were measured over a velocity range that was decided based on each line profile or line width. In the Appendix in Figure~\ref{fig:zmoment}, the color scale and contours show the integrated intensity maps, and the spectra are the average spectra for each line.

For a single transition CLASS provides four line parameters: the integrated intensity, W ({K km s$^{-1}$}), the local standard of rest velocity, V$_{LSR}$ ({km s$^{-1}$}), FWHM line width, $\Delta V$ ({km s$^{-1}$}), and the main beam temperature, T$_{MB}$ (K). On the other hand, \nthp~and \hcn~were fit with the hyperfine structure (HFS) method to derive A x $\tau$$_{m}$, $V_{LSR}$, $\Delta V$, and $\tau$, the sum of all the opacities of all hyperfine transitions. The \nthp~line has 15 hyperfine components, while the \hcn~line has three which makes it possible to calculate the optical depths since optical depth can be written as a function of the relative strengths of the hyperfine components. The line parameters from the Gaussian fits are given in the Appendix \ref{section:fittingresults} (Table~\ref{tab_linefit}) and \nthp~and \hcn~line parameters are given in the Appendix Table~\ref{tab_linefit_n2hp}. Intensity (W) for \nthp, and \hcn~were derived by integrating the line over a wavelength range that covers all hyperfine line components (20.5 km/s on average for all sources). In order to estimate the uncertainty of the integrated intensity, we first derived the channel-by-channel rms noise by fitting the baseline with the line-free channels, then we followed Equation 3 in \citet{gre09} taking into account the number of channels of the integrated intensity and baseline fitting. Observed spectra averaged from a 38$''$ x 38$''$ square region are shown for all sources along with the integrated intensity maps in the Appendix Figure~\ref{fig:zmoment}. We note that for almost half of the sources, a good fit to the \hcn~spectrum was not possible due to the existing hyperfine line anomalies. As discussed in \citet{lou12} hyperfine anomalies in the J=1$\rightarrow$0 line are observed especially in line widths as well as the line strengths in massive star-forming regions. 

\subsection{Optical depths and excitation temperatures}\label{sec:opt_dep}

In order to calculate the column densities, we first need to calculate the optical depths and excitation temperatures. Following the hyperfine line fitting for the \nthp~line, A $\times$ $\tau_{m}$ and $\tau$ were derived. By using these parameters, and Equation (1) below \citep[as described in ][]{san11t}, excitation temperature was estimated. 

\begin{equation}\label{eq:tex}
T_{ex} = \frac{h\nu/k}{ln[\frac{h\nu/k}{(A \tau_m/\tau)+J_\nu(T_{bg})}+1]},
\end{equation}

where h is the Planck constant, k is the Boltzmann constant, $\nu$ is the frequency of the observed transition, and $J_{\nu}$(T$_{bg}$) is the Planck radiation which was calculated by the following equation

\begin{equation}
J_{\nu}(T_{bg}) = \frac {h\nu/k} {e^{h\nu/kT_{bg}}-1}
\end{equation}

where the background temperature, T$_{bg}$, is assumed to be 2.74~K. 

T$_{ex}$ was used for the rest of the measurements if the error was smaller than 30\% \citep[as in][]{fon11}, otherwise 11~K and 22~K was used for D24 and D8 clumps, respectively, which were calculated as average T$_{ex}$ in \citet{gia14}. For the \hcop, \htcop, \sio, \hnc, \hntc, and \hcn~lines, we assumed they share the same T$_{ex}$ as \nthp. 

CLASS provides the optical depths for the lines that have hyperfine structure since there is a constant ratio between the hyperfine lines, assuming local thermodynamic equilibrium (LTE). While we provide the optical depths for the \nthp~line in Table~\ref{tab_linefit_n2hp}, we could not calculate the optical depths for half of the sources with the \hcn~line. In many such cases, we could not provide a good fit due to hyperfine line anomalies as mentioned in Section~\ref{sec:specfit}. The mean optical depths are 1.40$\pm$0.09, and 3.64$\pm$0.28 for the \nthp and \hcn~lines, respectively.

We derived the optical depths for lines of \hcop~and \hnc~and their isotopologues \htcop, and \hntc~by following the method described in detail in \citet{san12}. This method requires the main beam brightness temperature for both the main and the isotopologue lines and assumes a constant abundance ratio of 50 for [\hcop/\htcop] and [\hnc/\hntc] \citep[e.g.,][]{sav02}. For one source (G014.114-00.574) since the main beam brightness temperature for the isotopologue was larger than the main molecule (T$_{H^{13}CO^{+}}$ > T$_{HCO^{+}}$) possibly due to self absorption, we could not calculate the optical depths. For three other sources we could not fit \htcop~data so we do not provide optical depths for those sources as well. For the same reasons for nine sources we could not calculate the optical depths for the \hnc~and \hntc~lines. Optical depths are reported in the Appendix in Table~\ref{tab_od} and the mean values are 20.3 and 23.2 for the optically thick \hcop~and \hnc~lines, respectively, and 0.4 and 0.4 for the optically thin isotopologues \htcop and \hntc, respectively. 

\subsection{Column densities}\label{sec:coldens}

Following the calculations of T$_{ex}$ and the optical depths, the column densities were estimated for all the lines under the assumption of LTE. The following equation \citep{cas02} was used to calculate the column densities for optically thin lines, N$_{thin}$:

\begin{equation}\label{eq:col_dens}
N_{thin} = \frac{8\pi W}{\lambda^3 A} \frac{g_l}{g_u} \frac{1}{J_\nu(T_{ex})-J_\nu(T_{bg})} \frac{1}{1-e^{(-h\nu/kT_{ex})}} \frac{Q_{rot}}{g_l e^{(-E_l / kT_{ex})}},
\end{equation}

where W is the integrated intensity of the line calculated from fitting single line transitions (see the Appendix Table~\ref{tab_linefit}), while g$_{l}$ and g$_{u}$ are the statistical weights of the lower and upper levels, A is the Einstein coefficient, Q$_{rot}$ is the partition function, and E$_{l}$ is the lower energy level. Cataloged terms mentioned above are obtained for each line from the CDMS (The Cologne Database for Molecular Spectroscopy) and JPL (Jet Propulsion Laboratory, Molecular Spectroscopy) databases. 

When the line is optically thick ($\tau$ > 1), the following equation was used to estimate the column density

\begin{equation}
N_{thick} = N_{thin} \frac{\tau} {1-e^{-\tau}}
.\end{equation}

For sources where a line is not detected, we calculated the upper limits by assuming optically thin lines with a line width of 2 $km s^{-1}$, and a brightness temperature that is three times the rms of the spectra (T$_{mb}$ = 3T$_{rms}$).
Estimated column densities are listed in the Appendix Table~\ref{tab_columndens}. 

Mean, median, and standard deviation values of the column densities are listed at the end of the Table~\ref{tab_columndens}. The derived column densities for \nthp~range from 4.4 $\times$ 10$^{12}$ cm$^{-2}$ to 116.2 $\times$ 10$^{12}$ cm$^{-2}$ with an average value of 3.2 ($\pm$0.3) $\times$ 10$^{13}$ cm$^{-2}$. This mean is similar to the value (2.1 $\times$ 10$^{13}$ cm$^{-2}$) derived for a set of IRDCs by \citet{sak08} and smaller than the average value, 6.3 $\times$ 10$^{13}$ cm$^{-2}$, derived for a set of massive young stellar objects and \hii regions by \citet{yu15}. This mean is also higher than the 10$^{12}$ cm$^{-2}$ which is a typical value found in low-mass cores \citep{cas02}.

\section{Discussion}\label{sec:discussion}

\subsection{Infall and outflow candidates}

The low angular resolution of the single-dish data does not allow us to reveal the substructure of the clumps, but we can still for studying infall which is relevant to the high-mass star formation process. The infall candidates discussed in this study should be primary candidates to study the core structure of massive infrared clumps. For instance, they will give crucial information regarding the early phases of high-mass star formation, fragmentation and cluster formation processes when studied with radiative transfer modeling. As an example, \citet{con18} study the high-mass clump G331.372-00.116 with ALMA, which was detected by ATLASGAL and observed by MALT90 similarly to the high-mass clumps studied in this paper. By examining the \hcop~and \hnc~lines and blue-shifted profiles, they estimate a core infall rate and discuss core collapse and competitive accretion scenarios. 
Within this context, one of the goals of our study is to find infall and outflow candidates among the dark clumps by examining their emission line morphology. For this purpose, we used the line profiles of \hcop, \htcop, \hnc, and \hntc. First, we examined the double-peaked lines and if they have greater peak emission in their blue shifted sides, they were designated as infall candidates. G008.684-00.367, G008.706-00.414, G013.178+00.059, G014.114-00.574, G326.987-00.031, G329.029-00.206, G331.708+00.583\_1, G331.708+00.583\_2, G340.784-00.097, G343.756-00.164, G351.444+00.659, G353.066+00.452\_2, and G354.944-00.537 have clear infall signatures based on the strong blue peaks in their \hcop~lines. In order to determine the asymmetry of the line more quantitatively, we used the method given in \citet{mar97} and determined the normalized velocity difference ($\delta v$) between optically thick (\hcop~or \hnc) and thin (\htcop~or \hntc) components. The normalized velocity difference is given as

\begin{equation}
\delta v = \frac{V_{LSR}(thick)-V_{LSR}(thin)}{\Delta V(thin)}
.\end{equation}

If the \hcop~or \hnc~lines have a significant blue shift ($\delta v$ < -0.25), the source was flagged as blue and if either of them has a significant red shift ($\delta v$ > 0.25), the source was flagged as red as given in Table~\ref{shift}. Thirteen of the sources have blue profiles (shown in Table~\ref{shift}), while seven of these sources had been previously found with blue profiles in \citet{he15}. Among these 13 sources, nine have blue profiles in both the \hcop~and \hnc~lines. 

In addition to the infall candidates with blue profiles, there are six sources with red profiles which are thought to be outflow indicators as stated in \citet{che10}. One of these sources, G014.632-00.577, has red profiles in both lines and it is shown in Figure~\ref{fig:spitzerRGB}. In order to differentiate infall from outflow or rotation, the profile of the spectra should be investigated throughout the mapped region. Within this context, we examined the line profiles and overall the spectral map within each region. The spectral maps for the infall and outflow candidates are shown in Figures~\ref{fig:specmap1} and~\ref{fig:specmap2} if they have high signal-to-noise ratio (S/N). For instance for previously mentioned outflow candidate G14.632-00.577, the \hcop~spectra show a double-peaked line profile at the central region of the clump, however, within the mapped area it changes from blue-shifted profile to the red-shifted profile as seen in Figure~\ref{fig:specmap2}. While this can be a sign for outflow, we note that rotation can also produce these kind of asymmetries. In this case, the velocity gradients of optically thin lines can be studied, however, insufficient S/N with the current single-dish data do not allow us to do further investigation.

Among the targets, eleven sources have a significant \sio~detection (over 3$\sigma$). Among them, G014.194-00.194, G014.632-00.575, G331.708+00.583, and G351.444+00.659 have excess in the line wings of \hcop~spectra which is also a known effect of outflows \citep{raw04}.

Considering the high optical depths of the \hcop, and \hnc~lines, the lines should be highly affected by gradients in geometry, excitation temperature, and velocity fields along the clumps. Indeed, some studies show that the infall signatures may not be observable in all molecular tracers in some cases. For example, \citet{chi14} examined the line profiles of \hcop~and \hnc~for specific transitions by using radiative transfer calculations and showed that the infall signs seen in \hcop~line profiles do not exist in \hnc~profiles in some transitions. Therefore, a line investigation with radiative transfer modeling for specific sources is needed to understand the detailed physics behind the lines.

\begin{table*}
\caption{\label{shift} List of the sources with blue and red profiles}
\centering
\tiny
\begin{threeparttable}
\begin{tabular}{llllllllllll}
\hline\hline
\noalign{\smallskip}
Source & $V_{LSR}$(\hcop) & $V_{LSR}$(\htcop) & $V_{LSR}$(\hnc) & $V_{LSR}$(\hntc) & $\Delta V$ (\htcop) & $\Delta V$ (\hntc) & $\delta v$ (\hcop) & $\delta v$ (\hnc) & Profile\tnote{1} \\
(km~s$^{-1}$) & (km~s$^{-1}$) & (km~s$^{-1}$) & (km~s$^{-1}$) & (km~s$^{-1}$) & (km~s$^{-1}$) & (km~s$^{-1}$) & (km~s$^{-1}$) & \\
\hline
G008.684-00.367         &       34.04$\pm$0.04  &               36.94$\pm$0.10                 &       34.22$\pm$0.06  &       37.40$\pm$0.12          &       4.46$\pm$0.23         &       5.30$\pm$0.40   &       -0.65$\pm$0.04  &       -0.60$\pm$0.05                 &       B,B\tnote{2} \\
G008.706-00.414         &       38.56$\pm$0.04          &               39.09$\pm$0.12                 &       38.65$\pm$0.04          &       39.38$\pm$0.11          &       2.82$\pm$0.27         &       1.81$\pm$0.22   &       -0.19$\pm$0.05          &       -0.41$\pm$0.08                 &       N,B\tnote{2} \\
G013.178+00.059         &       48.31$\pm$0.05          &               49.06$\pm$0.17                 &       49.13$\pm$0.04          &       49.44$\pm$0.15          &       2.90$\pm$0.60         &       3.70$\pm$0.40   &       -0.25$\pm$0.08          &       -0.09$\pm$0.04           &       B,N\tnote{2} \\
G014.114-00.574         &       18.27$\pm$0.29          &               19.96$\pm$0.12                 &       19.38$\pm$0.08          &       20.90$\pm$0.21          &       3.25$\pm$0.23         &       3.50$\pm$0.50   &       -0.52$\pm$0.10          &       -0.44$\pm$0.09                 &       B,B \\
G014.194-00.194         &       40.23$\pm$0.05          &               39.52$\pm$0.10                 &       39.85$\pm$0.032         &       39.47$\pm$0.14          &       2.86$\pm$0.26         &       2.80$\pm$0.40   &       0.25$\pm$0.05           &       0.14$\pm$0.05                 &       R,N \\
G014.632-00.577         &       20.32$\pm$0.026 &                18.79$\pm$0.08         &       19.90$\pm$0.029         &       18.81$\pm$0.07          &       2.34$\pm$0.21         &       1.94$\pm$0.16   &       0.65$\pm$0.07           &       0.56$\pm$0.06                 &       R,R \\
G309.382-00.134         &       -49.06$\pm$0.05         &               -50.38$\pm$0.22         &       -50.37$\pm$0.07         &       \nodata &       3.40$\pm$0.70         &       \nodata &       0.39$\pm$0.10           &       \nodata &       R,N \\
G317.867-00.151         &       -38.87$\pm$0.09         &               -40.07$\pm$0.21         &       -38.92$\pm$0.29         &       \nodata &       3.90$\pm$0.50         &       \nodata &       0.30$\pm$0.07           &       \nodata &       R,N \\
G318.779-00.137         &       -36.45$\pm$0.08         &               -38.59$\pm$0.25         &       -37.87$\pm$0.09         &       -37.94$\pm$0.34         &       3.40$\pm$0.60         &       3.90$\pm$0.70   &       0.64$\pm$0.14           &       0.02$\pm$0.09                 &       R,N \\
G326.987-00.031         &       -59.91$\pm$0.05         &               -58.39$\pm$0.27         &       -59.72$\pm$0.19         &       -58.46$\pm$0.22         &       2.90$\pm$0.70         &       2.30$\pm$0.50   &       -0.53$\pm$0.16          &       -0.54$\pm$0.18                 &       B,B \\
G329.029-00.206     &   -46.47$\pm$0.04     &       -43.30$\pm$0.16     &   -46.31$\pm$0.05     &   -43.21$\pm$0.11     &   4.9$\pm$0.4     &   3.33$\pm$0.29   &   -0.64$\pm$0.06      &   -0.93$\pm$0.09      &   B,B \\
G331.708+00.583\_1      &       -69.33$\pm$0.06         &               -67.20$\pm$0.21         &       -68.75$\pm$0.10         &       -67.05$\pm$0.20         &       4.00$\pm$0.40         &       3.40$\pm$0.50   &       -0.53$\pm$0.08          &       -0.49$\pm$0.10                 &       B,B\tnote{2}  \\
G331.708+00.583\_2      &       -69.57$\pm$0.034        &               -67.14$\pm$0.14         &       -68.95$\pm$0.030        &       -67.17$\pm$0.14         &       4.08$\pm$0.33         &       2.63$\pm$0.29   &       -0.59$\pm$0.06          &       -0.68$\pm$0.09                 &       B,B\tnote{2}  \\
G340.784-00.097         &       -102.46$\pm$0.14        &               -100.87$\pm$0.10         &       -101.55$\pm$0.06        &       -101.16$\pm$0.25        &       1.39$\pm$0.28         &       2.70$\pm$1.70   &       -1.14$\pm$0.27          &       -0.14$\pm$0.13                 &       B,N \\
G342.484+00.182         &       -40.92$\pm$0.07         &               -41.53$\pm$0.08         &       -41.04$\pm$0.012        &       \nodata &       1.64$\pm$0.19         &\nodata &      0.38$\pm$0.08           &       \nodata &       R,N \\
G343.756-00.164         &       -29.87$\pm$0.07         &               -27.69$\pm$0.08         &       -29.12$\pm$0.034        &       -27.60$\pm$0.12         &       2.66$\pm$0.19         &       2.92$\pm$0.33   &       -0.82$\pm$0.07          &       -0.52$\pm$0.07                 &       B,B \\
G351.444+00.659         &       -5.68$\pm$0.02  &               -4.35$\pm$0.031         &       -4.46$\pm$0.014         &       -4.31$\pm$0.04          &       3.89$\pm$0.07         &       3.89$\pm$0.10   &       -0.34$\pm$0.011         &       -0.04$\pm$0.01         &       B,N\tnote{2} \\
G353.066+00.452\_1      &       2.25$\pm$0.04           &               1.62$\pm$0.14                 &       1.68$\pm$0.05           &       1.56$\pm$0.26           &       2.24$\pm$0.31         &       2.60$\pm$0.60   &       0.28$\pm$0.08           &       0.05$\pm$0.10                 &       R,N \\
G353.066+00.452\_2      &       -2.89$\pm$0.03  &               -1.80$\pm$0.08                 &       -2.25$\pm$0.04          &       -1.68$\pm$0.22          &       1.74$\pm$0.18         &       1.60$\pm$0.40   &       -0.63$\pm$0.08          &       -0.36$\pm$0.17                 &       B,B \\
G354.944-00.537         &       -6.85$\pm$0.07          &               -5.90$\pm$0.11                 &       -6.54$\pm$0.05          &       -6.14$\pm$0.23          &       1.98$\pm$0.22         &       1.08$\pm$0.40   &       -0.48$\pm$0.09          &       -0.22$\pm$0.14                 &       B,B\tnote{2} \\ 
\hline
\hline
\end{tabular}
\begin{tablenotes}
\item[1]{B, R, and N are used for blue-shifted, red-shifted and neither blue nor red shifted profiles for \hcop, and \hnc~line profiles, respectively.}
\item[2]{These sources are found with either \hcop or \hnc~blue-shifted profiles in \citet{he15} too.}
\end{tablenotes}
\end{threeparttable}
\end{table*}

\subsection{Mass infall rates}

In order to estimate the mass infall rate, we used \hcop~and \hnc~molecular line emission of the sources with blue-peaked profiles. A radiative transfer model "Hill5" \citep{dev05} from the PySpecKit3 spectroscopic analysis toolkit \citep{gin11} allowed us to reproduce the observed line profiles and derive an infall velocity ($v_{\rm{in}}$) which can be used to estimate a clump mass infall rate ($\dot{M}$).

First, we fitted the average spectrum of \hcop~and \hnc~for 13 infall candidates that are listed with blue profiles in Table~\ref{shift}. The Hill5 model uses the following free parameters for which you are required to provide initial guesses and also set limits: the optical depth of the center of the line ($\tau$), the local standard of rest velocity (v$_{\rm{LSR}}$), the infall velocity (v$_{\rm{in}}$), the velocity dispersion ($\sigma$), and the peak excitation temperature (T$_{\rm{peak}}$). When the infall velocity is smaller than the velocity dispersion or the line profile does not show a separated red shifted peak, the fit results are not reliable. In these cases, we could not obtain a reliable fit. In the other cases the fits are shown in Figure~\ref{fig:Hill5} and derived infall rates are listed in Table~\ref{massinfall}.

Secondly, we calculated the volume densities assuming spherical clumps by \citep{hey16}

\begin{equation}\label{eq:tex}
n(H_{2}) = \frac {3 M_{cl}} {4 \pi \mu m_{H} R_{cl}^{3}} ,
\end{equation}

where $\mu$=2.8, m$_{\rm{H}}$ is the mass of the hydrogen atom, and M$_{\rm{cl}}$ and R$_{\rm{cl}}$ are the mass and radius of the clumps (from Table~\ref{malt90sourcetable}), respectively. By using this density value, and the infall velocity, v$_{\rm{in}}$, from the "Hill5" model fit, we calculated a mass infall rate for each clump by

\begin{equation}\label{eq:tex}
\dot{M}= 4\pi R_{cl}^2 n(H_{2}) \mu m_{H} v_{in} .
\end{equation}

Table~\ref{massinfall} shows the volume densities and the mass infall rates for the ten infall candidates. Mass infall rates range between 0.2 $\times$ 10$^{-3}$ M$_\odot$yr$^{-1}$ $-$ 1.8 $\times$10$^{-2}$ M$_\odot$yr$^{-1}$ which are similar to the value, 1.3 $\times$10$^{-2}$ M$_\odot$yr$^{-1}$, found by \citet{liut13} for the protocluster G10.6-0.4. We note that these infall rates refer to the gas accreted by the clump at larger scales, not at the smaller core scale. Previous mass infall rates found in high-mass protostellar regions range between 10$^{-5}$ $-$ 10$^{-3}$ M$_\odot$yr$^{-1}$ \citep[e.g.,][]{liu17,per13} whose high end is similar to the values we found. These high infall rates indicate that the clumps are undergoing a global infall and feed the possible high-mass protostellar and dense cores, we cannot, however estimate how much gas is accreted on to the cores themselves with the current data.

\begin{figure*}[ht]

\begin{subfigure}[t]{0.5\textwidth}
\includegraphics[width=0.9\textwidth]{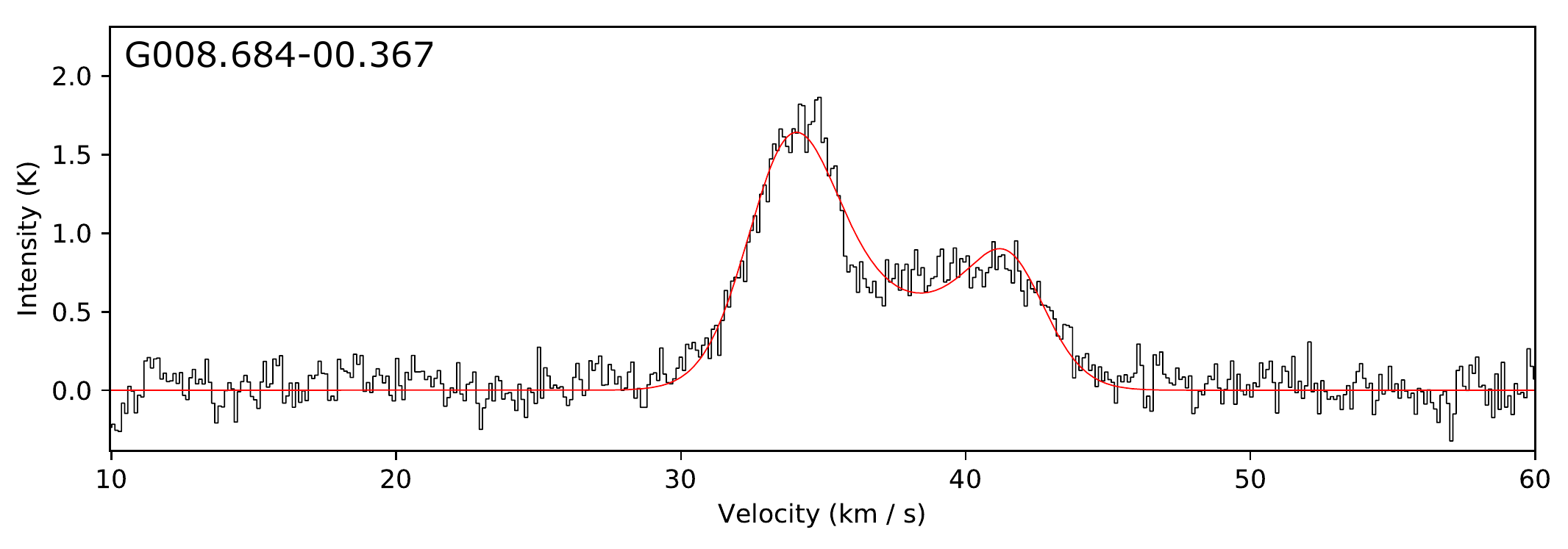}
\includegraphics[width=0.9\textwidth]{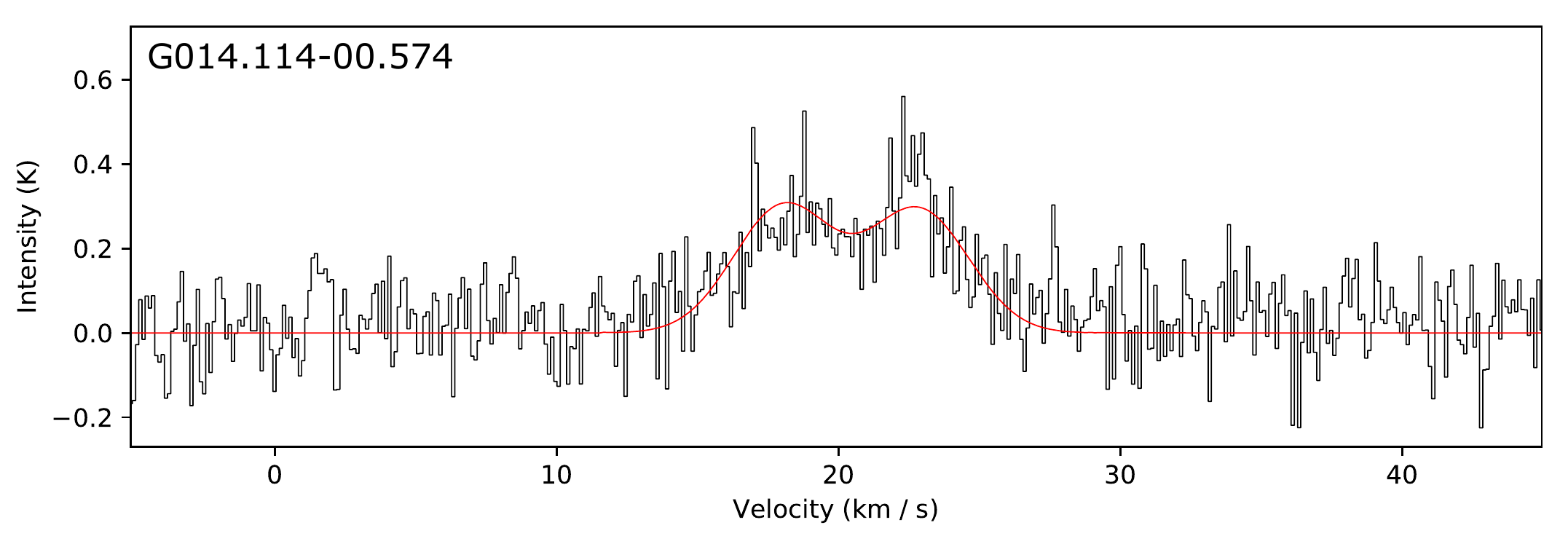}
\includegraphics[width=0.9\textwidth]{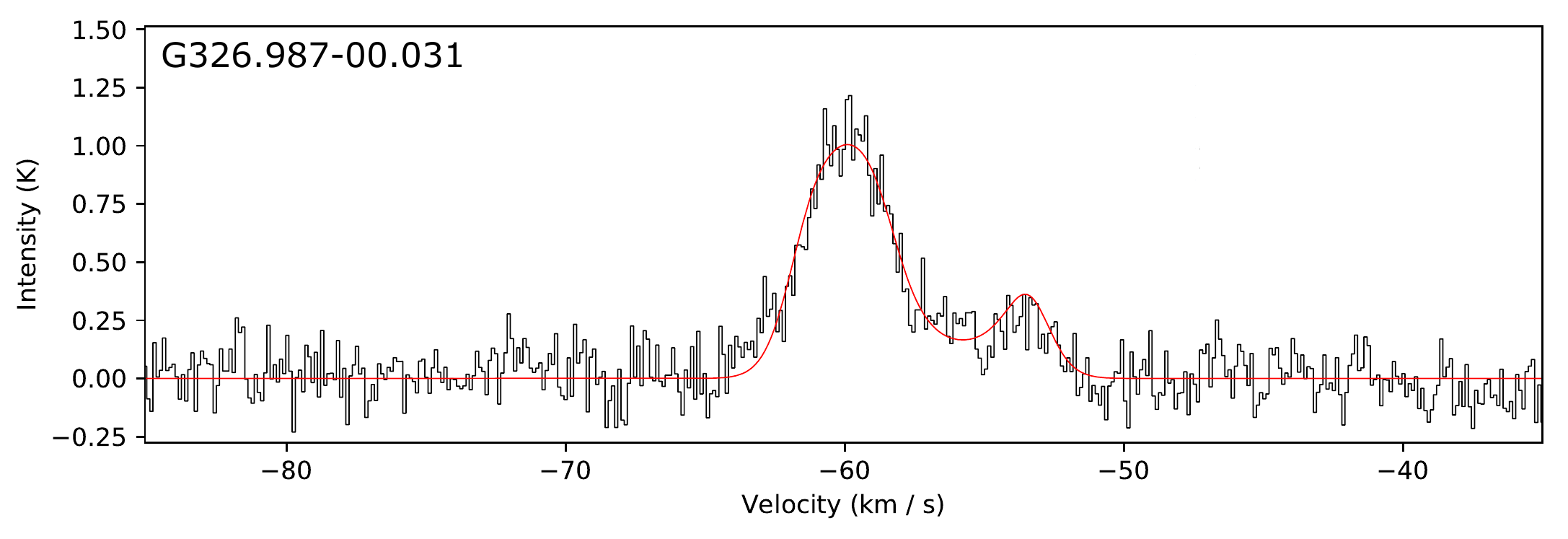}
\includegraphics[width=0.9\textwidth]{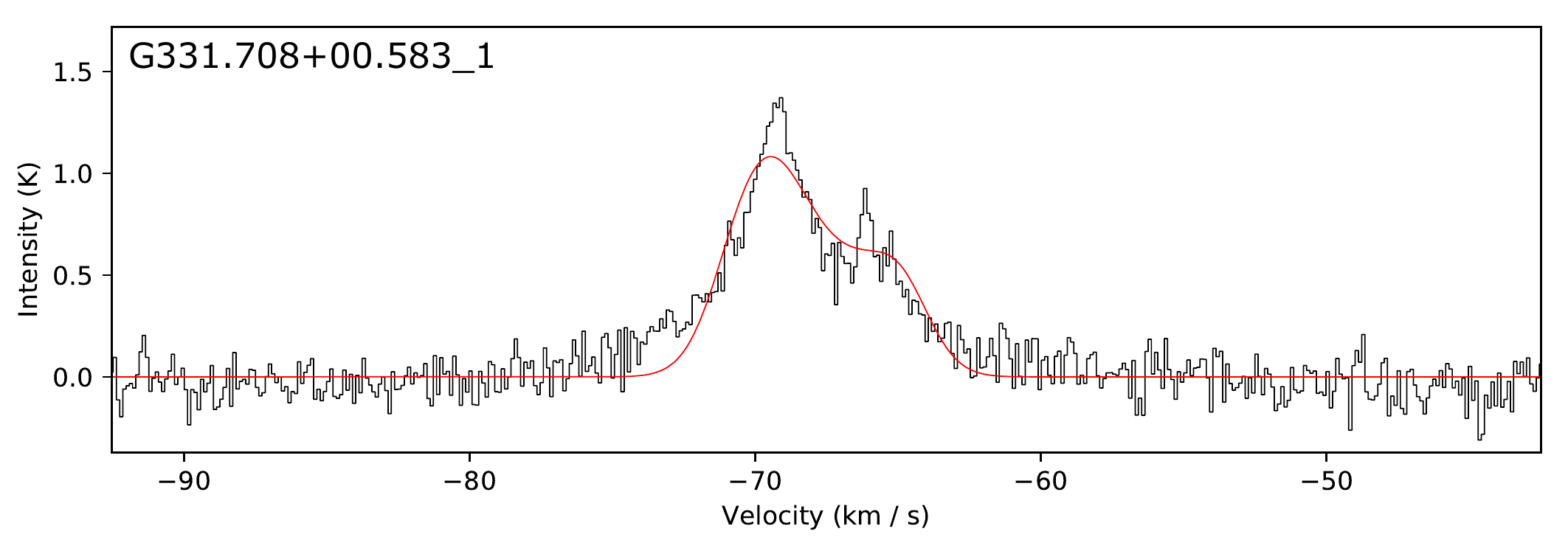}
\includegraphics[width=0.9\textwidth]{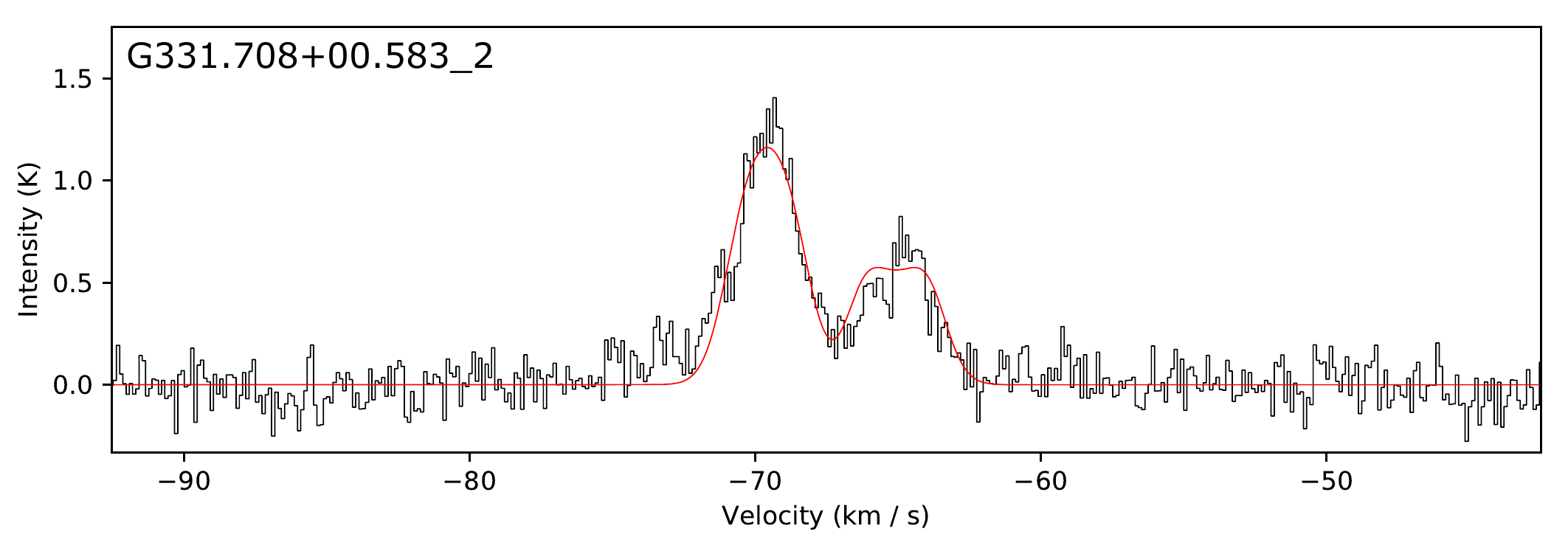}
\includegraphics[width=0.9\textwidth]{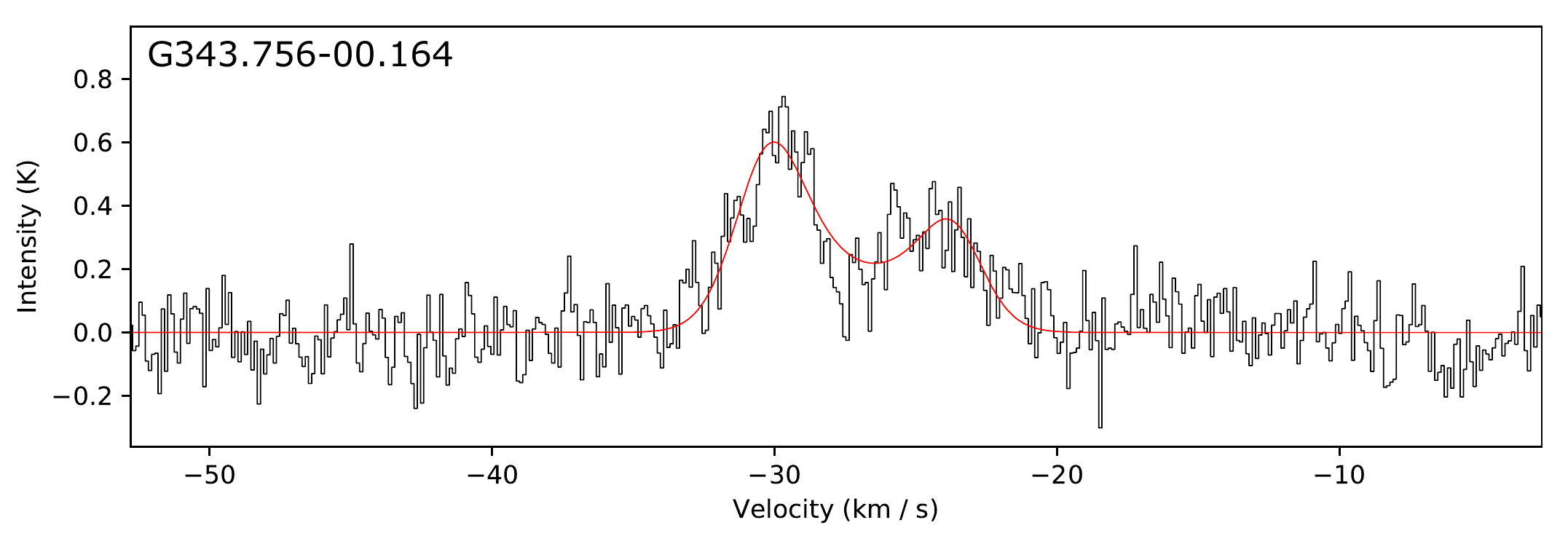}
\includegraphics[width=0.9\textwidth]{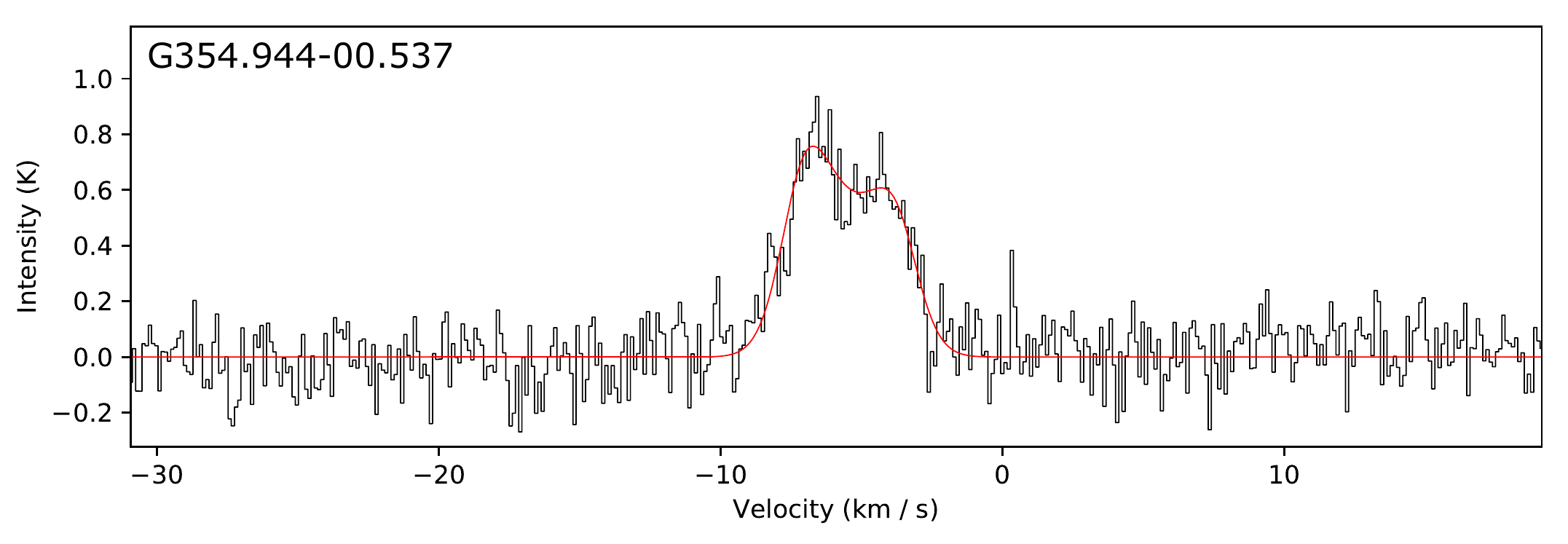}

\caption{\hcop(1-0) spectrum}
\label{fig:subim1}
\end{subfigure}
\begin{subfigure}[t]{0.5\textwidth}
\includegraphics[width=0.9\textwidth]{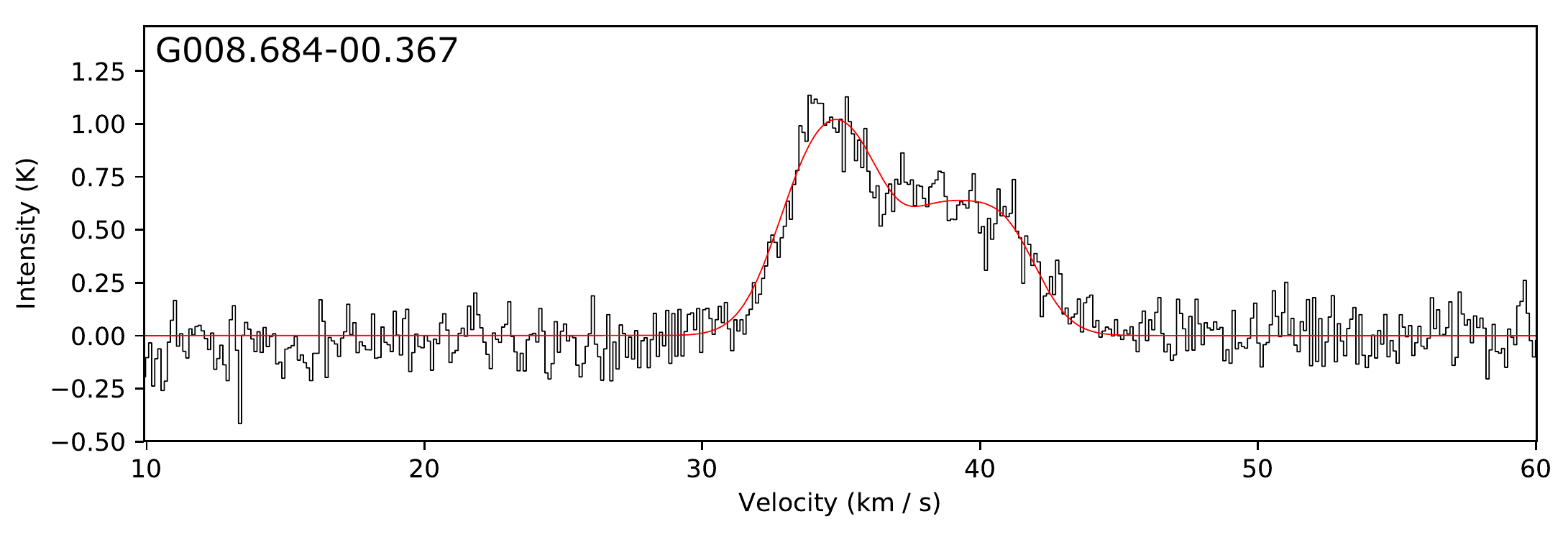}
\includegraphics[width=0.9\textwidth]{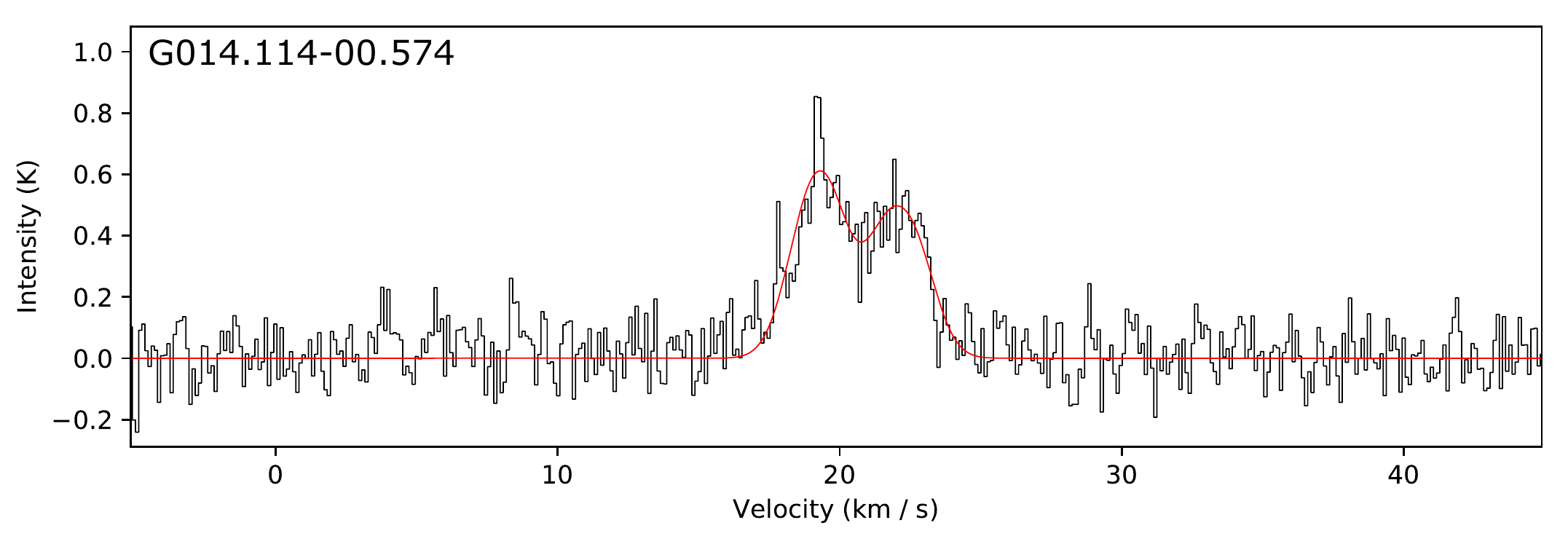}
\includegraphics[width=0.9\textwidth]{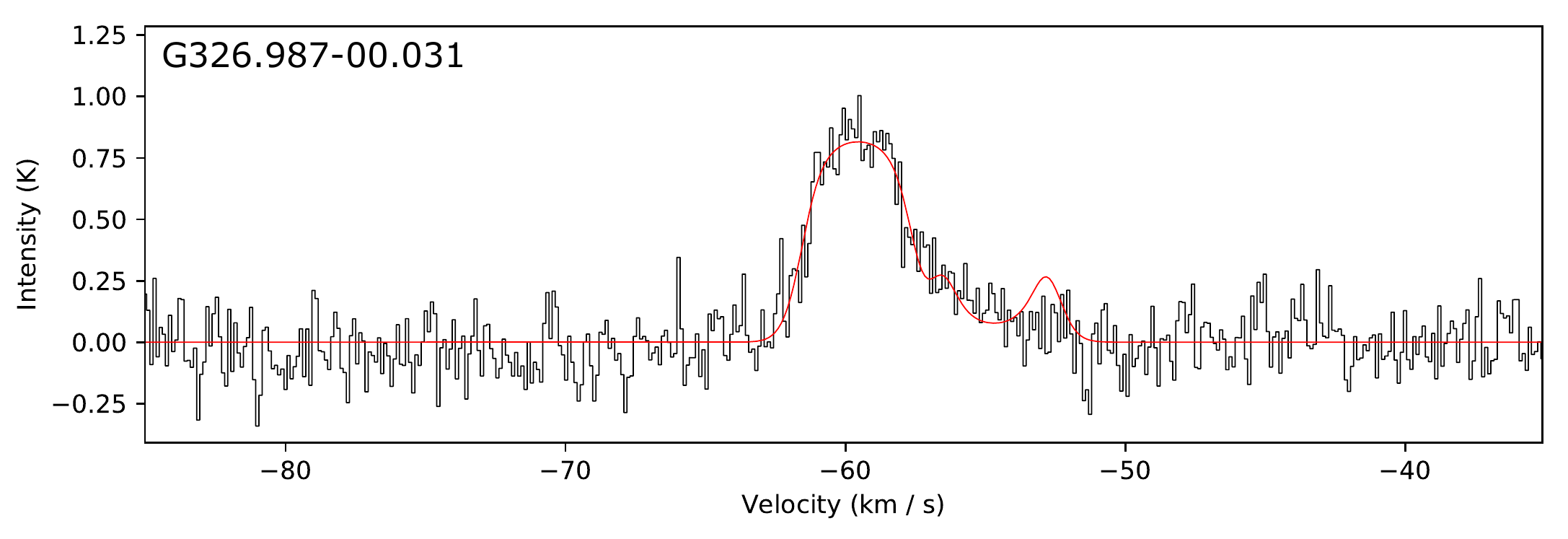}
\includegraphics[width=0.9\textwidth]{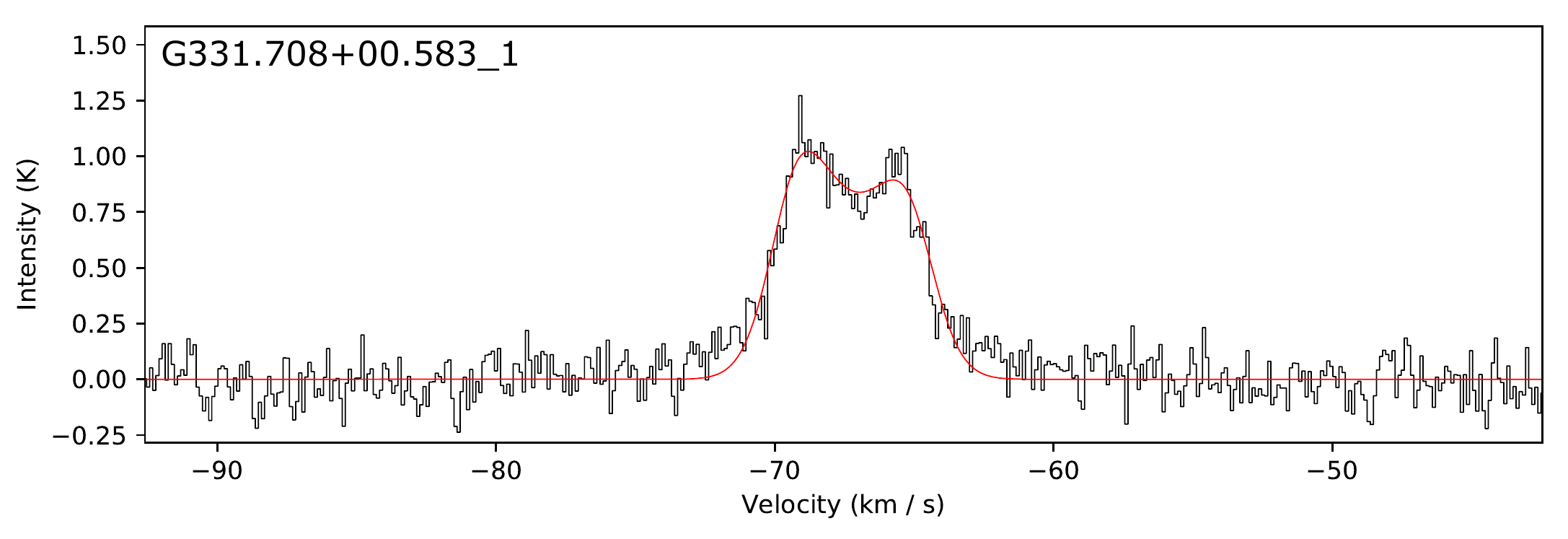}
\includegraphics[width=0.9\textwidth]{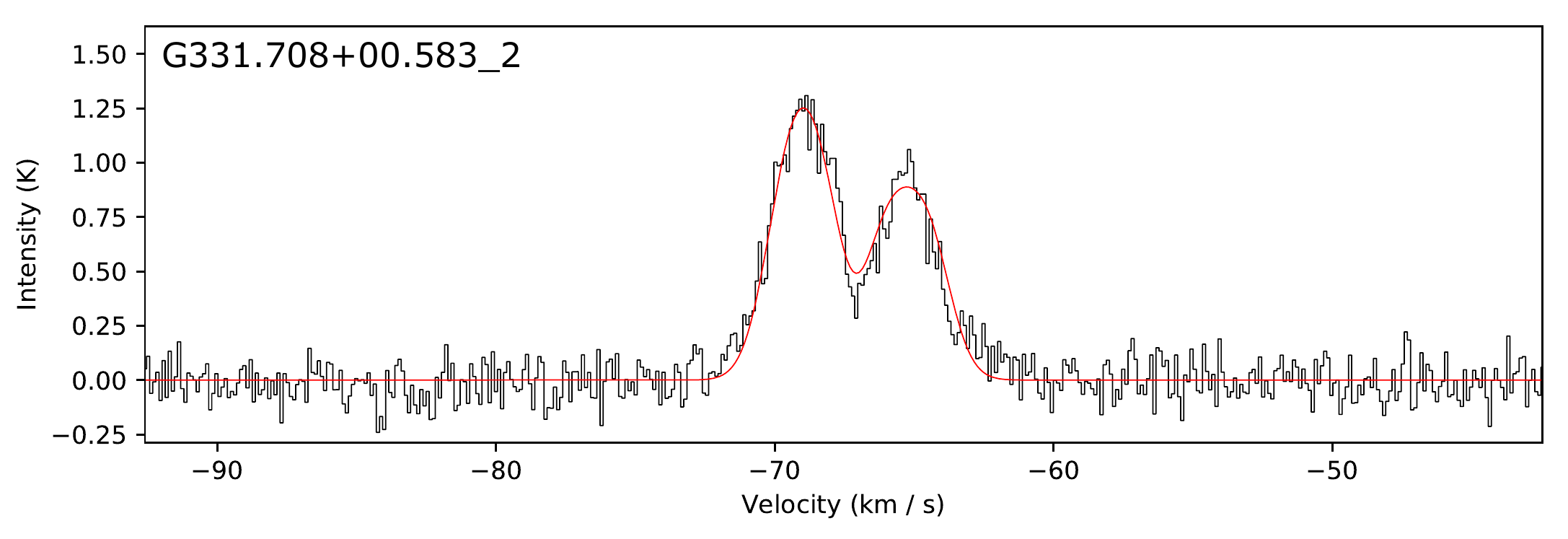}
\includegraphics[width=0.9\textwidth]{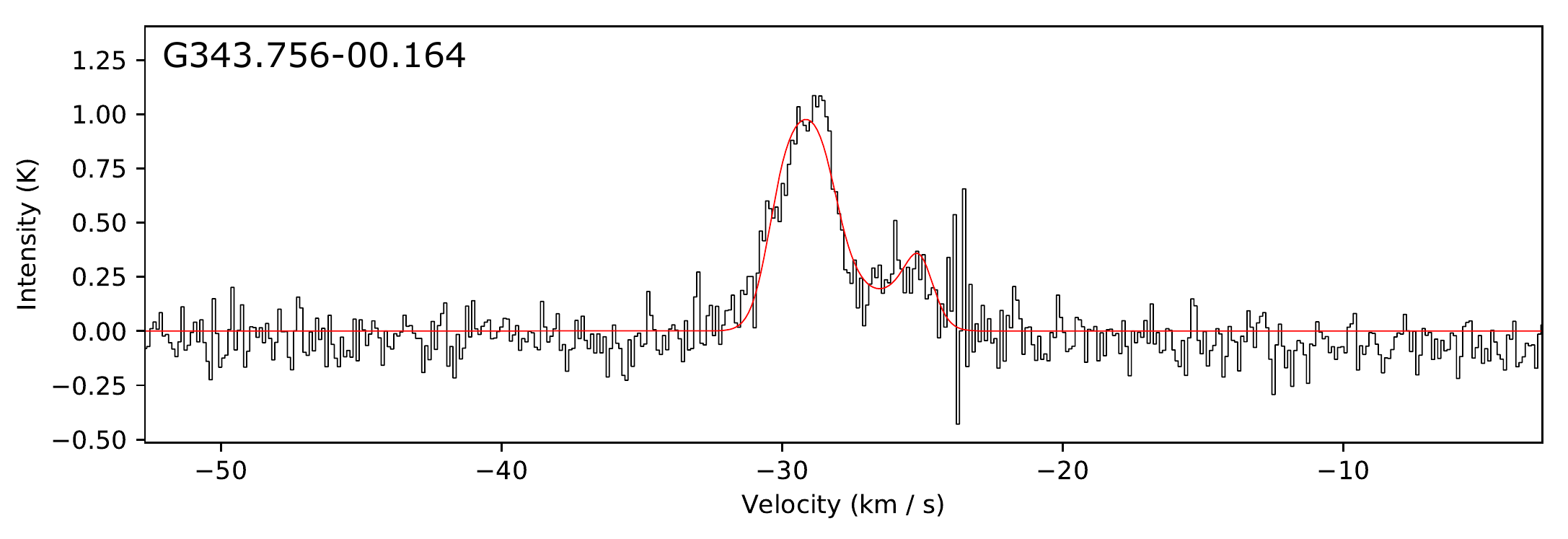}
\includegraphics[width=0.9\textwidth]{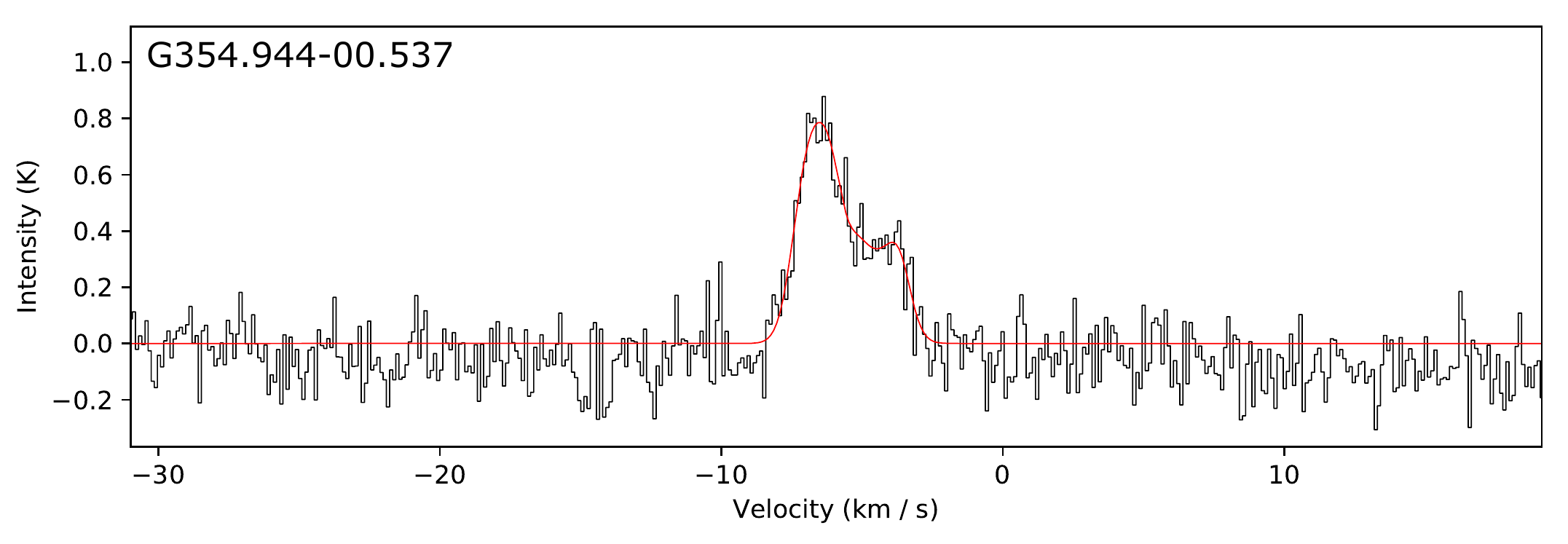}
\caption{\hnc(1-0) spectrum}
\label{fig:subim2}
\end{subfigure}
\caption{Average \hcop~(on left) and \hnc~spectra (on right) for the infall candidates are shown in black and the "Hill5" fit in red.}\label{fig:Hill5}
\label{fig:image2}

\end{figure*}

\addtocounter{figure}{-1}

\begin{figure*}[ht]
\begin{subfigure}[t]{0.5\textwidth}
\centering
\includegraphics[width=0.9\textwidth]{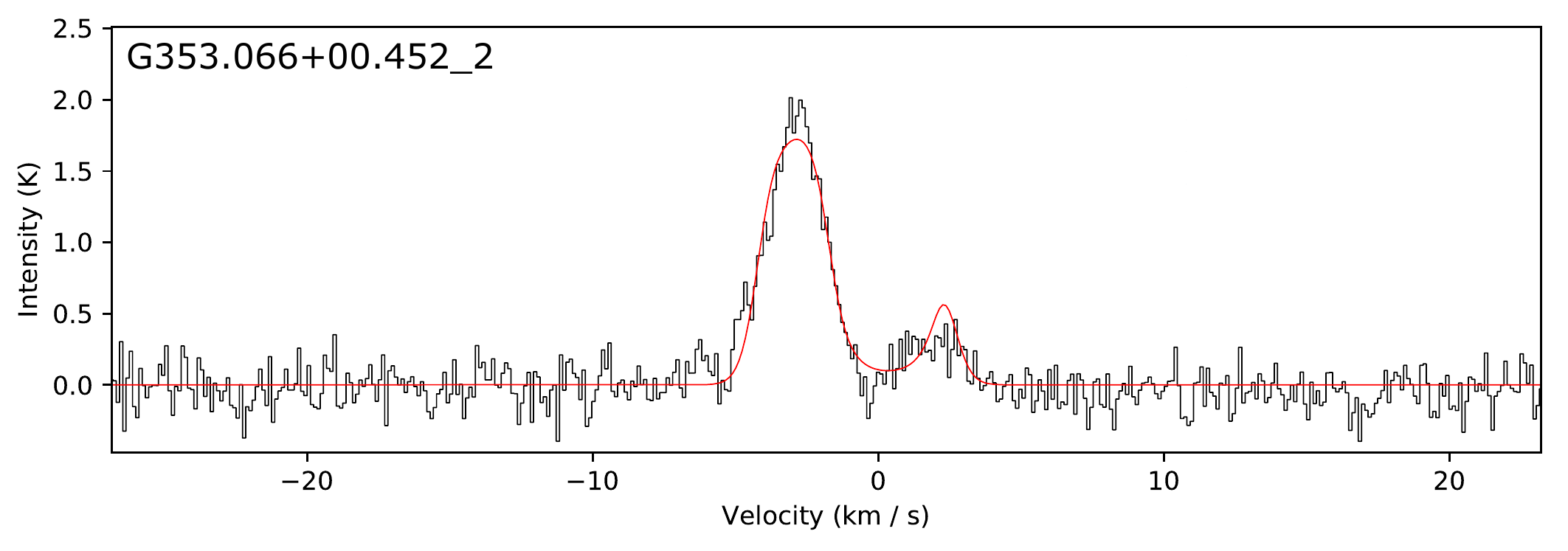}
\includegraphics[width=0.9\textwidth]{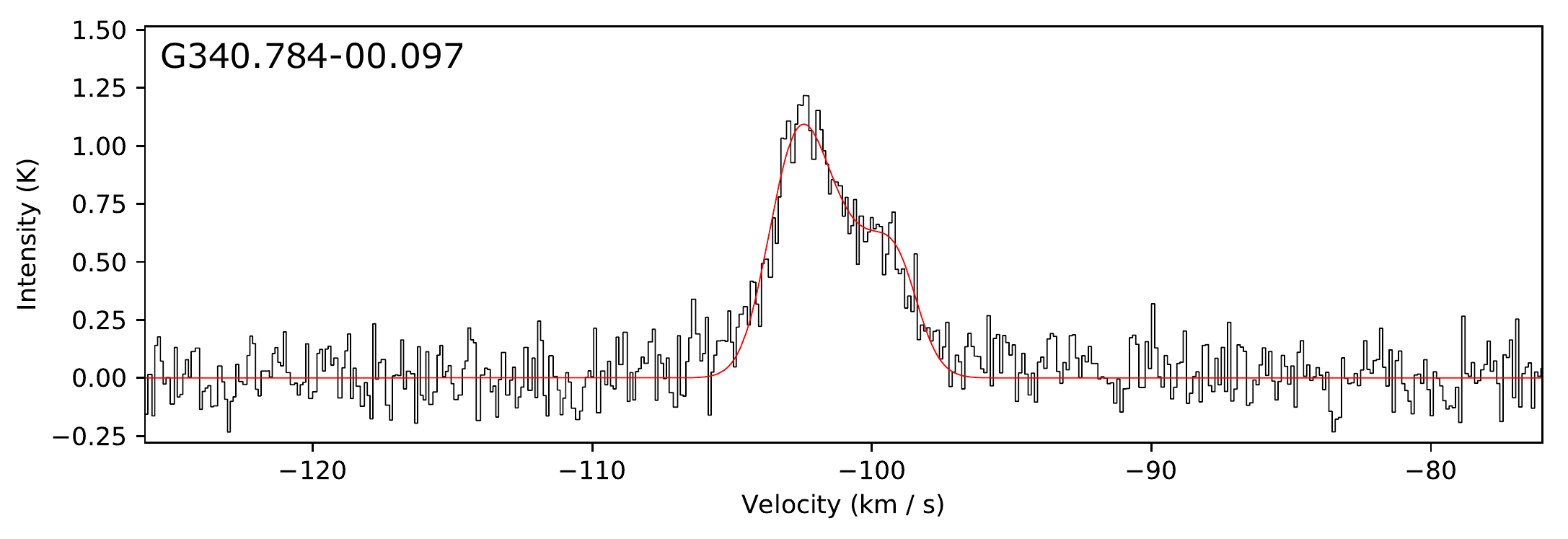}
\caption{\hcop(1-0) spectra}
\end{subfigure}
\begin{subfigure}[t]{0.5\textwidth}
\includegraphics[width=0.9\textwidth]{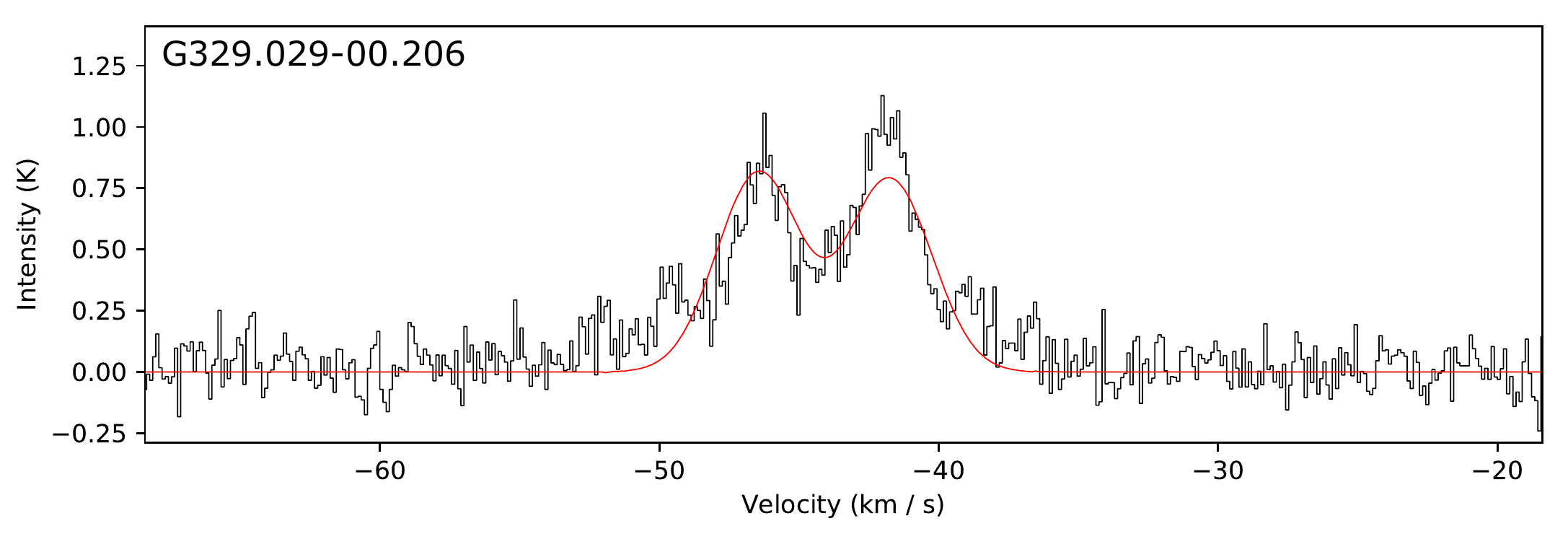}
\caption{\hnc(1-0) spectra}
\end{subfigure}
\caption{Continued}\label{fig:Hill5}
\end{figure*}

\begin{table*}[ht]
\caption{\label{massinfall} Mass infall rates.}
\centering
\tiny
\begin{tabular}{lllllll}
\hline\hline
\noalign{\smallskip}
No & Name & n(H$_{2}$) & v$_{\rm{in}}$ (\hcop) & v$_{\rm{in}}$ (\hnc) & $\dot{M}$ (\hcop) & $\dot{M}$ (\hnc)\\
 & & (10$^{3}$ cm$^{-3}$) & (km s$^{-1}$) & (km s$^{-1}$) & (10$^{-3}$ M$_\odot$yr$^{-1}$) & (10$^{-3}$ M$_\odot$yr$^{-1}$)\\
\hline
1 & G008.684-00.367 & 1.5 & 1.1$\pm$0.5 & 2.35$\pm$0.49 & 4.4$\pm$2.1 & 9.32$\pm$1.94 \\
2 & G014.114-00.574 & 1.3 & 2.0$\pm$1.0 & 1.39$\pm$0.44 & 3.8$\pm$1.9 & 2.62$\pm$0.83\\
3 & G326.987-00.031 & 2.6 & 1.8$\pm$0.8 & 2.43$\pm$0.52 & 4.9$\pm$2.4 & 6.77$\pm$1.45\\
4 & G329.029-00.206 & 1.4 & \nodata & 2.34$\pm$0.35 & \nodata & 8.00$\pm$3.84\\
5 & G331.708+00.583\_1 & 0.4 & 1.2$\pm$2.3 & 0.23$\pm$0.75 & 9.5$\pm$18.1 & 1.81$\pm$5.92 \\
6 & G331.708+00.583\_2 & 0.4 & 2.3$\pm$0.3 & 1.86$\pm$0.24 & 17.8$\pm$2.5 & 14.68$\pm$1.89\\
7 & G340.784-00.097 & 0.3 & 0.9$\pm$2.0 & \nodata & 3.0$\pm$6.5 & \nodata\\
8 & G343.756-00.164 & 1.3 & 0.7$\pm$1.2 & 1.18$\pm$0.83 & 2.0$\pm$3.2 & 3.13$\pm$2.20\\
9 & G353.066+00.452\_2 & 2.0 & 1.2$\pm$0.3 & \nodata & 1.1$\pm$0.2 & \nodata \\
10 & G354.944-00.537 & 1.6 & 1.2$\pm$0.4 & 1.01$\pm$0.67 & 0.2$\pm$0.5 & 0.52$\pm$0.34\\
\hline
\hline
\end{tabular}
\end{table*}

\subsection{Nature of some of the infall candidates}\label{fig:nature}

Some of the sources with infall and/or outflow signatures in MALT90 data were already studied by other authors. These studies confirmed them as infall and/or outflow candidates and revealed filamentary structures and multiple cores. This shows that MALT90 clumps are excellent targets to further study the early star formation process with higher quality data. Here we summarize the properties of three of these clumps while we will not give detailed discussion for all the clumps. In order to visually inspect the infrared appearance of the environment of these clumps, we generated the Spitzer images shown in Figure~\ref{fig:spitzerRGB}. Spitzer images of the other clumps can be found in the ATLASGAL catalog\footnote[6]{https://atlasgal.mpifr-bonn.mpg.de/cgi-bin/ATLASGAL$\_$DATABASE.cgi}.

G008.684-00.367 - An extended source is seen in the integrated intensity maps in the upper panels of Figure~\ref{fig:zmoment}. The \nthp~emission shows a single elongated structure, where two cores can be seen in the \hntc~integrated emission line map as well as in the dust continuum map. The bottom panels of Figure~\ref{fig:zmoment} show each spectral line. The \hcop~and \hnc~lines have double peaked-profiles, while the \hntc~line does not have a second component. Although the \htcop~line has a weak second component, it was not possible to obtain a good fit due to its low S/N. In order to examine the line profile change along the clump region, we generated the spectral maps as seen in Figure~\ref{fig:specmap1}. The top-left panel shows the double-peaked profile in the \hcop~spectra in the clump region. A strong blue peaked profile in \hcop~and \hnc~spectra is a typical infall signature (Table~\ref{shift}). 

There is no infrared source detected at the central coordinates (see Figure~\ref{fig:spitzerRGB}), and the low value of T$_{ex}$ (5.11$\pm$0.25~K) and high column density of \nthp~(55.8$\pm$9.7 $\times$ 10$^{12}$~cm$^{-2}$) indicate that the clump is at a very early stage of star formation. Extended \sio~emission and high column density for \sio~(1.4$\pm$0.2 $\times$ 10$^{12}$ cm$^{-2}$) indicate that there might be shocks in this clump due to outflows. 

The clump might be part of the star-forming complex IRAS 18032-2137 where we see an embedded stellar cluster BDS2003-3 \citep{bic03} and an UC\hii region G8.67-0.36 \citep{woo89}. \citet{lon11} studied this clump using SMA data and revealed three continuum sources separated by a minimum of 1.2$''$ and 2.2$''$ ($\sim$6200 AU and $\sim$9700 AU) and a bipolar outflow seems associated with one of the cores. Although the cores they studied have masses of 4~M$_\odot$, 7~M$_\odot$, and 10~M$_\odot$, \citet{lon11} conclude that G008.684-00.367 can still form O-type stars through large-scale infall and accretion from the very massive gas reservoir surrounding it. 

\textit{G014.114-00.574} - An extended-filamentary source in all integrated intensity maps and dust continuum emission is shown in the upper panels of Figure~\ref{fig:zmoment}. The \hcop~ and \hnc~lines show double-peaked profiles and blue profiles (see Table~\ref{shift}). It shows a small velocity gradient over $\sim$1~pc in \nthp~emission with $\sim$1~km s$^{-1}$ as seen in Figure~\ref{fig:velocitymap}. Filamentary structure, multiple cores in optically thin line emission maps, and the velocity gradient make this clump an interesting target for the study of early fragmentation in high-mass star formation. 

The clump is located within the giant molecular cloud (GMC) M17 SWex (Figure~\ref{fig:spitzerRGB}), which extends $\sim$50~pc southwest from the prominent Galactic \hii region M17 \citep[d=1.98~kpc,][]{wu14}, contains more than ten thousand YSO candidates, and represents a proto-OB association \citep{pov10}. \citet{pov16} suggested that the whole GMC is a good example of a distributed mode of star formation which is dominated by intermediate-mass stars that will produce a large OB association with few massive clusters, or the massive cores will produce massive clusters since they are still accreting mass. \citet{bus16} revealed 13 fragments within this clump (Hub-S), arranged in a manner that is more consistent with thermal Jeans fragmentation than a turbulent supported scenario.  

\textit{G331.708+00.583} - Two clumps are seen in all molecular emission lines as well as in dust continuum emission (Figure~\ref{fig:zmoment}) and in 24~$\mu$m infrared emission (Figure~\ref{fig:spitzerRGB}). One is at the center, and the other is in the northwest corner. The central clump is associated with two EGOs which are in general thought to be tracing the jets/outflow from the MYSOs \citep{cyg13}. 

The high column density of \nthp~of each clump indicates that they are both at early stages, and the \nthp~column density is higher in the second clump (33.5$\pm$7.1 $\times$10$^{12}$ cm$^{-2}$) which might indicate that it is younger than the central clump, G331.708+00.583\_1, (16.1$\pm$2.5 $\times$ 10$^{12}$ cm$^{-2}$). The spectra for both clumps have multiple line components in optically thick lines and blue profiles in both the \hcop~and \hnc~lines as seen in Figure~\ref{fig:specmap1} and Table~\ref{shift} which indicate that both clumps are possible infall targets. Furthermore, G331.708+00.583\_1 shows line wings in the \hcop~and \hnc~lines which might be outflow indicators as shown in \citet{ces97}. 

The infrared cloud that hosts both clumps is located in the fourth Galactic quadrant and the kinematic distance is either 4.0~kpc or 10.7~kpc. \citet{yu15} also studied this source with MALT90 data, finding the blue profiles in the \hcop~line indicating infall, as well as a red profile in the \hnc~line indicating an outflow activity. Both clumps are promising targets with which to investigate in detail the possible cores and outflows with high-sensitivity and high-resolution data. 

\begin{figure*}[ht]
\centering
\includegraphics[width=0.38\textwidth]{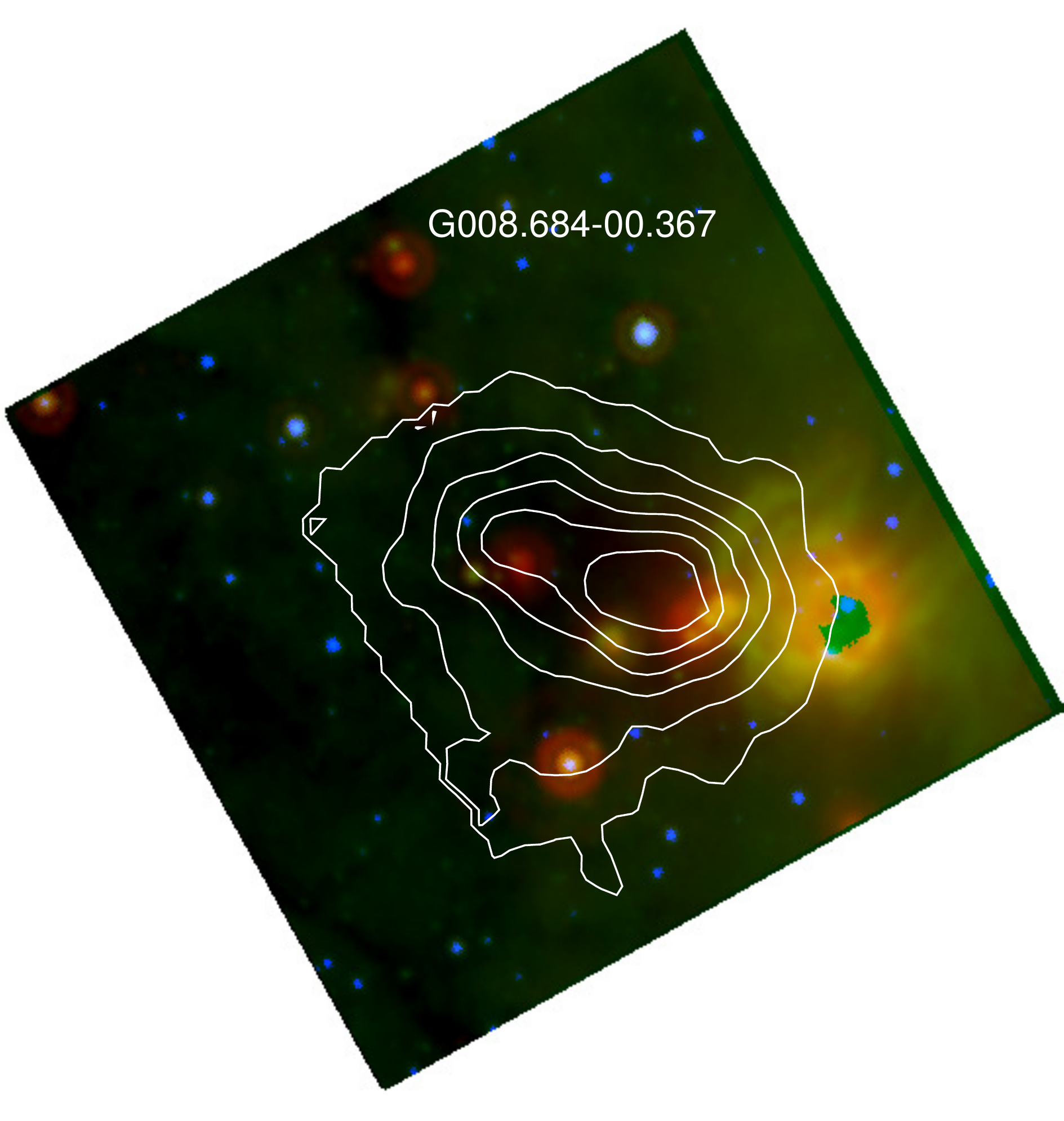}\vspace{0.5cm}
\includegraphics[width=0.48\textwidth]{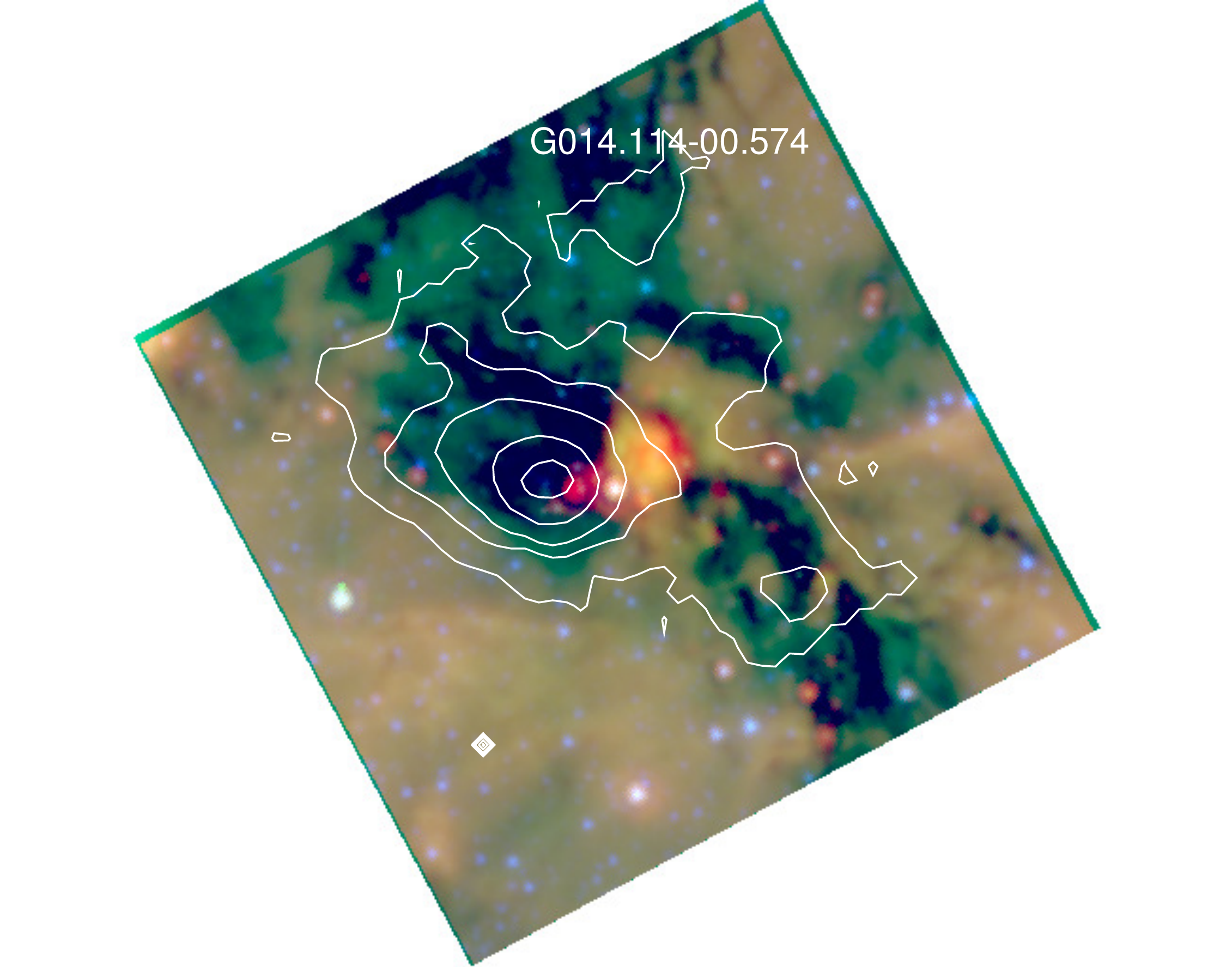}\\
\includegraphics[width=0.4\textwidth]{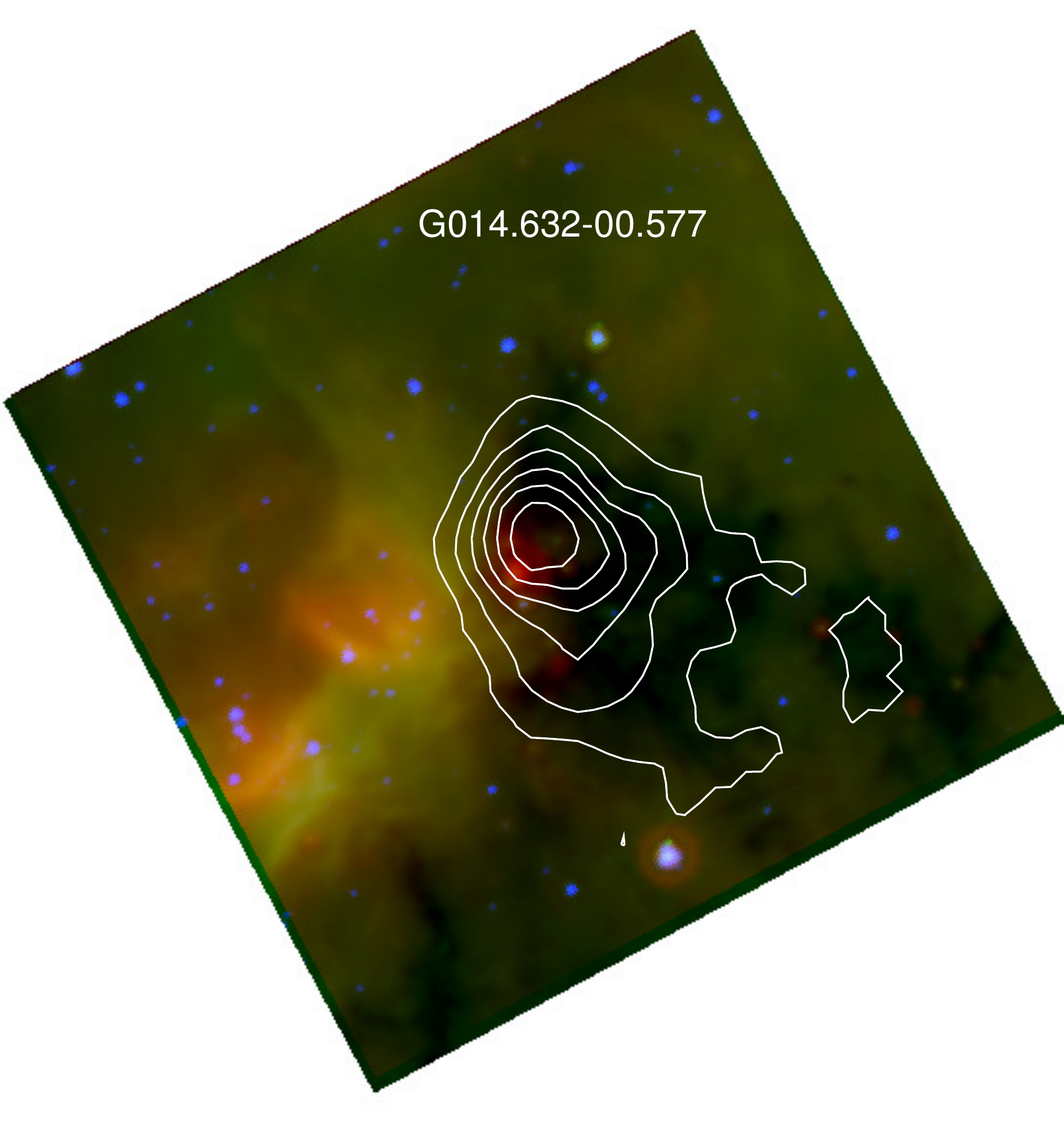}\vspace{0.5cm}
\includegraphics[width=0.5\textwidth]{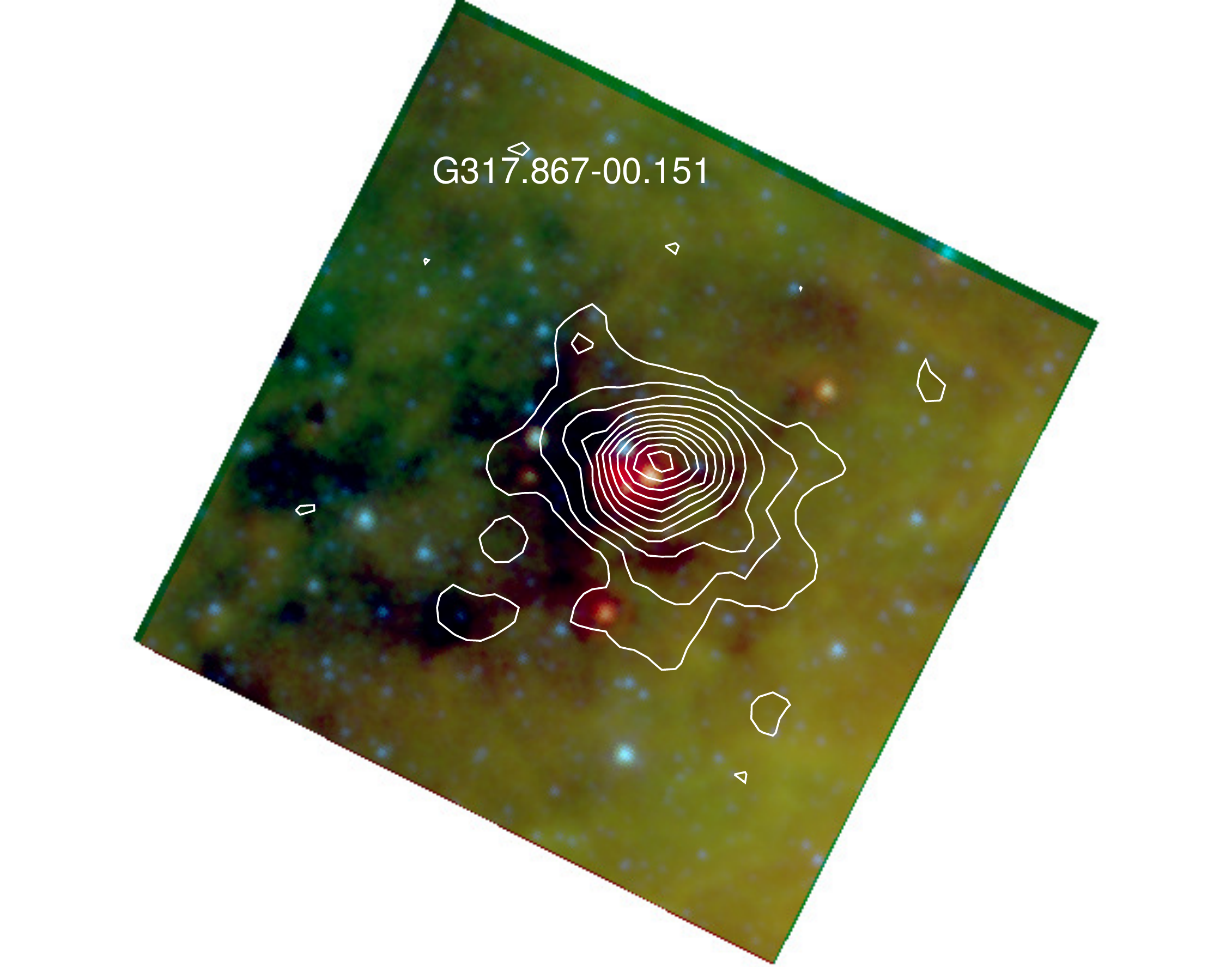}\\
\includegraphics[width=0.48\textwidth]{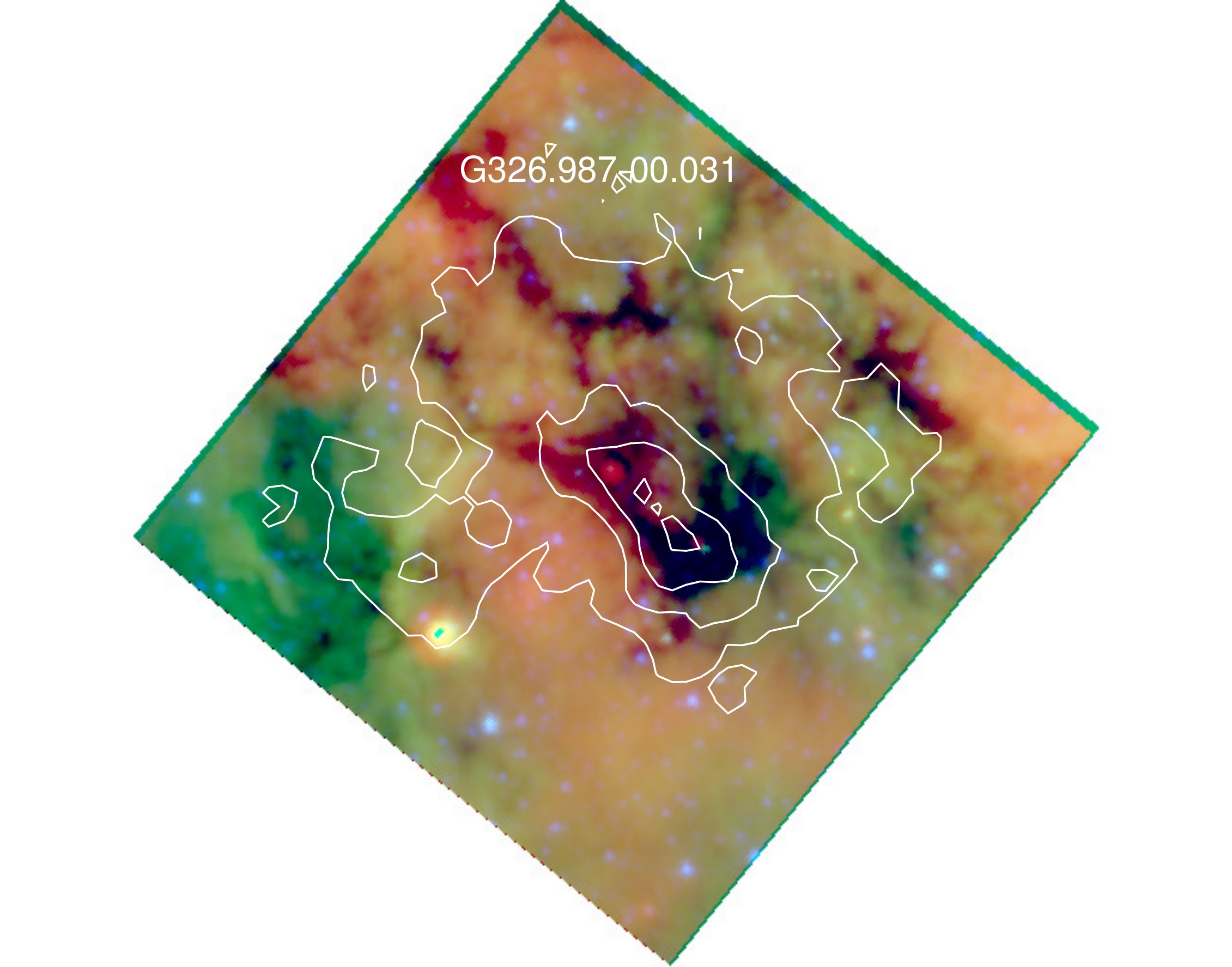}\vspace{0.5cm}
\includegraphics[width=0.5\textwidth]{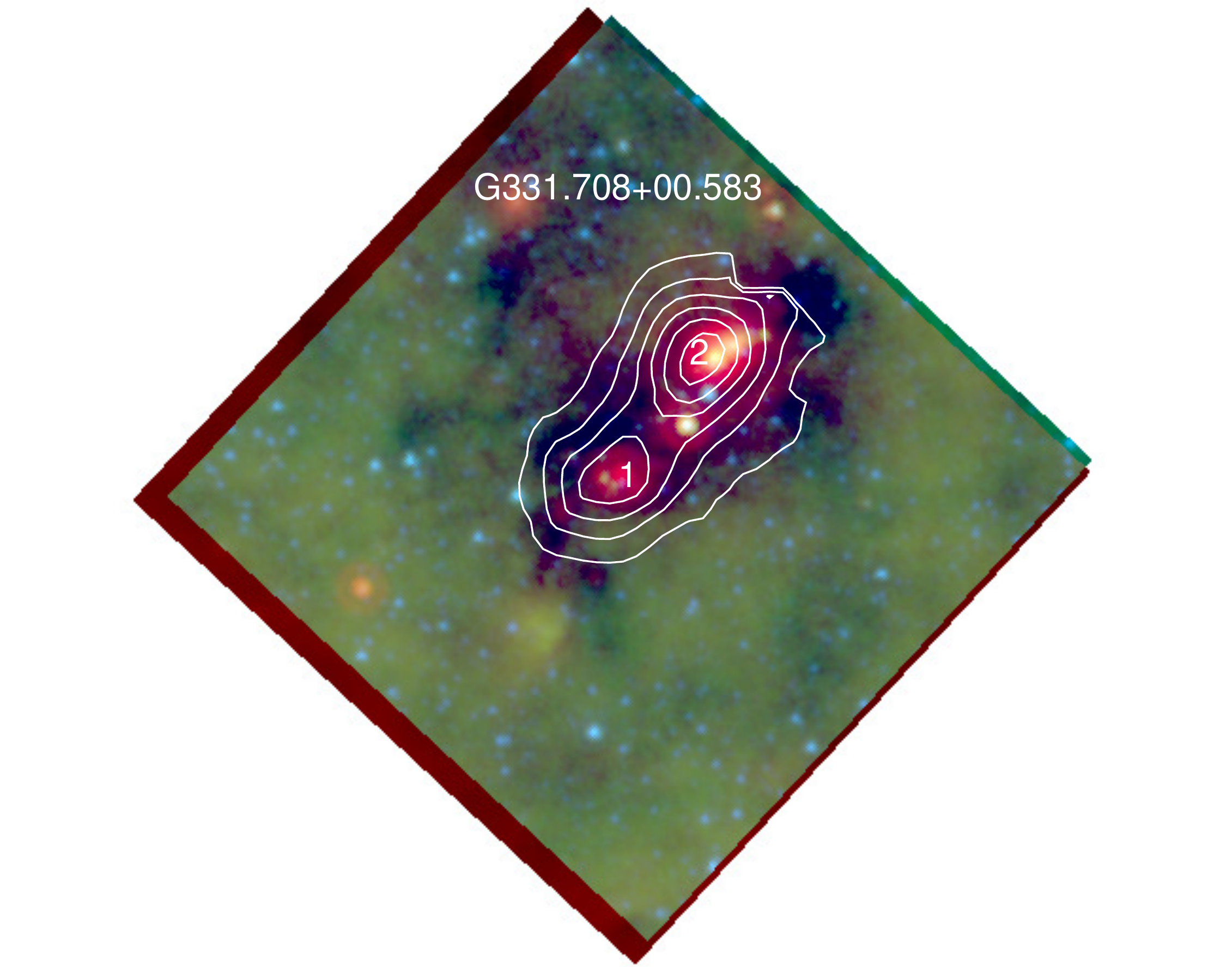}\\
\caption{\textit{Spitzer} RGB images for some of the infrared dark clumps (blue: 3.6~$\mu$m, green: 8~$\mu$m, red: 24~$\mu$m) with \nthp~emission overlaid with the contour levels of 3$\sigma$, 6$\sigma$, 9$\sigma$, etc.}\label{fig:spitzerRGB}
\end{figure*}

\addtocounter{figure}{-1}

\begin{figure*}[ht]
\caption{Continued}
\centering
\includegraphics[width=0.48\textwidth]{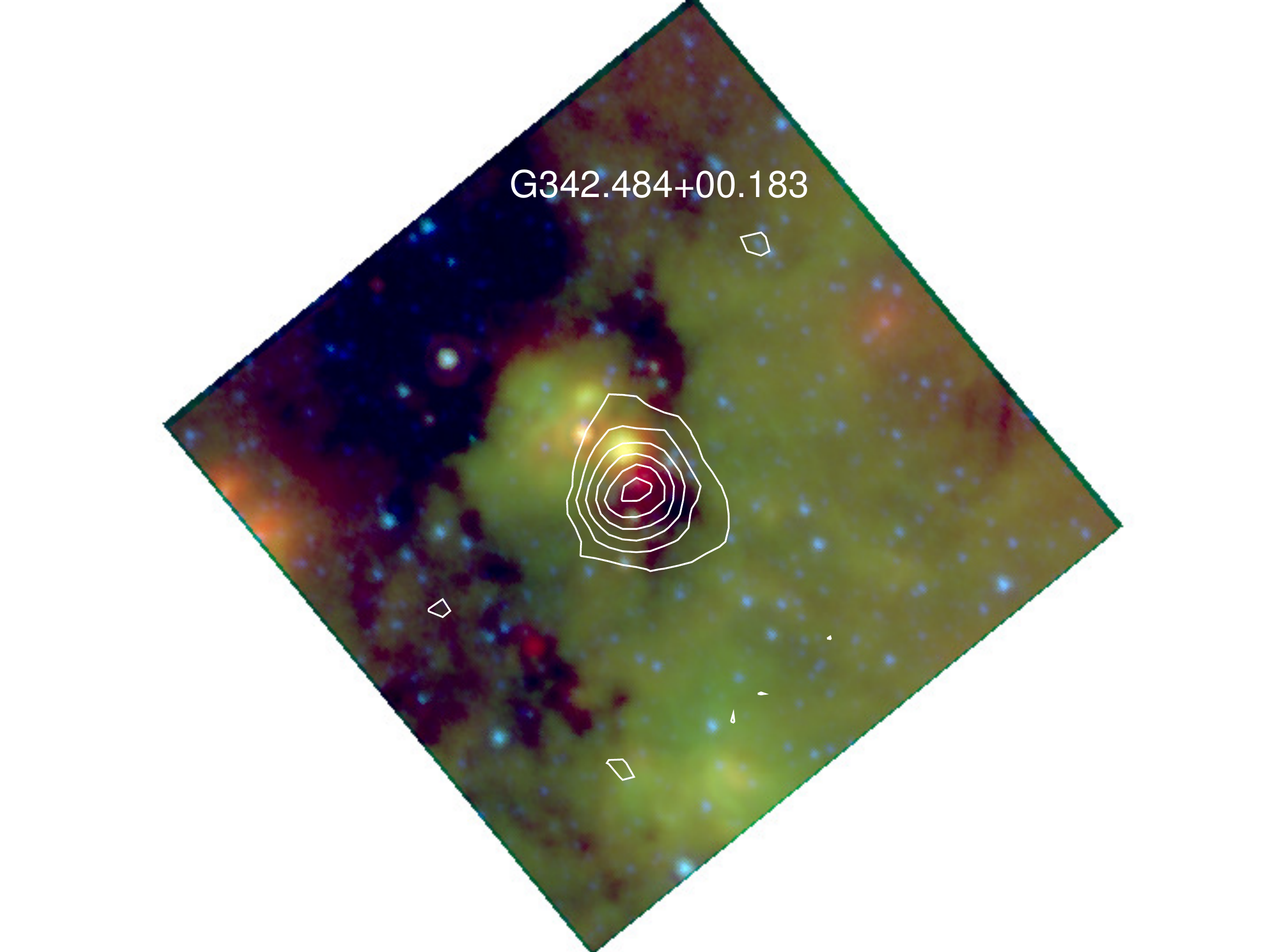}\vspace{0.5cm}
\includegraphics[width=0.48\textwidth]{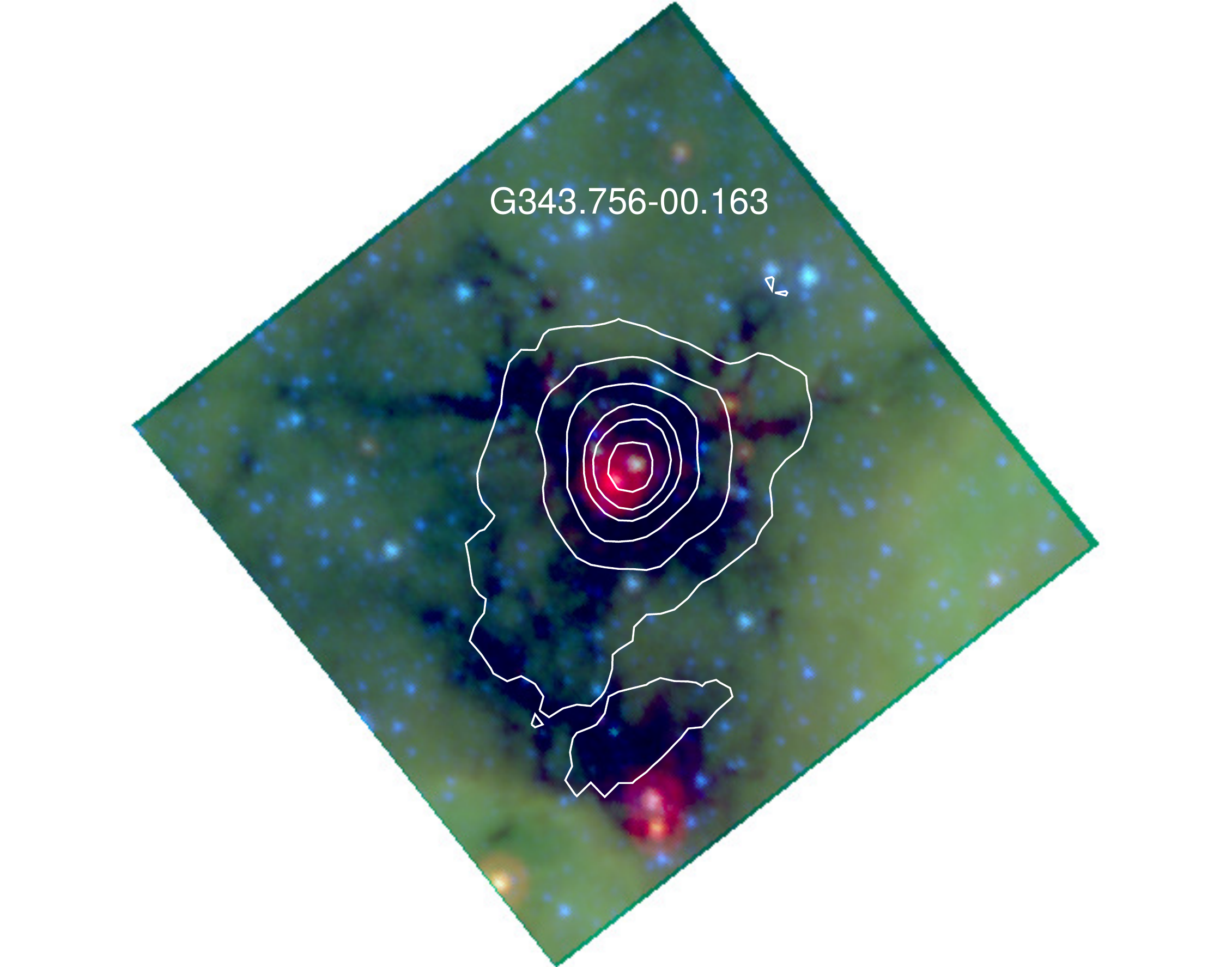}\\
\includegraphics[width=0.40\textwidth]{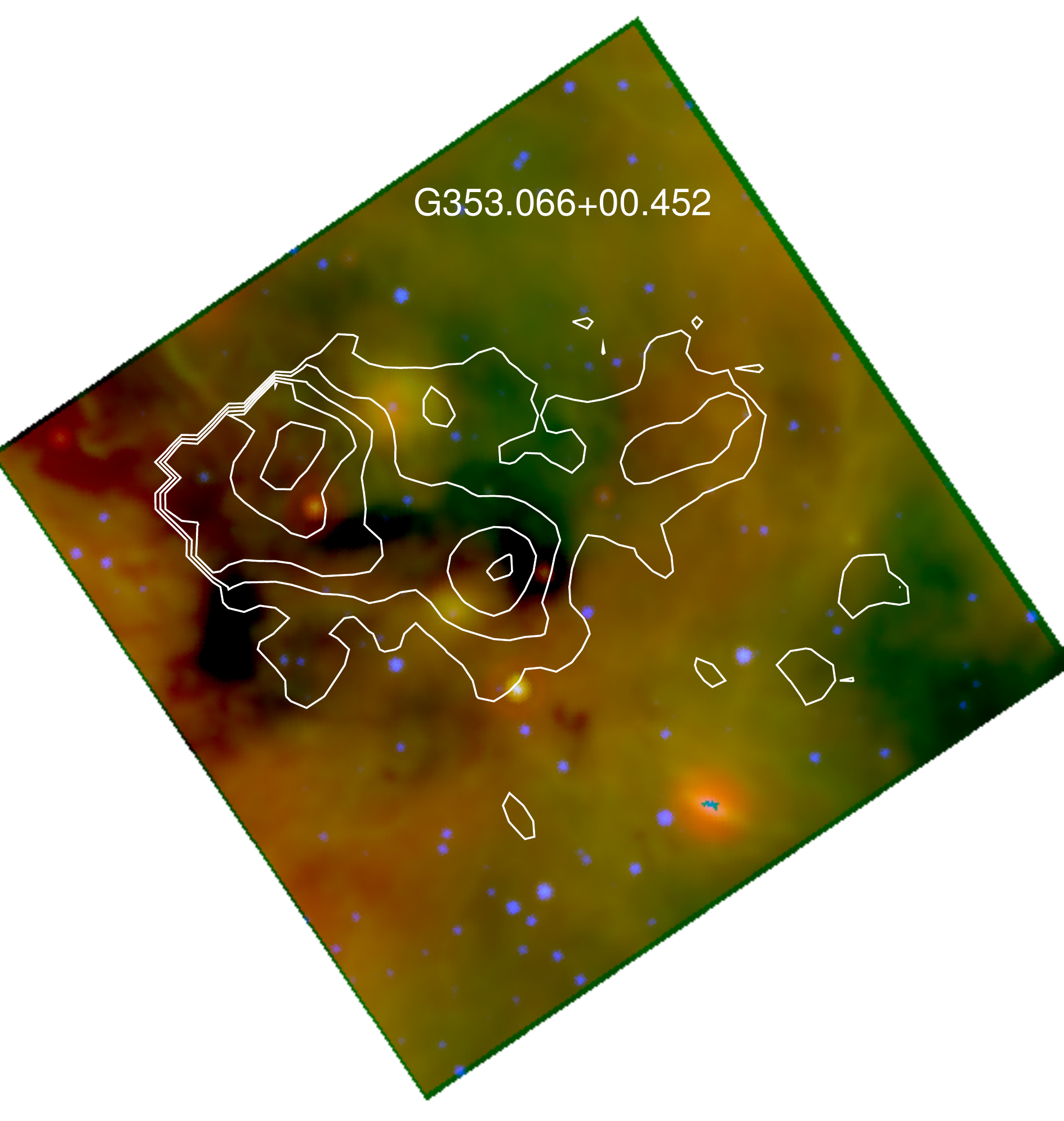}\vspace{0.5cm}
\includegraphics[width=0.40\textwidth]{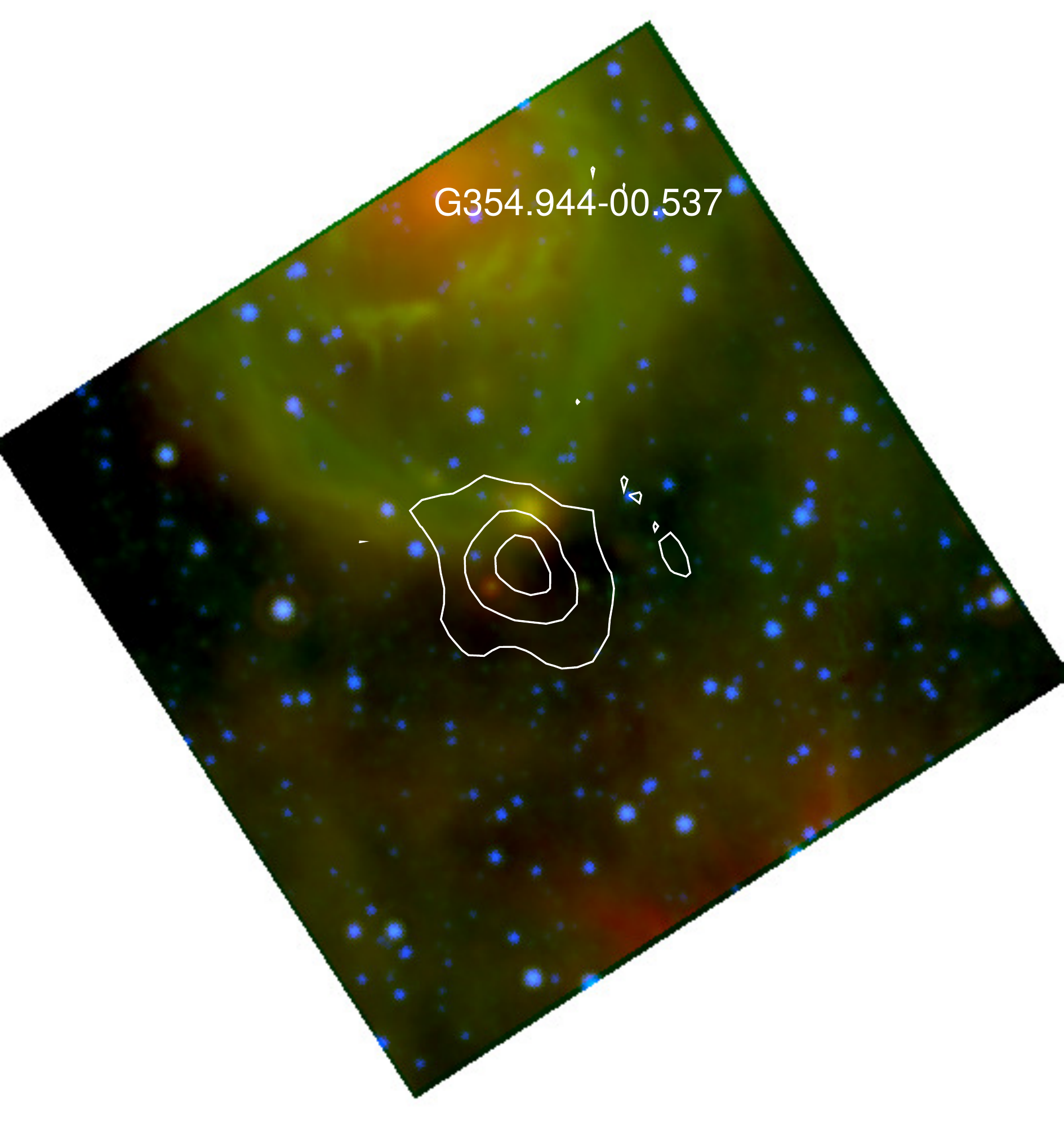}\\
\end{figure*}

\begin{figure*}
\centering
\includegraphics[width=0.45\textwidth]{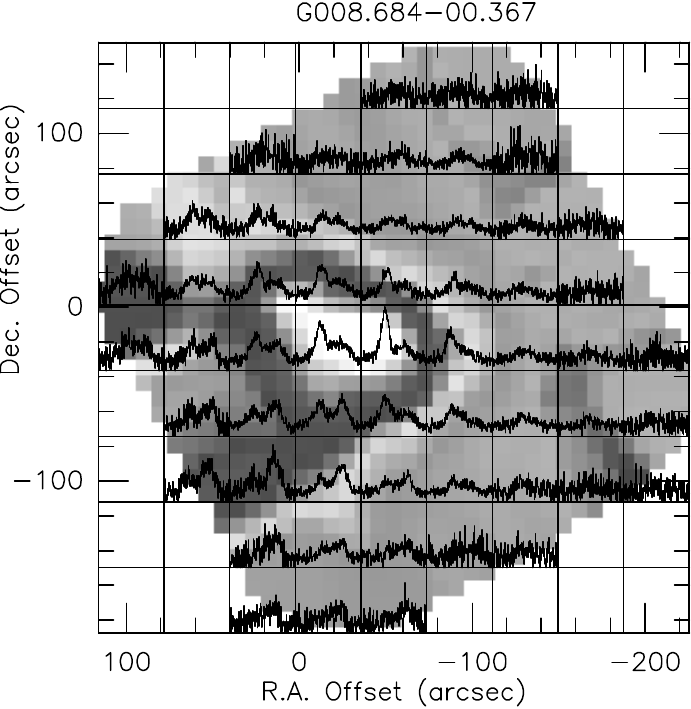}\hspace{0.3cm}
\includegraphics[width=0.45\textwidth]{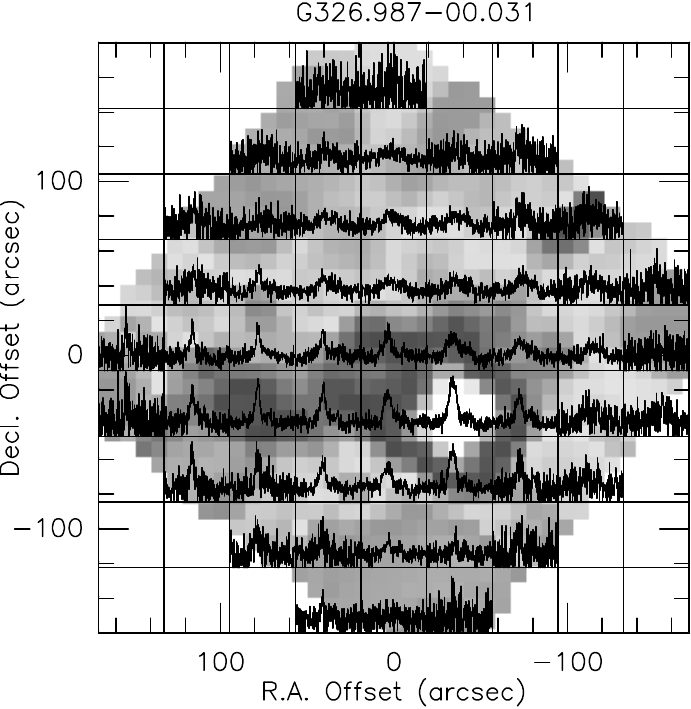}\vspace{0.3cm}
\includegraphics[width=0.45\textwidth]{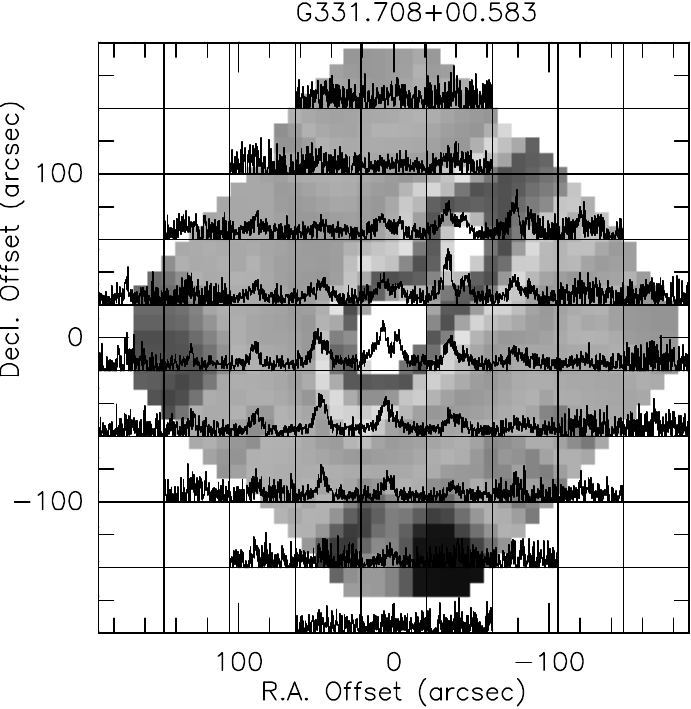}\hspace{0.3cm}
\includegraphics[width=0.45\textwidth]{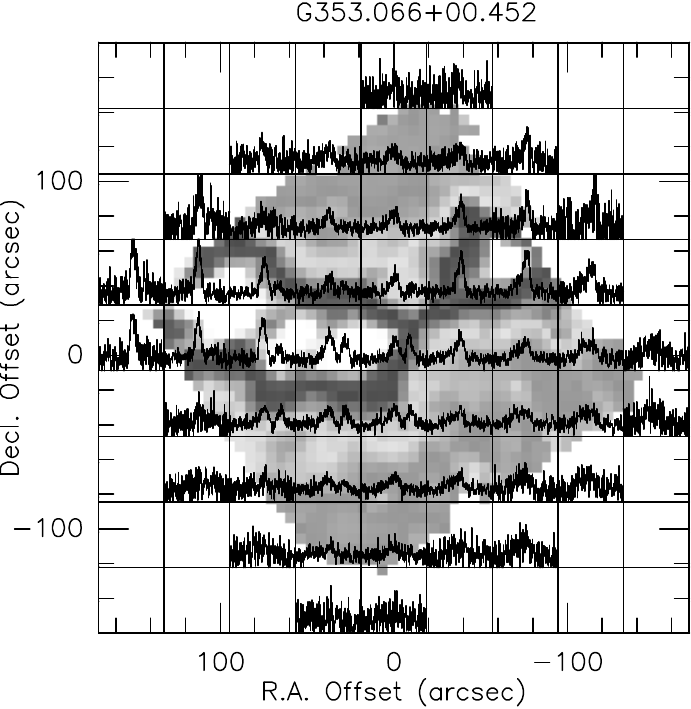}\vspace{0.3cm}
\caption{\hcop~map grid of four possible infall and outflow candidates (gridded to beam size) overlaid on the \hcop~ integrated intensity map.}\label{fig:specmap1}
\end{figure*}

\begin{figure*}
\centering
\includegraphics[width=0.45\textwidth]{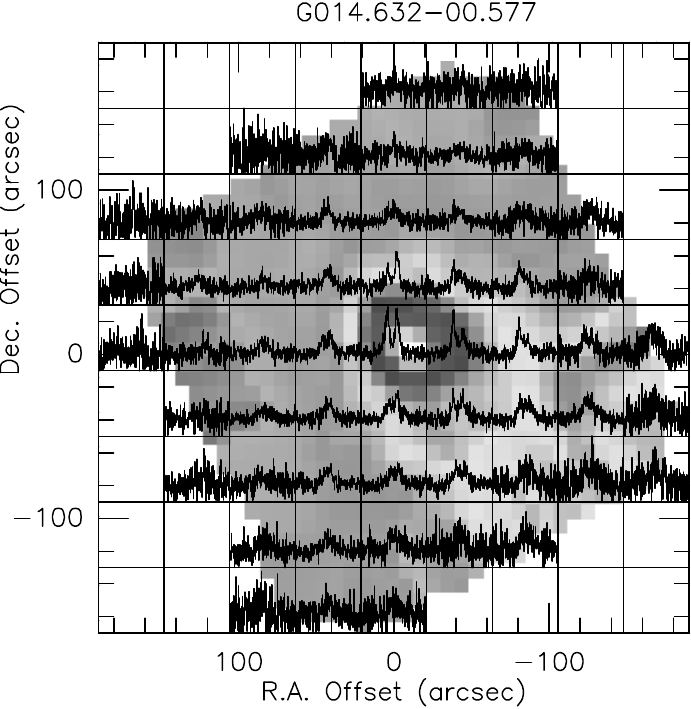}\hspace{0.3cm}
\includegraphics[width=0.45\textwidth]{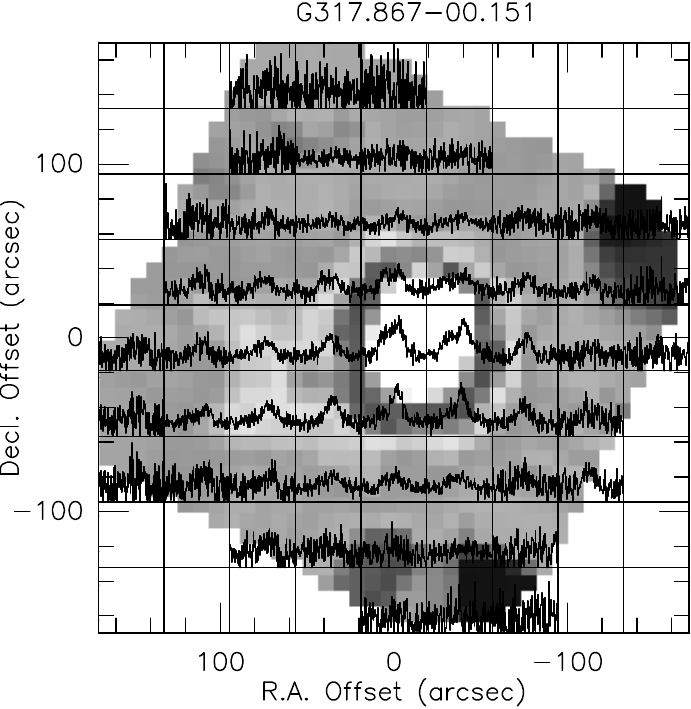}\vspace{0.3cm}
\caption{\hcop~map grid of two possible outflow candidates (gridded to beam size) overlaid on the \hcop~ integrated intensity map.}\label{fig:specmap2}
\end{figure*}

\subsection{Kinematics}\label{sec:kinematics}

\begin{figure*}
\centering
\includegraphics[width=0.4\textwidth]{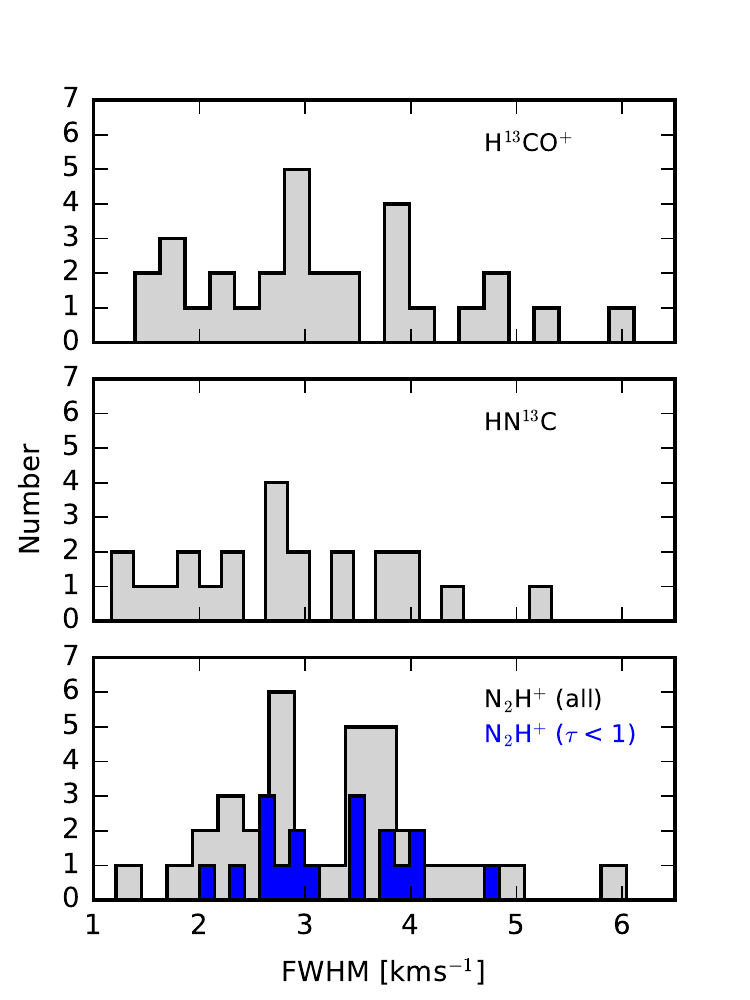}
\includegraphics[width=0.4\textwidth]{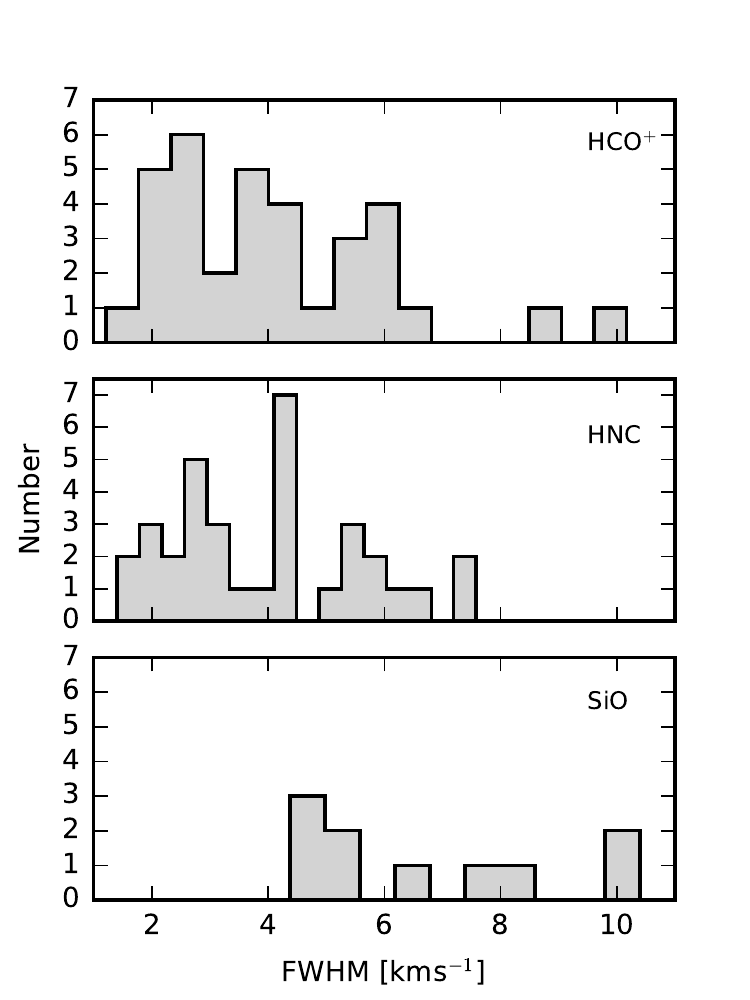}
\caption{Histograms of the number distributions of line widths. \emph{Left}: For the \htcop, \hntc, and \nthp~lines. Optically thin \nthp~lines are plotted in blue. \emph{Right}: For the \hcop, \hnc, and \sio~lines.}\label{fig:linewidths_hist}
\end{figure*}

\begin{figure*}
\centering
\includegraphics[width=0.4\textwidth]{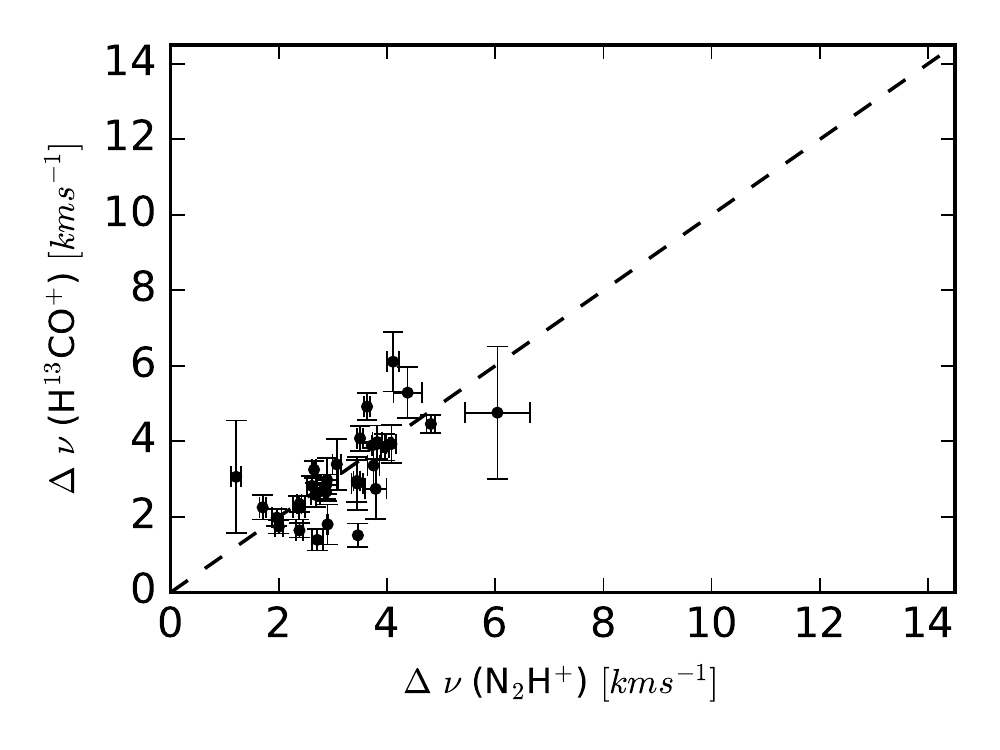}
\includegraphics[width=0.4\textwidth]{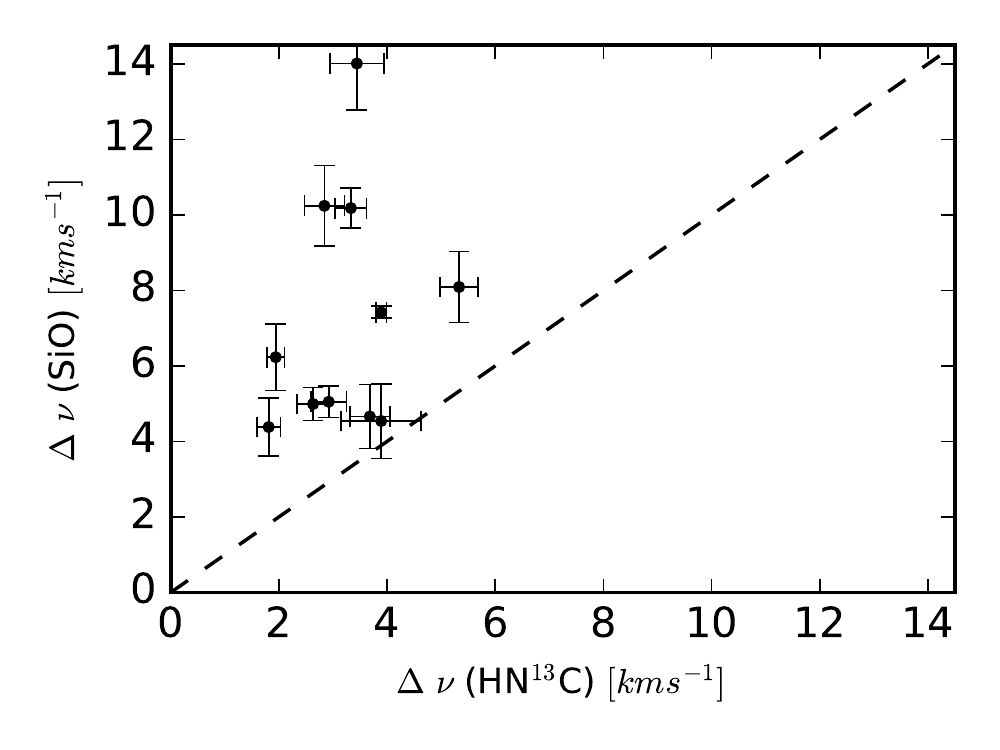}
\caption{Velocity widths of several molecular lines are compared. A ratio of one is shown as a black dashed line.}\label{fig:linewidths}
\end{figure*}

In order to obtain information about the gas kinematics inside the molecular clumps, line widths can be used. Line broadening in optically thin lines is commonly observed in more active clumps due to increased turbulence with star formation activity as proposed by \citet{cha09} and observed in some studies \citep[e.g.,][]{sak08,vas11,san12}. In order to see if there is any trend in the line widths, we first chose lines which are arguably optically thin, since opacity can also cause line broadening \citep{bel05}. The line widths are measured from the averaged spectra in 38$''$ x 38$''$ box (MOPRA beam size) around the clump's \nthp peak emission. To see any trends in the measured line widths of different spectral lines, we plot the number distributions of the line widths of the optically thin \htcop, \hntc~and \nthp~lines in the left panel of Figure~\ref{fig:linewidths_hist}.  The line widths for optically thin ($\tau$ < 1.0) \htcop~range between 1.4 \kms{} and 6.1 \kms{} with a median value of 3.2 \kms{}. Similarly \hntc~line widths range between 1.2 \kms{} and 4.5 \kms{} with a mean value of 2.8 \kms{}. For half of the \nthp~lines which are optically thin, the line widths range between 2.0 \kms{} and 4.8 \kms{}, with a mean value of 3.3 \kms{}, and for all \nthp~lines including the optically thick ones the mean value is 3.2 \kms{}. We conclude that there is a smooth distribution which peaks around 3 \kms{}, while the line widths for all three molecules are smaller than 6~\kms. In addition, the left panel of the Figure~\ref{fig:linewidths} shows a tight correlation between the velocity widths of \nthp~and \hntc~lines which might be a sign that these lines come from the same region. 

These \nthp~values are also similar to those found for 159 clumps (quiescent to active) in \citet{san12}, while the median \nthp~line width is similar to the median values they found for more active (3.0 \kms{}) and red clumps (3.4 \kms{}) rather than quiescent (2.8 \kms{}) or intermediate clumps (2.7 \kms{}). Considering the small number of targets in our study, we can say that this difference is not significant. Furthermore, \citet{sak08} studied 29 MSX objects, which are classified as mostly quiescent and intermediate or active clumps by \citet{san12}, and found that the mean \nthp~line width is about 2.4 \kms{} which is lower than the value we found. Similarly, \citet{vas11} studied 37 clumps (varying from quiescent to active and red) and found a mean value of 1.8 \kms{} and \citet{pir03} found the mean \nthp~line width value for 35 dense molecular cores to be 2.4 \kms{}. On the other hand, for low-mass starless cores the average value is found to be $\sim$0.5 km s$^{-1}$ \citep{ben98}.

\begin{figure}
\centering
\includegraphics[width=0.4\textwidth]{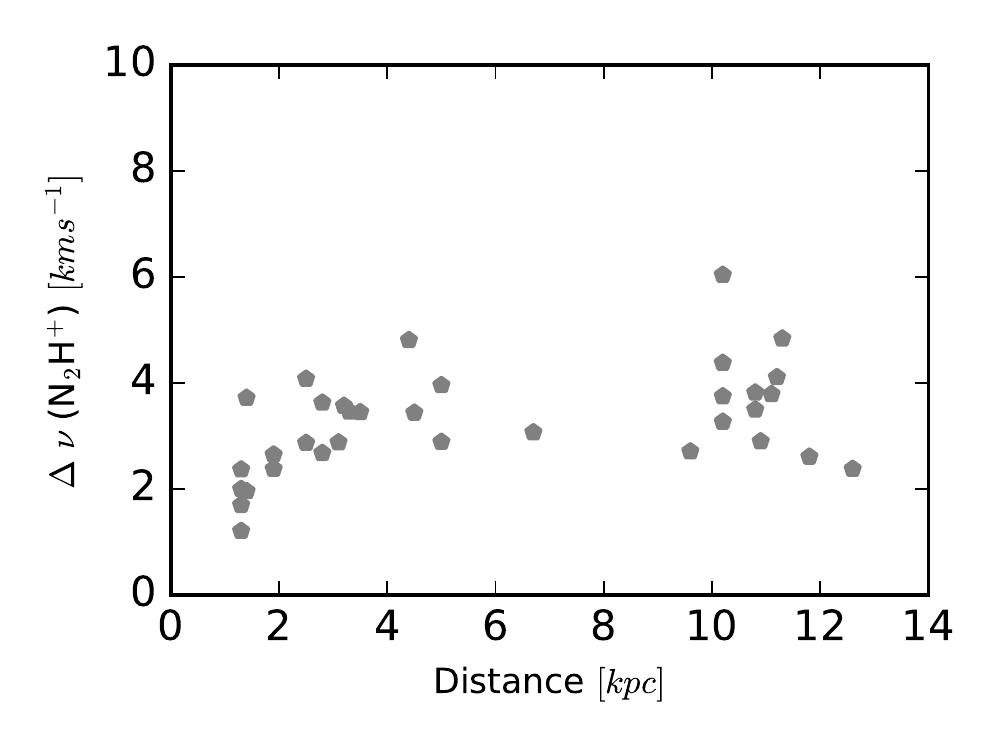}
\caption{\nthp~line width distribution of the clumps according to their kinematic distances.}\label{fig:nthp_distance}
\end{figure}

Considering that our large beam size corresponds to 0.18 pc at a distance of 1 kpc and 1.85 pc at a distance of 10 kpc, we might be covering a larger area for clumps at larger distances which might increase the line width as expected from Larson's laws \citep{lar81}. To check if this was the case for our sample, we examined how the \nthp~line widths are distributed by the distances. As can be seen in Figure~\ref{fig:nthp_distance}, large line widths cannot be explained by large distances. Therefore, the larger \nthp~line widths might indicate initially turbulent clumps which favors the turbulence supported cores theory \citep{mck03} for massive star formation. However, since we do not know if these clumps are starless or not, higher resolution data are needed to study the cores within these clumps.

Figure~\ref{fig:linewidths_hist} shows the line width distributions for the optically thick \hcop~and \hnc~lines. Line widths range between 1.2 \kms{} and 10.2 \kms{} and 1.4 \kms{} and 7.6 \kms{} and the mean values are 4.1 \kms{} and 4.0 \kms{} for the \hcop~and \hnc~lines, respectively. The larger line widths compared to the \nthp~lines can be explained with larger optical depths. Also as can be clearly seen in some sources (e.g., G014.194-00.194, G331.708+00.583) line wings can cause larger line widths possibly originating from outflows. G014.194-00.194, G331.708+00.583\_1, and G343.756-00.164 with \sio~signal within the range of 9$\sigma$, G329.029-00.206 with 12$\sigma$, and G351.444+00.659 with 15$\sigma$ are the most promising regions to look for outflows.

In general, the \sio~lines, whose widths are also shown in Figure~\ref{fig:linewidths_hist} are broader than the \htcop~and \hntc~lines and range between 4.4 \kms{} and 14.0 \kms{} with a mean value of 7.2 \kms{}, as expected when there are shocks or outflows (see the right panel of Figure~\ref{fig:linewidths}). The majority of the sources, however, show no strong outflow signatures in other molecular lines which indicates that the \sio~emission might not be due to the shocks in outflows but instead are due to colliding flows. {\sio} has been studied toward the mini-starburst region W43 where extended {\sio} emission is observed with line widths >6 km s$^{-1}$. The emission is not connected to any specific outflow activities and is proposed to originate from large-scale interactions during the cloud formation process \citep{ngu13,lou16}. Therefore, extended {\sio} emission might not be affected by star formation activity in all cases and it might represent remnant emission from the cloud formation process. In our study, only eleven sources have \sio~detections over 3$\sigma$ and only five of them are over 6$\sigma$. Although the line widths are large enough to give clues about outflows, it is difficult to derive conclusions about the nature of the \sio~emission in these clumps without higher sensitivity observations, considering there is no detection of velocity gradients detected for the \sio~line.

\begin{figure*}
\includegraphics[width=0.5\textwidth]{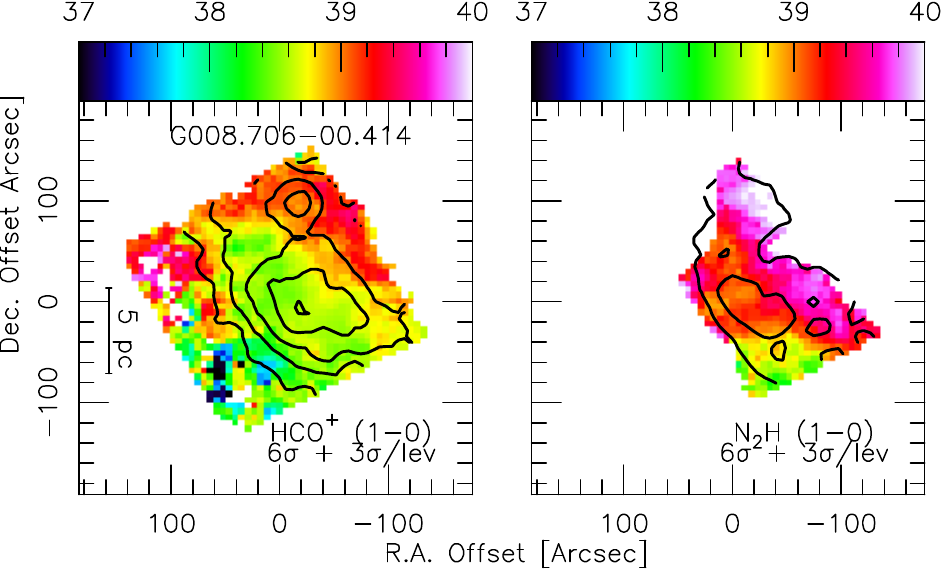}\vspace{0.5cm}
\includegraphics[width=0.5\textwidth]{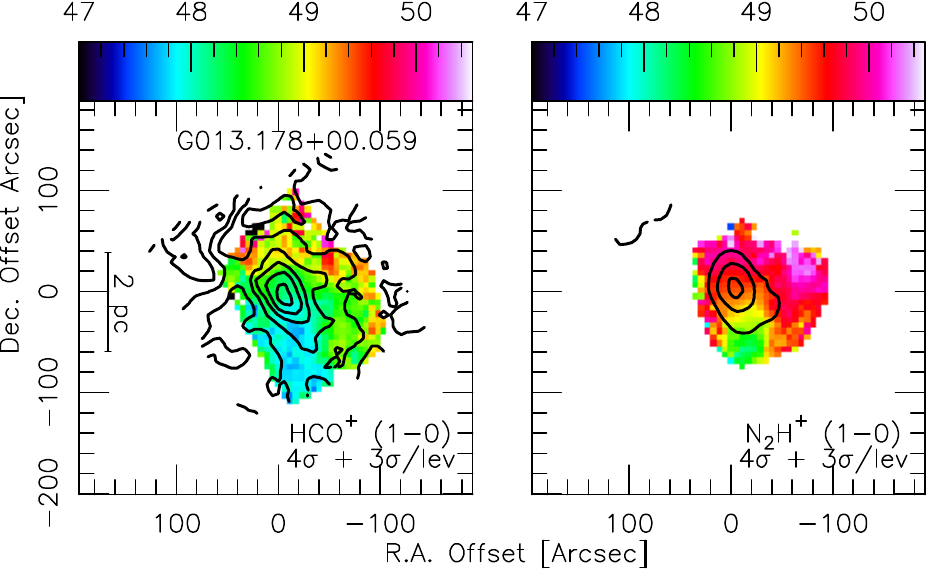}\\
\includegraphics[width=0.5\textwidth]{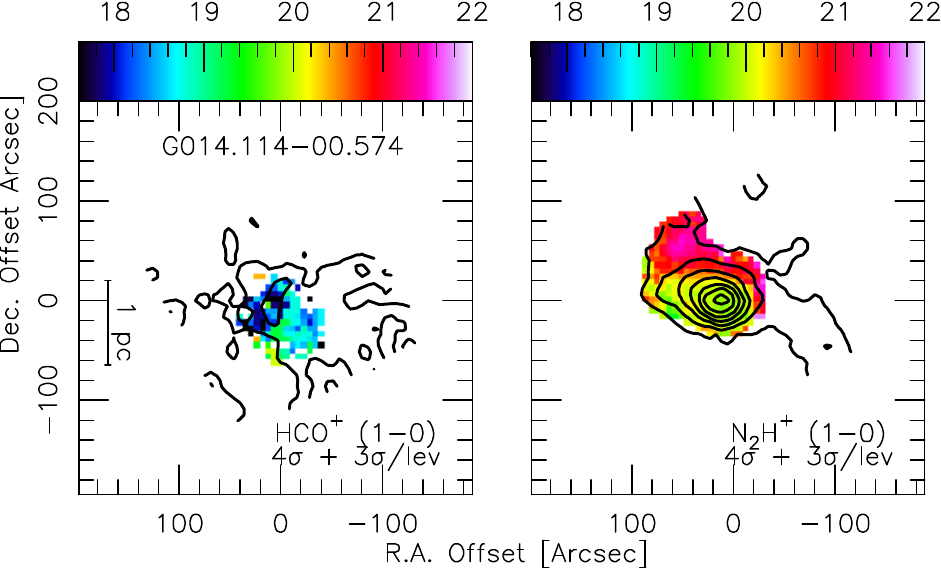}\vspace{0.5cm}
\includegraphics[width=0.5\textwidth]{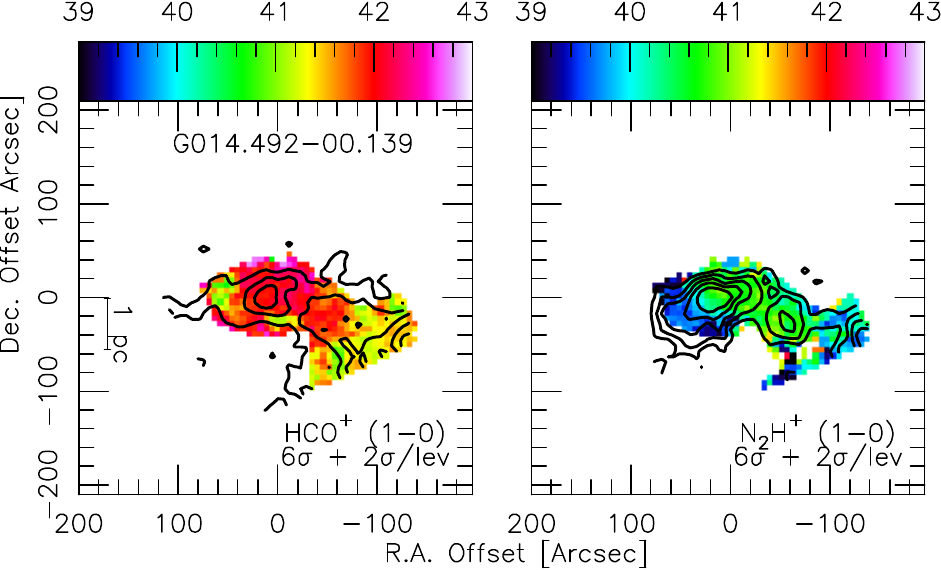}\\
\caption{Velocity (moment 1) maps overlaid with the intensity contours of the same line showing the velocity gradients in each clump. The top color bar is in units of \kms{}. Contour levels are shown on each image. All the velocity maps are generated with 3$\sigma$ level of the average rms of \hcop and \nthp channel maps. We show only a few examples here; maps for all the sources can be found in the electronic version.\label{fig:velocitymap}}
\end{figure*}

To check the velocity gradients, we constructed velocity maps of the sources from the \hcop~and \nthp~lines as shown in Figure~\ref{fig:velocitymap}. There are 25 sources which show only small velocity gradients of up to $\sim$2~km/s across each clump (a few parsecs) which are consistent with the velocity gradient values observed in other clouds \citep{bel13}. High-resolution interferometric observations at core scales could reveal $\sim$5 times more velocity gradients \citep{bel13} in these regions and reveal the outflows originating from individual cores.

Due to the different gas-grain chemistry of the molecular species, we can use different lines to trace specific motions such as infalls outflow, rotation or turbulent motions. Studies show that outflows from intermediate- and high-mass protostars can generate parsec-scale velocity gradients \citep[e.g.,][]{ben04}, and rotation of the cloud and cores and turbulence can also affect the observed velocity gradients \citep[e.g.,][]{bel13}. For instance, \hcop~enhancement is known to occur in ionized gas due to strong shocks \citep{dic80,whi87,raw04,gir05} and high-velocity \hcop~emission, line wings, and velocity gradients are already observed in many star-forming regions with outflows \citep[e.g.,][]{san82,gar92,jor04}. The variation of the \nthp~line velocities is mostly used to trace the accretion flows or rotation of the dense gas \citep[e.g.,][]{dif01,kir13,dha18}. For this study, with the limited resolution of single-dish observations, however, we note that the observed velocity gradients are generated by the bulk motion of the common envelope, rather than individual cores within the clump.

\subsection{Integrated intensities}\label{sec:intensity}

The relationships between the intensities of the different molecules can be used to understand the chemical evolution in star-forming regions. We find the mean integrated intensities to be 5.12 \Kkms{}, 9.79 \Kkms{}, 4.22 \Kkms{}, and 6.92 \Kkms{} for the \hcop, \nthp, \hnc, and \hcn~lines, respectively, as shown in Table~\ref{intensity}. Since \nthp~is a convenient molecule to study the kinematics of cold, dense star-forming regions \citep{cas02} due to the resistance of depletion under these conditions, and also traces the gas with a density one order of magnitude higher than what \hcop~does \citep{shi15}, we compared the mean integrated intensity of this line to other star-forming regions in order to derive information about the evolutionary stages of these clumps. With the mean integrated intensity of 9.79$\pm$0.28 \Kkms{} (median is 6.62$\pm$1.63 \Kkms{}), the selected sample is at similar evolutionary levels with the quiescent and intermediate clumps studied in \citet{vas11}. They studied 15 IRDCs which were classified as quiescent, mid, and active with mean integrated intensities of 4.7 \Kkms{}, 7.1 \Kkms{}, and 10.6 \Kkms{}, respectively. 

\begin{figure*}
\centering
\includegraphics[width=0.4\textwidth]{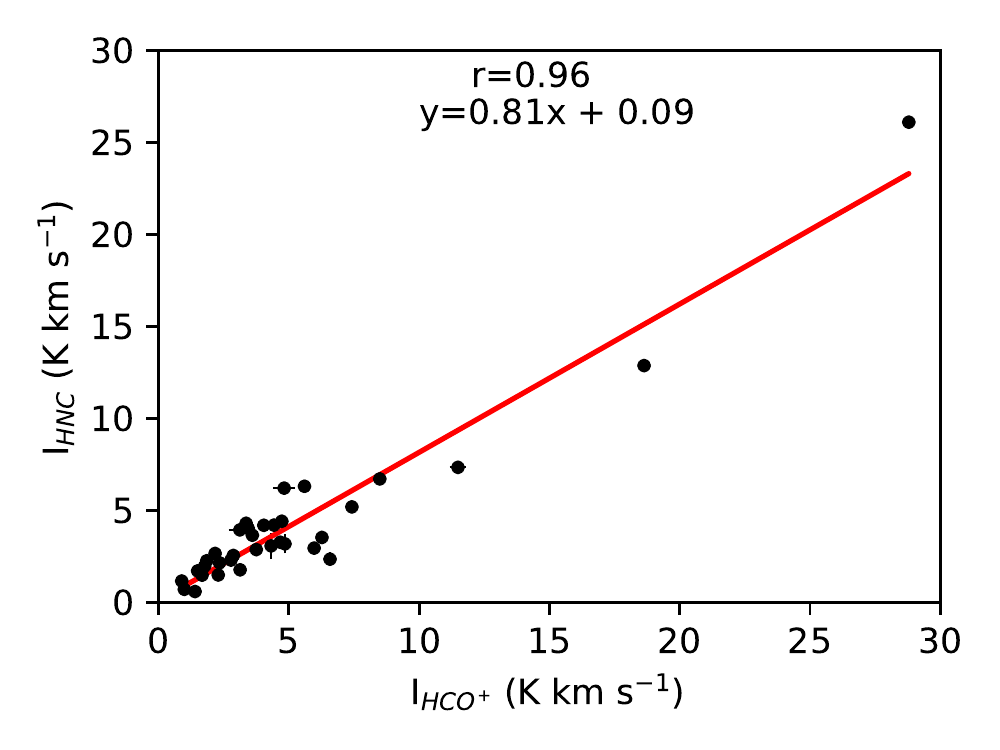}
\includegraphics[width=0.4\textwidth]{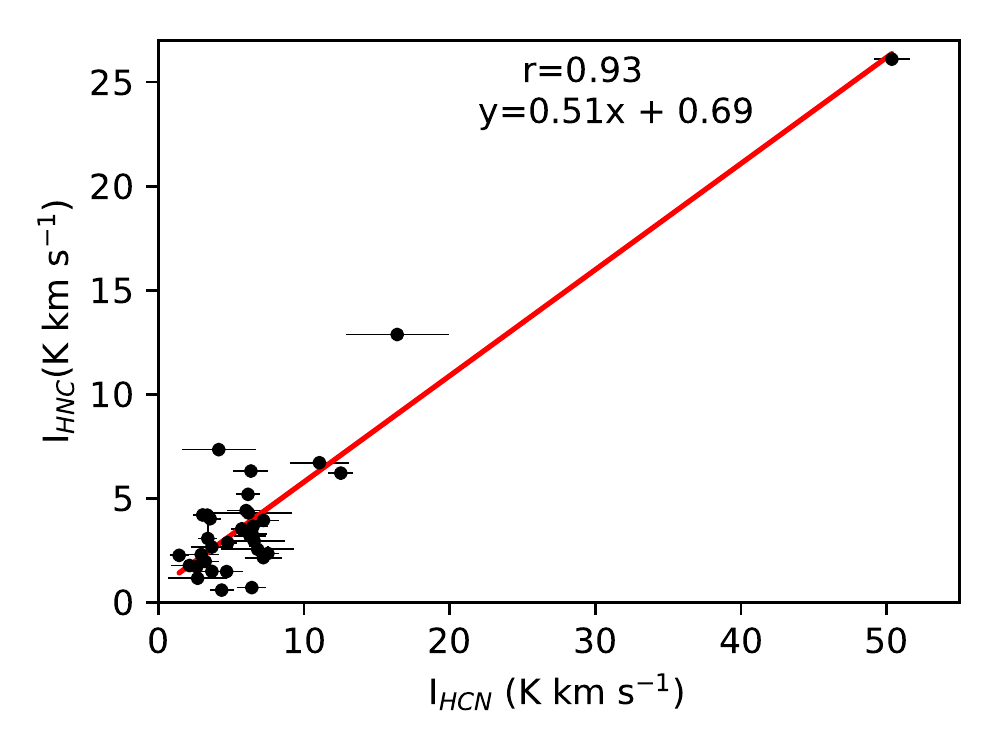}
\includegraphics[width=0.4\textwidth]{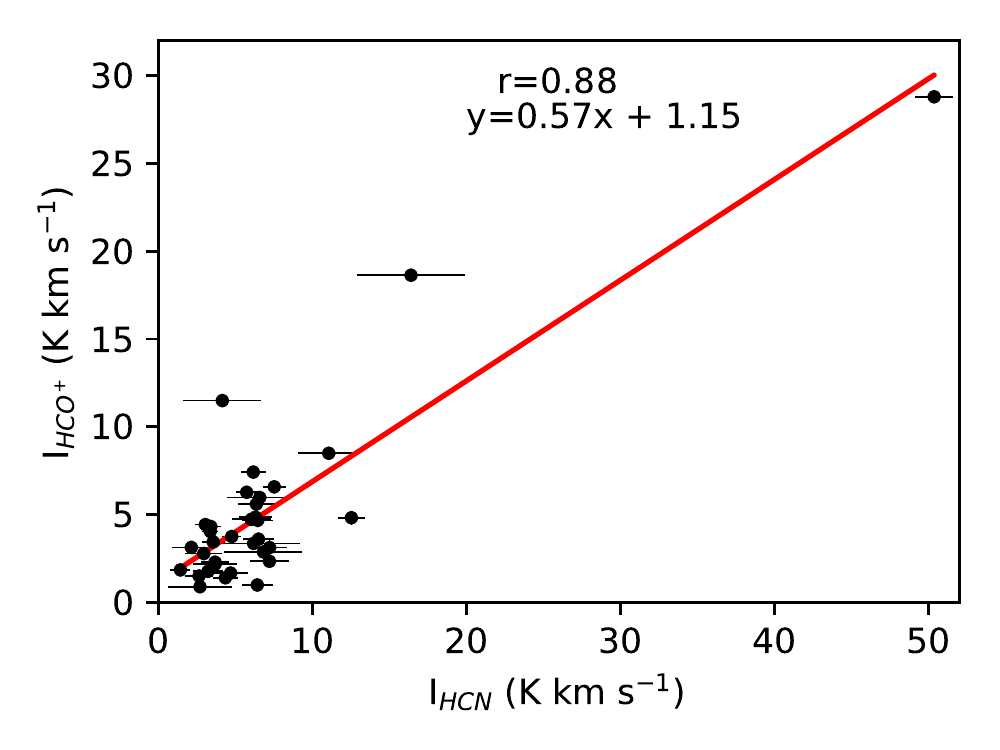}
\includegraphics[width=0.4\textwidth]{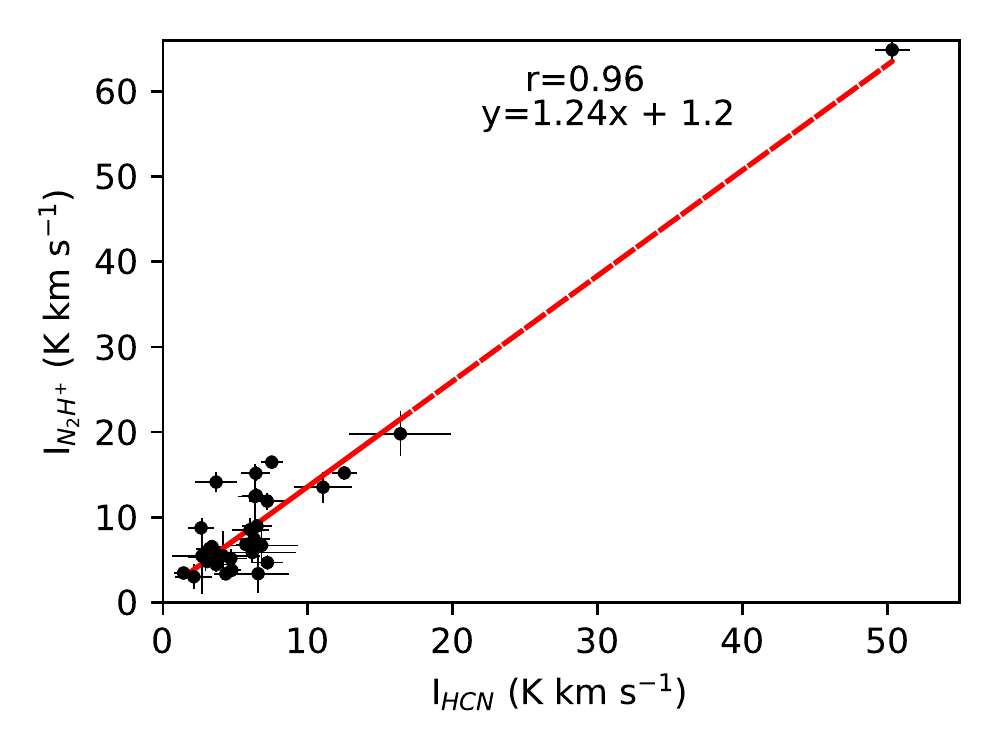}
\caption{Integrated intensity relations between \hcop, \hnc, and \hcn. The red line represent the linear fit.}\label{fig:int_intensity}
\end{figure*}

In order to see how the intensities of the different molecules are correlated we plotted the intensities of \hcop-\hnc, \hnc-\hcn, \hcop-\hcn, and \nthp-\hcn~in Figure~\ref{fig:int_intensity}. The linear fit results and r-values (shown on the panels) indicate that there is a tight correlation between emission from these molecules during the evolution of the clumps. We note, however, that the tight correlation is also caused by the data points corresponding to G351.444+00.659 which has large intensity values. Excluding this data point gives still a tight correlation (r=91) between \hnc~and \hcop, while the correlation between other molecules is decreases (r$\sim$ 0.70). Similar correlations between emission from these molecules had been found for other IRDCs \citep[e.g.,][]{liux13}.

The median intensity ratios are 1.30$\pm$0.17, 1.40$\pm$0.56, and 0.61$\pm$0.11 for I$_{HCN}$/I$_{HCO^{+}}$, I$_{N_{2}H^{+}}$/I$_{HCN}$, and I$_{HNC}$/I$_{HCN}$, respectively. The value found for the ratio of I$_{HCN}$/I$_{HCO^{+}}$ is similar to the value of 1.67$\pm$0.83 found in the starburst ring of NGC 1097 \citep{hsi12}, while it is bigger than the mean value of 0.67 found for 14 IRDCs in \citet{liu13}. This behavior indicates that the I$_{HCN}$/I$_{HCO^{+}}$ ratio is not dependent on evolution but rather other mechanisms such as X-ray ionization by \hii~regions. Massive stars in the vicinity of the IRDCs and clumps may also play a role by affecting the intensities as also stated in \citet{hsi12} and \citet{liu13}. On the other hand, the ratio of I$_{N_{2}H^{+}}$/I$_{HCN}$ was found to be 1.40$\pm$0.56, which is similar to the value of $\sim$1.5 found in \citet{liux13}, which might indicate that the IRDCs examined in both studies are at similar evolutionary stages.

\begin{table}
\caption{\label{intensity} Statistical properties of the intensities.}
\centering
\tiny
\begin{tabular}{lllll}
\hline\hline
\noalign{\smallskip}
 &  $I_{HCO^{+}}$ & $I_{N_{2}H^{+}}$ & $I_{HNC}$ & $I_{HCN}$ \\
\hline
Mean & 5.12(0.03) & 9.79(0.28) & 4.22(0.04) & 6.92(0.37) \\
Median & 3.67(0.09) & 6.62(1.63) & 3.10(0.40) & 5.87(1.67)\\
Std & 5.30 & 10.52 & 4.45 & 8.13 \\
\hline
\hline
\end{tabular}
\end{table}

\subsection{Column density ratios}\label{sec:abundance}

The Appendix Table~\ref{ratio} shows the column density ratios of different molecules. The mean column density ratios of N(HNC)/N(HCO$^{+}$), N(N$_{2}$H$^{+}$)/N(HCO$^{+}$), N(N$_{2}$H$^{+}$)/N(HNC), N(HNC)/N(HCN), and N(HCN)/N(HCO$^{+}$) are 2.3$\pm$0.1, 0.8$\pm$0.1, 1.3$\pm$0.7, 8.9$\pm$1.0, and 0.8$\pm$0.23, respectively (median and standard deviation values are also shown in Table~\ref{ratio}). 

The median ratio of N(HNC)/N(HCO$^{+}$), 1.2$\pm$0.2, is similar to the value, 1.7$\pm$0.2 found by \citet{liux13} for 14 IRDCs. On the other hand, for massive complex star-forming regions in the Galactic Plane (e.g., W49, W51, G10.62-0.39, and G34.3+0.1) this ratio was found to be only 0.5$\pm$0.3 by \citet{god10} which indicates that the ratio of N(HNC)/N(HCO$^{+}$) might be affected by the ambient environment. Hence, it is difficult to use it as an indicator for early evolution. 

On the other hand, the column density ratio of N(HCN)/N(HCO$^{+}$) with a median of 0.2 seems similar for most of the sources except G014.492-00.039, G338.786+00.476, and G353.417-00.079\_2 which have quite high ratios (3.2, 4.8, 9.4, respectively) compared to others. This behavior is a result of lower limit column density estimations for \hcop~since we were not be able to calculate optical depths for these sources. Similarly, due to the emission line anomalies of \hcn, and not being able to estimate an optical depth for half of the sources, we might also be underestimating the \hcn~column densities. This effect will result in a systematic error of an order of magnitude in the determined column densities. With these constraints, we can say that for all the sources except the three mentioned above, N(HCN)/N(HCO$^{+}$) is almost same within the errors in all clumps. Considering that some of the clumps are associated with protostellar sources, and YSOs while some others with massive star forming (MSF) regions, we can say that this column density ratio is not strongly affected by the different evolutionary phases.

Similarly, we do not see any significant difference in the column density ratios of other molecules. These results, however, are based on small number of sources and considering the possible systematic errors especially for \hcn~column densities, and in some cases for \hcop~column densities as well, the column density ratios cannot be used to claim clear evolutionary phase differences between the studied clumps.

\section{Summary} \label{sec:malt90summary}

The physical and chemical properties of 30 infrared-dark high-mass clumps at 8/24 $\mu$m from the ATLASGAL survey were investigated using MALT90 Survey data. The molecular lines \hcop(1-0), \hnc(1-0), and \nthp(1-0) were used as density tracers, along with their isotopologues \htcop(1-0) and \hntc(1-0) and, \sio(2-1) was used as an outflow tracer. Velocity maps and spectral line profiles for 30 sources were investigated. Column densities and the ratios of column densities were calculated in order to determine the nature of these sources which are at early stages of potential high-mass star formation. 

1. We find 13 sources with blue profiles in both/either \hcop~and/or \hnc~lines which might indicate infall and estimated clump mass infall rates ranging between 0.2 $\times$ 10$^{-3}$ M$_\odot$yr$^{-1}$ $-$ 1.8 $\times$10$^{-2}$ M$_\odot$yr$^{-1}$. 

2. The integrated intensities of the \hcop, \nthp, \hnc, and \hcn~lines have tight correlations with each other which indicate that they might share a similar chemical evolution during the early stages of IRDC formation. Large line widths for the \nthp~line might indicate turbulence, and large widths of the \hcop, \hnc, and \sio~lines might indicate outflow activities.

3. \hcop and \hcn~line emission is found to be optically thick, with median values of 18.08, and 15.50, respectively, while line emission from their isotopologues \htcop~and \hntc~is optically thin with median values of 0.35 and 0.30, respectively. Sixteen sources have optically thin \nthp~lines, while the median value for all sources is 1.00$\pm$0.18.

4. The excitation temperature, T$_{ex}$ is calculated as <24~K for all sources as expected in the cold dense regions within IRDCs.

5. Eleven sources have a significant \sio~detection (above 3$\sigma$) but the signal-to-noise ratio is not sufficient enough to do further analysis.

6. We used the \hcop~and \nthp~lines to investigate the kinematics within the clumps, and only a few sources (G008.706$-$00.414, G013.178$+$00.059, G014.114$-$00.574) show small velocity gradients over the clumps with a maximum of 2 \kms in both lines. 

7. Some sources might have multiple cores, outflow, and infall activities based on the line profiles and kinematics. In order to understand the true nature of these clumps, however, it is necessary to obtain high angular resolution and high sensitivity data and obtain radiative transfer models to reproduce the line profiles.

\begin{acknowledgements}

This work is supported by the STARFORM Sinergia Project CH:CRSII2\_160759 funded by the Swiss National Science Foundation. The authors also thank Dr. Joseph Hora for careful reading of the manuscript and valuable comments that helped improve the paper. 
This work has made use of the data products from the Millimetre Astronomy Legacy Team 90-GHz (MALT90) survey and the APEX Telescope Large Area Survey of the Galaxy (ATLASGAL), which is a collaboration between the Max-Planck-Gesellschaft, the European Southern Observatory (ESO) and the Universidad de Chile; and also used observations made with the \textit{Spitzer} Space Telescope, which is operated by the Jet Propulsion Laboratory, California Institute of Technology, under contract with NASA. 
This research has made use of NASA’s Astrophysics Data System Abstract Service; and the SIMBAD data base, operated at CDS, Strasbourg, France. We thank the anonymous referee for providing valuable comments and suggestions.

\end{acknowledgements}

\bibliographystyle{bibtex/aa}
\bibliography{bibtex/biblio.bib}

\begin{appendix}

\section{Fitting Results}
\label{section:fittingresults}

\begin{table*}
\caption{\label{tab_linefit} Fitting results of the spectral lines.}
\centering
\tiny
\begin{threeparttable}
\begin{tabular}{lllllll}
\hline\hline
\noalign{\smallskip}
Source & Line Properties & \multicolumn{5}{c}{Lines}  \\ 
       & & HCO$^{+}$(1-0) & H$^{13}$CO$^{+}$(1-0) & SiO(2-1) & HNC(1-0) & HN$^{13}$C(1-0)\\
\hline
& W (K km~s$^{-1}$)\tnote{a} & 6.58(0.18)\tnote{1} & 2.41(0.11) & 1.48(0.15) & 2.35(0.38)\tnote{1} & 1.98(0.12) \\
G008.684-00.367 & $V_{LSR}$ (km~s$^{-1}$) & 34.04(0.04)\tnote{1} & 36.94(0.10) &
37.71(0.42) & 34.22(0.06)\tnote{1} & 37.40(0.17) \\ 
 & $\Delta V$ (km~s$^{-1}$) & 3.70(0.10)\tnote{1} & 4.46(0.23) & 8.09(0.94) & 2.72(0.24)\tnote{1} & 5.33(0.35) \\ 
 & $T_{mb}$ (K) & 1.67\tnote{1} & 0.51 & 0.17 & 0.81\tnote{1} & 0.35 \\
\hline   
& W (K km~s$^{-1}$) & 7.42(0.11) & 0.96(0.08) & 0.82(0.11) & 5.19(0.10) & 0.58(0.07) \\
G008.706-00.414 & $V_{LSR}$ (km~s$^{-1}$) & 38.56(0.04) & 39.09(0.12) & 39.35(0.32) & 38.65(0.04) & 39.38(0.11) \\ 
 & $\Delta V$ (km~s$^{-1}$) & 5.30(0.09) & 2.82(0.27) & 4.38(0.77) & 4.20(0.10) & 1.81(0.22) \\ 
& $T_{mb}$ (K) & 1.32 & 0.32 & 0.18 & 1.16 & 0.30 \\
\hline 
& W (K km~s$^{-1}$) & 3.13(0.16)\tnote{2} & \nodata & \nodata & 0.94(0.12)\tnote{2} & \nodata \\
G010.444-00.017 & $V_{LSR}$ (km~s$^{-1}$) & 74.00(0.14)\tnote{2} & \nodata & \nodata & 66.17(0.22)\tnote{2} & \nodata \\
 & $\Delta V$ (km~s$^{-1}$) & 6.69(0.47)\tnote{2} & \nodata & \nodata & 4.11(0.61)\tnote{2} & \nodata \\ 
& $T_{mb}$ (K) & 0.44\tnote{2} & \nodata & \nodata & 0.21\tnote{2} & \nodata \\
\hline
& W (K km~s$^{-1}$) & 5.60(0.14) & 0.83(0.12) & 1.12(0.16) & 6.31(0.11) & 1.12(0.09) \\
G013.178+00.059 & $V_{LSR}$ (km~s$^{-1}$) & 48.31(0.05) & 49.06(0.17) & 48.54 (0.44) & 49.13(0.04) & 49.45(0.15) \\ 
 & $\Delta V$ (km~s$^{-1}$) & 4.41(0.13) & 2.95(0.55) & 7.32(1.53) & 5.10(0.10) & 3.68(0.37) \\ 
& $T_{mb}$ (K) & 1.19 & 0.26 & 0.14 & 1.16 & 0.29 \\
\hline
& W (K km~s$^{-1}$) & 1.50(0.23)\tnote{1} & 1.18(0.08) & \nodata & 1.71(0.13)\tnote{1} & 0.64(0.08) \\
G014.114-00.574 & $V_{LSR}$ (km~s$^{-1}$) & 18.27(0.29)\tnote{1} & 19.96(0.12) & \nodata & 19.38(0.08)\tnote{1} & 20.90(0.21) \\ 
 & $\Delta V$ (km~s$^{-1}$) & 5.12(0.85)\tnote{1} & 3.25(0.24) & \nodata & 2.68(0.25)\tnote{1} & 3.45(0.51) \\ 
& $T_{mb}$ (K) & 0.28\tnote{1} & 0.34 & \nodata & 0.60\tnote{1} & 0.17 \\
\hline
& W (K km~s$^{-1}$) & 8.49 (0.16) & 1.13(0.08) & 2.65(0.19) & 6.71(0.01) & 0.82(0.09) \\
G014.194-00.194 & $V_{LSR}$ (km~s$^{-1}$) & 40.23(0.05) & 39.52(0.10) & 40.26(0.31) & 39.85(0.03) & 39.47(0.14) \\ 
& $\Delta V$ (km~s$^{-1}$) & 5.77(0.14) & 2.86(0.26) & 10.24(1.07) & 4.35(0.06) & 2.84(0.37)  \\ 
& $T_{mb}$ (K) & 1.38 & 0.37 & 0.24 & 1.44 & 0.27  \\
\hline
& W (K km~s$^{-1}$) & 2.78 (0.12) & \nodata & \nodata & 2.30(0.08) & 0.29(0.05) \\ 
G014.492-00.139 & $V_{LSR}$ (km~s$^{-1}$) & 41.96(0.07) & \nodata & \nodata & 40.84(0.05) & 41.02(0.11) \\ 
& $\Delta V$ (km~s$^{-1}$) & 3.64(0.23) & \nodata & \nodata & 3.00 (0.13) & 1.17(0.24)  \\ 
& $T_{mb}$ (K) & 0.72 & \nodata & \nodata & 0.72 & 0.23  \\
\hline
& W (K km~s$^{-1}$) & 2.34(0.07)\tnote{1}  & 1.04(0.07)  & 1.49(0.15) & 2.15(0.08)\tnote{1} & 0.91(0.06) \\ 
G014.632-00.577 & $V_{LSR}$ (km~s$^{-1}$) & 20.32(0.03)\tnote{1}  & 18.79(0.08)  & 17.76(0.30) & 19.90(0.03)\tnote{1}  & 18.81(0.07) \\ 
& $\Delta V$ (km~s$^{-1}$) & 1.94(0.07)\tnote{1}  & 2.34(0.21)  & 6.86(1.03) & 1.85(0.08)\tnote{1}  & 1.94(0.16)  \\ 
& $T_{mb}$ (K) & 1.13\tnote{1} & 0.42  & 0.20 & 1.09\tnote{1}  & 0.44  \\
\hline
& W (K km~s$^{-1}$) & 18.63(0.22) & 1.60(0.12) & \nodata & 12.87(0.14) & 1.56(0.12) \\ 
G018.876-00.489 & $V_{LSR}$ (km~s$^{-1}$) & 66.29(0.04) & 66.52(0.14) & \nodata & 66.67(0.03) & 66.30(0.14) \\ 
& $\Delta V$ (km~s$^{-1}$) & 6.22(0.09) & 3.84(0.35) & \nodata & 5.64(0.08) & 3.73(0.37)  \\ 
& $T_{mb}$ (K) & 2.81 & 0.39 & \nodata & 2.14 & 0.39  \\
\hline
& W (K km~s$^{-1}$) & 3.36(0.13)\tnote{2} & 0.81(0.12) & \nodata & 4.30(0.13) & \nodata \\ 
G309.382-00.134 & $V_{LSR}$ (km~s$^{-1}$) & -49.06(0.05)\tnote{2} & -50.38(0.22) & \nodata & -50.37(0.07) & \nodata \\ 
& $\Delta V$ (km~s$^{-1}$) & 2.82(0.13)\tnote{2} & 3.39(0.68) & \nodata & 5.34(0.19) & \nodata  \\ 
  & $T_{mb}$ (K) & 0.74\tnote{2} & 0.23 & \nodata & 0.76 & \nodata  \\
\hline
& W (K km~s$^{-1}$) & 4.85(0.25)\tnote{2} & 1.07(0.12) & \nodata & 3.18(0.52)\tnote{2} & \nodata \\ 
G317.867-00.151  & $V_{LSR}$ (km~s$^{-1}$) & -38.87(0.09)\tnote{2} & -40.07(0.21) & \nodata & -38.92(0.29)\tnote{2} & \nodata \\ 
& $\Delta V$ (km~s$^{-1}$) & 4.18(0.25)\tnote{2} & 3.93(0.51) & \nodata & 4.30(0.51)\tnote{2} & \nodata  \\ 
& $T_{mb}$ (K) & 1.09\tnote{2} & 0.26 & \nodata & 0.70\tnote{2} & \nodata  \\
\hline
& W (K km~s$^{-1}$) & 3.44(0.14)\tnote{2} & 0.65(0.10) & 0.70(0.13) & 4.02(0.12) & 0.59(0.10) \\
G318.779-00.137  & $V_{LSR}$ (km~s$^{-1}$) & -36.45(0.08)\tnote{2} & -38.59(0.25) & -38.06(0.41) & -37.87(0.09) & -37.94(0.33) \\ 
& $\Delta V$ (km~s$^{-1}$) & 4.03(0.22)\tnote{2} & 3.36(0.62) & 4.54(0.99) & 5.86(0.21) & 3.89(0.74)  \\ 
& $T_{mb}$ (K) & 0.80\tnote{2} & 0.18 & 0.14 & 0.65 & 0.14  \\
\hline
& W (K km~s$^{-1}$) & 6.27(0.14) & 0.93(0.10) & \nodata & 3.53(0.12) & \nodata \\
G320.881-00.397  & $V_{LSR}$ (km~s$^{-1}$) & -44.51(0.06) & -44.48(0.13) & \nodata & -44.83(0.08) & \nodata \\ 
& $\Delta V$ (km~s$^{-1}$) & 5.21(0.13) & 2.58(0.31) & \nodata & 4.45(0.17) & \nodata  \\ 
& $T_{mb}$ (K) & 1.13 & 0.34 & \nodata & 0.75 & \nodata  \\
\hline
& W (K km~s$^{-1}$) & 4.32(0.13)\tnote{1} & 0.56(0.10) & \nodata & 3.87(0.12)\tnote{1} & 3.07(0.72) \\
G326.987-00.031 & $V_{LSR}$ (km~s$^{-1}$) & -59.91(0.05)\tnote{1} & -58.39(0.27) & \nodata & -59.39(0.06)\tnote{1} & -59.72(0.19) \\ 
& $\Delta V$ (km~s$^{-1}$) & 3.82(0.15)\tnote{1} & 2.88(0.70) & \nodata & 4.16(0.17)\tnote{1} & 3.38(0.28)  \\ 
& $T_{mb}$ (K) & 1.06\tnote{1} & 0.18 & \nodata & 0.87\tnote{1} & 0.85  \\
\hline
& W (K km~s$^{-1}$) & 3.16(0.17)\tnote{3} & 1.78(0.11) & 4.49(0.18) & 0.71(0.08)\tnote{2} & 1.36(0.10) \\
G329.029-00.206 & $V_{LSR}$ (km~s$^{-1}$) & -39.78(0.19)\tnote{3} & -43.30(0.16) & -44.45(0.19) & -46.31(0.05)\tnote{2} & -43.21(0.11) \\ 
& $\Delta V$ (km~s$^{-1}$) & 7.49(0.48)\tnote{3} & 4.92(0.36) & 10.18(0.53) & 1.40(0.13)\tnote{2} & 3.33(0.29)  \\ 
& $T_{mb}$ (K) & 0.40\tnote{3} & 0.34 & 0.41 & 0.45\tnote{2} & 0.38  \\
\hline
\end{tabular}
\begin{tablenotes}
\item[a] Integrated intensity of the line.
\item[1] Spectral lines show two line components. The properties of the first component with the velocity close to the V$_{LSR}$ are given here. \item[2] Spectral lines show two line components. The properties of the second component with the velocity close to the V$_{LSR}$ are given here. \item[3] Spectral line could not be fitted with single or double Gaussian distribution due to self absorption. The properties of the third component are given here.
\end{tablenotes}
\end{threeparttable}
\end{table*}

\addtocounter{table}{-1}

\begin{table*}
\caption{Continued}
\centering
\tiny
\begin{threeparttable}
\begin{tabular}{lcccccc}
\hline\hline
\noalign{\smallskip}
Source & Line properties & \multicolumn{5}{c}{Lines}  \\ 
       & & HCO$^{+}$(1-0) & H$^{13}$CO$^{+}$(1-0) & SiO(2-1) & HNC(1-0) & HN$^{13}$C(1-0)\\
\hline
& W (K km~s$^{-1}$) & 4.67(0.19)\tnote{1} & 1.01(0.10) & 2.64(0.18) & 3.27(0.26)\tnote{1} & 0.73(0.09) \\
G331.708+00.583$-{1}$ & $V_{LSR}$ (km~s$^{-1}$) & -69.33(0.06)\tnote{1} & -67.20(0.21) & -67.14(0.45) & -68.75(0.10)\tnote{1} & -67.05(0.20) \\ 
& $\Delta V$ (km~s$^{-1}$) & 3.96(0.21)\tnote{1} & 3.98(0.45) & 14.01(1.24) & 3.03(0.22)\tnote{1} & 3.44(0.51)  \\ 
& $T_{mb}$ (K) & 1.11\tnote{1} & 0.24 & 0.18 & 1.02\tnote{1} & 0.20  \\
\hline
& W (K km~s$^{-1}$) & 3.60(0.10)\tnote{1} & 1.30(0.09) & 1.51(0.11) & 3.65(0.09)\tnote{1} & 0.84(0.08) \\
G331.708+00.583$-{2}$ & $V_{LSR}$ (km~s$^{-1}$) & -69.57(0.03)\tnote{1} & -67.14(0.15) & -67.11(0.17) & -68.95(0.03)\tnote{1} & -67.17(0.14) \\ 
& $\Delta V$ (km~s$^{-1}$) & 2.80(0.10)\tnote{1} & 4.08(0.33) & 4.99(0.44) & 2.72(0.08)\tnote{1} & 2.63(0.29)  \\ 
& $T_{mb}$ (K) & 1.21\tnote{1} & 0.30 & 0.28 & 1.26\tnote{1} & 0.30  \\
\hline
& W (K km~s$^{-1}$) & 1.85(0.14) & 0.73(0.11) & \nodata & 2.26(0.13) & \nodata \\
G333.656+00.059 & $V_{LSR}$ (km~s$^{-1}$) & -84.20(0.16) & -84.71(0.23) & \nodata & -84.59(0.12) & \nodata \\ 
& $\Delta V$ (km~s$^{-1}$) & 4.38(0.39) & 2.98(0.57) & \nodata & 4.29(0.30) & \nodata  \\ 
& $T_{mb}$ (K) & 0.40 & 0.23 & \nodata & 0.50 & \nodata \\
\hline
& W (K km~s$^{-1}$) & 4.82(0.41)\tnote{2} & 0.51(0.14)\tnote{2} & \nodata & 6.21(0.22)\tnote{2} & 1.32(0.10)\tnote{2} \\
G335.789+00.174 & $V_{LSR}$ (km~s$^{-1}$) & -49.30(0.02)\tnote{2} & -49.45(0.13)\tnote{2} & \nodata & -49.71(0.07)\tnote{2} & -49.73(0.10)\tnote{2} \\ 
& $\Delta V$ (km~s$^{-1}$) & 3.95(0.16)\tnote{2} & 1.51(0.31)\tnote{2} & \nodata & 4.48(0.12)\tnote{2} & 2.63(0.20)\tnote{2}  \\ 
& $T_{mb}$ (K) & 1.15\tnote{2} & 0.32\tnote{2} & \nodata & 1.12\tnote{2} & 0.20\tnote{2} \\
\hline
& W (K km~s$^{-1}$) & 3.75(0.14) & 0.28(0.07) & \nodata & 2.87(0.10) & 0.40(0.07) \\
G336.958-00.224 & $V_{LSR}$ (km~s$^{-1}$) & -71.54(0.09) & -71.18(0.20) & \nodata & -71.83(0.06) & -71.34(0.20) \\ 
& $\Delta V$ (km~s$^{-1}$) & 5.69(0.28) & 1.80(0.53) & \nodata & 4.01(0.17) & 2.18(0.45)  \\ 
& $T_{mb}$ (K) & 0.62 & 0.15 & \nodata & 0.67 & 0.17 \\
\hline
& W (K km~s$^{-1}$) & 4.04(0.17)\tnote{3} & 0.96(0.10) & \nodata & 4.19(0.14)\tnote{2} & \nodata \\
G337.176-00.032 & $V_{LSR}$ (km~s$^{-1}$) & -69.59(0.09) & -68.15(0.32) & \nodata & -68.77(0.09)\tnote{2} & \nodata \\ 
& $\Delta V$ (km~s$^{-1}$) & 5.82(0.34) & 6.12(0.79)\tnote{2} & \nodata & 6.11(0.30) & \nodata  \\ 
& $T_{mb}$ (K) & 0.65 & 0.15 & \nodata & 0.65\tnote{2} & \nodata \\
\hline
& W (K km~s$^{-1}$) & 4.43(0.15) & 0.43(0.10) & \nodata & 4.20(0.14) & 0.33(0.09) \\
G337.258-00.101 & $V_{LSR}$ (km~s$^{-1}$) & -68.47(0.14) & -67.93(0.29) & \nodata & -67.92(0.12) & -68.30(0.29) \\ 
& $\Delta V$ (km~s$^{-1}$) & 8.51(0.34) & 2.74(0.79) & \nodata & 7.25(0.31) & 2.30(0.71)  \\ 
& $T_{mb}$ (K) & 0.49 & 0.15 & \nodata & 0.54 & 0.14 \\
\hline
& W (K km~s$^{-1}$) & 1.40(0.28)\tnote{2} & \nodata & \nodata & 0.59(0.15)\tnote{2} & 0.60(0.11) \\
G338.786+00.476 & $V_{LSR}$ (km~s$^{-1}$) & -66.33(0.06)\tnote{2} & \nodata & \nodata & -66.26(0.11)\tnote{2} & -64.11(0.40) \\ 
& $\Delta V$ (km~s$^{-1}$) & 2.21(0.27)\tnote{2} & \nodata & \nodata & 1.85(0.39)\tnote{2} & 4.44(0.86)  \\ 
& $T_{mb}$ (K) & 0.59\tnote{2} & \nodata & \nodata & 0.30\tnote{2} & 0.13 \\
\hline
& W (K km~s$^{-1}$) & 3.12(0.43)\tnote{1} & 0.47(0.08) & \nodata & 3.94(0.12) & 0.55(0.19) \\
G340.784-00.097 & $V_{LSR}$ (km~s$^{-1}$) & -102.46(0.15)\tnote{1} & -100.87(0.10) & \nodata & -101.55(0.06) & -101.16(0.25) \\ 
& $\Delta V$ (km~s$^{-1}$) & 2.69(0.25)\tnote{1} & 1.9(0.32) & \nodata & 4.41(0.17) & 2.73(1.72)  \\ 
& $T_{mb}$ (K) & 1.09\tnote{1} & 0.32 & \nodata & 0.84 & 0.19 \\
\hline
& W (K km~s$^{-1}$) & 2.29(0.14)\tnote{2} & 0.66(0.07) & \nodata & 1.49(0.01)\tnote{2} & \nodata \\
G342.484+00.182 & $V_{LSR}$ (km~s$^{-1}$) & -40.92(0.07)\tnote{2} & -41.53(0.08) & \nodata & -41.04(0.01)\tnote{2} & \nodata \\ 
& $\Delta V$ (km~s$^{-1}$) & 2.53(0.23)\tnote{2} & 1.64(0.19) & \nodata & 2.09(0.08)\tnote{2} & \nodata  \\ 
& $T_{mb}$ (K) & 0.52\tnote{2} & 0.38 & \nodata & 0.67\tnote{2} & \nodata \\
\hline
& W (K km~s$^{-1}$) & 2.17(0.09)\tnote{1} & 1.27(0.08) & 1.43(0.10) & 2.66(0.08)\tnote{1} & 0.98(0.09) \\
G343.756-00.164 & $V_{LSR}$ (km~s$^{-1}$) & -29.87(0.07)\tnote{1} & -27.69(0.08) & -27.41(0.18) & -29.12(0.03)\tnote{1} & -27.60(0.12) \\ 
& $\Delta V$ (km~s$^{-1}$) & 3.32(0.18)\tnote{1} & 2.66(0.19) & 5.05(0.42) & 2.43(0.09)\tnote{1} & 2.92(0.33)  \\ 
& $T_{mb}$ (K) & 0.61\tnote{1} & 0.45 & 0.27 & 1.03\tnote{1} & 0.31 \\
\hline
& W (K km~s$^{-1}$) & 28.79(0.20)\tnote{1} & 6.06(0.10) & 8.62(0.12) & 26.10(0.12) & 4.73(0.11) \\
G351.444+00.659 & $V_{LSR}$ (km~s$^{-1}$) & -5.69(0.02)\tnote{1} & -4.35(0.03) & -3.98(0.06) & -4.47(0.01) & -4.31(0.04) \\ 
& $\Delta V$ (km~s$^{-1}$) & 5.46(0.05)\tnote{1} & 3.89(0.07) & 7.42(0.16) & 6.03(0.03) & 3.89(0.10)  \\ 
& $T_{mb}$ (K) & 4.95\tnote{1} & 1.46 & 1.09 & 4.06 & 1.14 \\
\hline
& W (K km~s$^{-1}$) & 1.77(0.06)\tnote{2} & 0.99(0.10) & \nodata & 1.97(0.08) & 0.38(0.06) \\
G351.571+00.762 & $V_{LSR}$ (km~s$^{-1}$) & -2.58(0.06)\tnote{2} & -3.12(0.10) & \nodata & -3.06(0.06) & -2.79(0.10) \\ 
& $\Delta V$ (km~s$^{-1}$) & 2.10(<0.01)\tnote{2} & 2.25(0.32) & \nodata & 2.83(0.13) & 1.30(0.23)  \\ 
& $T_{mb}$ (K) & 0.41\tnote{2} & 0.41 & \nodata & 0.65 & 0.28 \\
\hline
& W (K km~s$^{-1}$) & 2.87(0.09)\tnote{2} & 0.82(0.10) & \nodata & 2.55(0.08)\tnote{2} & 0.53(0.11) \\
G353.066+00.452$-{1}$ & $V_{LSR}$ (km~s$^{-1}$) & 2.25(0.04)\tnote{2} & 1.62(0.14) & \nodata & 1.68(0.05)\tnote{2} & 1.56(0.26) \\ 
& $\Delta V$ (km~s$^{-1}$) & 2.50(<0.01)\tnote{2} & 2.24(0.31) & \nodata & 2.70(<0.01)\tnote{2} & 2.63(0.63)  \\ 
& $T_{mb}$ (K) & 1.08\tnote{2} & 0.34 & \nodata & 0.89\tnote{2} & 0.19  \\
\hline
& W (K km~s$^{-1}$) & 4.73(0.11)\tnote{1} & 1.00(0.09) & \nodata & 4.41(0.09)\tnote{1} & 0.36(0.09) \\
G353.066+00.452$-{2}$ & $V_{LSR}$ (km~s$^{-1}$) & -2.89(0.03)\tnote{1} & -1.80(0.08) & \nodata & -2.25(0.04)\tnote{1} & -1.68(0.22) \\ 
& $\Delta V$ (km~s$^{-1}$) & 2.36(0.06)\tnote{1} & 1.74(0.18) & \nodata & 3.04(0.02)\tnote{1} & 1.58(0.42) \\ 
& $T_{mb}$ (K) & 1.89\tnote{1} & 0.54 & \nodata & 1.36\tnote{1} & 0.21  \\
\hline
& W (K km~s$^{-1}$) & 5.74(0.12) & 0.68(0.08) & \nodata & 2.88(0.10) & \nodata\\
G353.066+00.452$-{3}$ & $V_{LSR}$ (km~s$^{-1}$) & -3.11(0.03) & -2.75(0.09) & \nodata & -2.86(0.04) & \nodata \\
& $\Delta V$ (km~s$^{-1}$) & 2.94(0.08) & 1.48(0.23) & \nodata & 2.32(0.09) & \nodata \\
& $T_{mb}$ (K) & 1.84 & 0.43 & \nodata & 1.16 & \nodata\\
\hline
& W (K km~s$^{-1}$) & 11.49(0.32) & 1.82(0.18)  & \nodata & 7.34(0.25) & 0.40(0.10)  \\
G353.417-00.079\_1 & $V_{LSR}$ (km~s$^{-1}$) & -50.52(0.13) & -49.76(0.23)  & \nodata & -49.84(0.12) & -50.24(0.34)  \\ 
& $\Delta V$ (km~s$^{-1}$) & 10.17(0.42) & 5.29(0.68) & \nodata & 7.58(0.35) & 2.38(0.55) \\ 
& $T_{mb}$ (K) & 1.06 & 0.32 & \nodata & 0.91 &  0.16 \\
\hline 
\hline
\end{tabular}
\end{threeparttable}
\end{table*}

\addtocounter{table}{-1}

\begin{table*}
\caption{Continued}
\centering
\tiny
\begin{threeparttable}
\begin{tabular}{lcccccc}
\hline\hline
\noalign{\smallskip}
Source & Line Properties & \multicolumn{5}{c}{Lines}  \\ 
       & & HCO$^{+}$(1-0) & H$^{13}$CO$^{+}$(1-0) & SiO(2-1) & HNC(1-0) & HN$^{13}$C(1-0)\\
            \hline
& W (K km~s$^{-1}$) & 0.89(0.17)\tnote{2} & 0.46(0.14)  & \nodata & \nodata & \nodata  \\
G353.417-00.079\_2 & $V_{LSR}$ (km~s$^{-1}$) & -58.30(0.14)\tnote{2} & -57.23(0.90)  & \nodata & \nodata & \nodata  \\ 
& $\Delta V$ (km~s$^{-1}$) & 2.27(0.39)\tnote{2} & 4.76(1.75)  & \nodata & \nodata & \nodata   \\ 
& $T_{mb}$ (K) & 0.37\tnote{2} & 0.09  & \nodata & \nodata & \nodata   \\
\hline
& W (K km~s$^{-1}$) & 1.67(0.13)\tnote{1} & 0.68(0.07) & \nodata & 1.48(0.09)\tnote{1} & 0.32(0.07) \\
G354.944-00.537 & $V_{LSR}$ (km~s$^{-1}$) & -6.85(0.07)\tnote{1} & -5.90(0.11) & \nodata & -6.54(0.05)\tnote{1} & -6.14(0.23) \\ 
& $\Delta V$ (km~s$^{-1}$) & 2.23(0.18)\tnote{1} & 1.98(0.22) & \nodata & 1.74(0.12)\tnote{1} & 1.78(0.44)  \\ 
& $T_{mb}$ (K) & 0.70\tnote{1} & 0.32 & \nodata & 0.80\tnote{1} & 0.17 \\
\hline
\end{tabular}
\end{threeparttable}
\end{table*}

\begin{sidewaystable*}
\caption{\label{tab_linefit_n2hp} Fitting results of the \nthp and \hcn.} 
\centering
\tiny
\begin{threeparttable}
\begin{tabular}{llllll|lllll}
\hline\hline
Source & \multicolumn{5}{c}{\nthp(1-0)} & \multicolumn{5}{c}{\hcn(1-0)} \\
      & W\tnote{a} & A $\times$ $\tau$$_{m}$\tnote{b} & $V_{LSR}$ & $\Delta v$ & $\tau$\tnote{b} & W\tnote{a} & A $\times$ $\tau$$_{m}$\tnote{b} & $v_{LSR}$ & $\Delta v$\tnote{c} & $\tau$\tnote{b,c}\\
      & (K km~s$^{-1}$) & & (km~s$^{-1}$) & (km~s$^{-1}$) & & (K km~s$^{-1}$) & & (km~s$^{-1}$) & (km~s$^{-1}$) & \\[0.5ex]
\hline
G008.684-00.367  & 16.47(0.55) & 4.62(0.20)     & 36.80(0.04)   &   4.81(0.08) & 2.19(0.22)  & 7.51(0.75) & 2.08(0.07) & 34.50(0.14)  & 4.07 & 2.47  \\[0.5ex]
G008.706-00.414 & 6.81(0.75) & 2.77(0.22)       & 39.00(0.03)   &   2.61(0.09) & 0.71(0.38) & 6.15(0.83) & 2.78(0.36)  & 38.20(0.12)   &   4.23(0.23) & 5.54(1.02) \\[0.5ex]
G010.444-00.017 &  3.01(1.47) & 1.34(0.27)      & 74.60(0.13)   &   3.27(0.31) & 4.20(1.43) & 2.13(1.28) & \nodata     & \nodata     & \nodata     & \nodata     \\[0.5ex]
G013.178+00.059   & 12.44(3.81) & 4.03(0.20)    & 49.30(0.03)   &   3.44(0.06) & 1.00(<0.01) & 6.35(1.20) & 2.78(0.63)  & 38.20(0.12)     & 4.23(0.23)     & 5.54(1.02)     \\[0.5ex]
G014.114-00.574   & 8.75(0.71) & 0.17(<0.01)    & 19.90(0.02)   &   2.65(0.04) & 0.10(<0.01) & 2.63(0.91) & \nodata     & \nodata     & \nodata     & \nodata     \\[0.5ex]
G014.194-00.194   & 13.53(1.80) & 0.24(<0.01)   & 39.70(0.02)   &   2.88(0.03) & 0.10(<0.01) & 11.05(2.01) & \nodata     & \nodata     & \nodata     & \nodata     \\[0.5ex]
G014.492-00.139   & 5.32(1.68) & 1.65(0.19)     & 40.20(0.06)   &   3.47(0.14) & 1.20(0.57) & 2.94(1.20) & 0.84(0.08)     & 41.10(0.20)     & 4.40(0.61)     & 2.40     \\[0.5ex]
G014.632-00.577   & 11.89(0.99) & 5.74(0.24)    & 18.60(0.01)   &   2.38(0.04) & 1.45(0.23) & 7.20(1.28) & 3.31(0.59)     & 20.10(0.13)    & 2.60     & 3.69(1.38)     \\[0.5ex]
G018.876-00.489   & 19.80(2.65) & 5.55(0.24)    & 66.10(0.03)   &   3.96(0.07) & 0.71(0.20) & 16.39(3.53) & 8.00(0.05)  & 66.60(0.08)  & 4.00 & 5.75(0.12) \\
G309.382-00.134   & 5.86(1.15) & 2.05(0.05)     & -50.60(0.04)  &       3.07(0.08) & 0.10(0.30) & 6.17(3.01) & \nodata     & \nodata     & \nodata     & \nodata     \\[0.5ex]
G317.867-00.151   & 7.46(1.09) & 1.89(0.04)     & -40.50(0.05)  &       4.08(0.09) & 0.10(0.21)  & 6.29(1.09) & \nodata     & \nodata     & \nodata     & \nodata     \\[0.5ex]
G318.779-00.137   & 5.97(1.17) & 1.60(0.06)     & -38.60(0.07)  &       3.75(0.11) & 0.13(0.10) & 3.55(0.74) & 0.68(0.03)  & -37.50(0.33) & 6.67(0.51)  & 2.38(0.09) \\[0.5ex]
G320.881-00.397   & 6.80(0.71) & 2.90(0.21)     & -44.60(0.03)  &       2.68(0.08) & 1.01(0.37) & 5.72(0.71) & 1.38(0.15)  & -45.50(0.10) & 4.51(0.26)  & 0.57(0.49) \\[0.5ex]
G326.987-00.031   & 6.54(0.71) & 2.03(0.17)     & -58.80(0.04)  &       3.45(0.11) & 0.52(0.39) & 3.39(0.65) & 0.77(0.05)  & -60.00(0.14)  & 3.50(0.25) & 0.10(0.21) \\[0.5ex]
G329.029-00.206   & 15.16(0.89) & 7.45(0.31)    & -43.40(0.03)  &       3.63(0.06) & 4.90(0.31) & 6.41(1.00) & \nodata     & \nodata     & \nodata     & \nodata     \\[0.5ex]
G331.708+00.583\_1  & 12.56(0.91) & 2.80(0.17)  & -67.10(0.04)  & 3.82(0.09) & 0.98(0.30)  & 6.44(1.01) & \nodata     & \nodata     & \nodata     & \nodata     \\[0.5ex]
G331.708+00.583\_2  & 8.96(0.81) & 5.15(0.24)  & -67.5(0.03)  & 3.50(0.06) & 2.48(0.27)  & 6.49(1.01) & \nodata     & \nodata     & \nodata     & \nodata     \\[0.5ex]
G333.656+00.059   & 3.45(0.59) & 1.55(0.28)     & -84.90(0.08)  &       2.89(0.21) & 1.75(1.06) & 1.42(0.65) & \nodata     & \nodata     & \nodata     & \nodata     \\[0.5ex]
G335.789+00.174   & 15.20(0.85) & 4.27(0.04)    & -50.00(<0.01) &       3.46(<0.01) & 0.12(<0.01) & 12.52(0.86) & \nodata     & \nodata     & \nodata     & \nodata     \\[0.5ex]
G336.958-00.224   & 3.78(0.65) & 1.39(0.04)     & -71.60(<0.01) & 2.90(0.01)  & 0.16(<0.01) & 4.75(0.62) & 1.99(0.36)  & -72.30(0.24) & 4.41(0.54)  & 6.01(1.33) \\[0.5ex]
G337.176-00.032   & 5.68(0.58) & 1.45(0.04)     & -68.50(0.06) & 4.11(0.11)  & 0.10(0.14) & 3.36(0.53) & \nodata     & \nodata      & \nodata     & \nodata     \\[0.5ex]
G337.258-00.101   & 4.80(0.78) & 1.65(0.23)     & -68.20(0.09) & 3.79(0.21)  & 1.52(0.71)  & 3.04(0.65) & \nodata     & \nodata      & \nodata     & \nodata     \\[0.5ex]
G338.786+00.476   &  3.35(0.76) & 0.75(0.02)    & -64.00(<0.01) & 4.84(0.02)  & 0.11(<0.01) & 4.34(0.83) & \nodata     & \nodata      & \nodata     & \nodata     \\[0.5ex]
G340.784-00.097  & 4.67(0.90) & 1.87(0.05)  & -101.00(0.04) & 2.71(0.10)  & 0.10(0.14) & 7.21(1.10) & \nodata     & \nodata      & \nodata     & \nodata     \\[0.5ex]
G342.484+00.182   & 4.44(0.91) & 1.95(0.05)     & -41.70(0.03) & 2.38(0.07)  & 0.10(0.29) & 3.67(0.91) & 2.19(0.42)  & -41.90(0.09) &  2.89(0.20) & 5.44 (1.55) \\[0.5ex]
G343.756-00.164   & 14.13(1.20) & 5.82(0.20)    & -27.80(0.02) & 2.87(0.04)  & 1.02(0.17) & 3.67(1.44) & 1.33(0.06)     & -29.80(0.11)      & 2.87     & 1.20     \\[0.5ex]
G351.444+00.659   & 64.89(1.16) & 24.00(0.02)   & -4.46(0.01)  & 3.72(0.01)  & 2.91(0.01) & 50.35(1.22) & \nodata     & \nodata      & \nodata     & \nodata     \\[0.5ex]
G351.571+00.762   & 6.30(1.06) & 4.44(0.34)     & -3.04(0.02)  & 1.70(0.06)  & 1.89(0.45) & 3.21(0.96) & 3.06(0.63)  & -3.51(0.09)  & 2.52(0.18)  & 8.63(2.20) \\[0.5ex]
G353.066+00.452\_1 & 6.69(2.41) & 3.23(0.34) &  1.60(0.03)  & 2.37(0.11)  & 1.10(0.57) & 6.81(2.53) &  3.05(0.31) & -2.89(0.09)  & 3.85(0.30)  & 3.53(0.60) \\[0.5ex]
G353.066+00.452\_2 &  8.54(1.34) & 4.24(0.29) & -1.85(0.02)  & 2.00(0.07)  & 0.67(0.34) & 6.02(1.28) & 2.05(0.06) & -2.55(0.04)  & 2.61(0.08)  & 0.10(0.13) \\[0.5ex]
G353.066+00.452\_3 &  3.37(2.22) & 3.02(0.40) & -2.89(0.03)  & 1.21(0.09)  & 1.95(0.73) & 6.57(2.10) & 2.19(0.06) & -3.10(0.04)  & 2.68(0.08)  & 0.10(0.09) \\[0.5ex]
G353.417-00.079\_1 & 5.46(2.87) & 1.93(0.32) & -49.60(0.12) & 4.29(0.28)  & 2.17(0.88) & 4.14(2.53) & \nodata & \nodata & \nodata  & \nodata \\[0.5ex]
G353.417-00.079\_2 & 5.46(4.43) & 2.02(0.38) & -54.90(0.26) & 6.04(0.60) & 4.41(1.11)  & 2.69(2.05) & \nodata & \nodata & \nodata & \nodata \\[0.5ex]
G354.944-00.537    & 5.16(1.06) & 3.72(0.39)    & -5.92(0.03)  & 1.96(0.10)  & 2.98(0.67)  & 4.68(1.11) & \nodata     & \nodata      & \nodata     & \nodata     \\[0.5ex]
\hline
\end{tabular}
\begin{tablenotes}
\item[a]{Integrated intensity is measured from the averaged spectra in order to estimate the column densities. The intensity is integrated over all hyperfine line components of \nthp, and \hcn (20.5 km/s in average for all sources) and this range is multiplied by RMS value of the spectra to estimate the error.}
\item[b]{A = f [J$_{\nu}$(T$_{ex}$) - J$_{\nu}$(T$_{bg}$)], and $\tau$ is the sum of all the opacities of the hyperfine transitions. Filling factor, f, is assumed to be one.}
\item[c]{The errors are not reported for line widths and optical depths for some of the \hcn~lines which show line anomalies, therefore the fit was applied with fixed fit parameters.}
\end{tablenotes}
\end{threeparttable}
\end{sidewaystable*}

\clearpage

\section{Optical depths and column densities}
\label{section:optdepthandcoldens}

\begin{table*}
\centering
\caption{\label{tab_od} Optical depth measurements for HCO$^{+}$, H$^{13}$CO$^{+}$, HNC, and HN$^{13}$C.}
\tiny
\begin{threeparttable}
\begin{tabular}{lllll}
\hline\hline
\noalign{\smallskip}
Source & $\tau$$_{\rm{HCO^+}}$ & $\tau$$_{\rm{H^{13}}CO^+}$\tnote{*} & $\tau$$_{\rm{HNC}}$ & $\tau$$_{\rm{HN^{13}}C}$\tnote{*} \\
\hline
G008.684-00.367         & 18.64(0.02)   &       0.36    & 103.15(0.20)  & 2.00 \\
G008.706-00.414         & 14.55(0.07)   &       0.28    & 15.81(0.11)   & 0.30\\
G010.444-00.017         & \nodata               &       \nodata     & \nodata           & \nodata  \\
G013.178+00.059         & 12.69(0.05)   &       0.25    & 15.00(0.08)   & 0.29\\
G014.114-00.574         & \nodata               &       \nodata         & 17.04(0.01)     & 0.33 \\
G014.194-00.194         & 15.94(<0.01)  &       0.31    & 10.62(<0.01)  & 0.21\\
G014.492-00.139         & \nodata               &       \nodata         & 19.46(0.11)     & 0.38\\
G014.632-00.577         & 23.91(0.06)   &       0.46    & 26.94(0.09)   & 0.52\\
G018.876-00.489         & 7.758(0.03)   &       0.15    & 10.63(0.05)   & 0.20\\
G309.382-00.134         & 11.36(0.05)   &       0.22    & \nodata               & \nodata\\
G317.867-00.151         & 14.15(0.06)   &       0.27    & \nodata               & \nodata\\
G318.779-00.137         & 13.24(0.06)   &       0.25    & 12.81(0.09)   & 0.24\\
G320.881-00.397         & 18.33(0.10)   &       0.36    & \nodata               & \nodata\\
G326.987-00.031         & 9.67(0.04)    &       0.19    & 11.79(0.08)   & 0.22\\
G329.029-00.206         & 30.09(0.03)   &       0.59    & 80.05(0.12)   & 1.57\\
G331.708+00.583\_1  & 12.60(0.06)       &       0.25    & 11.29(0.08)   & 0.22\\
G331.708+00.583\_2  & 14.52(0.03)       &       0.28    & 13.97(0.04)   & 0.27\\
G333.656+00.059         & 43.10(0.32)   &       0.86    & \nodata               & \nodata\\
G335.789+00.174         & 17.14(<0.01)  &       0.33    & 24.13(<0.01)  & 0.45\\
G336.958-00.224         & 14.41(0.01)   &       0.28    & 15.50(0.01)   & 0.29\\
G337.176-00.032         & 13.86(0.09)   &       0.26    & \nodata               & \nodata\\
G337.258-00.101         & 18.46(0.12)   &       0.36    & 15.21(0.14)   & 0.30\\
G338.786+00.476         & \nodata               &       \nodata         & 17.97(0.01) & 0.34\\
G340.784-00.097         & 18.05(0.08)   &       0.35    & 13.55(0.09)   & 0.26\\
G342.484+00.182         & 30.78(0.14)   &       0.59    & \nodata               & \nodata\\
G343.756-00.164         & 70.28(0.24)   &       1.34    & 18.52(0.04)   & 0.36\\
G351.444+00.659         & 18.16(<0.01)  &       0.35    & 17.43(<0.01)  & 0.33\\
G351.571+00.762         & 37.36(0.14)   &       0.73    & 29.03(0.16)   & 0.56\\
G353.066+00.452\_1  & 19.37(0.16)       &       0.38    & 12.43(0.15)   & 0.25\\
G353.066+00.452\_2  & 17.48(0.08)       &       0.34    & 8.86(0.06)    & 0.17\\
G353.066+00.452\_3  & 10.64(0.06)       &       0.21    & \nodata               & \nodata\\
G353.417-00.079\_1  & 18.08(0.07)       &       0.36    & 31.22(0.17)   & 0.62\\
G353.417-00.079\_2      & 13.95(0.04)   &       0.28    & \nodata       & \nodata\\
G354.944-00.537         & 30.91(0.10)   &       0.61    & 12.14(0.06)   & 0.24\\
\hline
Mean & 20.32(0.02) & 0.39(<0.01) & 23.19(0.02) & 0.45(<0.01) \\
Median & 18.08(<0.01) & 0.35(<0.01) & 15.50(0.01) & 0.30(<0.01) \\
Std & 12.29 & 0.24 & 14.02 & 0.27 \\
\hline
\hline
\end{tabular}
\begin{tablenotes}
\item[*]{The errors for optically thin lines are <0.01. For the error calculations the \textit{Uncertainties: a Python package for calculations with uncertainties} by Eric O. Lebigot was used (http://pythonhosted.org/uncertainties/). The errors on optical depths, T$_{ex}$, and integrated intensities are included in the error calculation of column densities.}
\end{tablenotes}
\end{threeparttable}
\end{table*}

\begin{table*}
\caption{\label{tab_columndens} Column density of HCO$^{+}$, H$^{13}$CO$^{+}$, SiO, HNC, HN$^{13}$C, N$_{2}$H$^{+}$, and HCN, and T$_{ex}$ calculated from N$_{2}$H$^{+}$ line.}
\centering
\tiny
\begin{threeparttable}
\begin{tabular}{lllllllll}
\hline\hline
\noalign{\smallskip}
Source & N(HCO$^{+}$) & N(H$^{13}$CO$^{+}$) & N(SiO) & N(HNC) & N(HN$^{13}$C) & N(N$_{2}$H$^{+}$) &  N(HCN) & T$_{ex}$(\nthp)\tnote{a} \\
.. & ($\times$10$^{14}$~cm$^{-2}$) & ($\times$10$^{12}$~cm$^{-2}$) & ($\times$10$^{12}$~cm$^{-2}$) & ($\times$10$^{14}$~cm$^{-2}$) & ($\times$10$^{12}$~cm$^{-2}$) & ($\times$10$^{12}$~cm$^{-2}$) & ($\times$10$^{12}$~cm$^{-2}$) & (K) \\
\hline
G008.684-00.367  &  1.4(0.1) &  3.6(0.2) &  1.4(0.2) &  3.2(0.5) &  11.7(0.8) &  55.8(6.1) &  30.0(3.0) &  5.1(0.2) \\
G008.706-00.414  &  1.2(0.1) &  1.3(0.2) &  0.6(0.1) &  1.3(0.2) &  1.4(0.2) &  9.3(1.4) &  59.7(16.1) &  7.0(2.2)\tnote{b} \\
G010.444-0<0.117  &  0.1(<0.1) &  5.2(\nodata) &  3.0(\nodata) &  0.1(<0.1) &  10.8(\nodata) &  56.3(44.6) &  8.0(5.3) &  3.1(0.1) \\
G013.178+0<0.159  &  0.7(<0.1) &  1.0(0.1) &  0.5(0.1) &  1.3(0.1) &  2.6(0.2) &  15.4(4.7) &  9.3(1.8) &  7.1(1.0) \\
G014.114-00.574  &  <0.1(<0.1) &  1.6(0.1) &  1.0(\nodata) &  0.4(<0.1) &  1.8(0.2) &  13.0(1.1) &  4.1(1.4) &  4.7(<0.1) \\
G014.194-00.194  &  1.5(<0.1) &  1.6(0.1) &  2.3(0.2) &  0.9(<0.1) &  2.0(0.2) &  18.0(2.4) &  16.1(2.9) &  5.4(<0.1) \\
G014.492-00.139  &  <0.1(<0.1) &  1.6(\nodata) &  1.0(\nodata) &  0.7(0.1) &  1.1(0.3) &  28.0(14.9) &  14.9(6.5) &  3.8(0.3) \\
G014.632-00.577  &  0.6(<0.1) &  1.4(0.1) &  1.0(0.1) &  0.8(<0.1) &  2.1(0.1) &  27.9(3.4) &  39.9(15.3) &  7.0(0.7) \\
G018.876-00.489  &  1.6(0.1) &  2.1(0.2) &  0.3(\nodata) &  2.2(0.3) &  3.9(0.4) &  26.9(4.3) &  165.5(40.4) &  11.0(2.3) \\
G309.382-00.134  &  0.4(<0.1) &  1.1(0.2) &  0.4(\nodata) &  1.8(0.3) &  1.6(\nodata) &  8.0(1.8) &  10.8(5.5) &  23.8(62.3)\tnote{b} \\
G317.867-00.151  &  0.8(0.1) &  1.5(0.2) &  0.6(\nodata) &  1.3(0.3) &  1.4(\nodata) &  10.1(1.9) &  11.0(2.5) &  22.1(40.4)\tnote{b} \\
G318.779-00.137  &  0.5(0.1) &  0.9(0.2) &  0.5(0.1) &  0.8(0.1) &  1.5(0.3) &  8.1(1.8) &  16.3(4.2) &  15.7(9.6)\tnote{b} \\
G320.881-00.397  &  1.2(0.1) &  1.3(0.2) &  0.7(\nodata) &  1.4(<0.1) &  1.7(\nodata) &  13.8(3.3) &  8.3(1.0) &  5.9(1.1) \\
G326.987-00.031  &  0.5(0.1) &  0.7(0.2) &  1.1(\nodata) &  0.6(0.2) &  1.0(0.2) &  8.9(1.4) &  5.9(1.4) &  7.0(3.0)\tnote{b} \\
G329.029-00.206  &  0.4(<0.1) &  3.4(0.2) &  4.9(0.3) &  0.8(0.1) &  7.9(0.6) &  116.2(12.5) &  10.2(1.6) &  4.5(0.1) \\
G331.708+00.583\_1  &  0.6(<0.1) &  1.3(0.1) &  2.1(0.3) &  0.5(<0.1) &  1.7(0.2) &  16.1(1.4) &  9.3(1.5) &  5.9(0.9) \\
G331.708+00.583\_2  &  0.6(<0.1) &  1.9(0.1) &  1.4(0.1) &  0.7(<0.1) &  2.2(0.2) &  33.5(4.9) &  9.6(1.5) &  5.1(0.3) \\
G333.656+00.059  &  1.4(0.6) &  2.1(0.9) &  1.2(\nodata) &  2.5(0.7) &  5.3(\nodata) &  15.1(11.8) &  19.6(14.6) &  3.8(0.6) \\
G335.789+00.174  &  2.1(0.2) &  1.6(0.5) &  1.9(\nodata) &  6.3(0.2) &  7.4(0.5) &  47.7(2.7) &  55.8(3.9) &  38.0(0.3) \\
G336.958-00.224  &  0.6(<0.1) &  0.4(0.1) &  0.9(\nodata) &  0.8(<0.1) &  1.0(0.2) &  5.3(0.9) &  8.8(1.2) &  12.1(0.3) \\
G337.176-00.132  &  0.6(0.1) &  1.3(0.2) &  0.3(\nodata) &  1.4(0.2) &  1.1(\nodata) &  7.7(1.2) &  5.9(1.3) &  17.7(20.8)\tnote{b} \\
G337.258-00.101  &  1.3(0.3) &  0.9(0.3) &  2.2(\nodata) &  1.0(0.2) &  1.2(0.4) &  17.1(9.5) &  5.4(1.6) &  4.0(0.6) \\
G338.786+00.476  &  <0.1(<0.1) &  0.2(\nodata) &  0.2(\nodata) &  0.2(<0.1) &  1.4(0.3) &  4.4(1.0) &  7.3(1.4) &  10.1(0.2) \\
G340.784-00.197  &  0.6(0.1) &  0.7(0.1) &  1.2(\nodata) &  0.9(0.1) &  1.4(0.5) &  6.3(1.4) &  12.6(2.7) &  21.9(25.9)\tnote{b} \\
G342.484+00.182  &  0.8(0.1) &  1.1(0.2) &  1.0(\nodata) &  0.9(0.1) &  1.7(\nodata) &  6.0(1.4) &  35.0(14.1) &  22.7(56.1)\tnote{b} \\
G343.756-00.164  &  1.6(0.1) &  2.6(0.2) &  1.0(0.1) &  0.7(<0.1) &  2.3(0.2) &  28.6(2.8) &  9.9(3.9) &  8.8(1.0) \\
G351.444+00.659  &  5.9(<0.1) &  8.8(0.1) &  6.4(0.1) &  7.5(<0.1) &  11.9(0.3) &  275.7(5.0) &  89.8(2.2) &  11.4(<0.1) \\
G351.571+00.762  &  0.7(0.1) &  1.7(0.2) &  1.1(\nodata) &  0.7(<0.1) &  0.9(0.2) &  18.7(5.2) &  40.5(15.9) &  5.4(0.6) \\
G353.066+00.452\_1  &  0.6(0.1) &  1.1(0.2) &  0.5(\nodata) &  0.4(<0.1) &  1.2(0.3) &  14.0(6.8) &  35.7(14.3) &  6.0(1.6) \\
G353.066+00.452\_2  &  0.9(0.1) &  1.4(0.2) &  0.5(\nodata) &  0.6(0.1) &  0.9(0.2) &  11.6(2.2) &  10.5(2.7) &  9.5(3.3)\tnote{b} \\
G353.066+00.452\_3  &  0.8(0.1) &  1.8(0.6) &  0.7(\nodata) &  0.8(0.1) &  2.2(\nodata) &  11.8(9.2) &  10.4(3.5) &  4.5(0.7) \\
G353.417-00.179\_1  &  3.8(0.8) &  4.4(1.0) &  1.6(\nodata) &  4.1(0.7) &  3.7(\nodata) &  35.0(24.1) &  8.5(5.4) &  3.7(0.3) \\
G353.417-00.179\_2  &  0.3(0.1) &  1.6(0.6) &  3.8(\nodata) &  <0.1(<0.1) &  8.0(\nodata) &  79.8(74.3) &  322.6(297.2) &  3.3(0.2) \\
G354.944-00.537  &  0.7(0.1) &  1.4(0.2) &  1.3(\nodata) &  0.3(<0.1) &  1.0(0.3) &  27.6(10.2) &  7.9(2.0) &  4.2(0.3) \\
\hline
Mean\tnote{d} & 1.1(<0.1) & 1.8(0.1) & 2.0(<0.1) & 1.4(<0.1) & 3.0(0.1) & 31.7(2.8) & 32.8(8.9) & 9.7(2.9) \\
Median & 0.7(0.1) & 1.4(0.1) & 1.4(0.2) & 0.8(0.1) & 1.6(0.2) & 15.7(2.5) & 10.6(3.1) & 6.5(1.7) \\
Std & 1.1 & 1.5 & 1.8 & 1.6 & 3.2 & 50.2 & 61.5 & 7.8 \\
\hline
\hline
\end{tabular}
\begin{tablenotes}
\item[a]{T$_{ex}$ is calculated for \nthp~lines as described in Section~\ref{sec:coldens}.}
\item[b]{Since the errors on T$_{ex}$ values calculated for these sources are larger than 30\%, in column density calculations 22~K or 11~K is used for D8, or D24 clumps, respectively.}
\item[c]{Column densities without errors represent the upper limits for non-detections. In these case, a $\Delta v$ = 2.0 km s$^{-1}$ was assumed and the brightness temperatures were set to three times the rms of the spectrum (T$_{mb}$ = 3T$_{rms}$). For these cases, no uncertainties are shown in the table.} 
\item[d]{The upper level column density values are excluded from the mean and median calculations.}
\item[e]{For the error calculations the \textit{Uncertainties: a Python package for calculations with uncertainties} by Eric O. Lebigot was used (http://pythonhosted.org/uncertainties/). The errors on optical depths, T$_{ex}$, and integrated intensities are included in the error calculation of column densities.}
\end{tablenotes}
\end{threeparttable}
\end{table*}

\begin{table*}
\caption{\label{ratio} Column density ratios.}
\centering
\tiny
\begin{tabular}{llllll}
\hline\hline
\noalign{\smallskip}
Source & N(HNC)/N(HCO$^{+}$) & N(N$_{2}$H$^{+}$)/N(HCO$^{+}$) & N(N$_{2}$H$^{+}$)/N(HNC) & N(HNC)/N(HCN) & N(HCN)/N(HCO$^{+}$) \\
\hline
G008.684-00.367  &  2.3(0.4) & 0.4(0.0)  & 0.2(0.0)  & 10.7(2.0)  & 0.2(0.0) \\
G008.706-00.414  &  1.1(0.2) & 0.1(0.0)  & 0.1(0.0)  & 2.2(0.7)  & 0.5(0.1)\\
G010.444-00.017  &  0.5(0.0) & 4.9(2.8)  & 9.8(5.9)  & 0.7(0.4)  & 0.7(0.4)\\
G013.178+00.059  &  1.7(0.1) & 0.2(0.1)  & 0.1(0.0)  & 13.6(2.6)  & 0.1(0.0)\\
G014.114-00.574  &  10.9(1.0) & 3.6(0.3)  & 0.3(0.0)  & 9.8(3.5)  & 1.1(0.4)\\
G014.194-00.194  &  0.6(0.0) & 0.1(0.0)  & 0.2(0.0)  & 5.8(1.0)  & 0.1(0.0)\\
G014.492-00.139  &  16.0(1.3) & 6.0(2.4)  & 0.4(0.2)  & 5.0(2.1)  & 3.2(1.3)\\
G014.632-00.577  &  1.4(0.1) & 0.5(0.1)  & 0.4(0.1)  & 1.9(0.7)  & 0.7(0.3)\\
G018.876-00.489  &  1.4(0.1) & 0.2(0.0)  & 0.1(0.0)  & 1.3(0.3)  & 1.0(0.2)\\
G309.382-00.134  &  4.3(0.8) & 0.2(0.0)  & 0.0(0.0)  & 17.1(9.1)  & 0.3(0.1)\\
G317.867-00.151  &  1.7(0.4) & 0.1(0.0)  & 0.1(0.0)  & 12.0(3.9)  & 0.1(0.0)\\
G318.779-00.137  &  1.6(0.3) & 0.2(0.0)  & 0.1(0.0)  & 5.1(1.6)  & 0.3(0.1)\\
G320.881-00.397  &  1.1(0.1) & 0.1(0.0)  & 0.1(0.0)  & 16.7(2.2)  & 0.1(0.0)\\
G326.987-00.031  &  1.3(0.4) & 0.2(0.0)  & 0.2(0.0)  & 9.9(3.7)  & 0.1(0.0)\\
G329.029-00.206  &  2.1(0.3) & 3.0(0.4)  & 1.4(0.2)  & 7.9(1.5)  & 0.3(0.0)\\
G331.708+00.583\_1  &  0.8(0.1) & 0.3(0.0)  & 0.3(0.0)  & 5.2(0.9)  & 0.1(0.0)\\
G331.708+00.583\_2  &  1.1(0.0) & 0.6(0.1)  & 0.5(0.1)  & 7.0(1.1)  & 0.2(0.0)\\
G333.656+00.059  &  1.8(0.3) & 0.1(0.0)  & 0.1(0.0)  & 12.7(8.6)  & 0.1(0.1)\\
G335.789+00.174  &  3.0(0.3) & 0.2(0.0)  & 0.1(0.0)  & 11.2(0.9)  & 0.3(0.0)\\
G336.958-00.224  &  1.2(0.1) & 0.1(0.0)  & 0.1(0.0)  & 8.7(1.2)  & 0.1(0.0)\\
G337.176-00.032  &  2.3(0.4) & 0.1(0.0)  & 0.1(0.0)  & 24.2(6.5)  & 0.1(0.0)\\
G337.258-00.101  &  0.8(0.1) & 0.1(0.0)  & 0.2(0.1)  & 18.4(4.0)  & 0.0(0.0)\\
G338.786+00.476  &  10.8(3.5) & 2.9(0.9)  & 0.3(0.1)  & 2.2(0.7)  & 4.8(1.3)\\
G340.784-00.097  &  1.4(0.3) & 0.1(0.0)  & 0.1(0.0)  & 6.8(1.8)  & 0.2(0.1)\\
G342.484+00.182  &  1.1(0.2) & 0.1(0.0)  & 0.1(0.0)  & 2.5(1.1)  & 0.4(0.2)\\
G343.756-00.164  &  0.4(0.0) & 0.2(0.0)  & 0.4(0.1)  & 7.2(2.8)  & 0.1(0.0)\\
G351.444+00.659  &  1.3(0.0) & 0.5(0.0)  & 0.4(0.0)  & 8.4(0.2)  & 0.2(0.0)\\
G351.571+00.762  &  1.0(0.1) & 0.3(0.1)  & 0.3(0.1)  & 1.8(0.7)  & 0.6(0.2)\\
G353.066+00.452\_1  &  0.7(0.1) & 0.2(0.1)  & 0.3(0.2)  & 1.2(0.5)  & 0.6(0.3)\\
G353.066+00.452\_2  &  0.7(0.1) & 0.1(0.0)  & 0.2(0.0)  & 6.0(1.8)  & 0.1(0.0)\\
G353.066+00.452\_3  &  1.0(0.1) & 0.1(0.1)  & 0.2(0.1)  & 7.5(2.4)  & 0.1(0.0)\\
G353.417-00.079\_1  &  1.1(0.1) & 0.1(0.1)  & 0.1(0.1)  & 48.1(29.4)  & 0.0(0.0)\\
G353.417-00.079\_2  &  0.1(0.0) & 2.3(2.0)  & 27.6(24.0)  & 0.0(0.0)  & 9.4(8.7)\\
G354.944-00.537  &  0.4(0.0) & 0.4(0.1)  & 1.0(0.3)  & 3.4(0.8)  & 0.1(0.0)\\
\hline
Mean & 2.3(0.1) & 0.8(0.1) & 1.3(0.7) & 8.9(1.0) & 0.8(0.3) \\
Median & 1.2(0.2) & 0.2(<0.1) & 0.2(<0.1) & 7.1(1.5) & 0.2(<0.1) \\
Std & 3.4 & 1.5 & 4.8 & 8.8 & 1.8 \\
\hline
\hline
\end{tabular}
\end{table*}

\clearpage

\section{Integrated intensity maps and line spectra}
\label{section:mapsandspectra}

\begin{figure*}
\centering
\tiny
\includegraphics[width=0.79\textwidth]{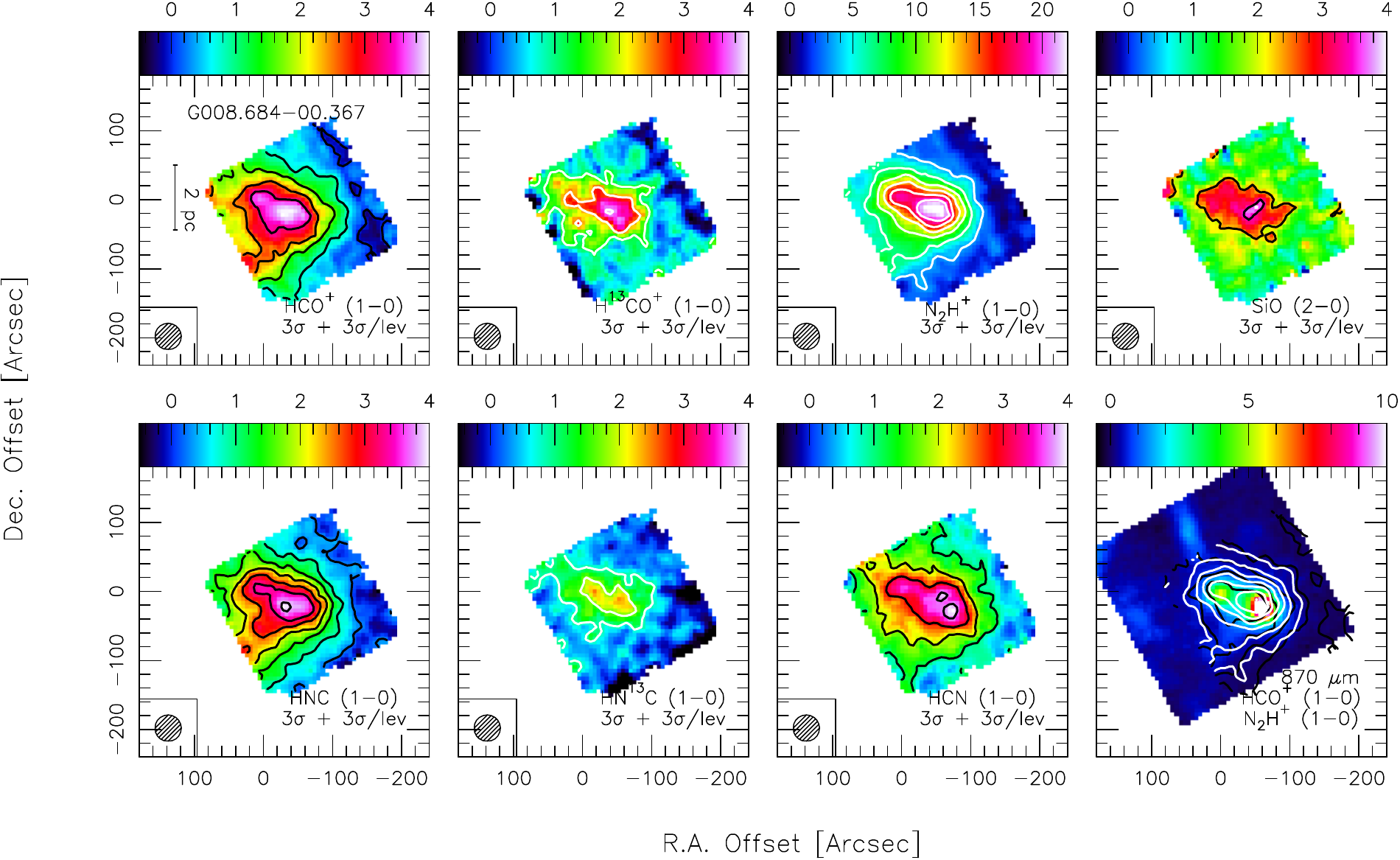}\vspace{0.50cm}
\includegraphics[width=0.49\textwidth]{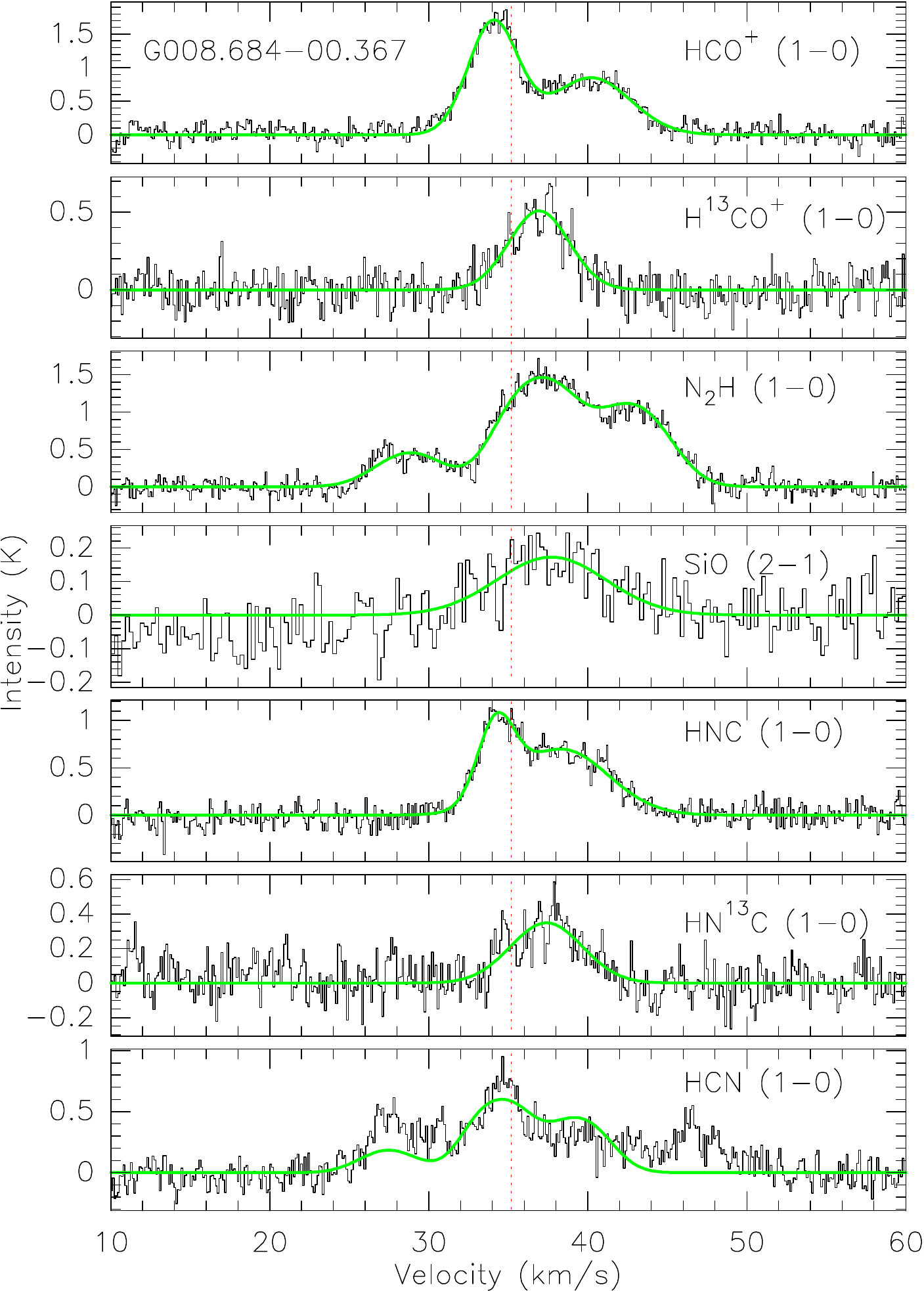}
\caption{\emph{Top}: Integrated intensity maps of the \hcop(1$-$0), \htcop(1$-$0), \sio, \nthp(1$-$0), \hnc(1$-$0), \hntc(1$-$0) and \hcn(1$-$0) emission for the high-mass ATLASGAL clumps. The last panel shows the zero moment map contours of the \hcop~(black) and \nthp~(white) line emissions overlaid on the ATLASGAL submm continuum emission at 870~$\mu$m. Contour levels are shown on each image. The Mopra telescope beam size is shown on the left corner of the maps. \emph{Bottom}: \hcop,\htcop,\nthp,\sio,\hnc,\hntc,\hcn~spectra for all sources. Velocity range is 50 \kms. The spectra is created by averaging the spectra in 38$''$ x 38$''$ box (MOPRA beam size) around the clump's N$_{2}$H$^{+}$ peak emission. The vertical red dotted line indicates the radial velocity of the ammonia line for the clumps given in \citet{wie15}. The single and double Gaussian or hyperfine line fits are shown with green lines.}\label{fig:zmoment}
\end{figure*}

\addtocounter{figure}{-1}

\begin{figure*}
\centering
\caption{Continued}
\includegraphics[width=0.80\textwidth]{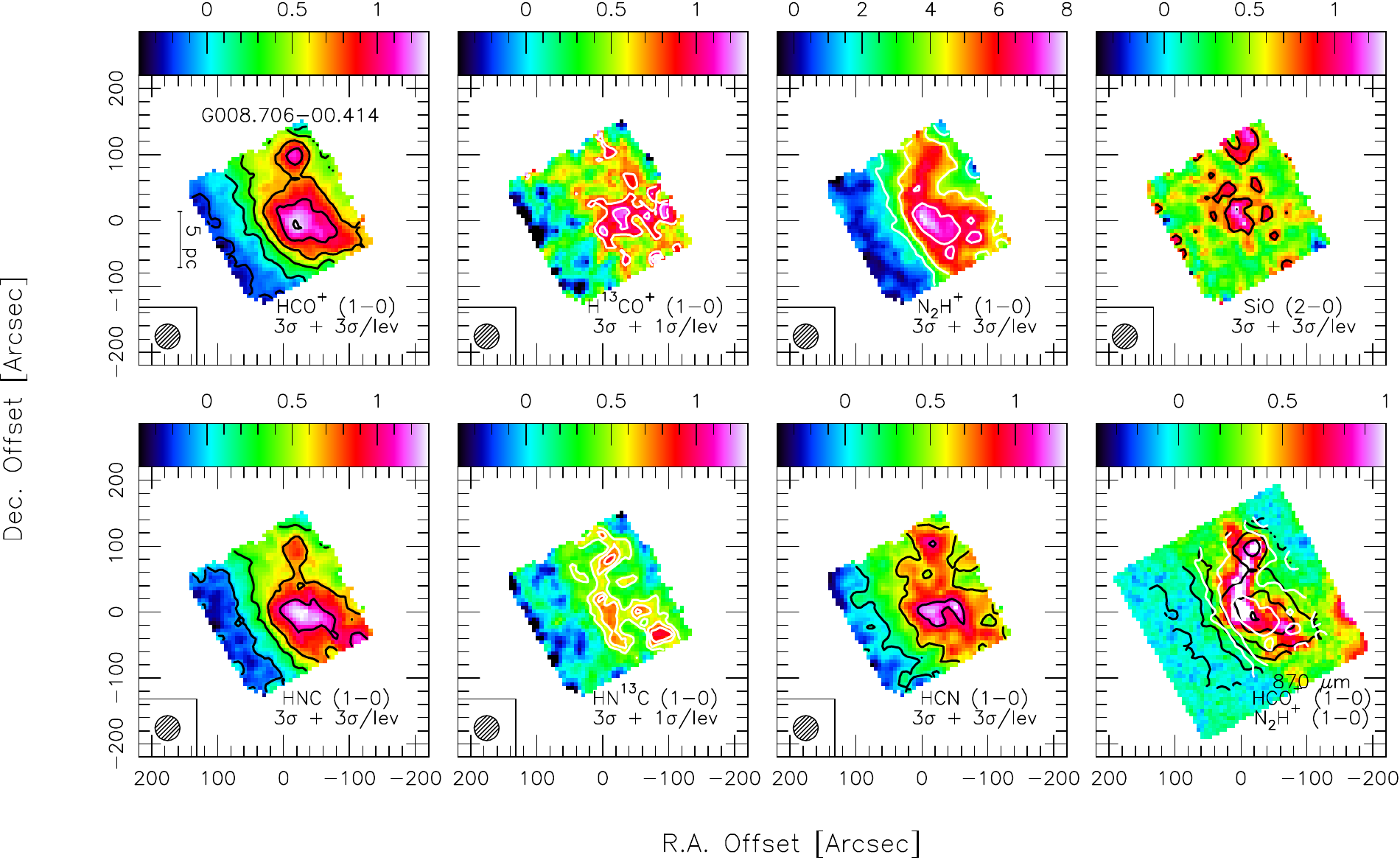}\vspace{0.5cm}
\includegraphics[width=0.50\textwidth]{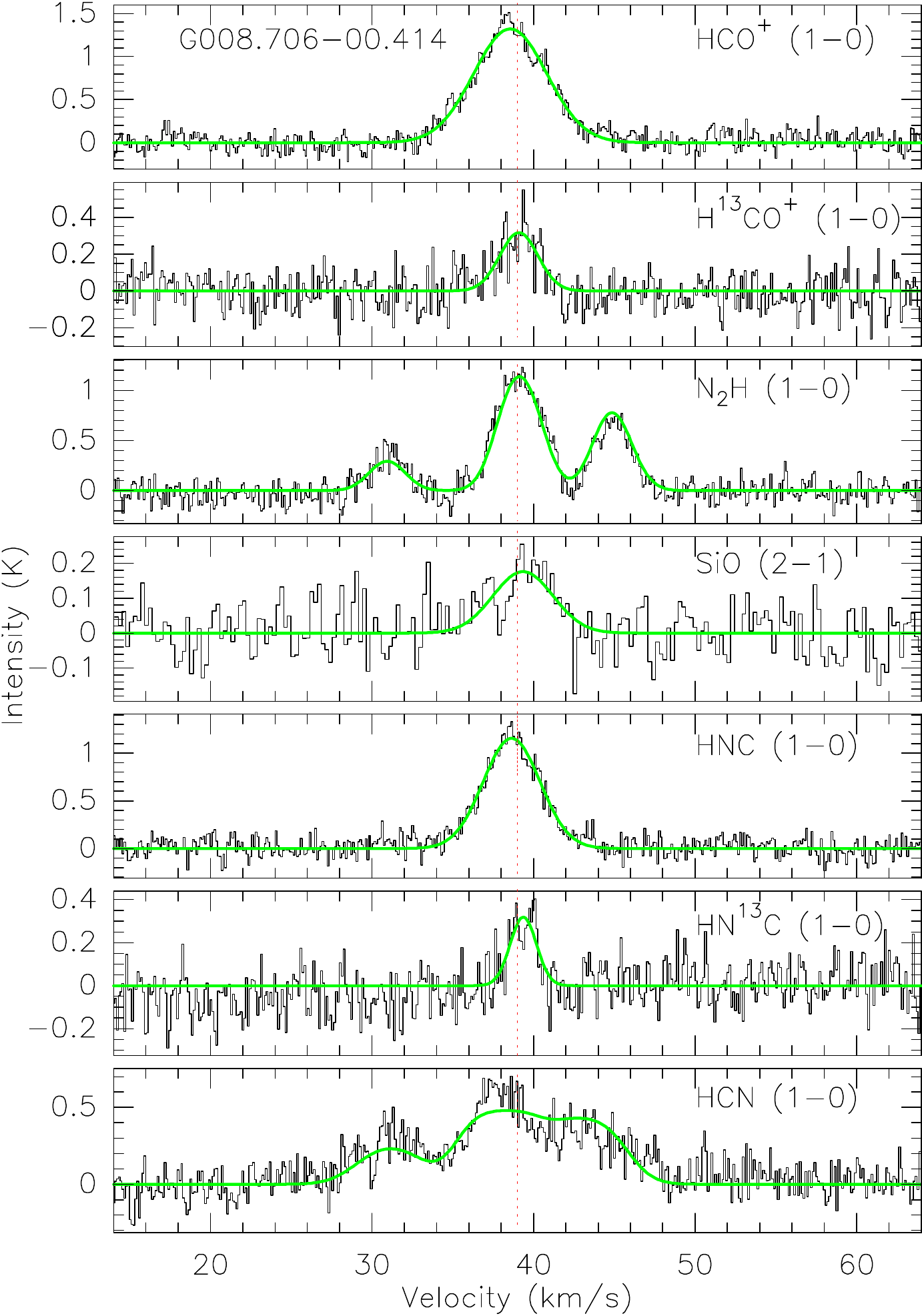}
\end{figure*}

\addtocounter{figure}{-1}
\begin{figure*}
\centering
\caption{Continued}
\includegraphics[width=0.80\textwidth]{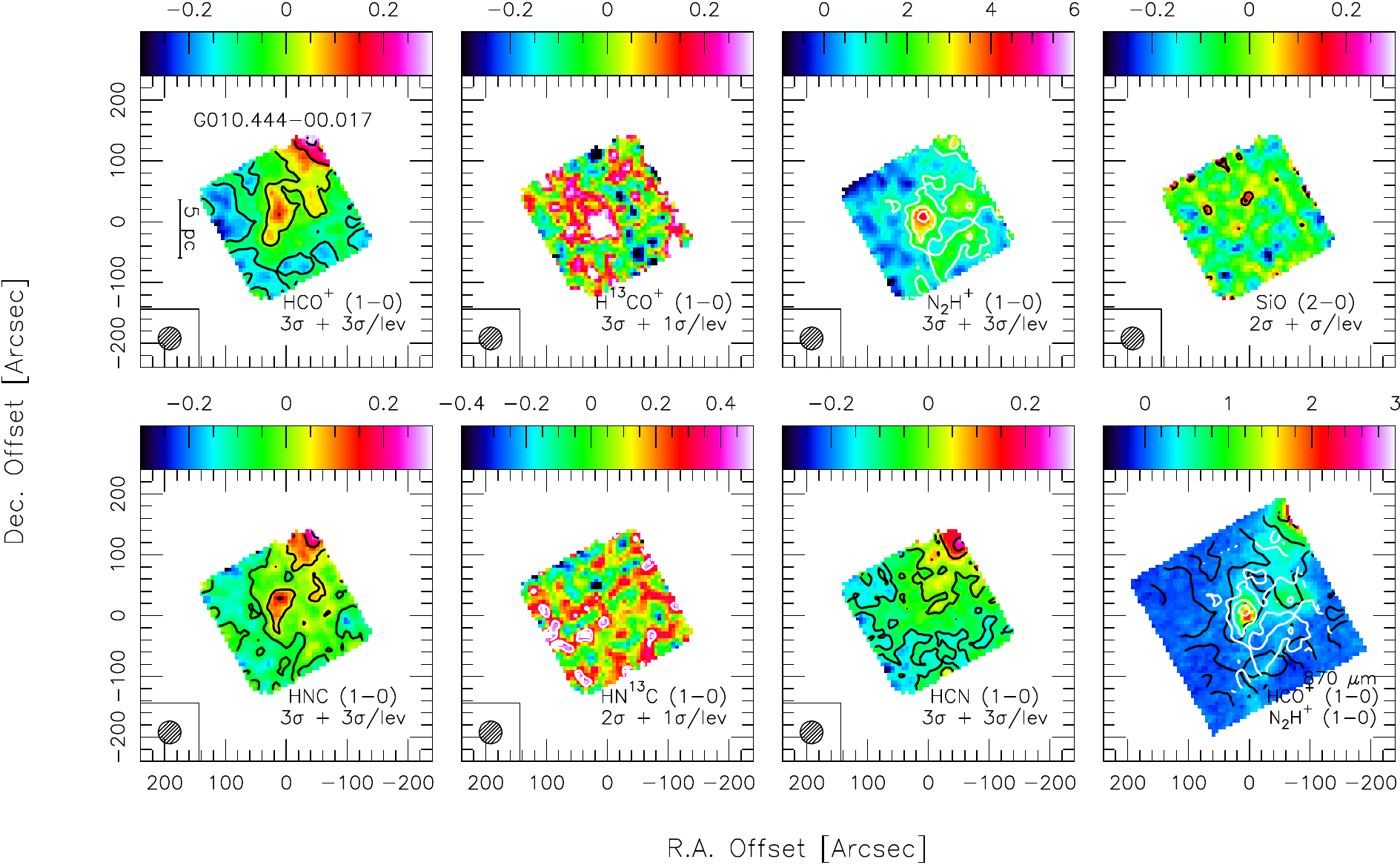}\vspace{0.5cm}
\includegraphics[width=0.50\textwidth]{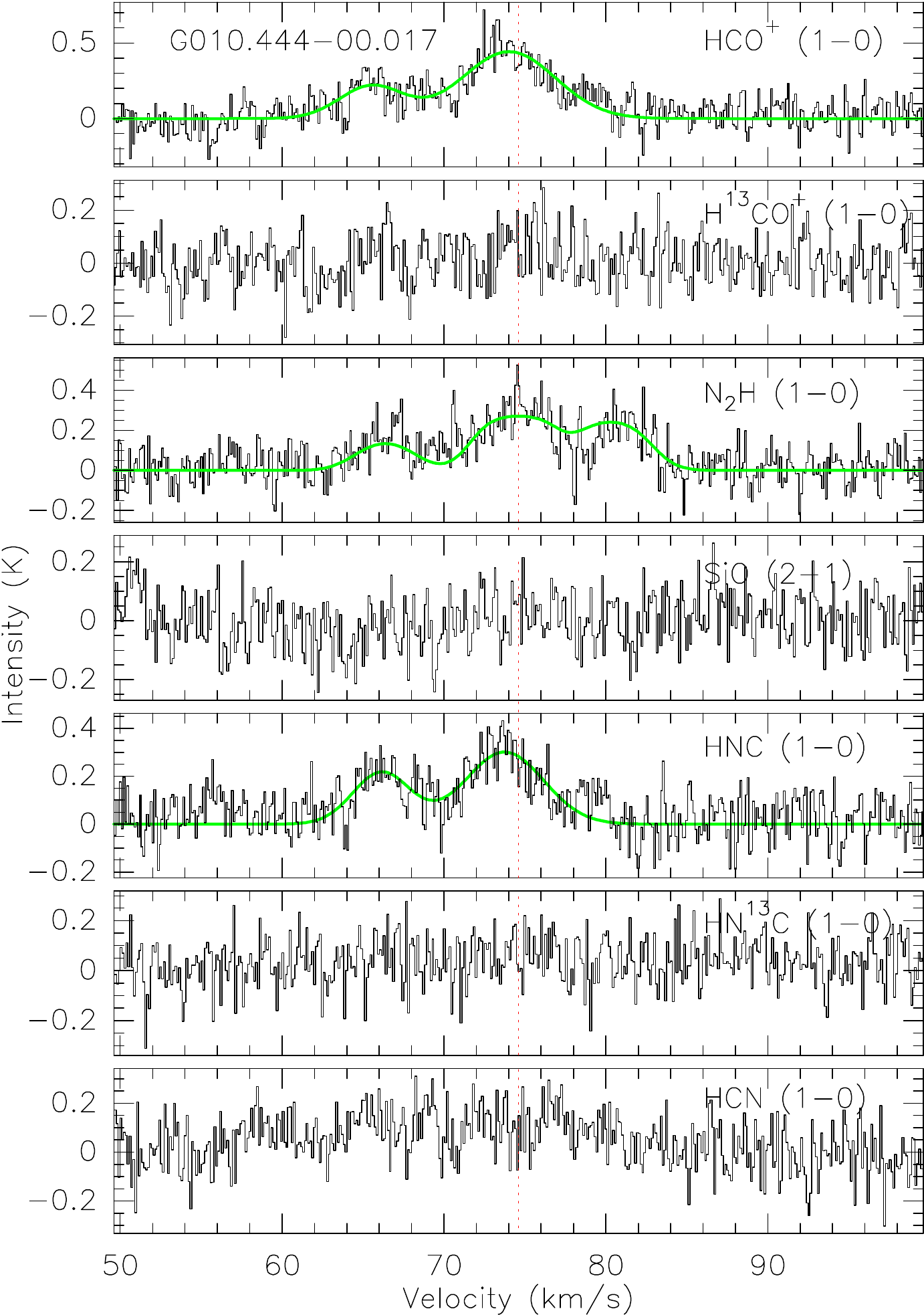}
\end{figure*}
%
\addtocounter{figure}{-1}
\begin{figure*}
\centering
\caption{Continued}
\includegraphics[width=0.80\textwidth]{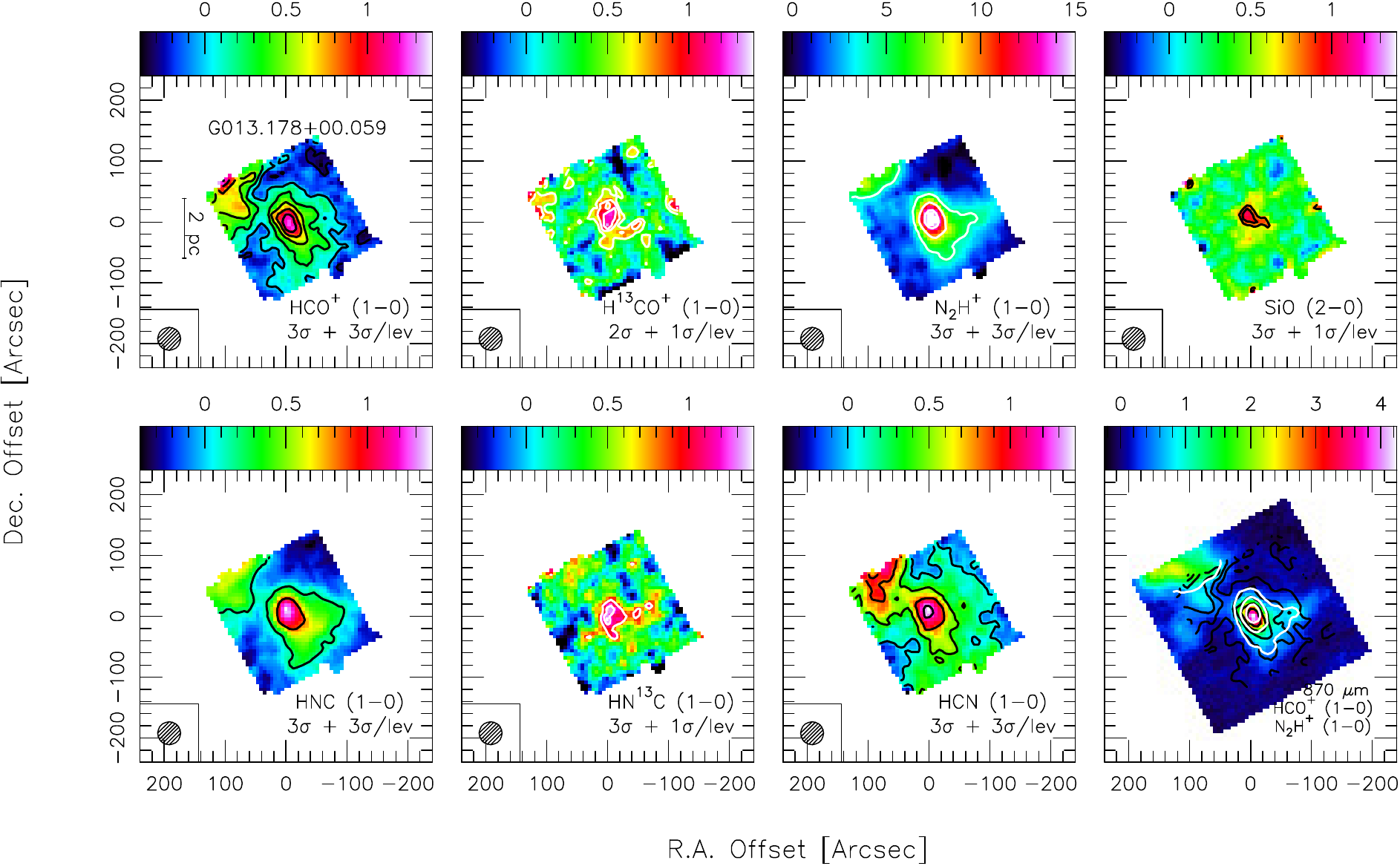}\vspace{0.5cm}
\includegraphics[width=0.50\textwidth]{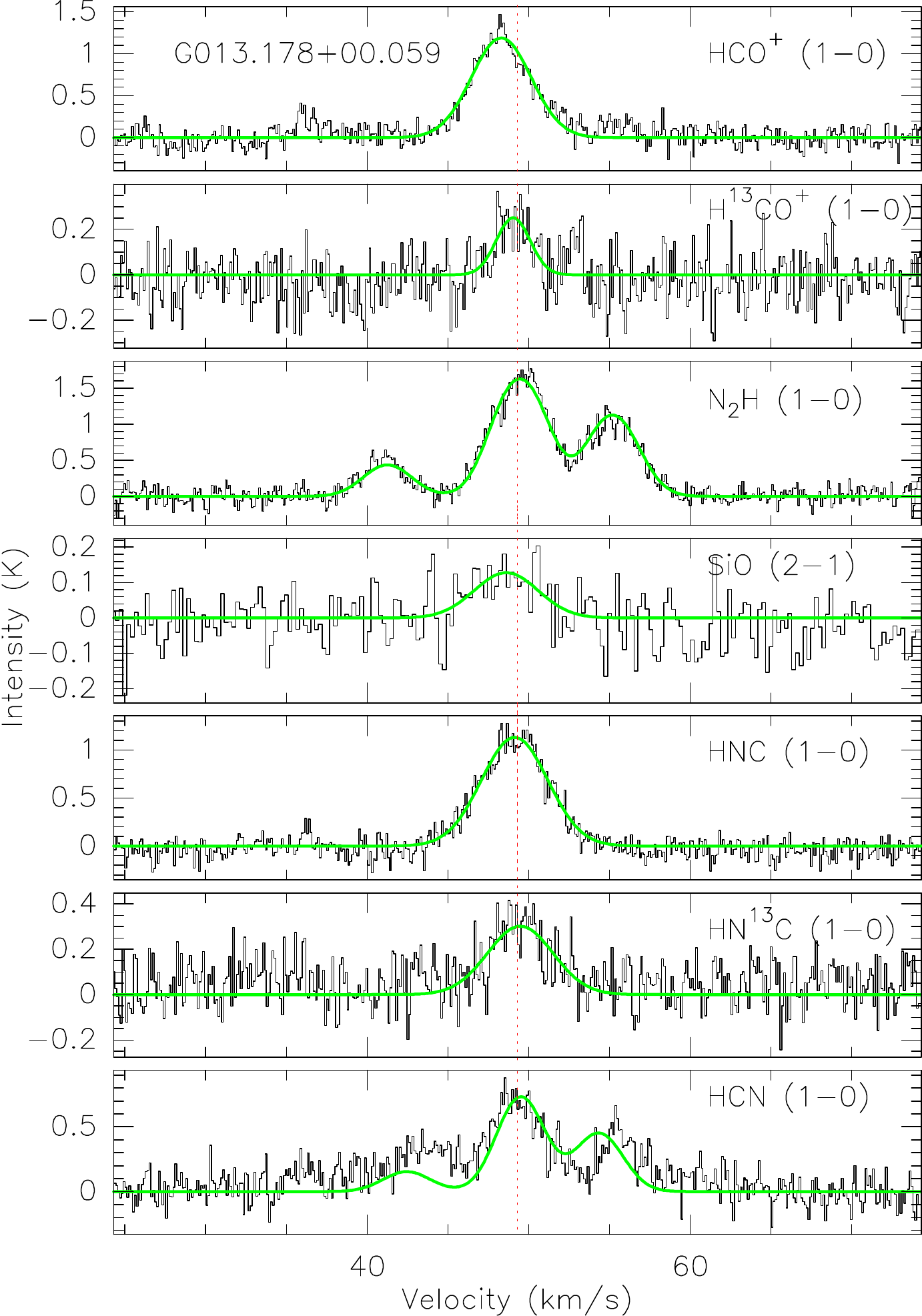}
\end{figure*}
\addtocounter{figure}{-1}
\begin{figure*}
\centering
\caption{Continued}
\includegraphics[width=0.80\textwidth]{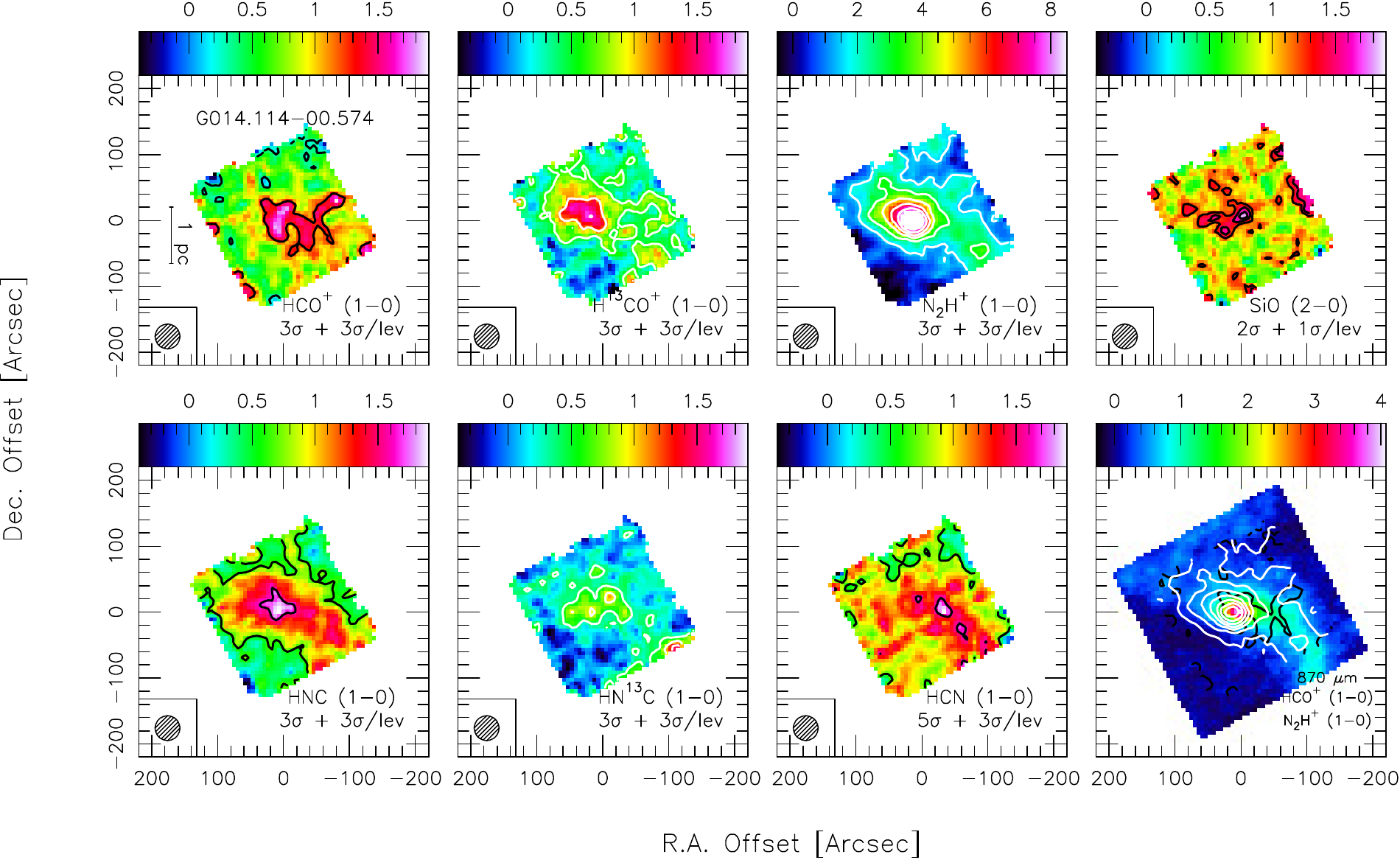}\vspace{0.5cm}
\includegraphics[width=0.50\textwidth]{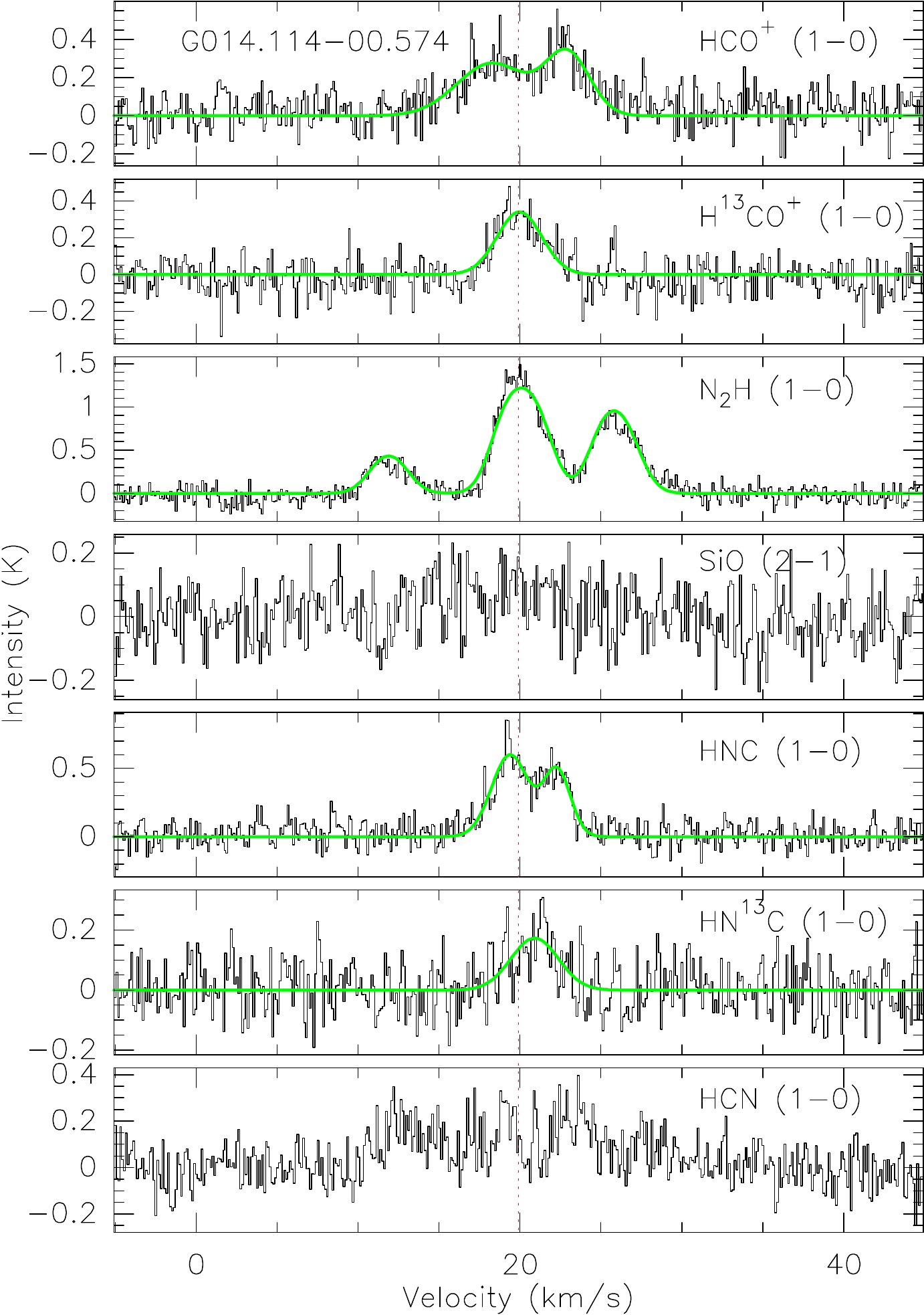}
\end{figure*}
\addtocounter{figure}{-1}
\begin{figure*}
\centering
\caption{Continued}
\includegraphics[width=0.80\textwidth]{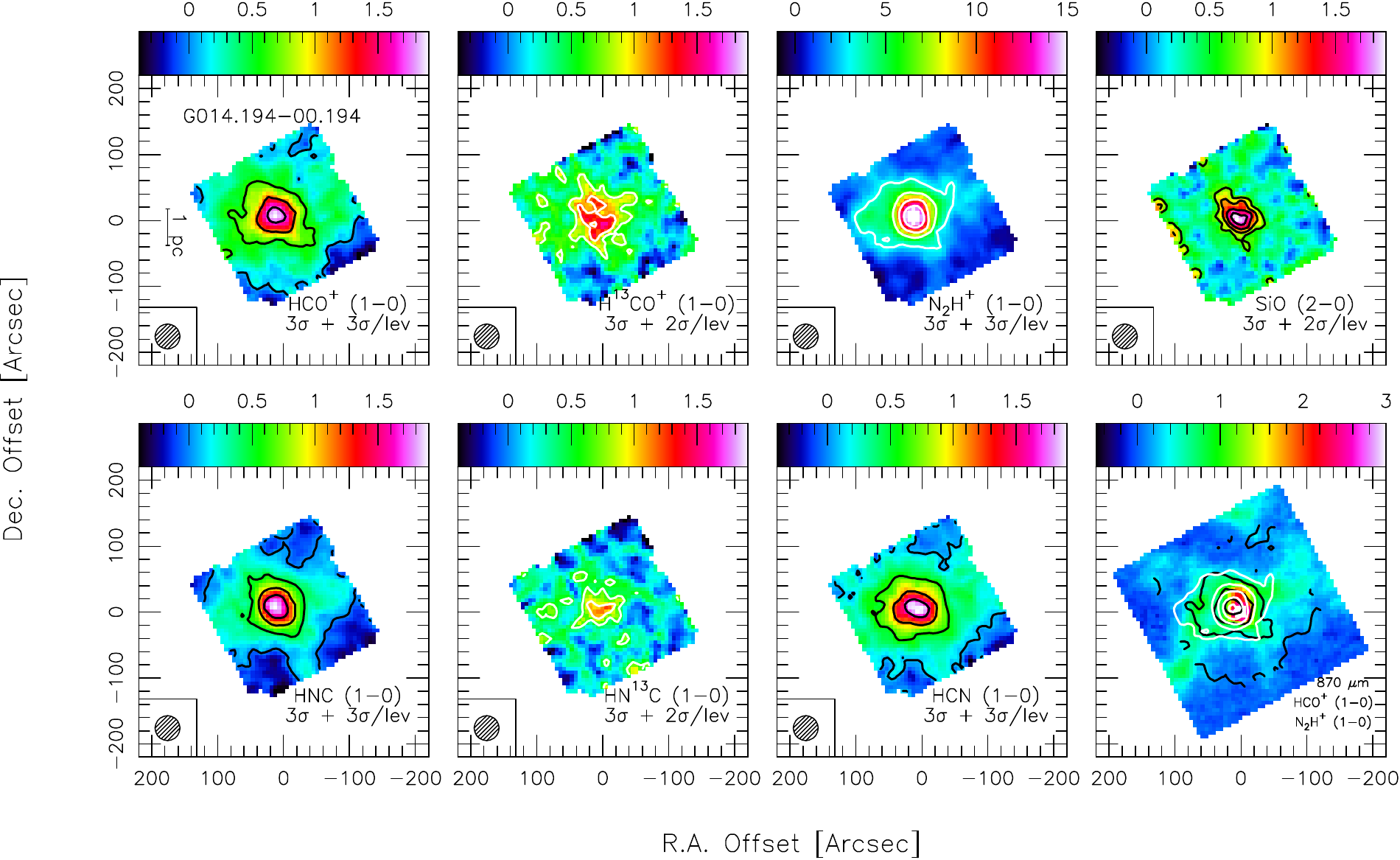}\vspace{0.5cm}
\includegraphics[width=0.50\textwidth]{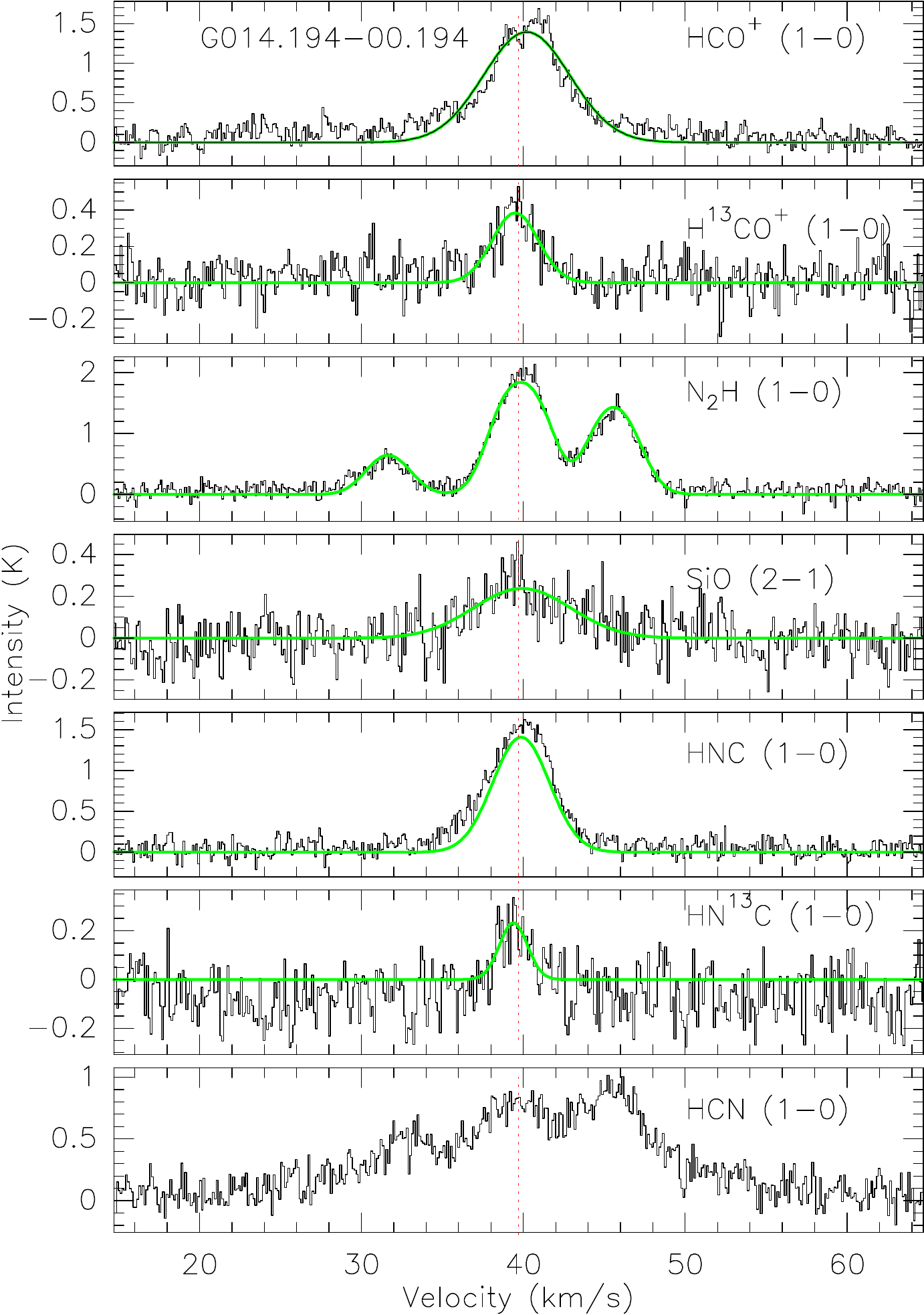}
\end{figure*}
\addtocounter{figure}{-1}
\begin{figure*}
\centering
\caption{Continued}
\includegraphics[width=0.80\textwidth]{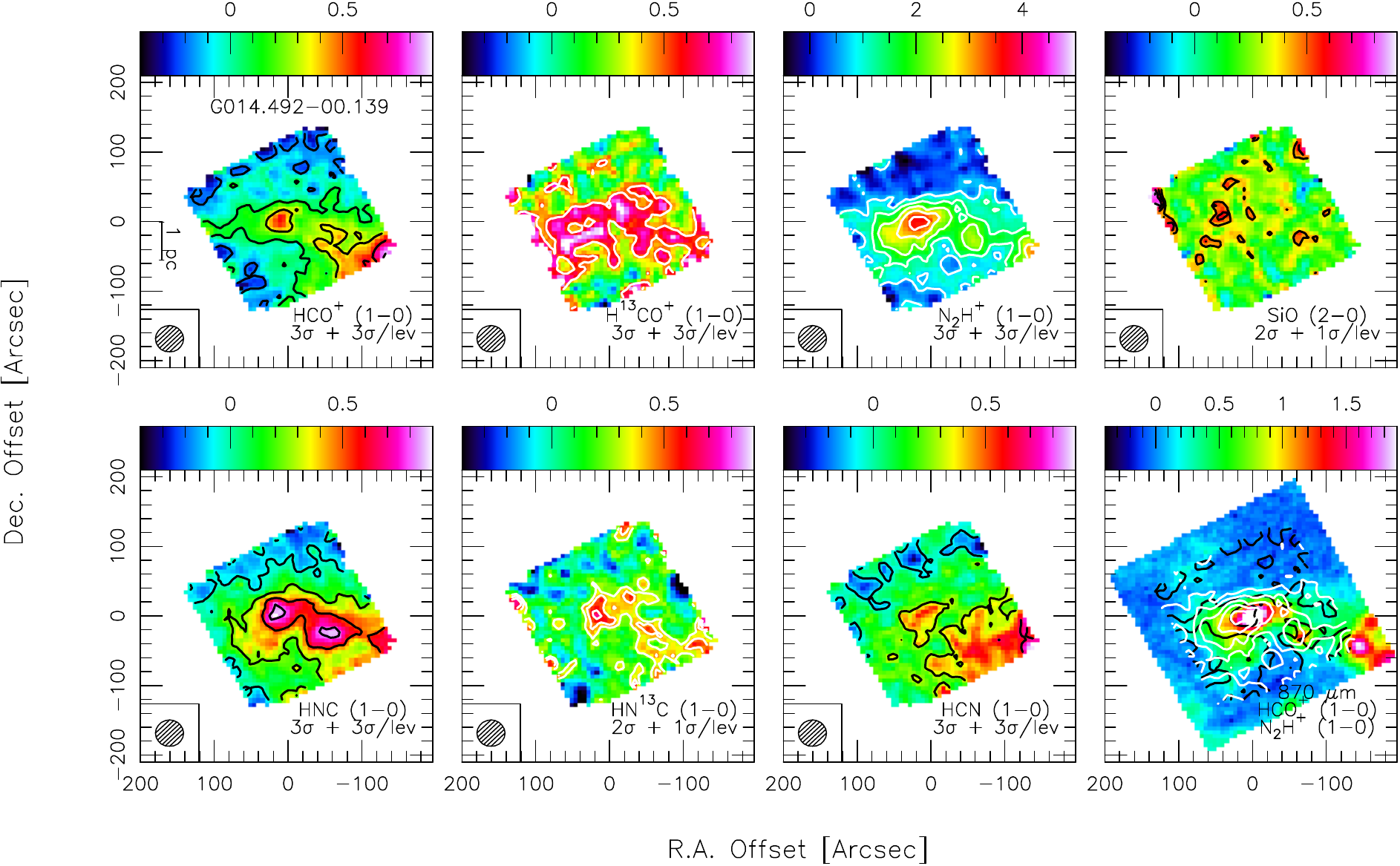}\vspace{0.5cm}
\includegraphics[width=0.50\textwidth]{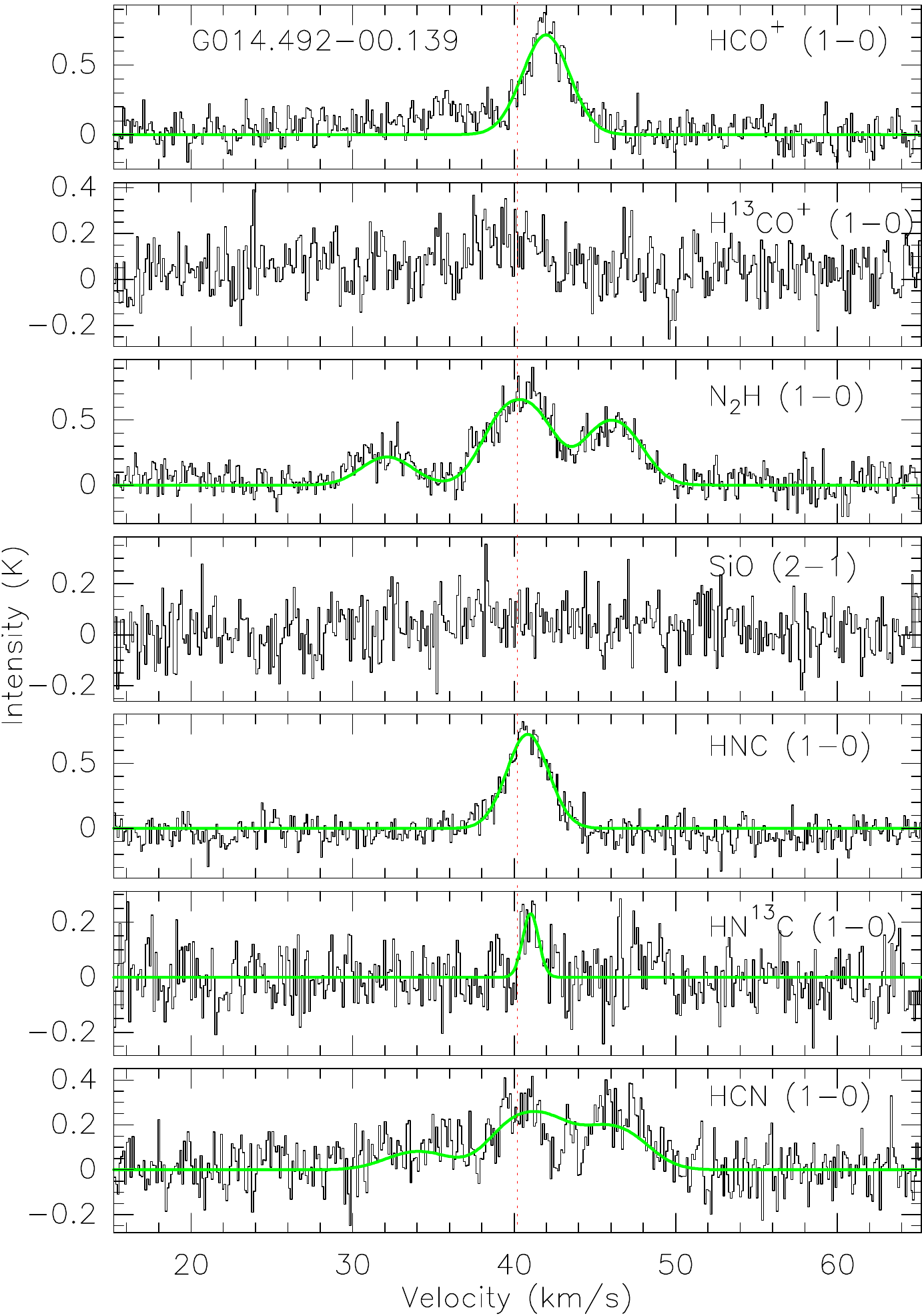}
\end{figure*}
\addtocounter{figure}{-1}
\begin{figure*}
\centering
\caption{Continued}
\includegraphics[width=0.80\textwidth]{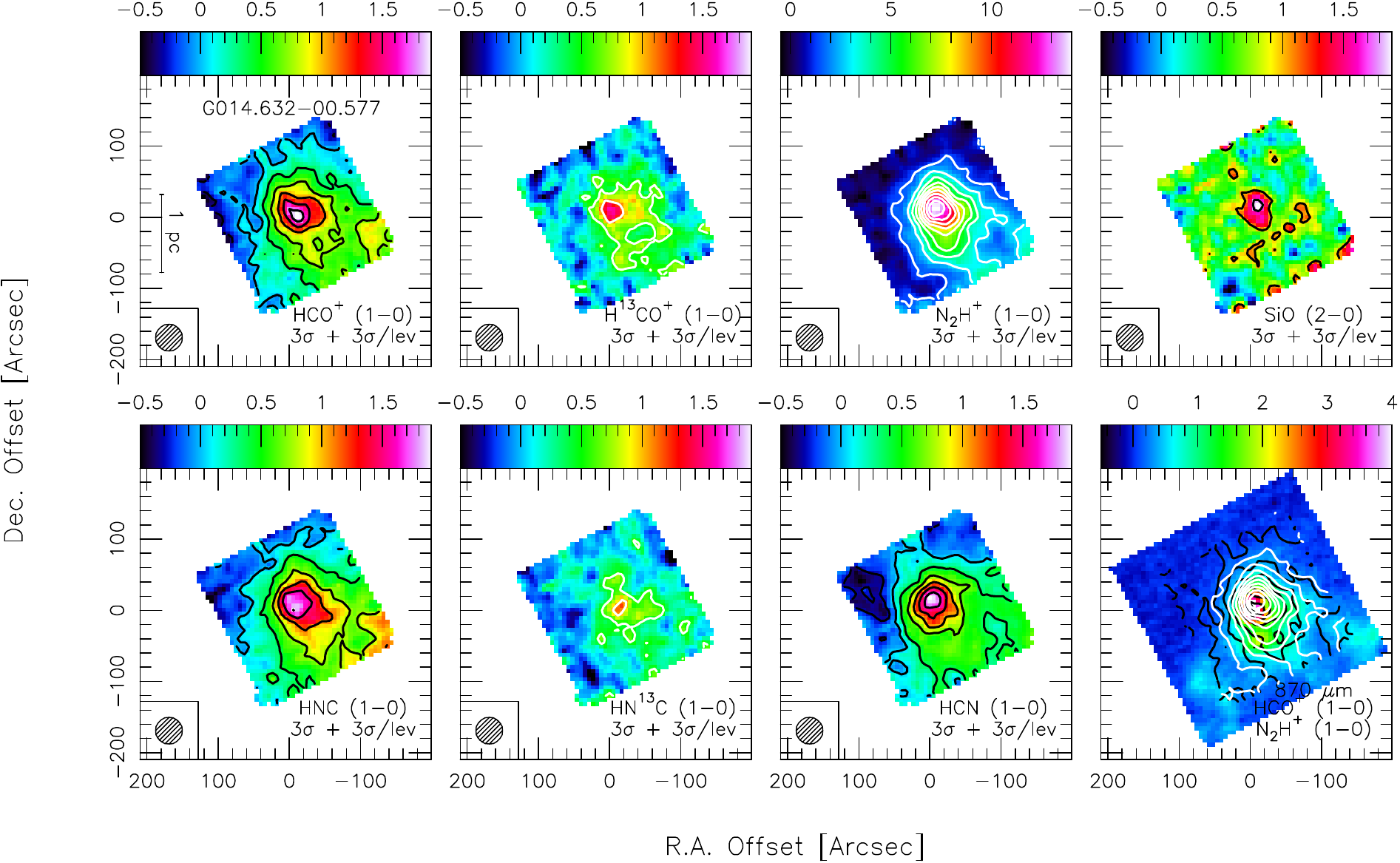}\vspace{0.5cm}
\includegraphics[width=0.50\textwidth]{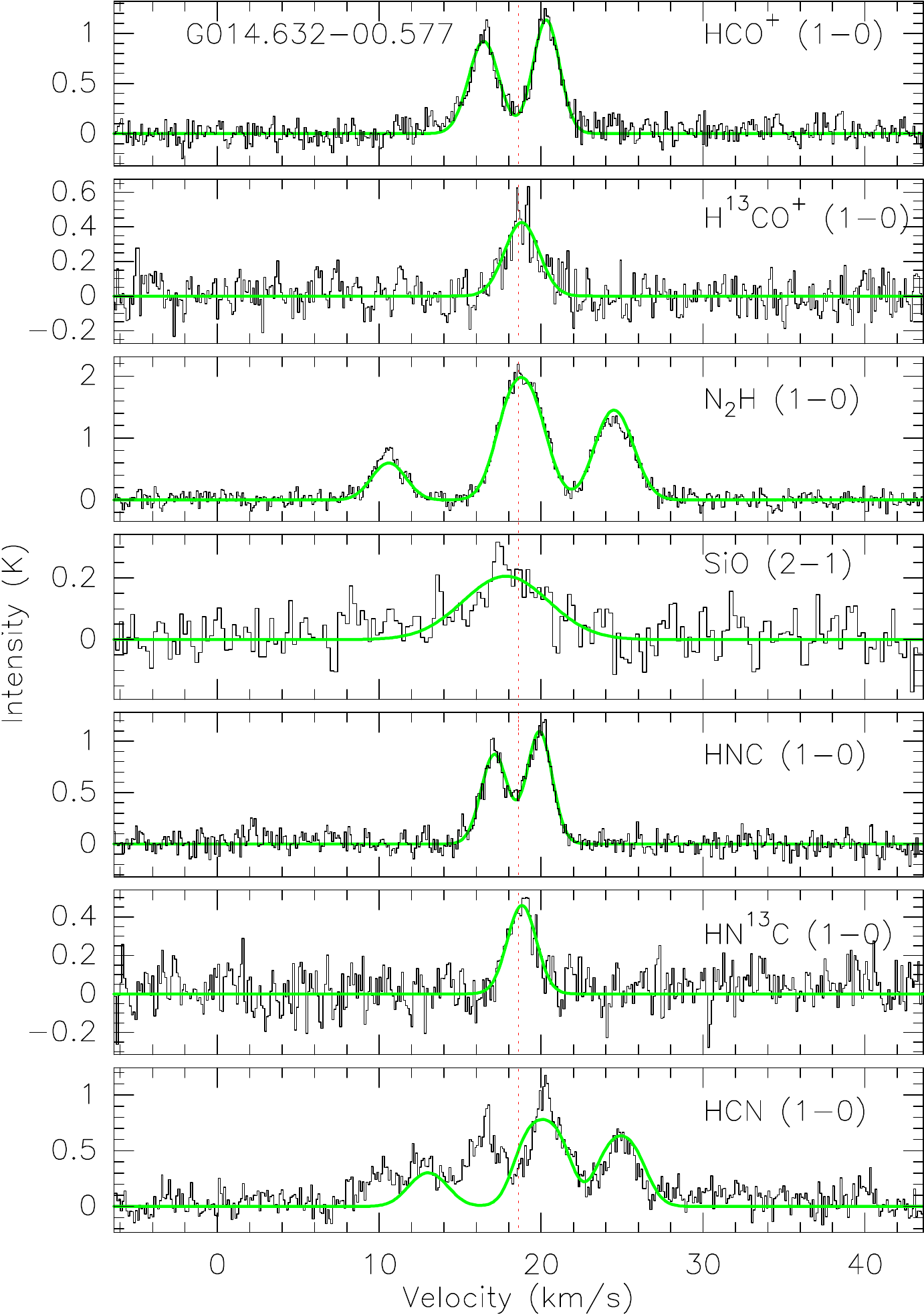}
\end{figure*}
\addtocounter{figure}{-1}
\begin{figure*}
\centering
\caption{Continued}
\includegraphics[width=0.80\textwidth]{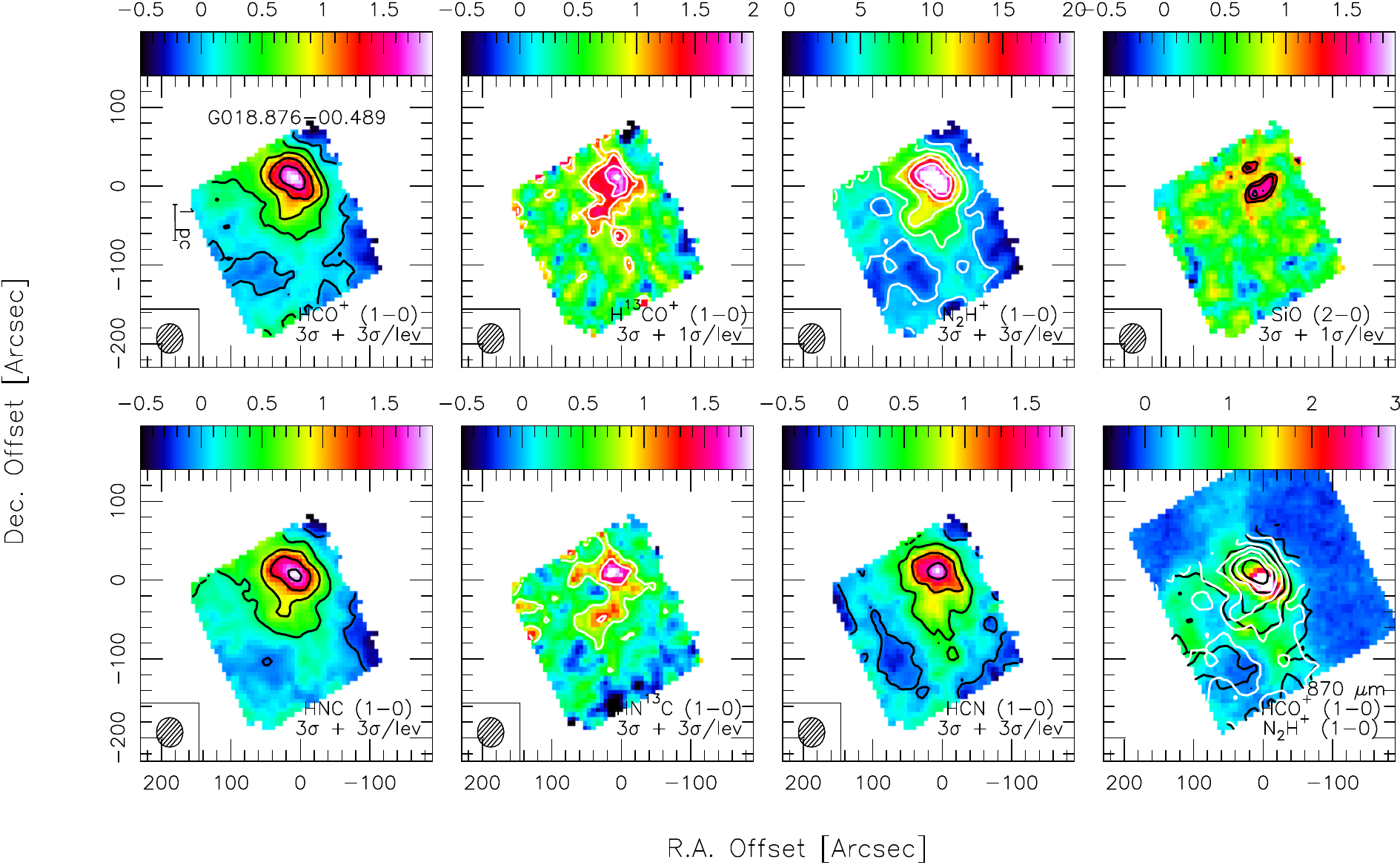}\vspace{0.5cm}
\includegraphics[width=0.50\textwidth]{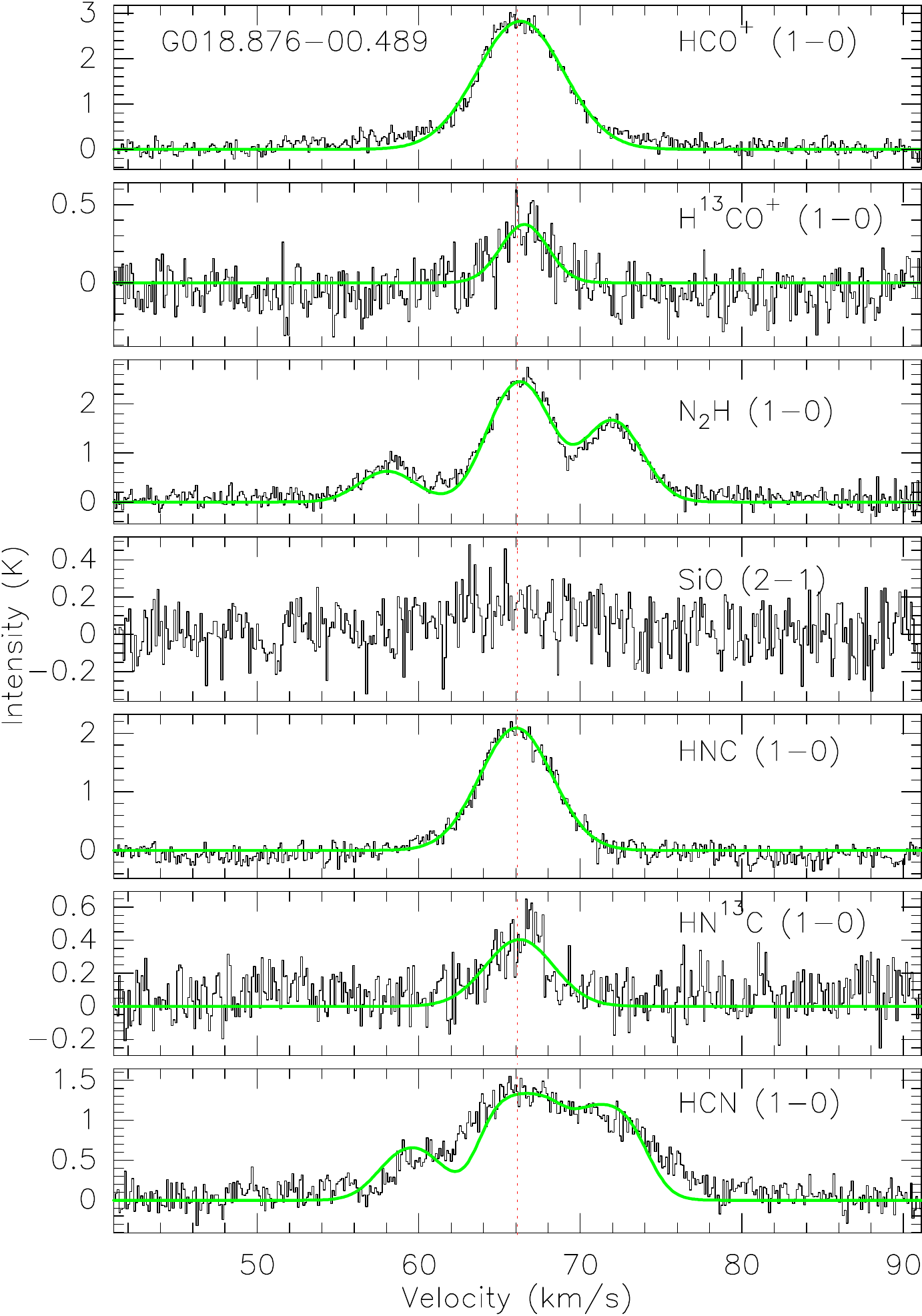}
\end{figure*}
\addtocounter{figure}{-1}
\begin{figure*}
\centering
\caption{Continued}
\includegraphics[width=0.80\textwidth]{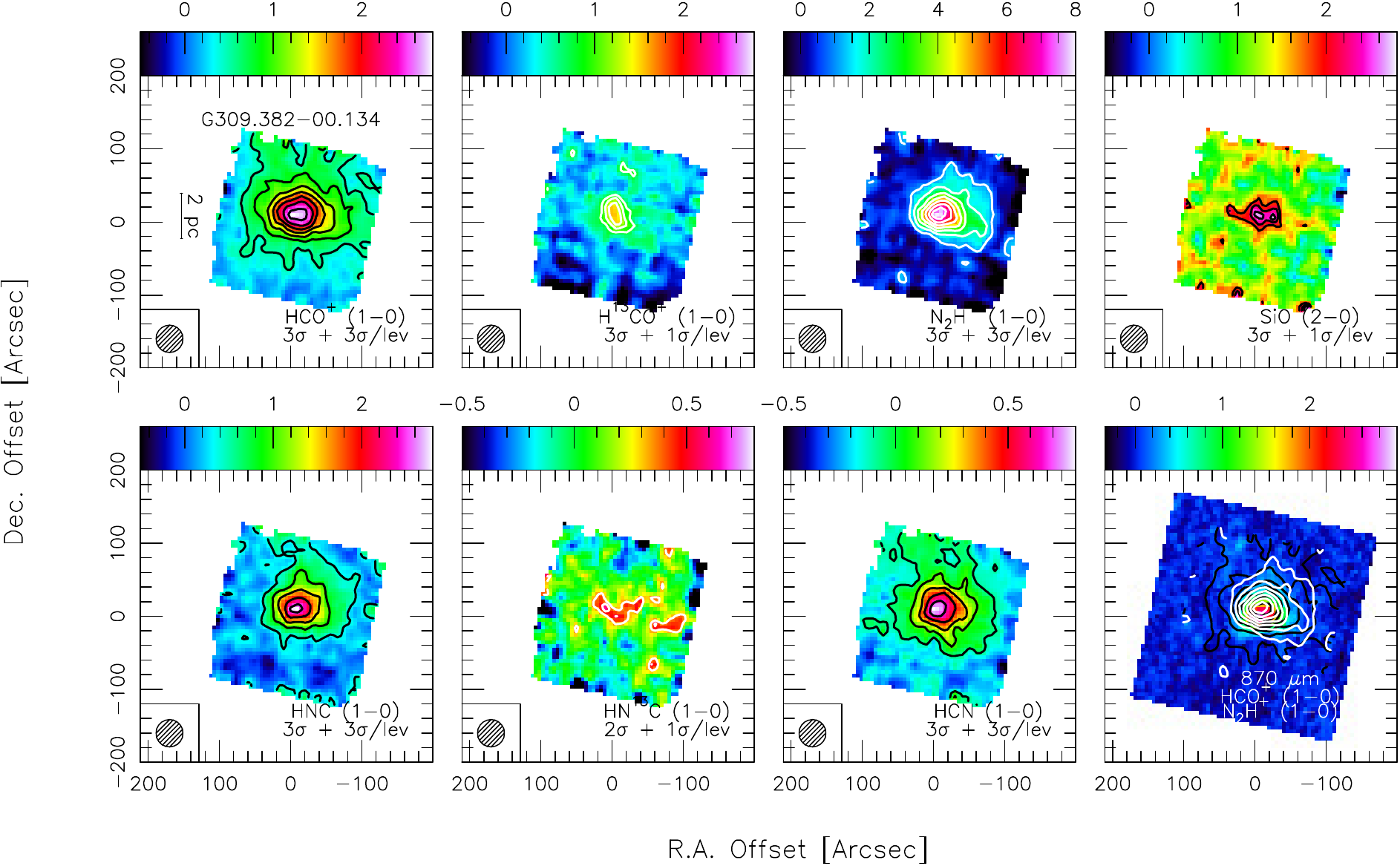}\vspace{0.5cm}
\includegraphics[width=0.50\textwidth]{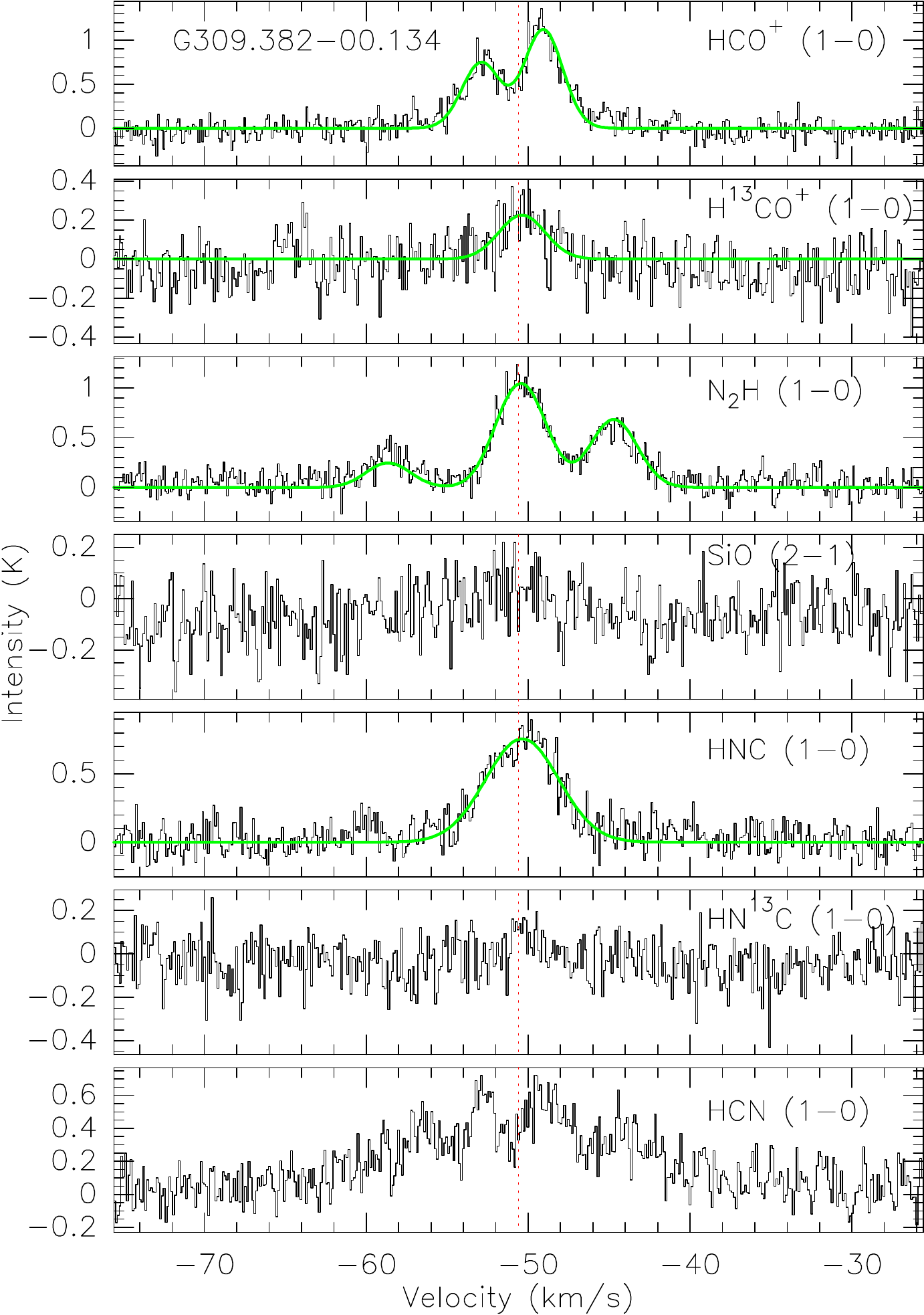}
\end{figure*}
\addtocounter{figure}{-1}
\begin{figure*}
\centering
\caption{Continued}
\includegraphics[width=0.80\textwidth]{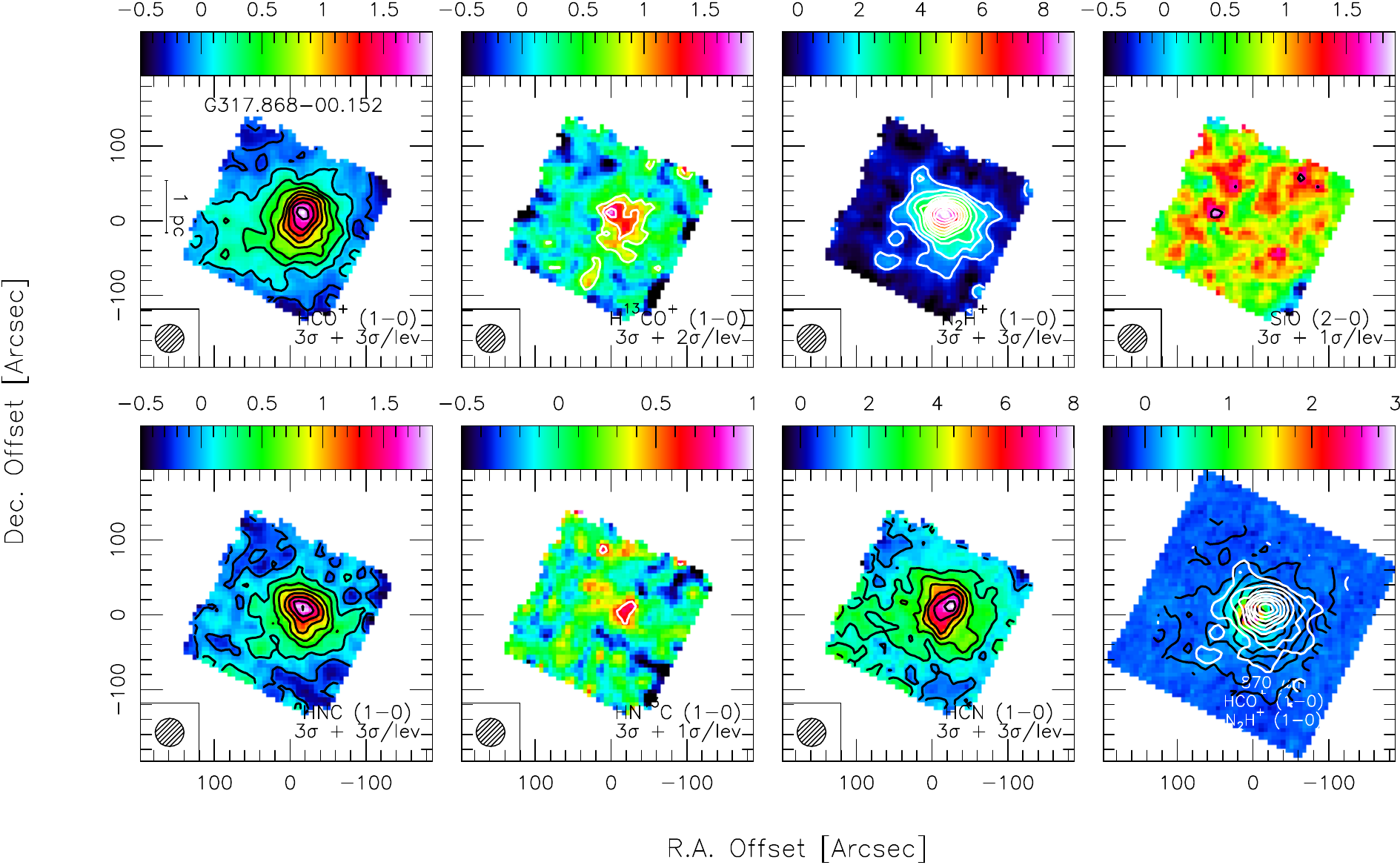}\vspace{0.5cm}
\includegraphics[width=0.50\textwidth]{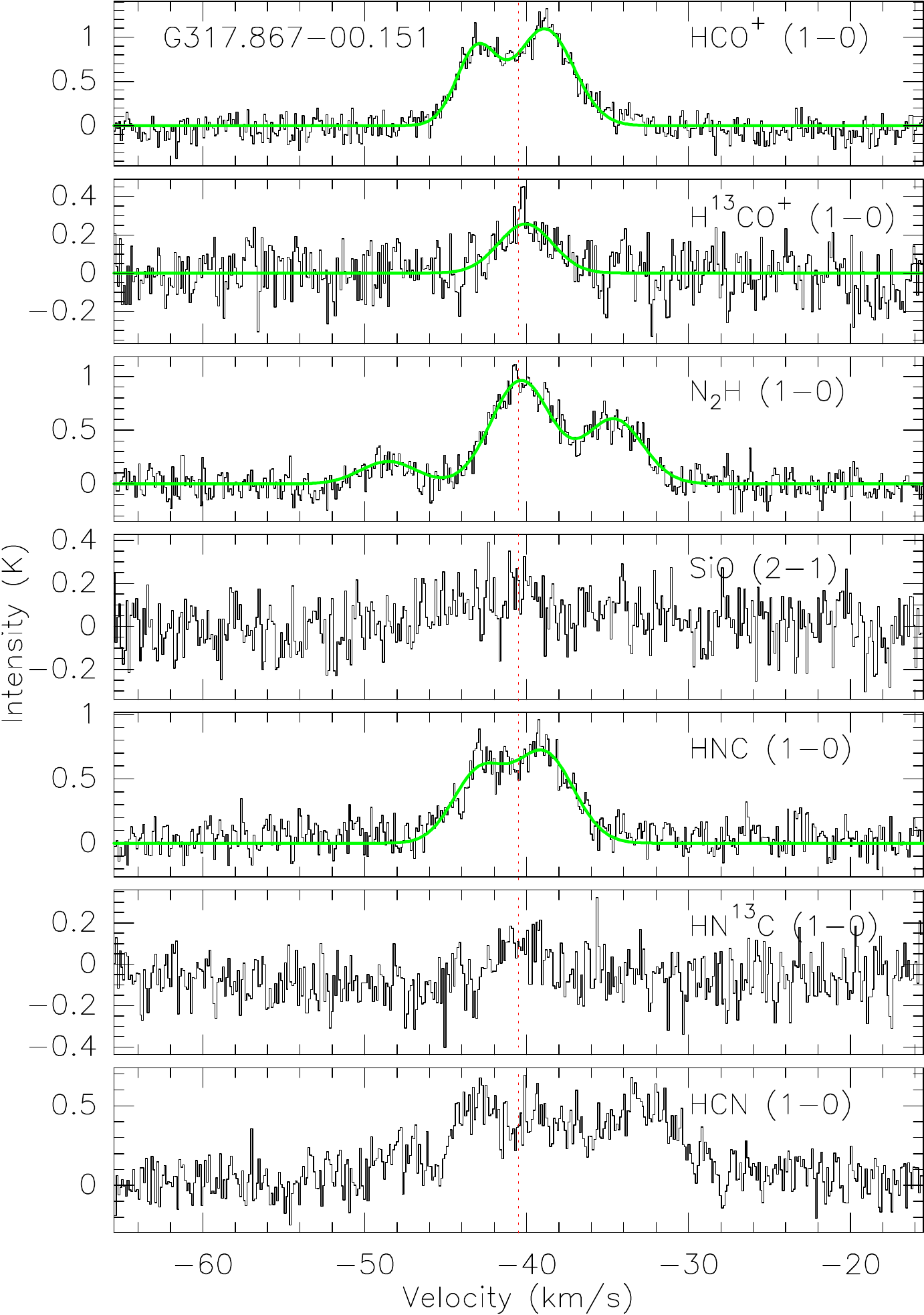}
\end{figure*}
\addtocounter{figure}{-1}
\begin{figure*}
\centering
\caption{Continued}
\includegraphics[width=0.80\textwidth]{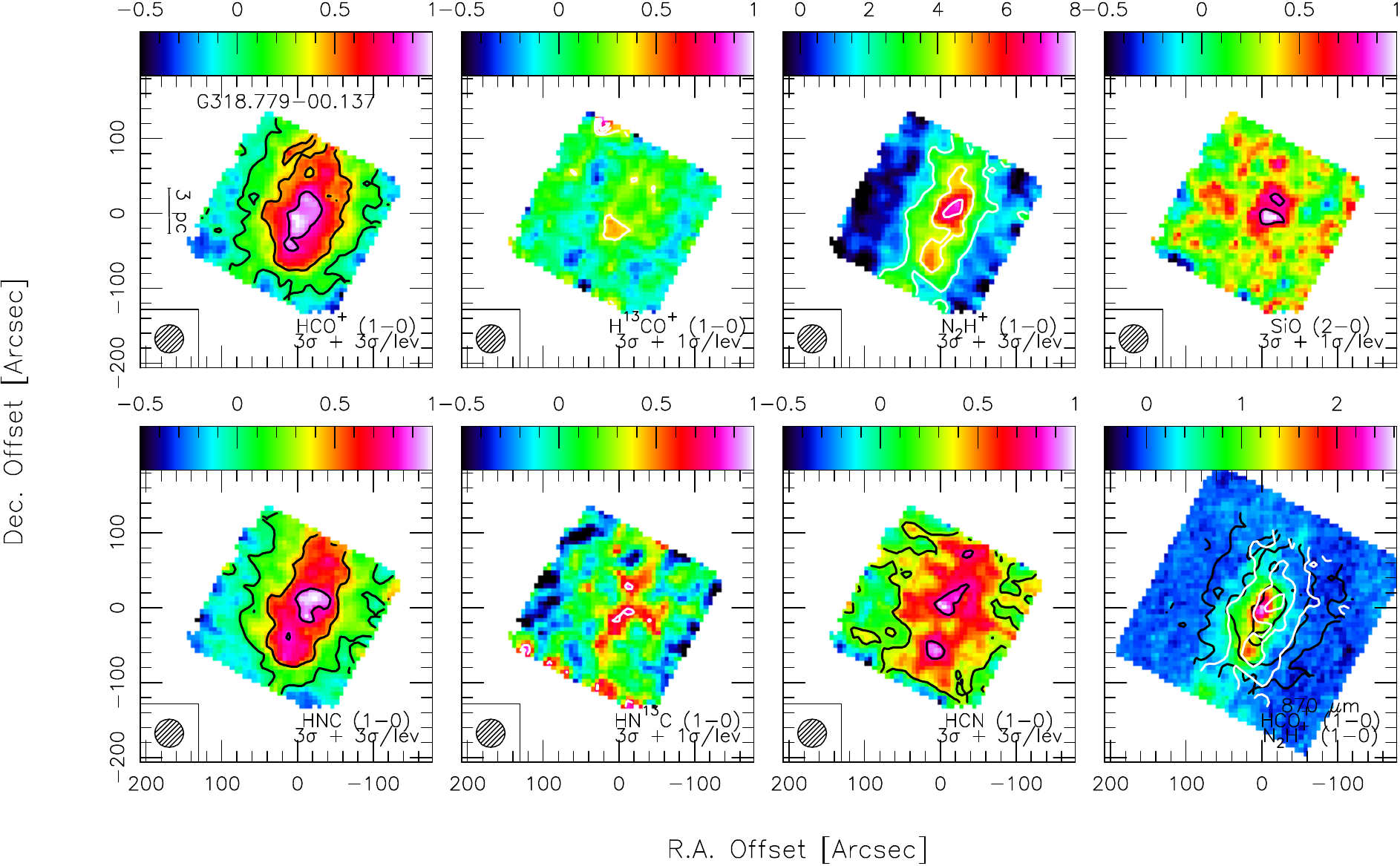}\vspace{0.5cm}
\includegraphics[width=0.50\textwidth]{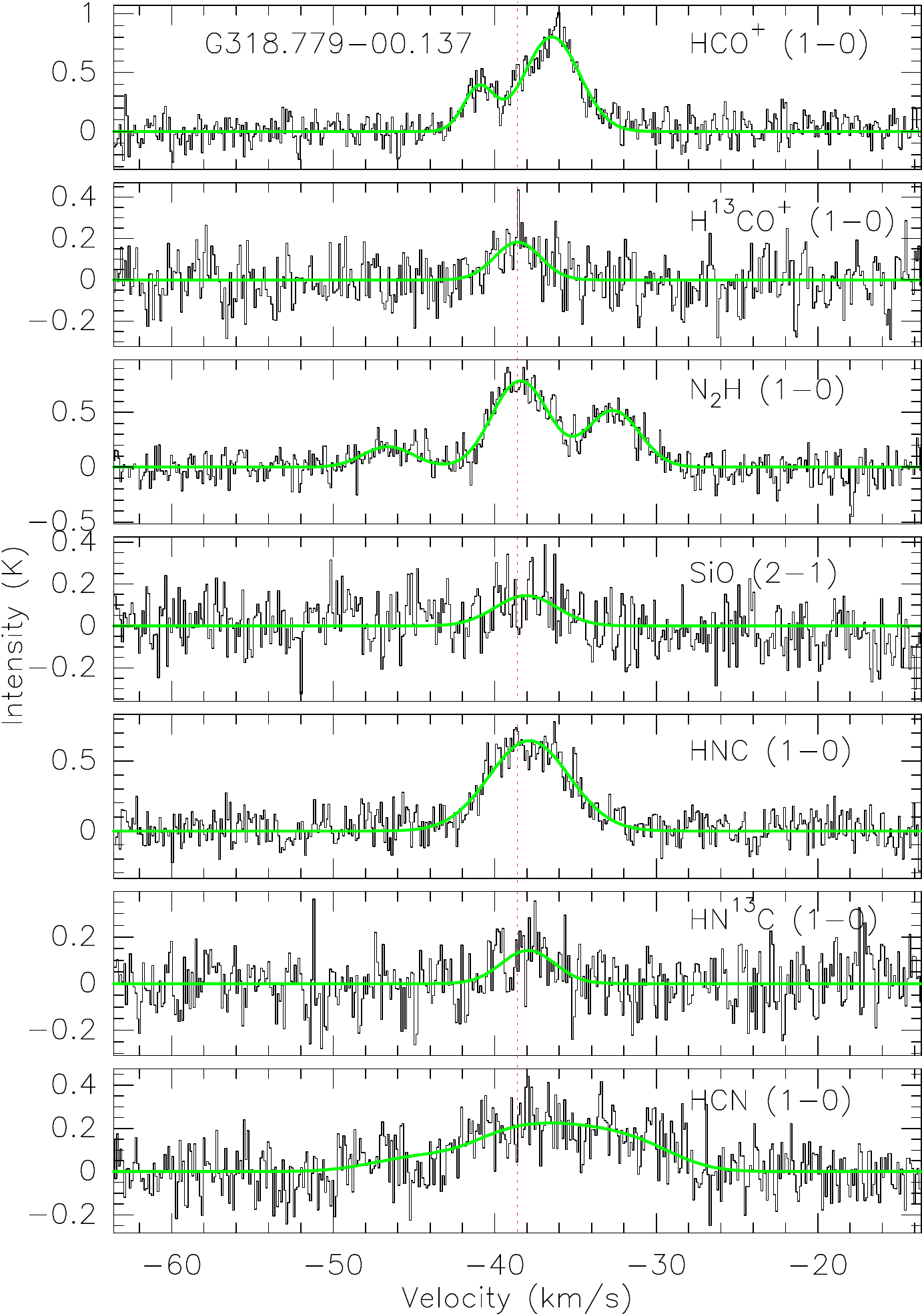}
\end{figure*}
\addtocounter{figure}{-1}
\begin{figure*}
\centering
\caption{Continued}
\includegraphics[width=0.80\textwidth]{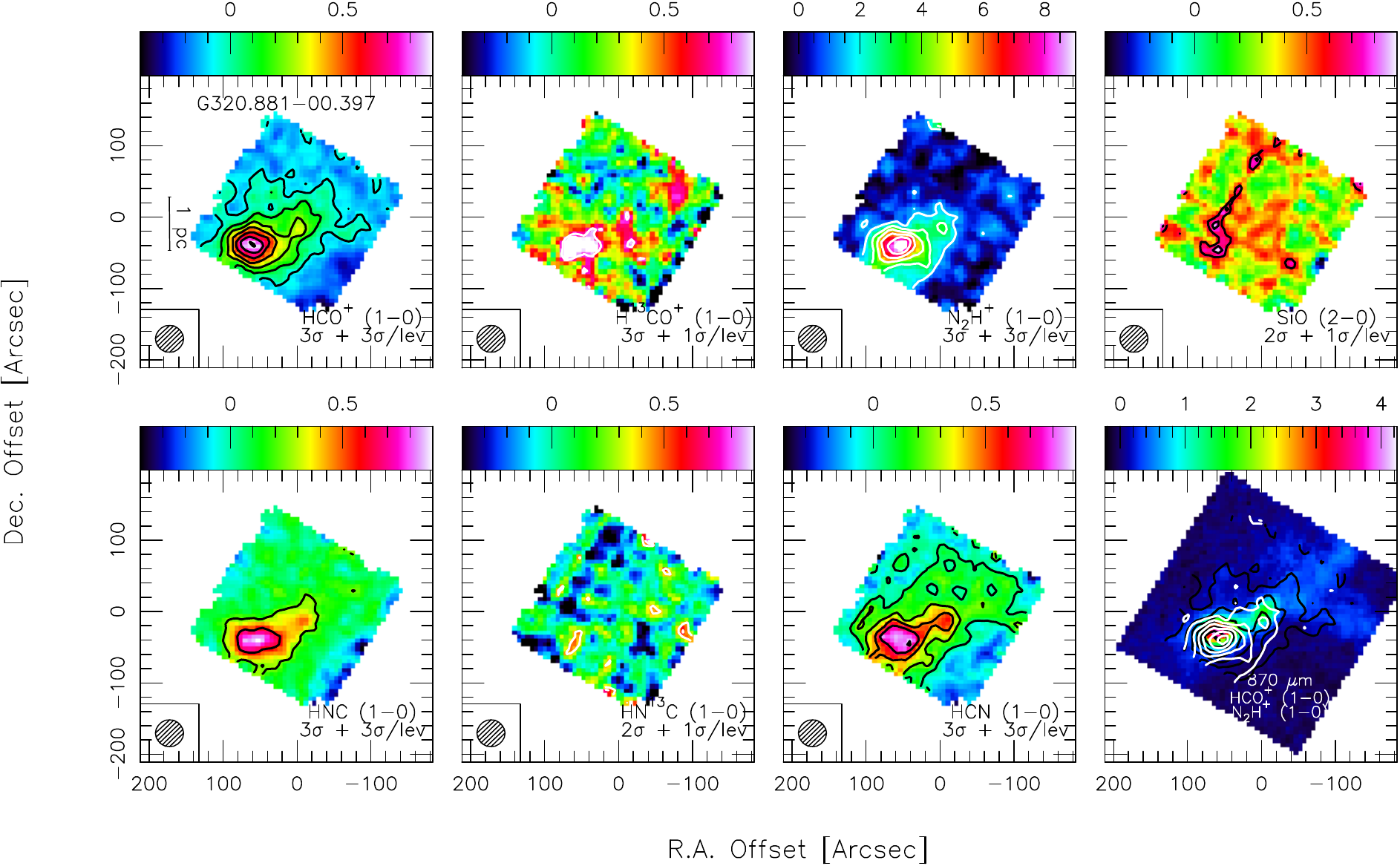}\vspace{0.5cm}
\includegraphics[width=0.50\textwidth]{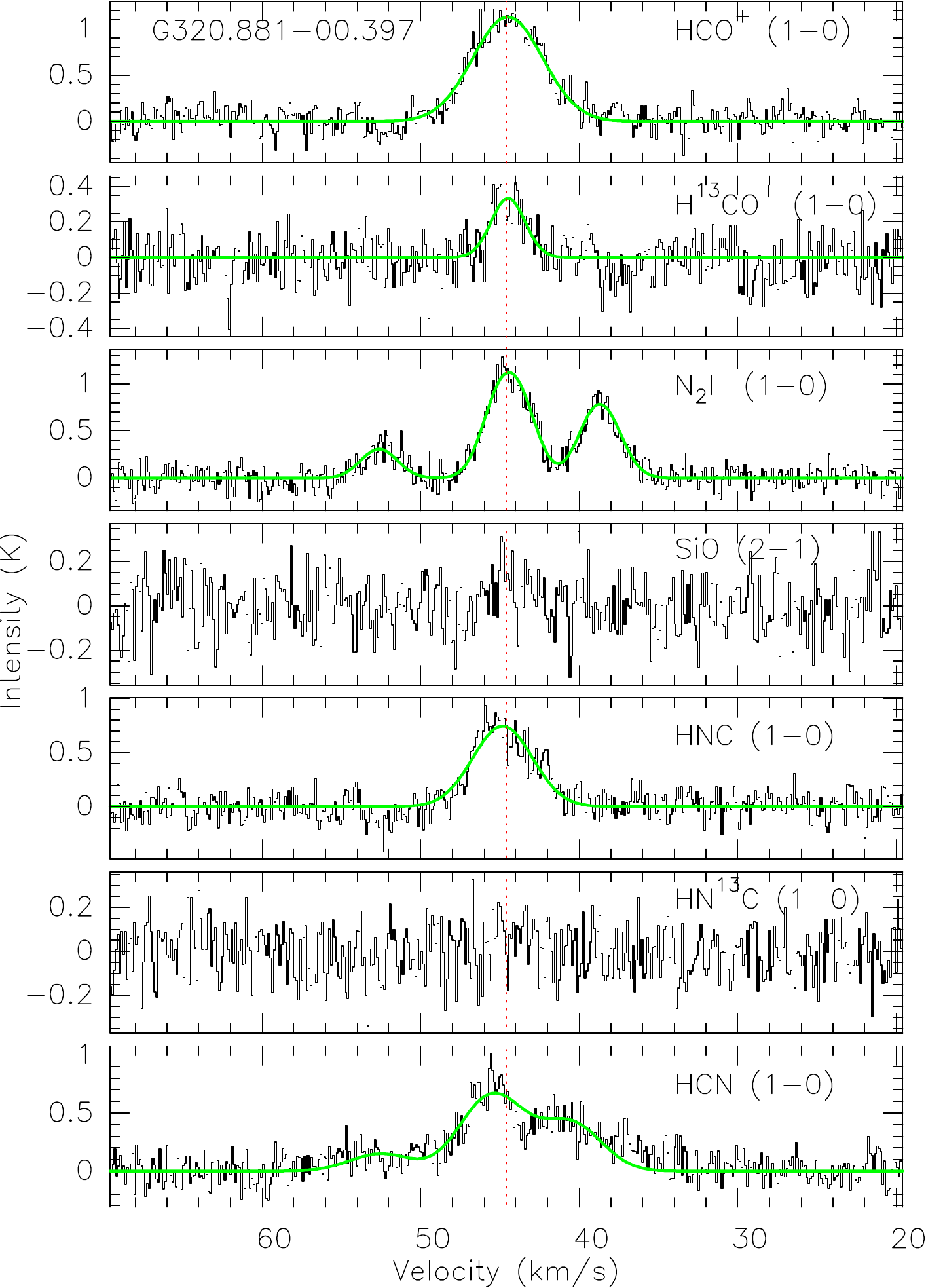}
\end{figure*}
\addtocounter{figure}{-1}
\begin{figure*}
\centering
\caption{Continued}
\includegraphics[width=0.80\textwidth]{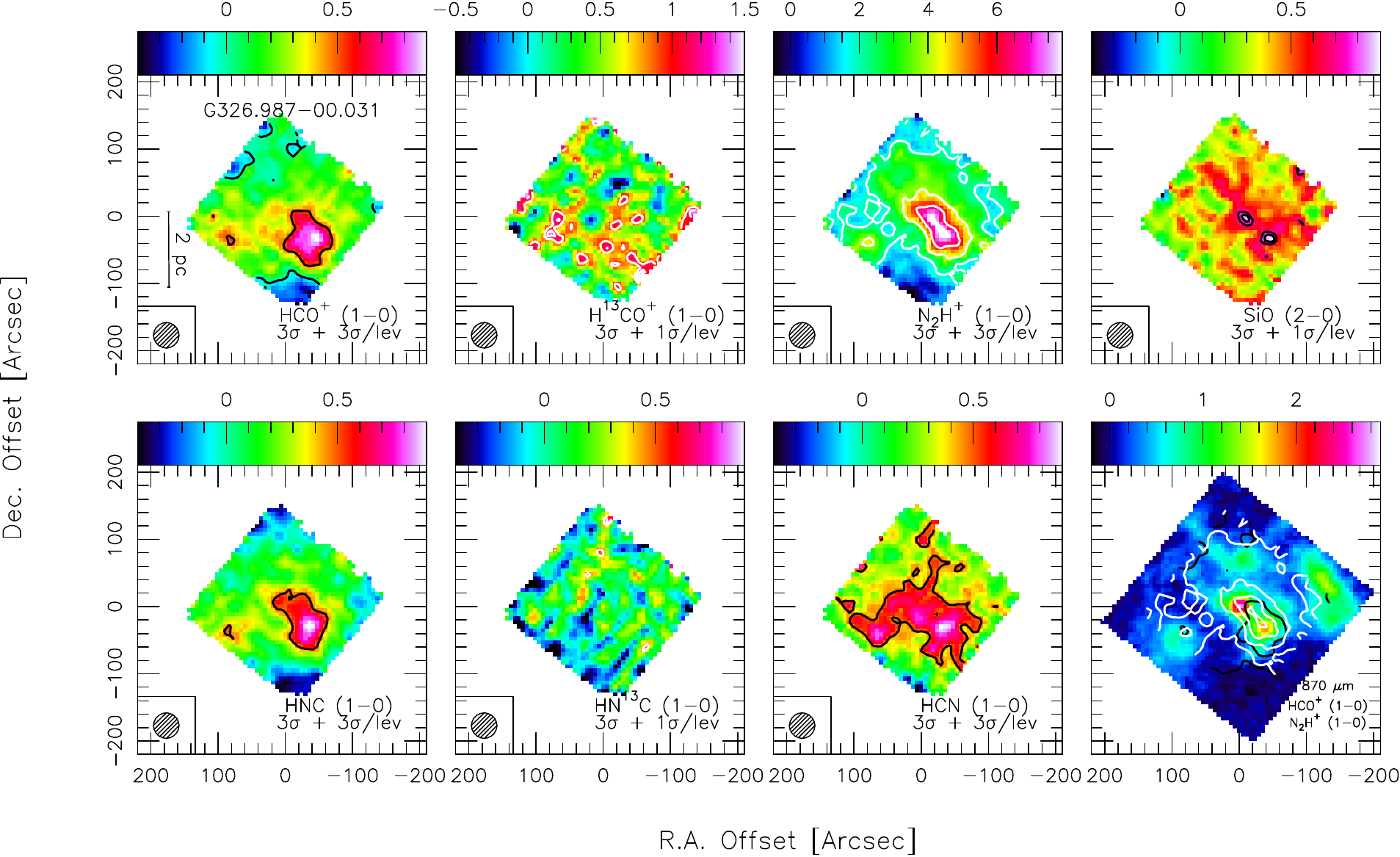}\vspace{0.5cm}
\includegraphics[width=0.50\textwidth]{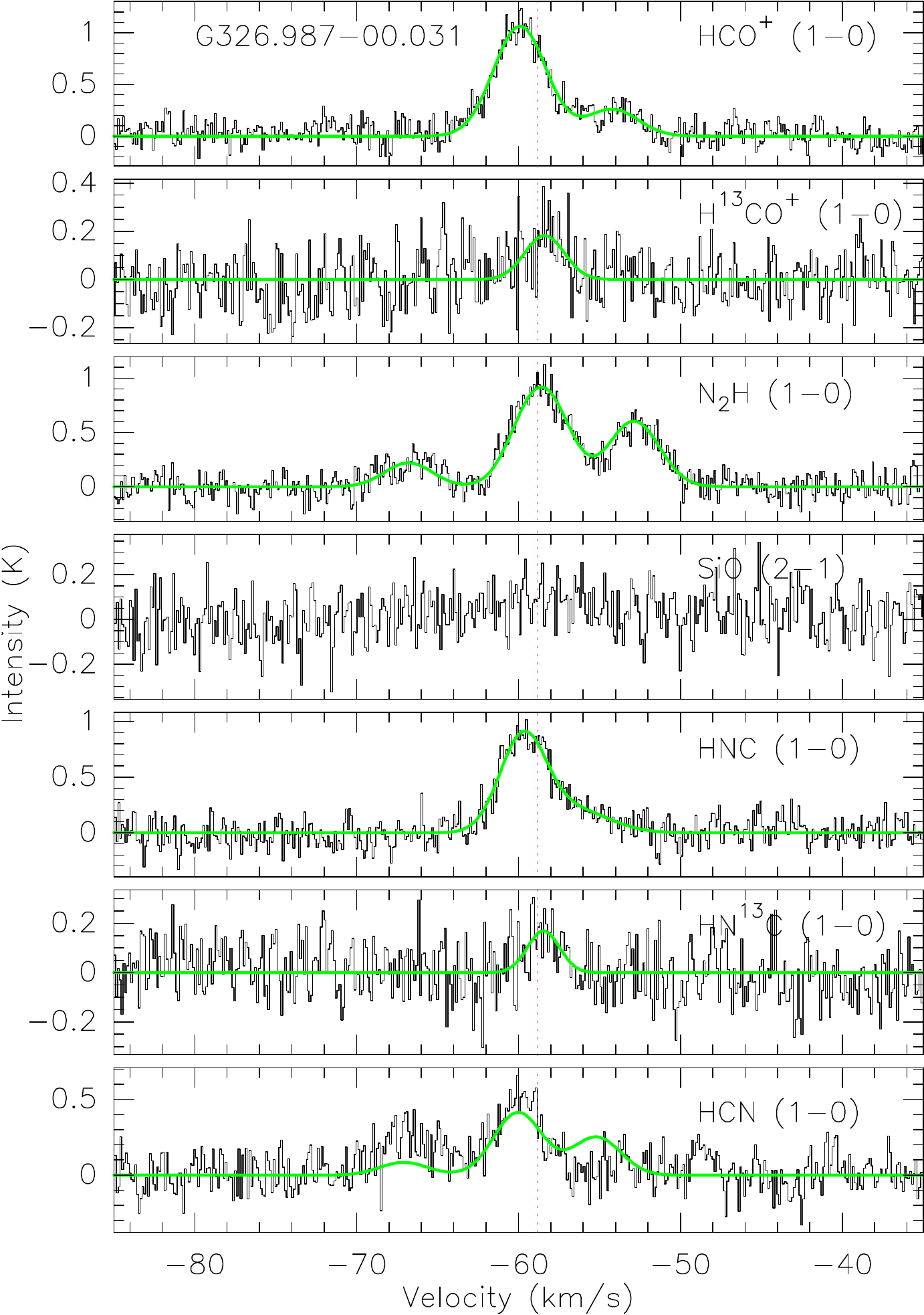}
\end{figure*}
\addtocounter{figure}{-1}
\begin{figure*}
\centering
\caption{Continued}
\includegraphics[width=0.80\textwidth]{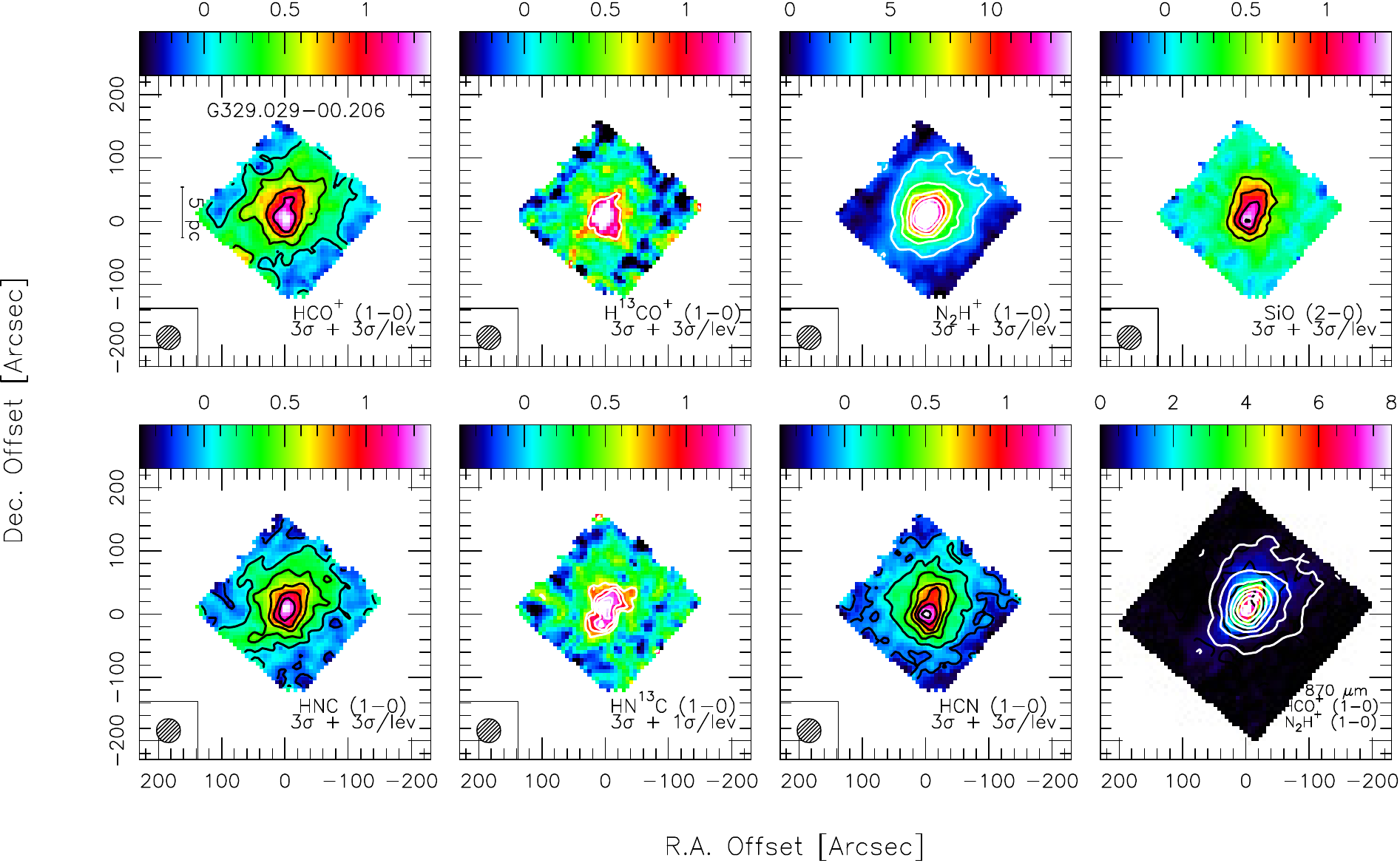}\vspace{0.5cm}
\includegraphics[width=0.50\textwidth]{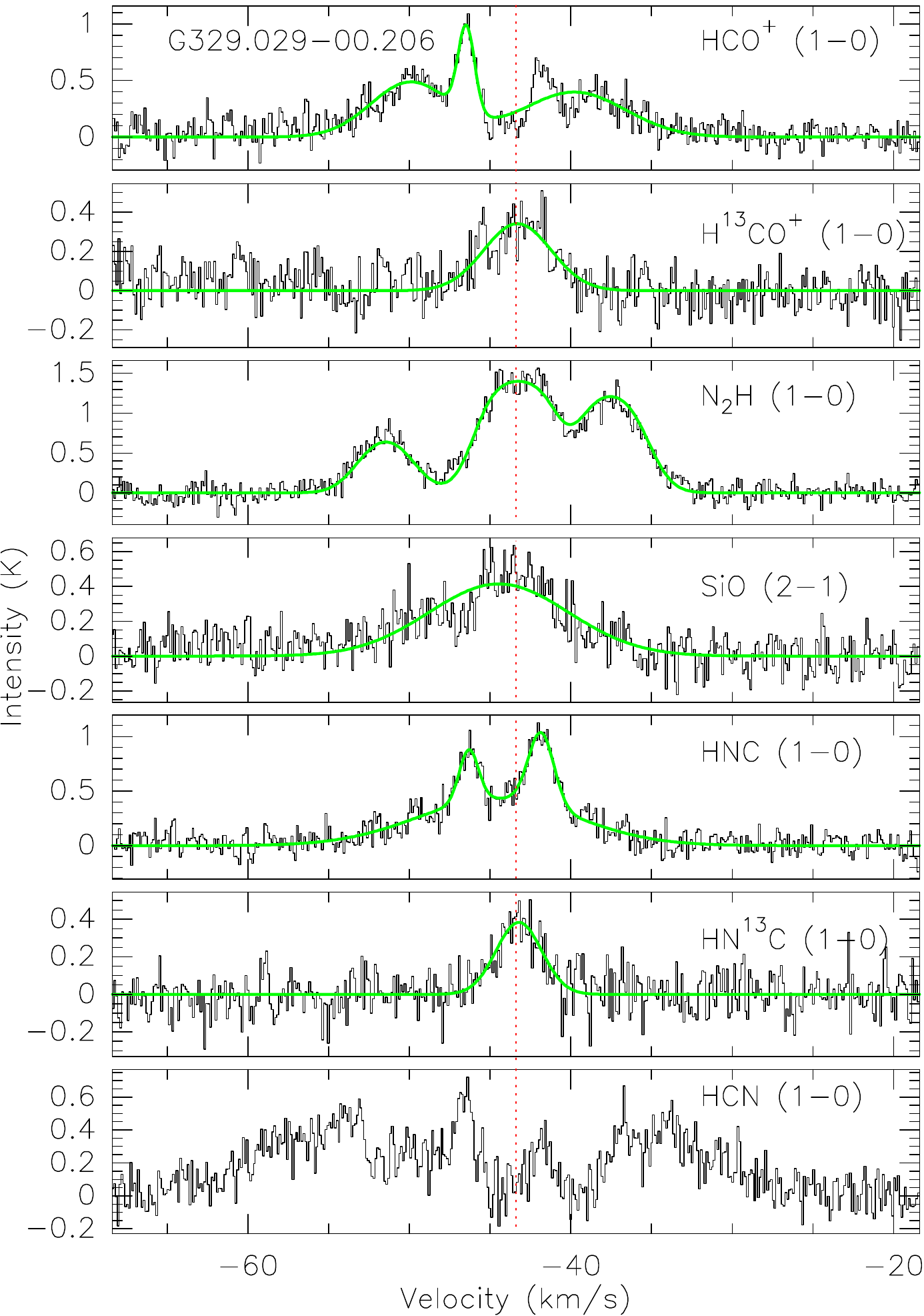}
\end{figure*}
\addtocounter{figure}{-1}
\begin{figure*}
\caption{Continued}
\centering
\includegraphics[width=0.80\textwidth]{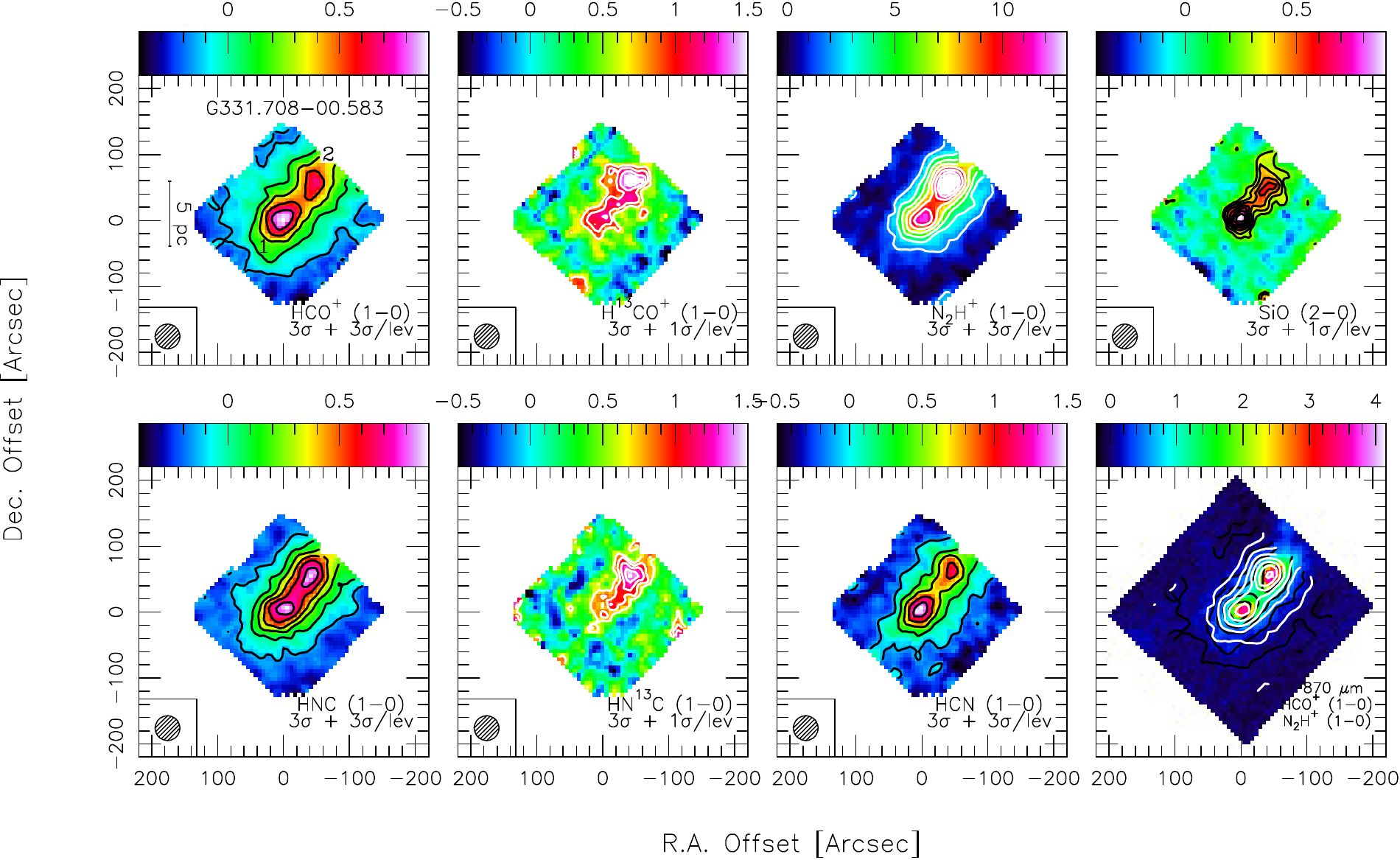}\vspace{0.5cm}
%
%
\centering
\includegraphics[width=0.45\textwidth]{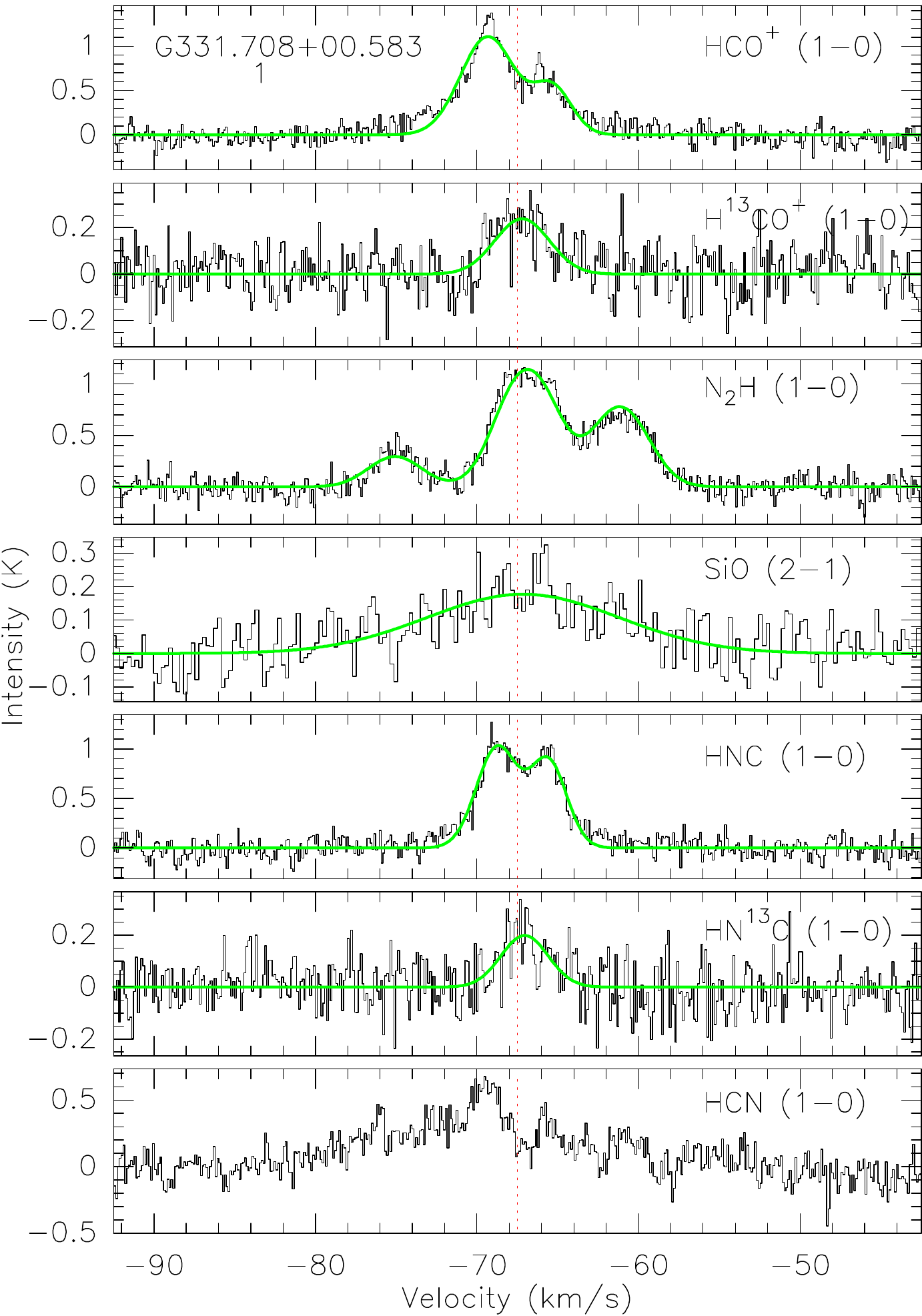}\hspace{0.5cm}
\includegraphics[width=0.45\textwidth]{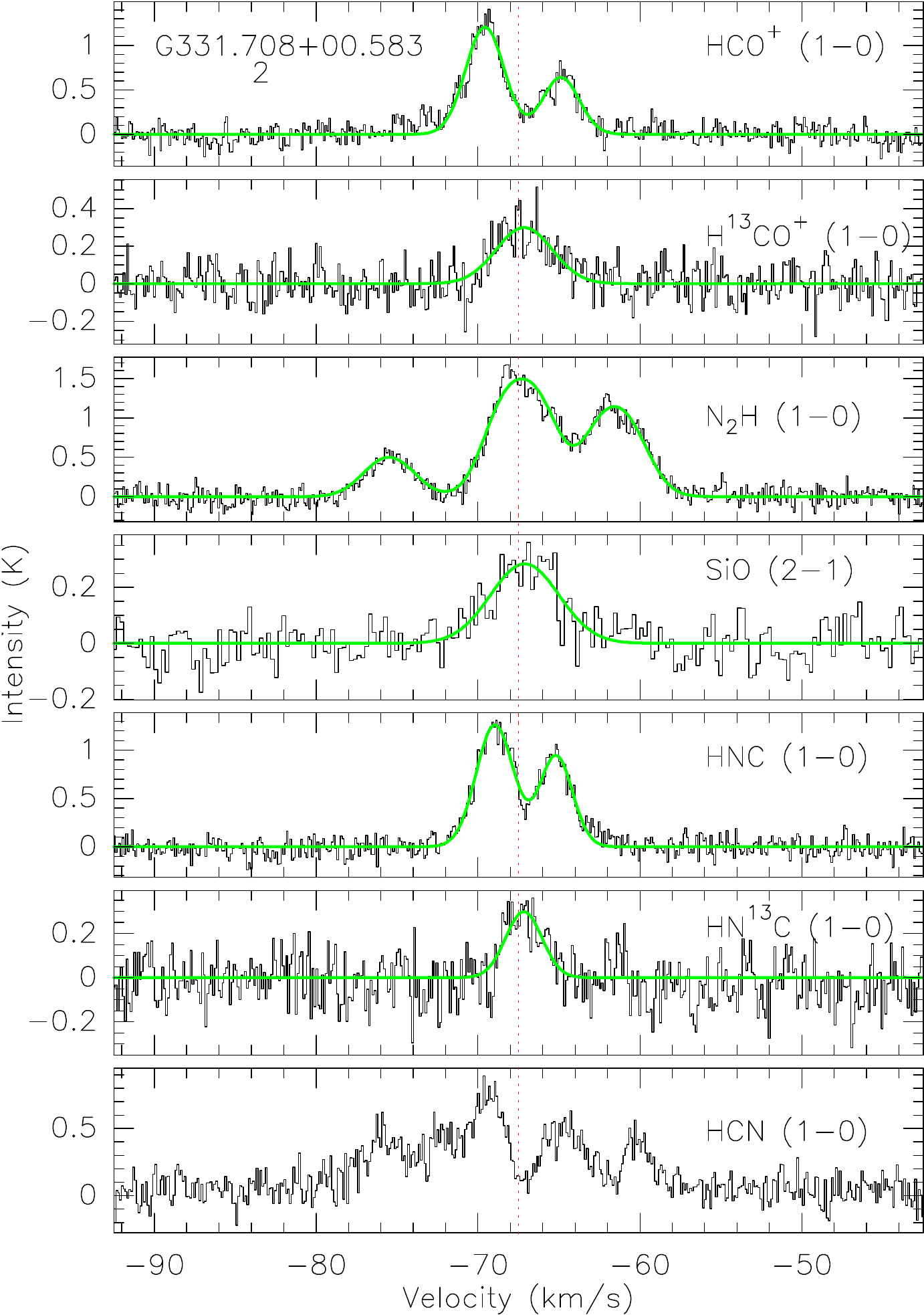}
\end{figure*}
\addtocounter{figure}{-1}

\begin{figure*}
\centering
\caption{Continued}
\includegraphics[width=0.80\textwidth]{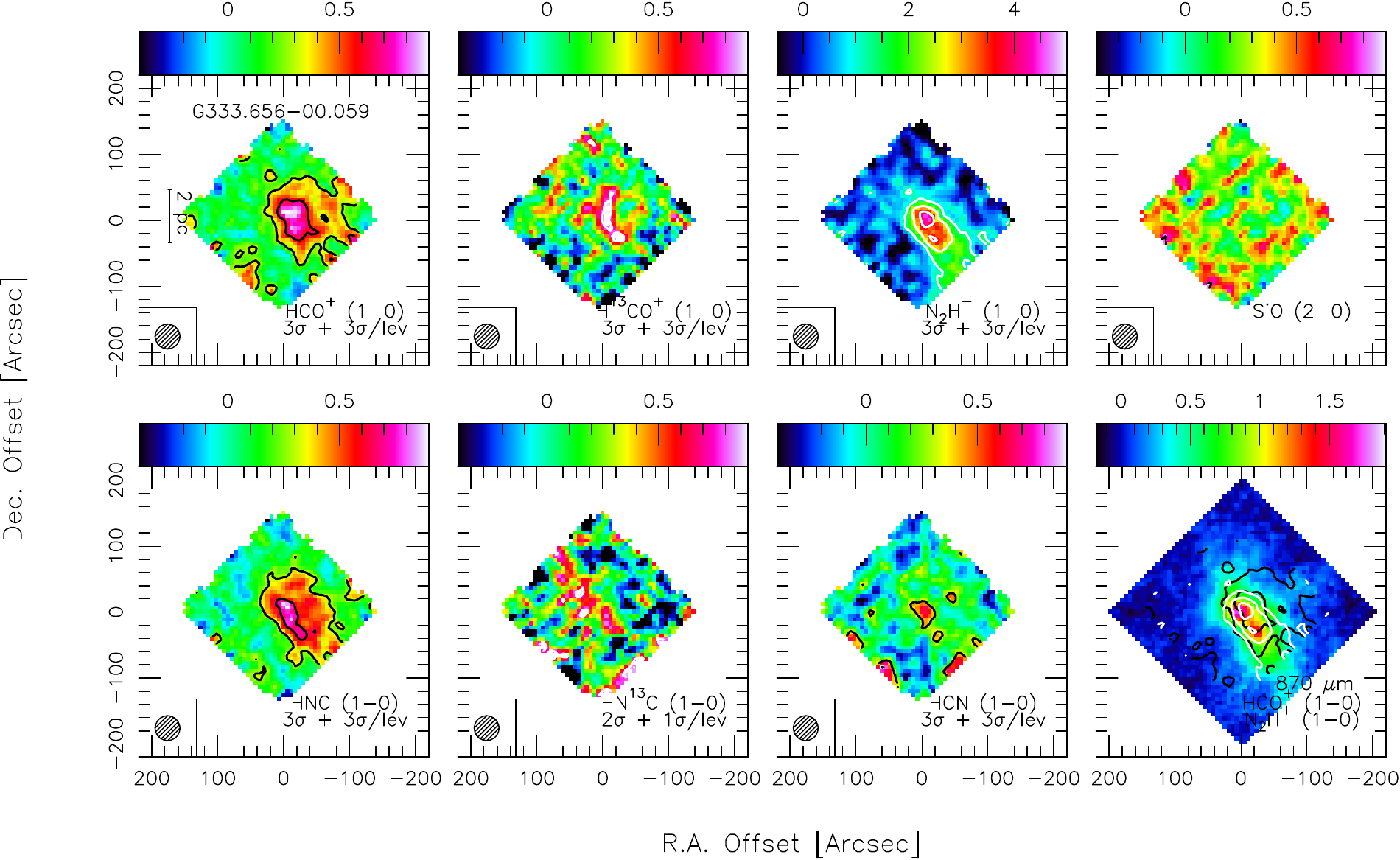}\vspace{0.5cm}
\includegraphics[width=0.50\textwidth]{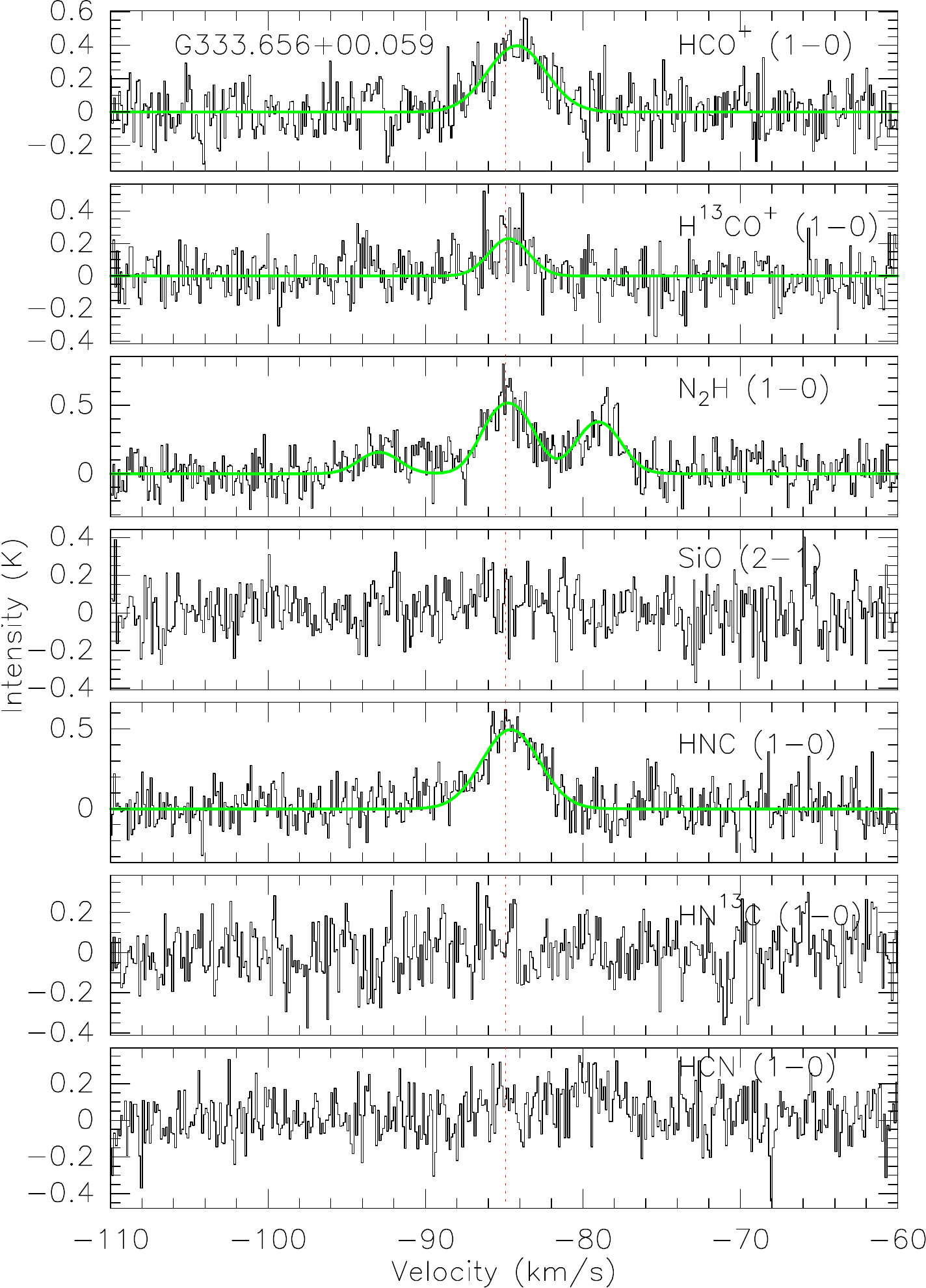}
\end{figure*}%

\addtocounter{figure}{-1}
\begin{figure*}
\centering
\caption{Continued}
\includegraphics[width=0.80\textwidth]{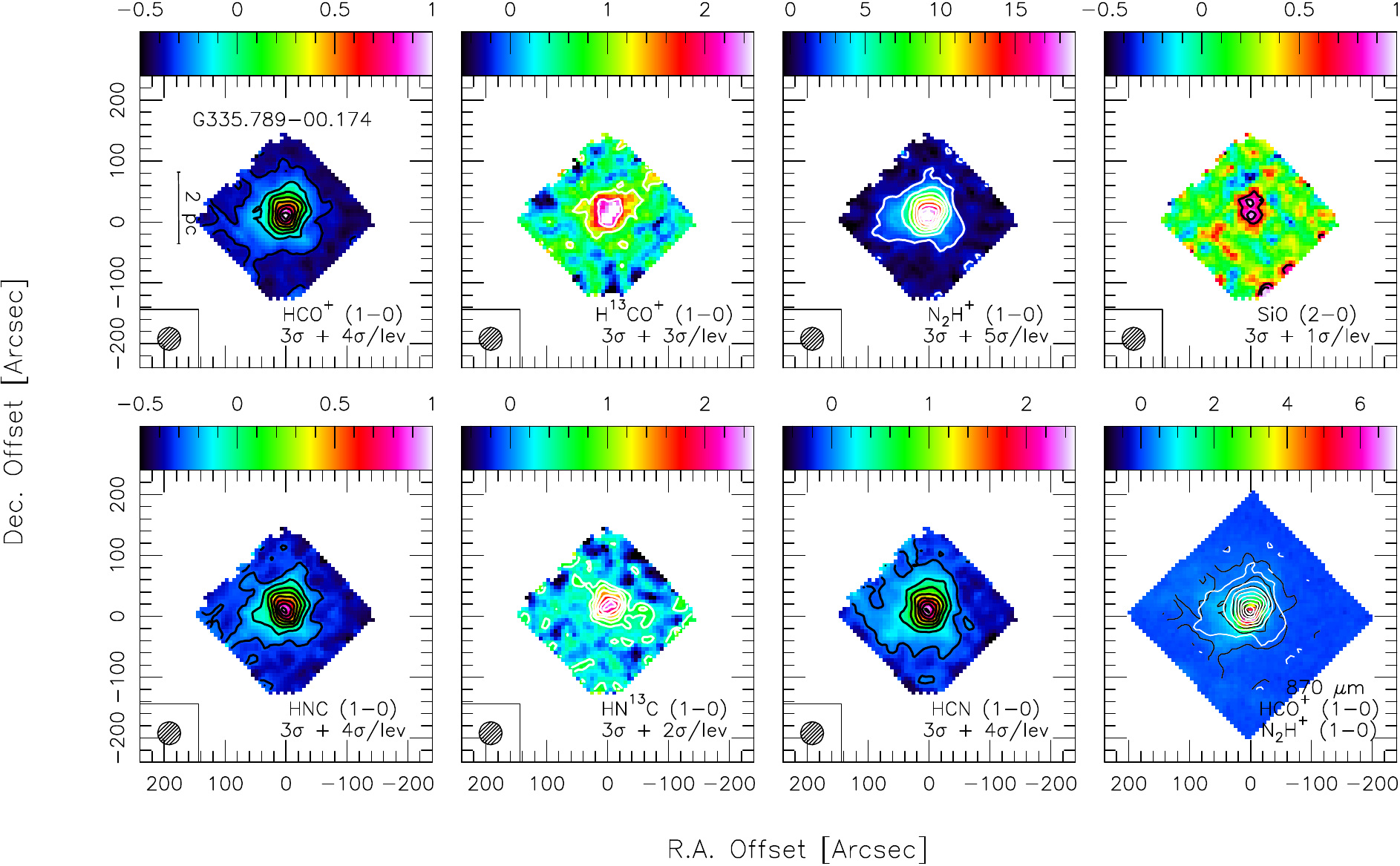}\vspace{0.5cm}
\includegraphics[width=0.50\textwidth]{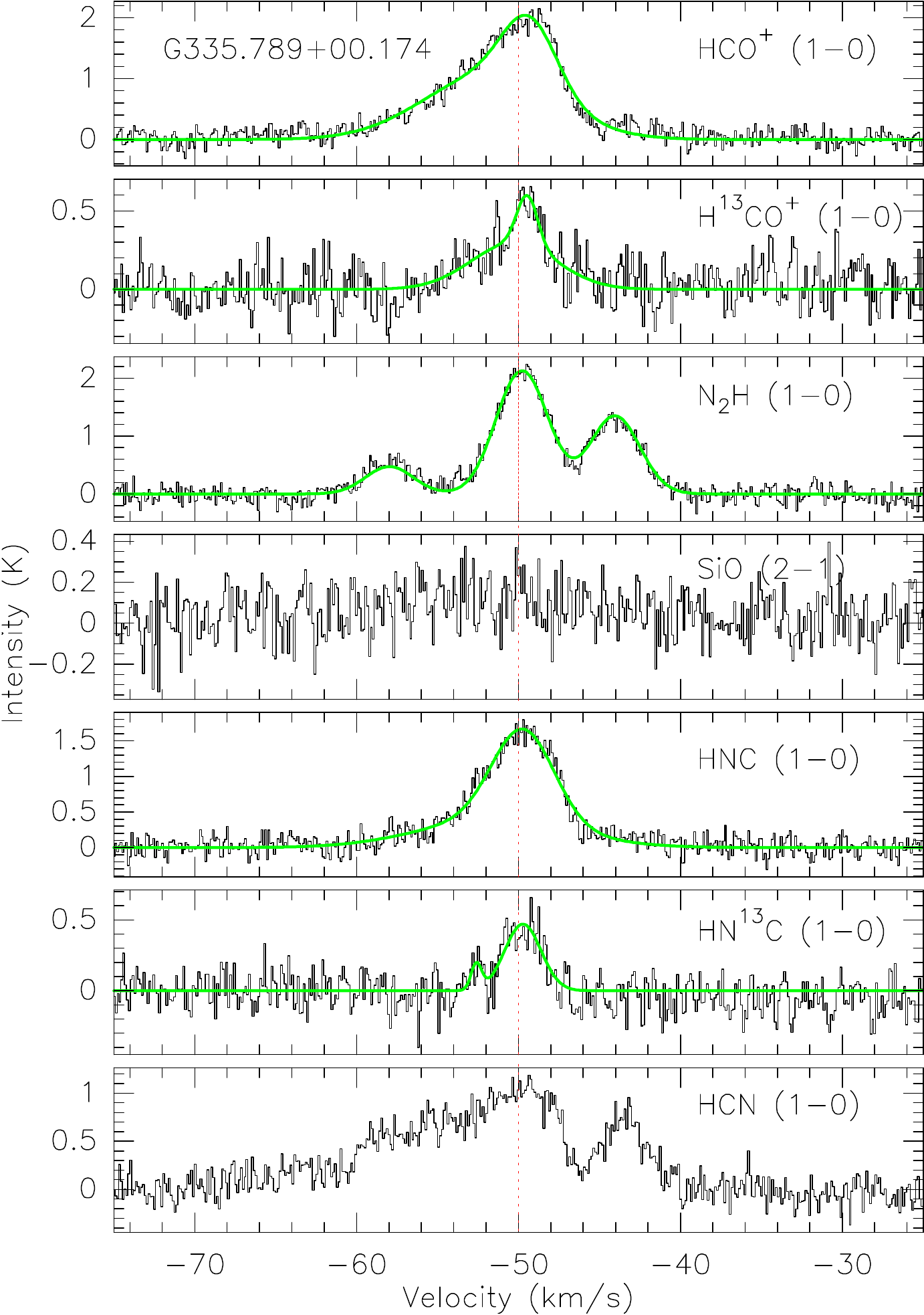}
\end{figure*}
\addtocounter{figure}{-1}

\begin{figure*}
\centering
\caption{Continued}
\includegraphics[width=0.80\textwidth]{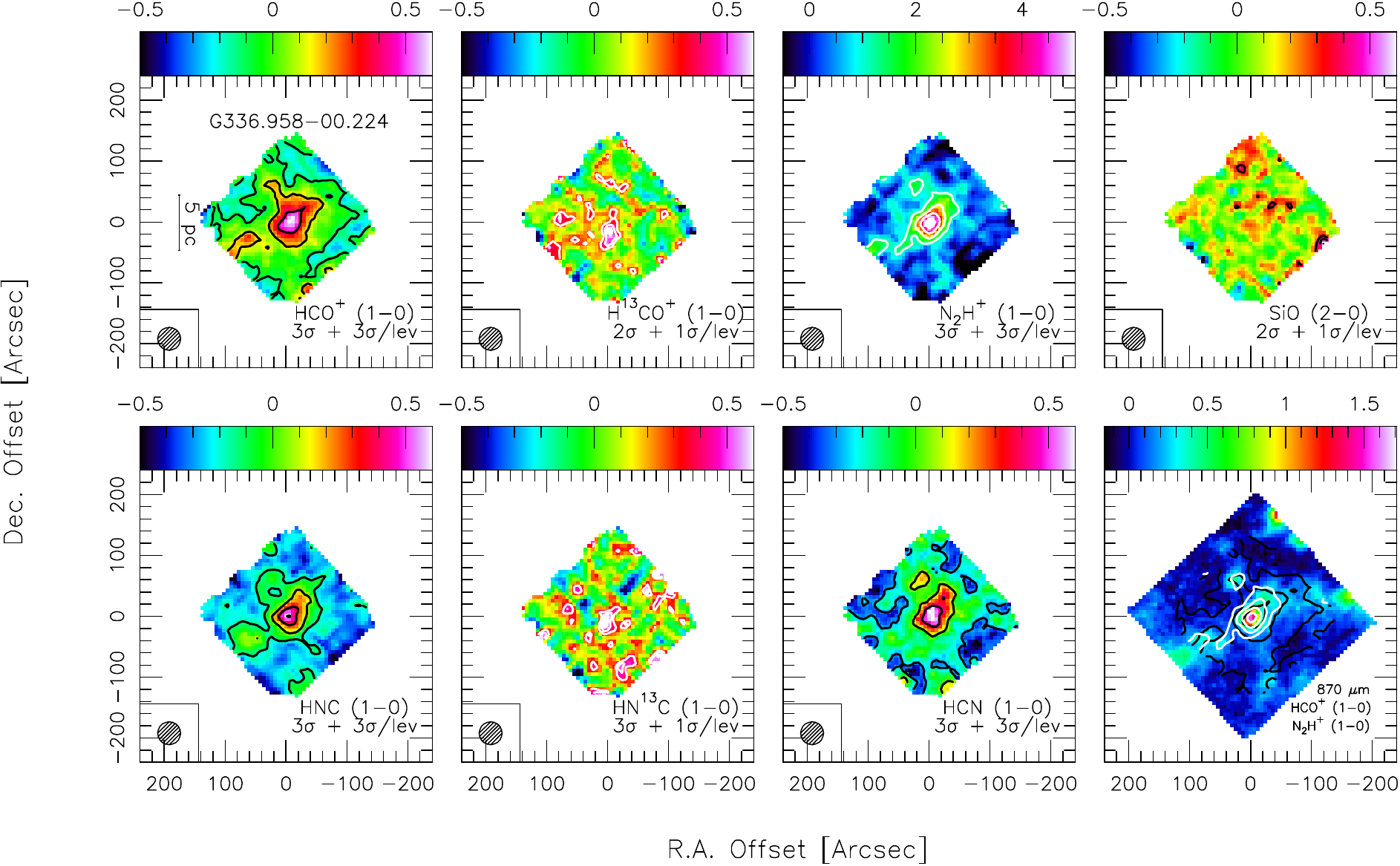}\vspace{0.5cm}
\includegraphics[width=0.50\textwidth]{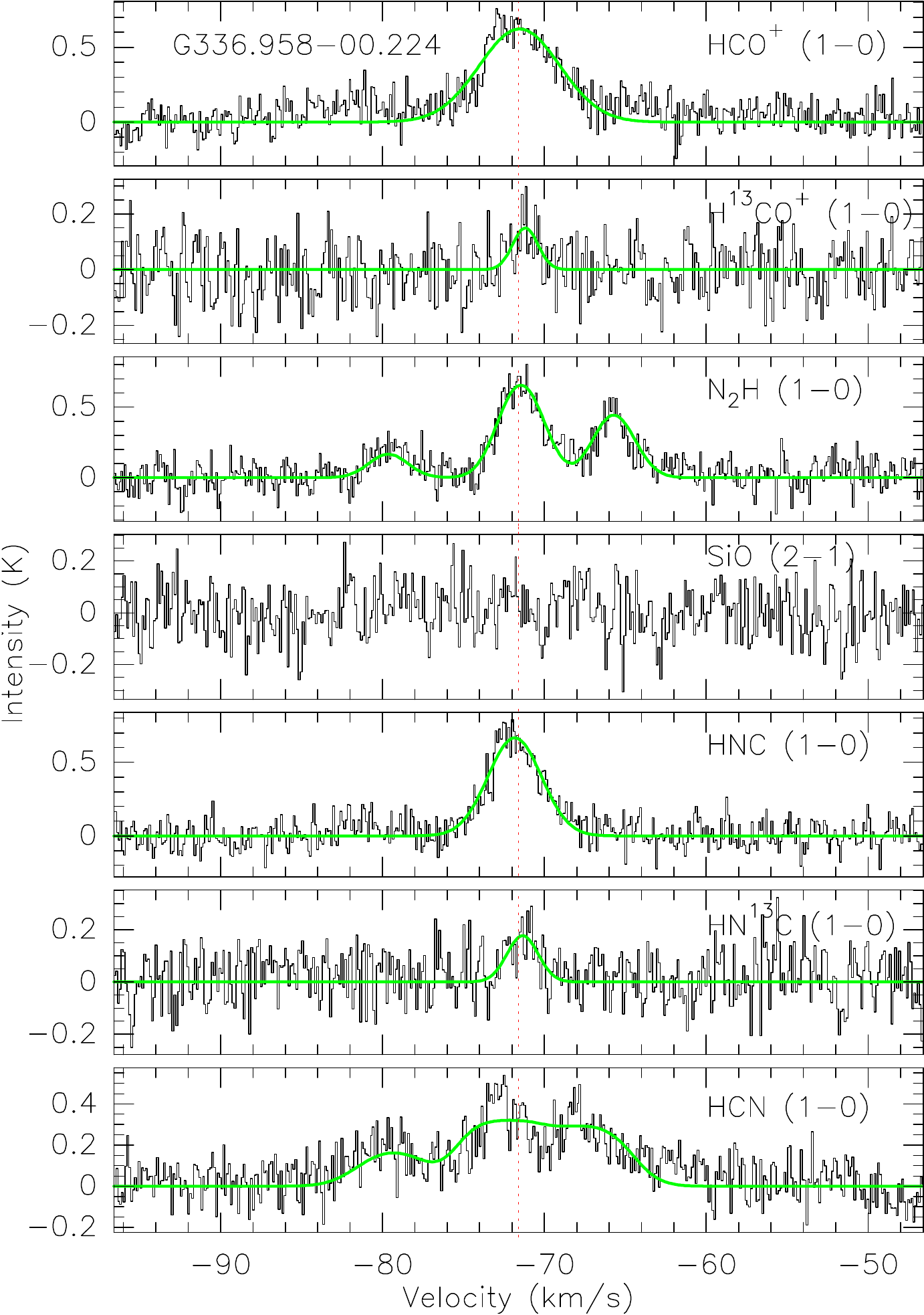}
\end{figure*}
\addtocounter{figure}{-1}
\begin{figure*}
\centering
\caption{Continued}
\includegraphics[width=0.80\textwidth]{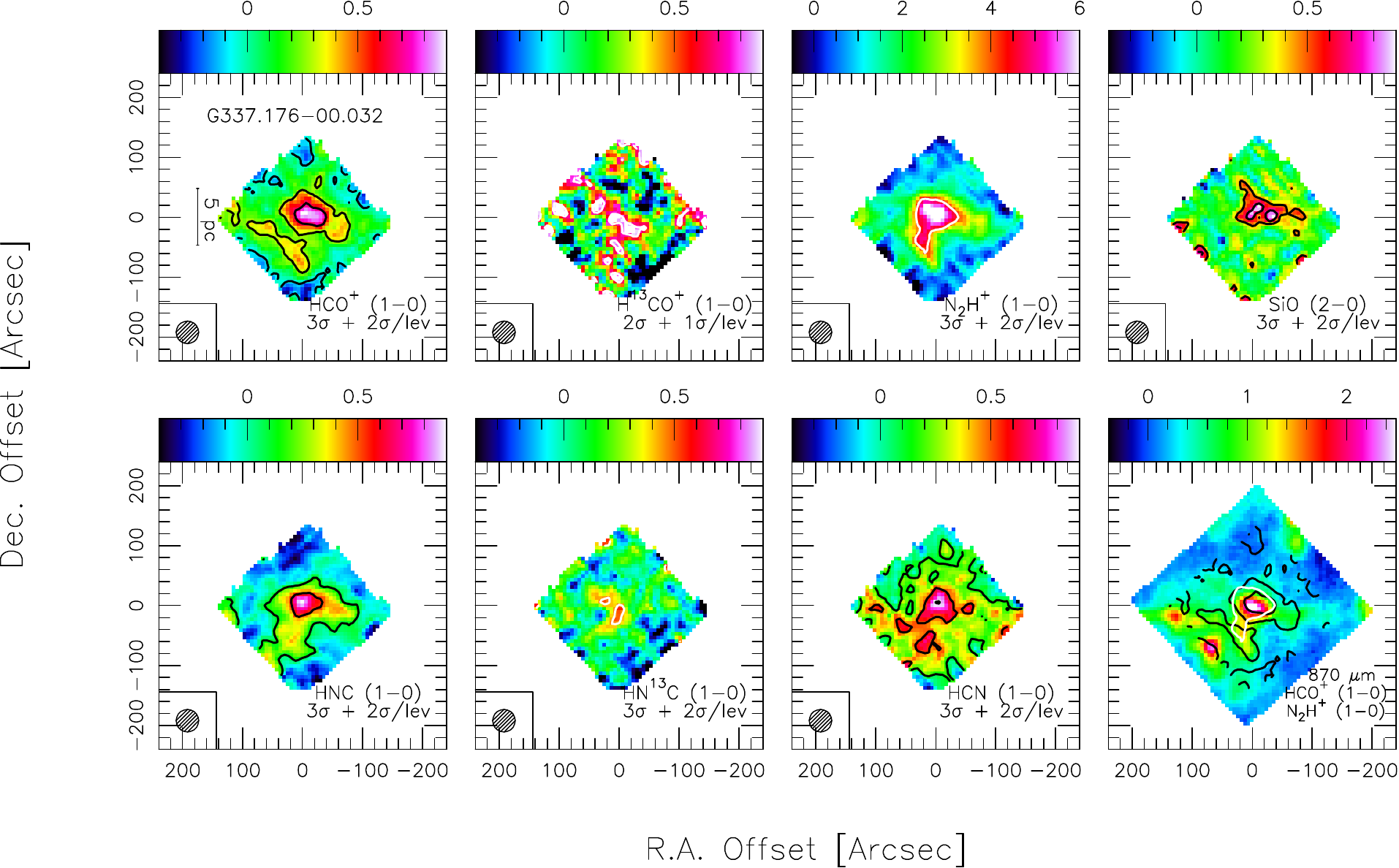}\vspace{0.5cm}
\includegraphics[width=0.50\textwidth]{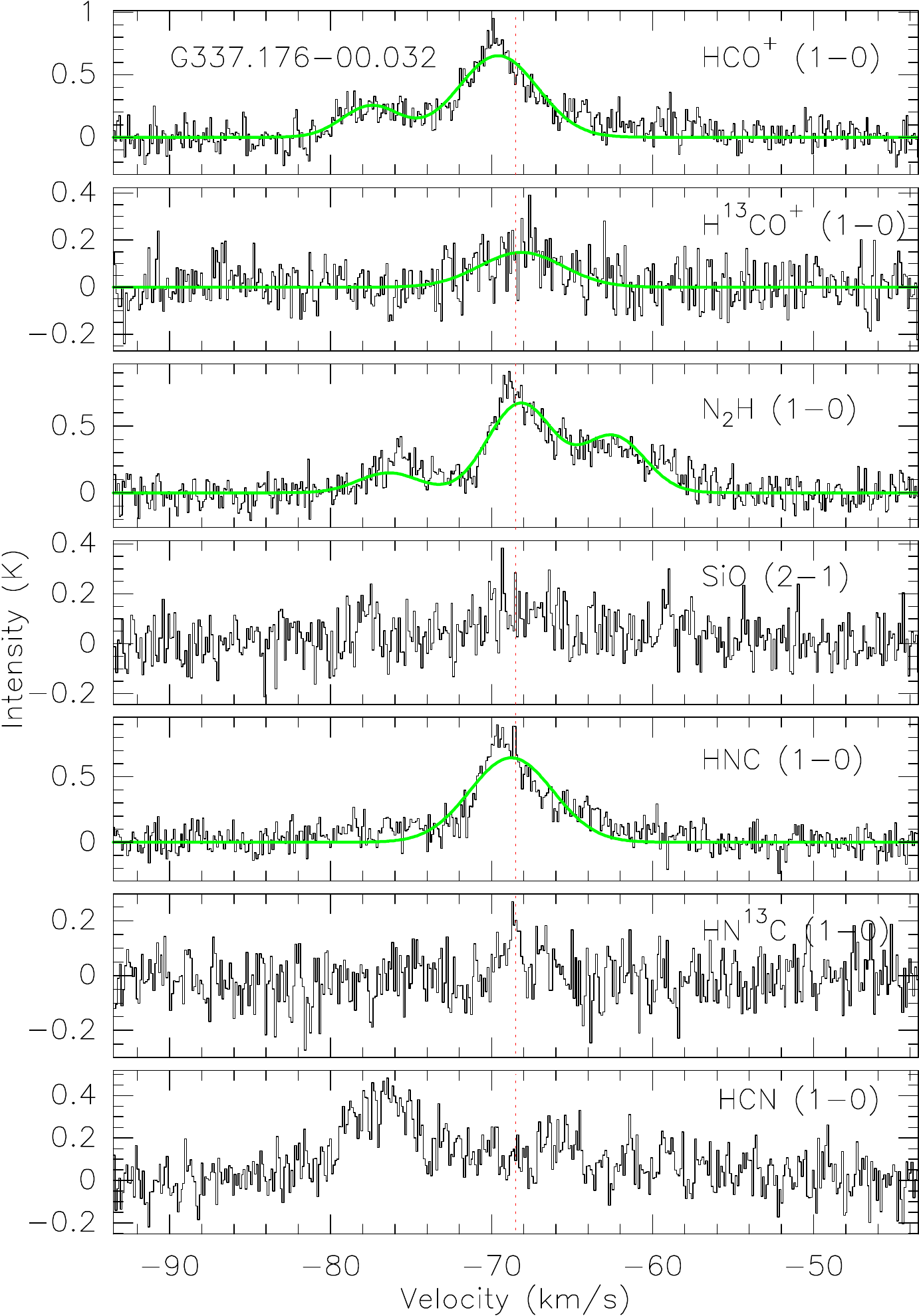}
\end{figure*}
\addtocounter{figure}{-1}
\begin{figure*}
\centering
\caption{Continued}
\includegraphics[width=0.80\textwidth]{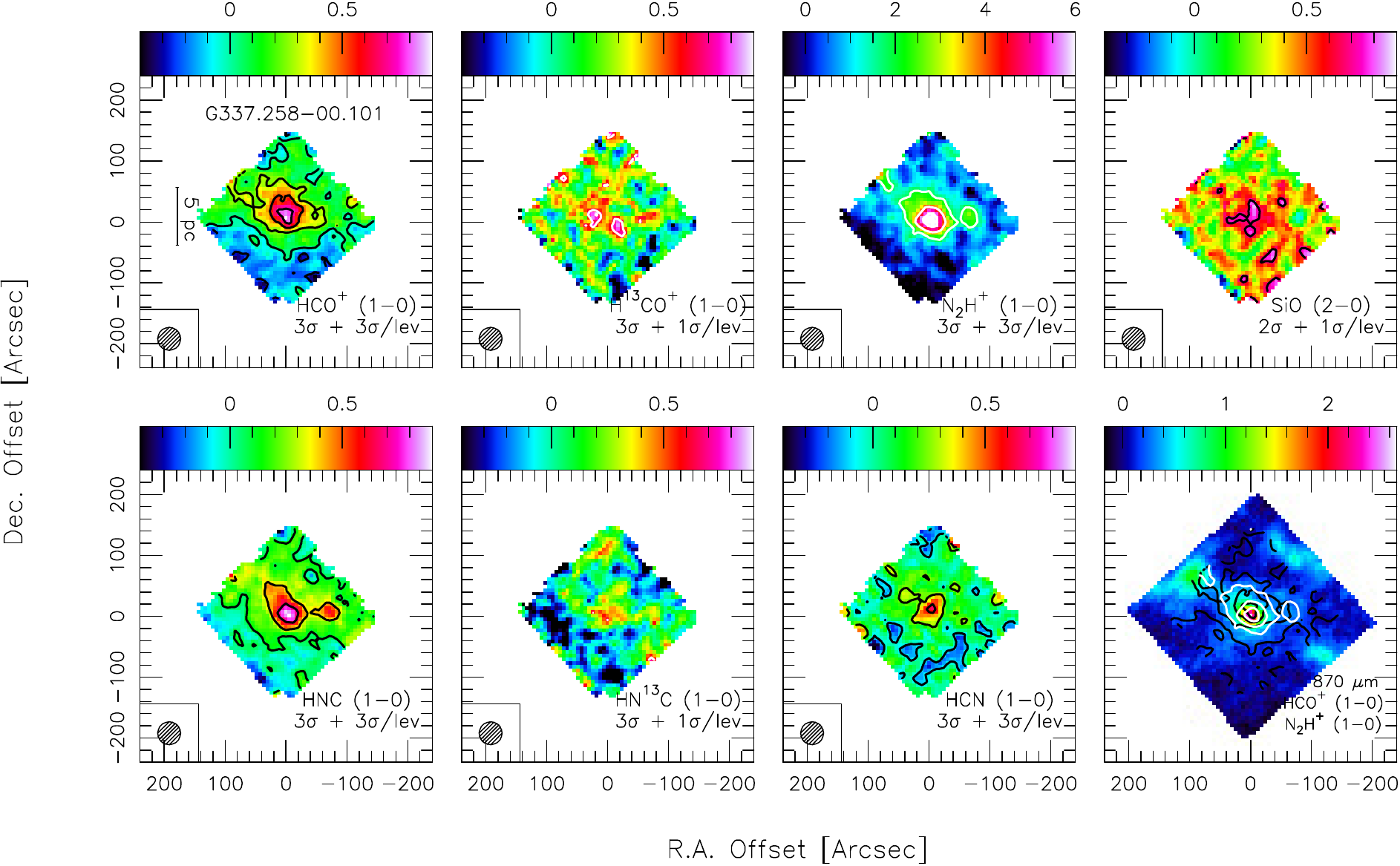}\vspace{0.5cm}
\includegraphics[width=0.50\textwidth]{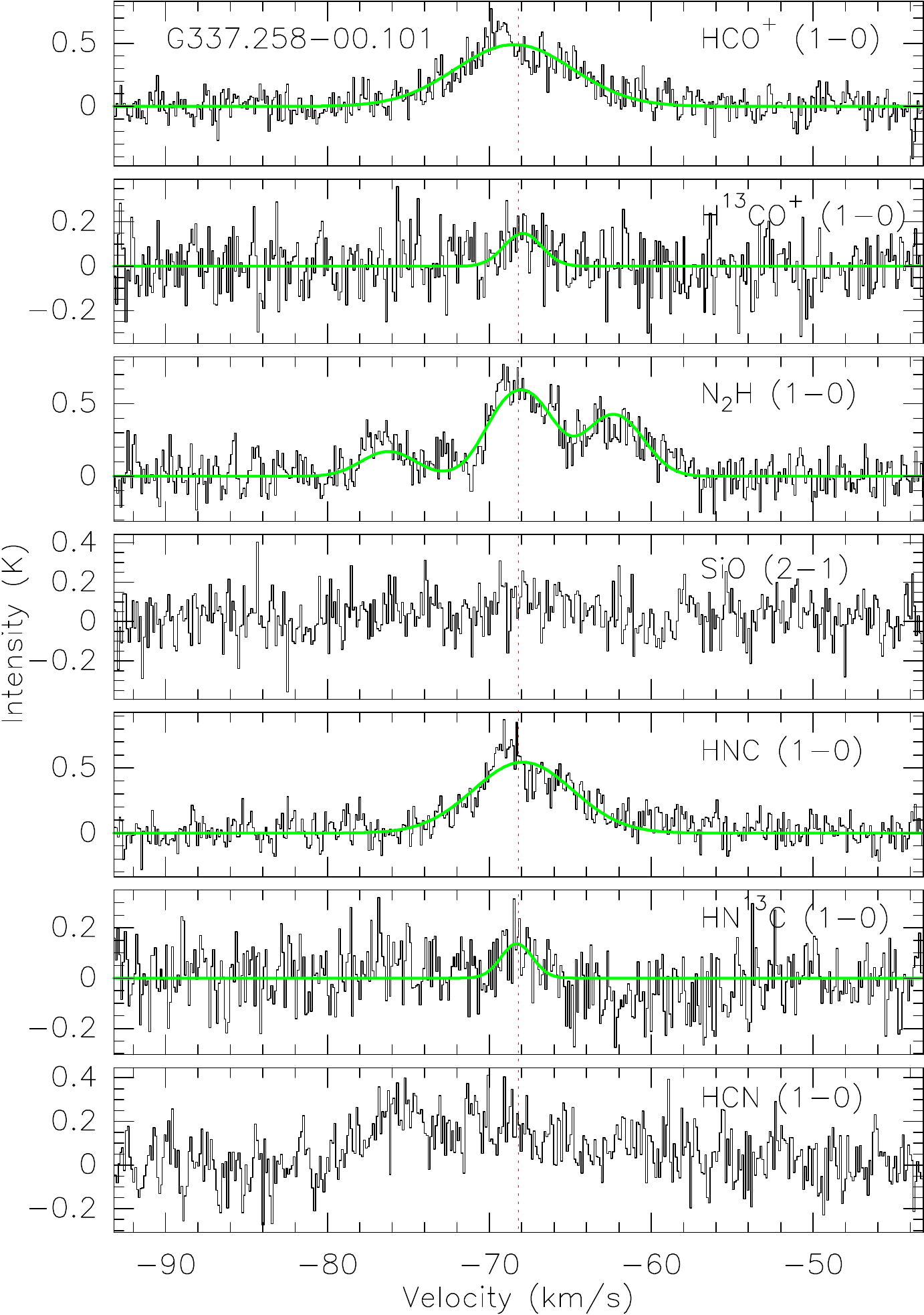}
\end{figure*}
\addtocounter{figure}{-1}
\begin{figure*}
\centering
\caption{Continued}
\includegraphics[width=0.80\textwidth]{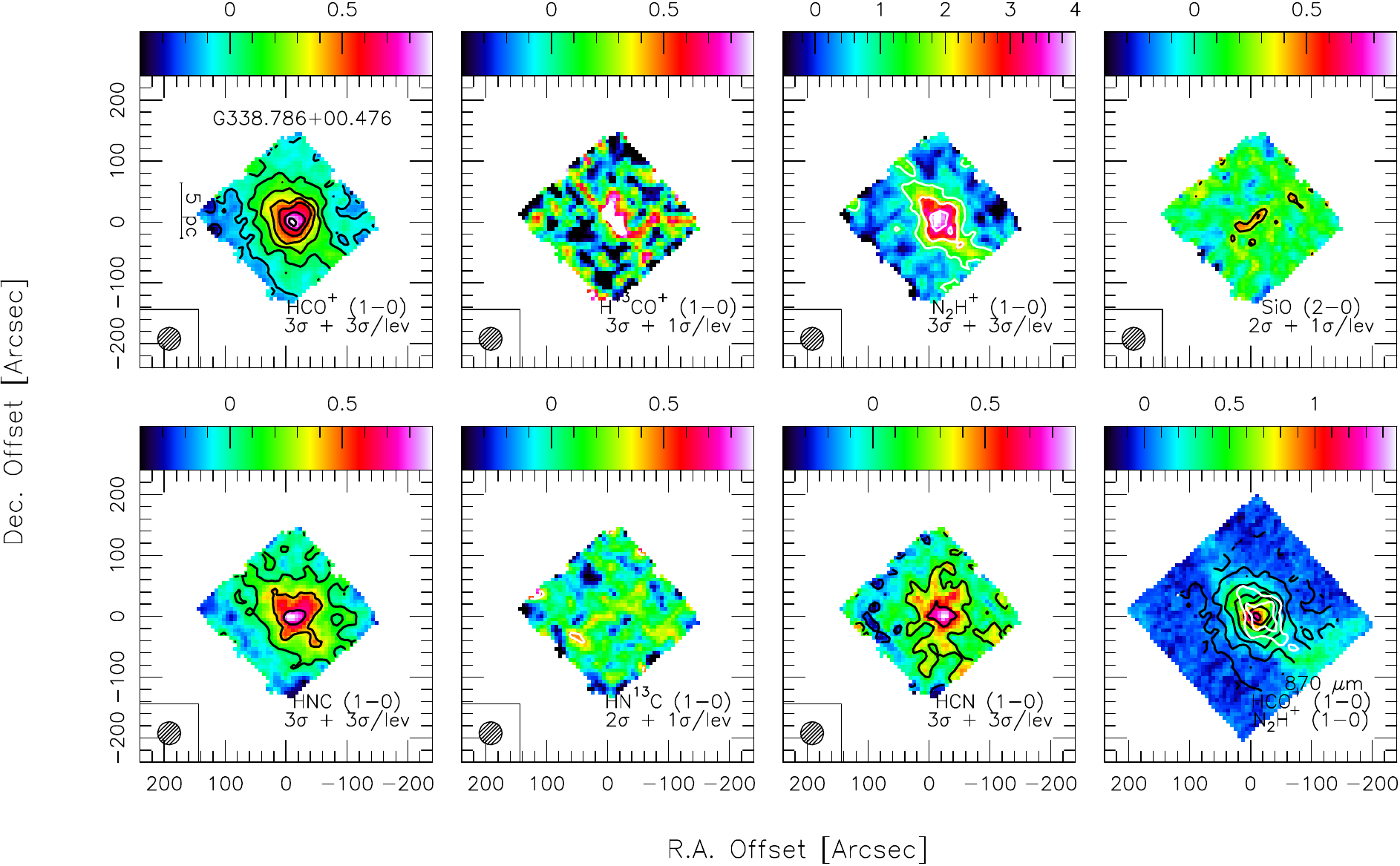}\vspace{0.5cm}
\includegraphics[width=0.50\textwidth]{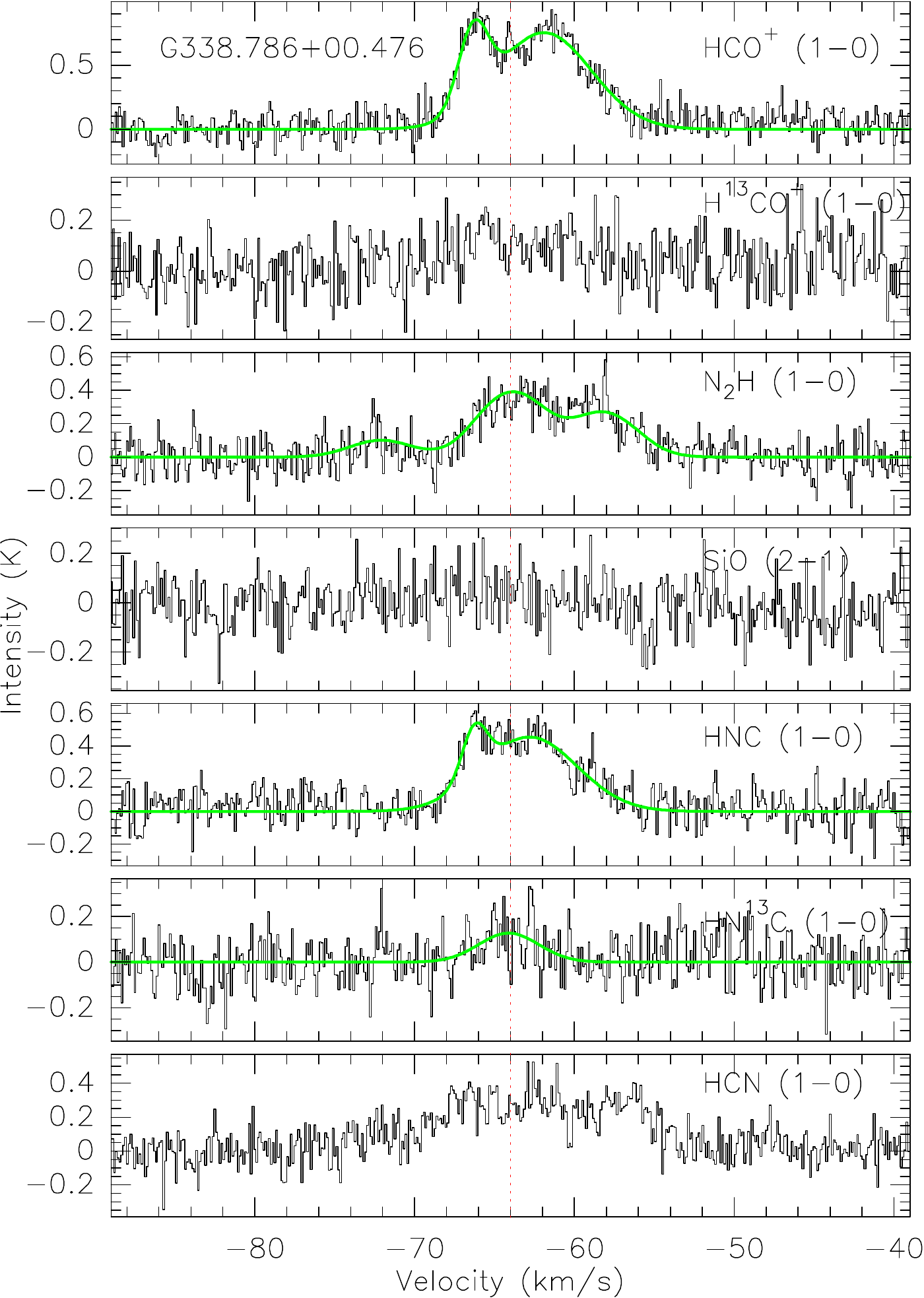}
\end{figure*}
\addtocounter{figure}{-1}
\begin{figure*}
\centering
\caption{Continued}
\includegraphics[width=0.80\textwidth]{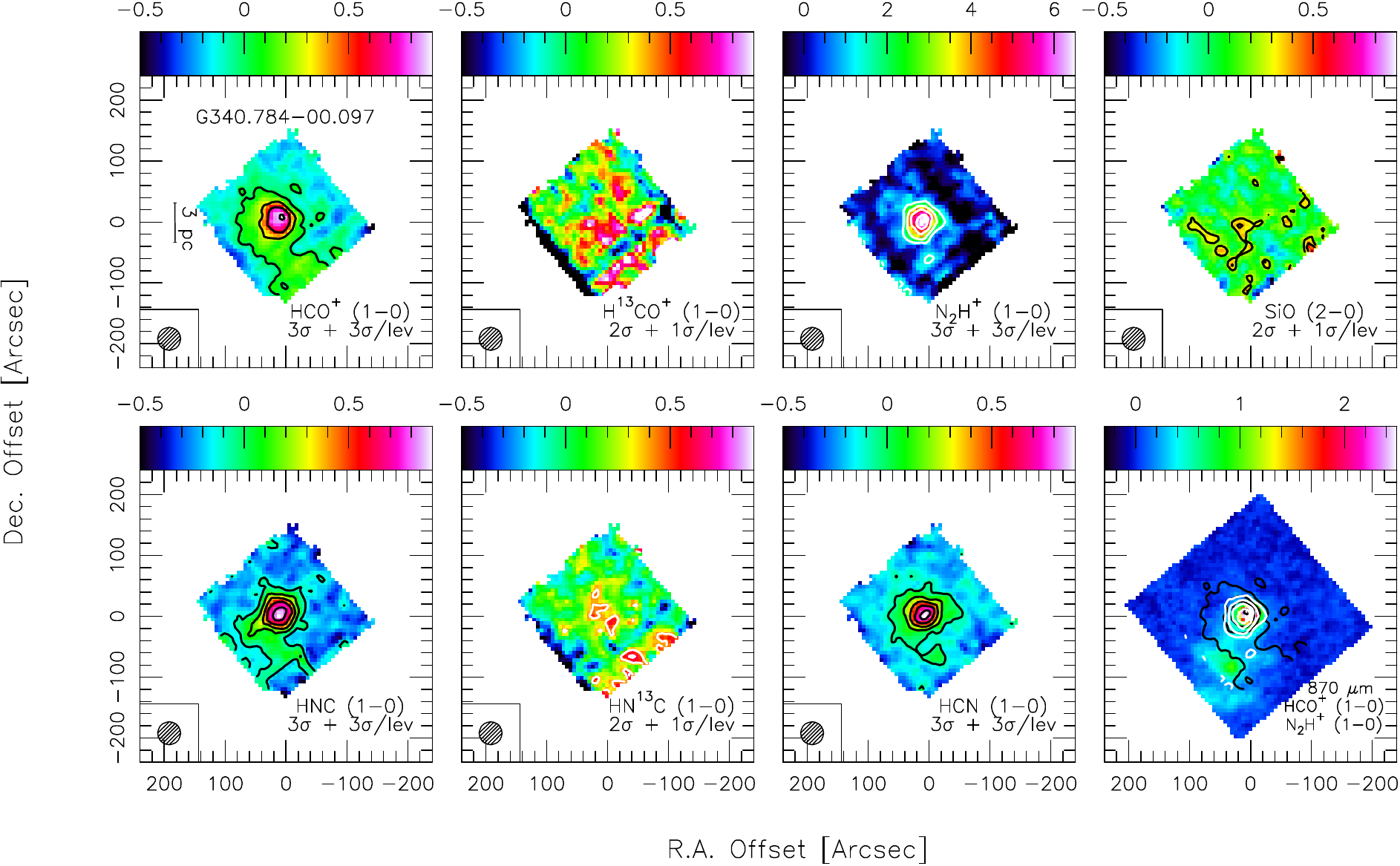}\vspace{0.5cm}
\includegraphics[width=0.50\textwidth]{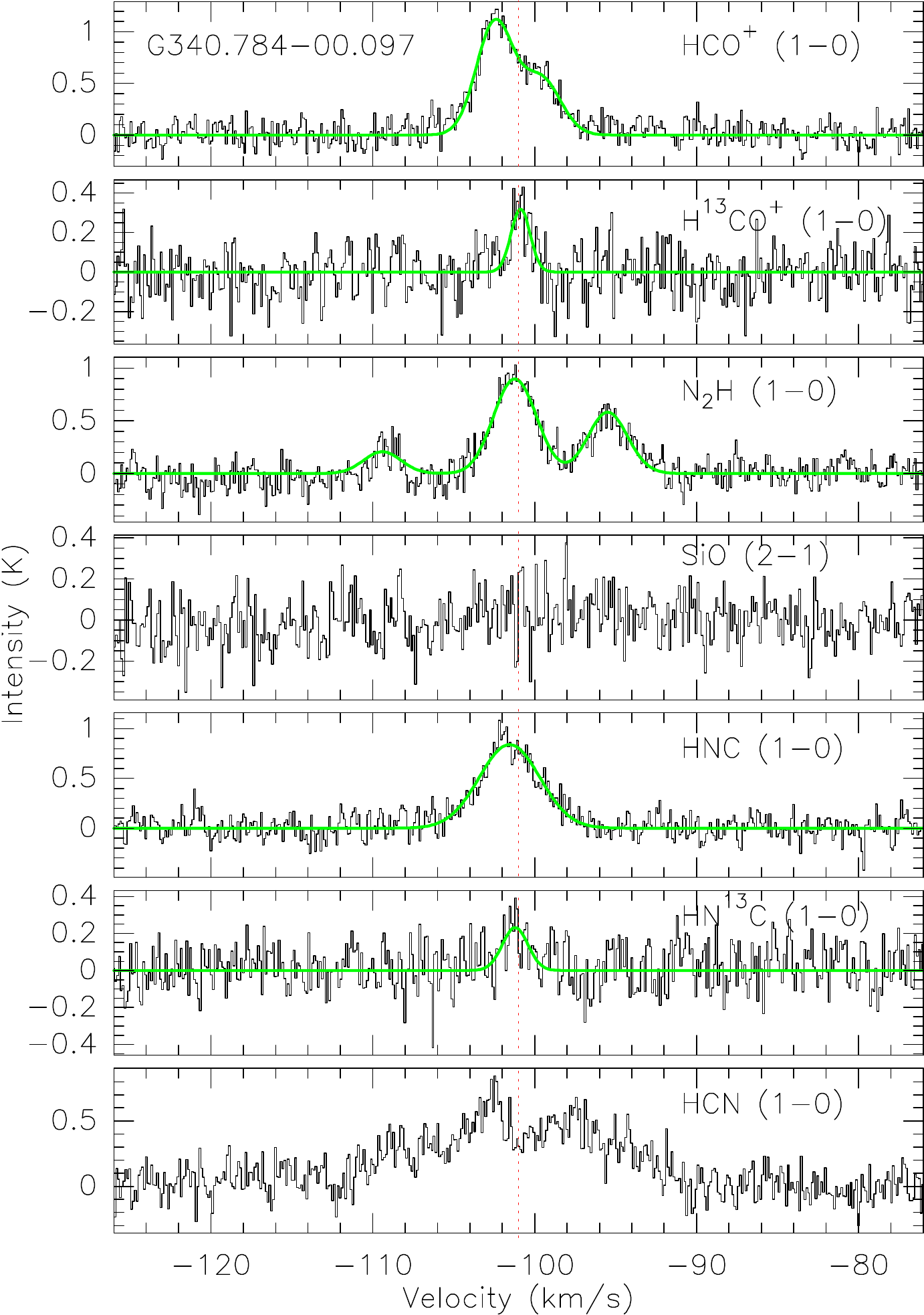}
\end{figure*}
\addtocounter{figure}{-1}
\begin{figure*}
\centering
\caption{Continued}
\includegraphics[width=0.80\textwidth]{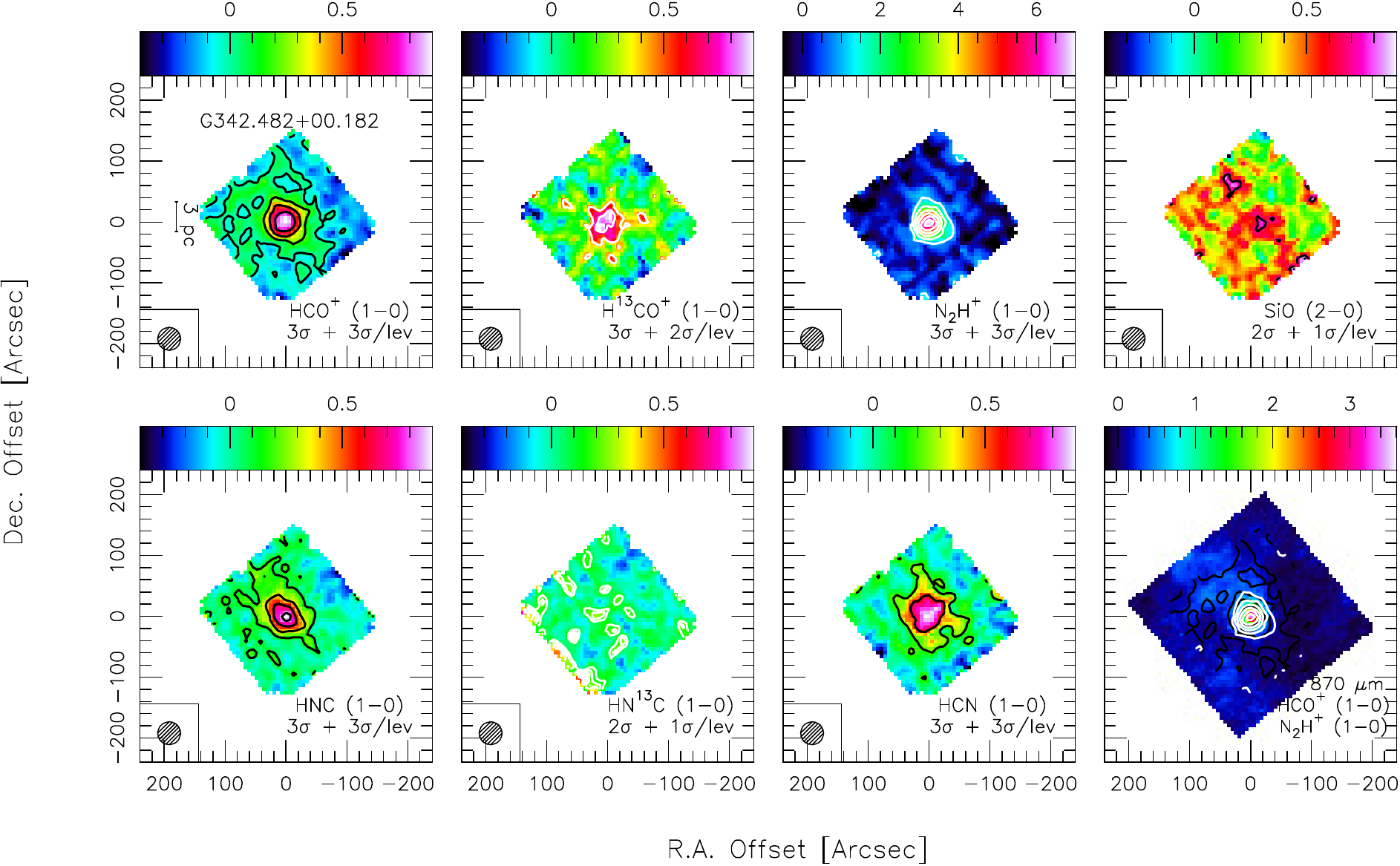}\vspace{0.5cm}
\includegraphics[width=0.50\textwidth]{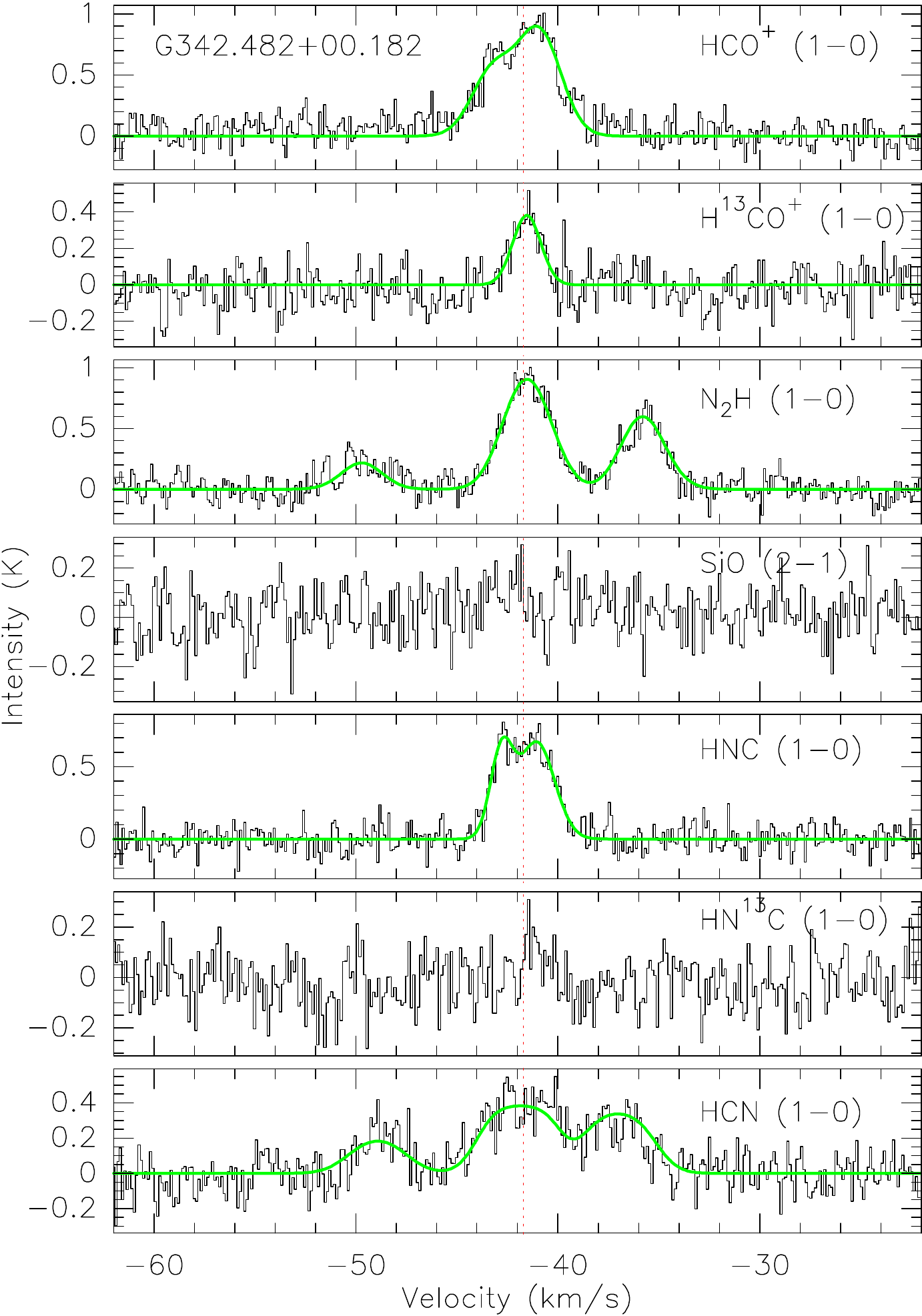}
\end{figure*}
\addtocounter{figure}{-1}
\begin{figure*}
\centering
\caption{Continued}
\includegraphics[width=0.80\textwidth]{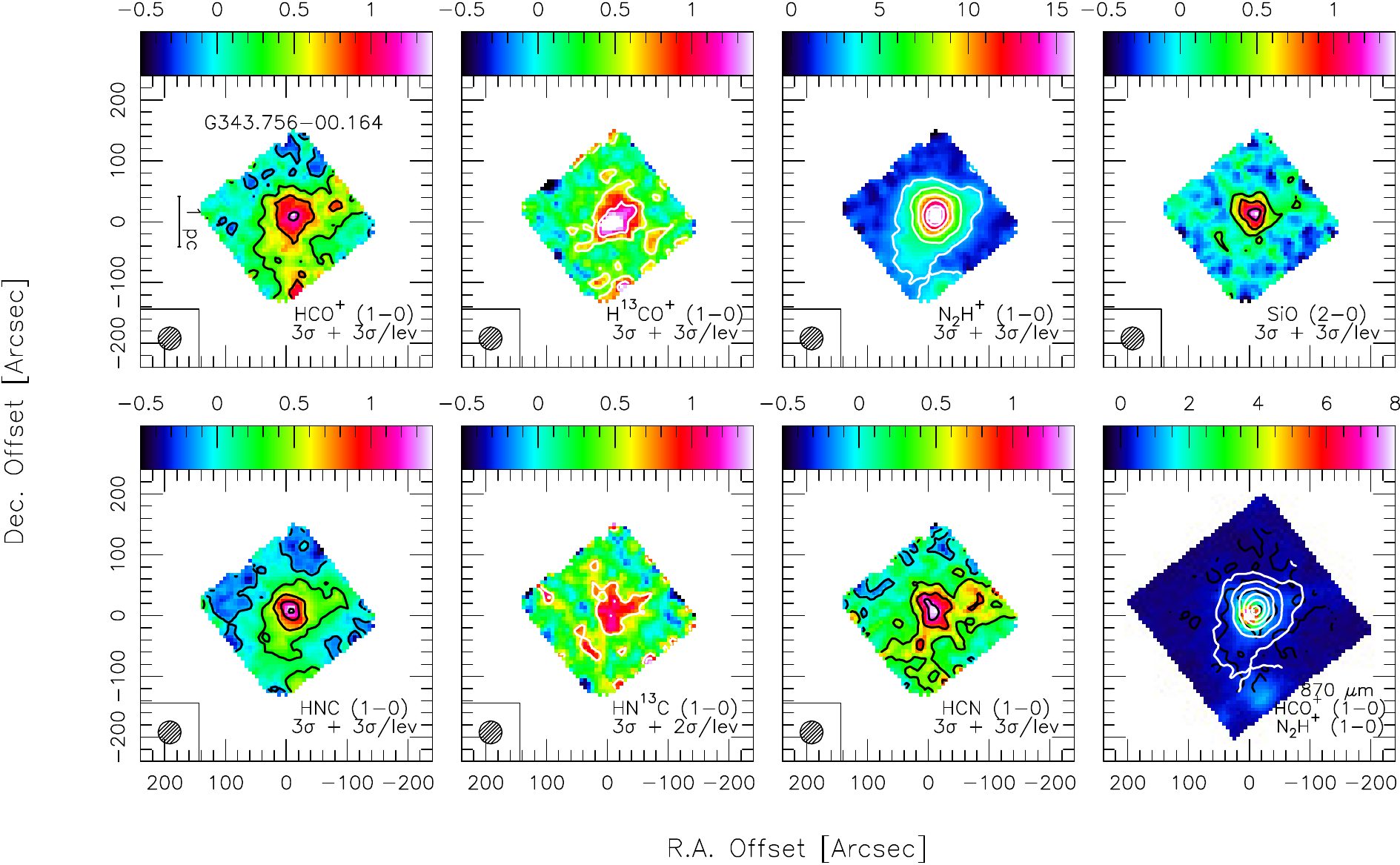}\vspace{0.5cm}
\includegraphics[width=0.50\textwidth]{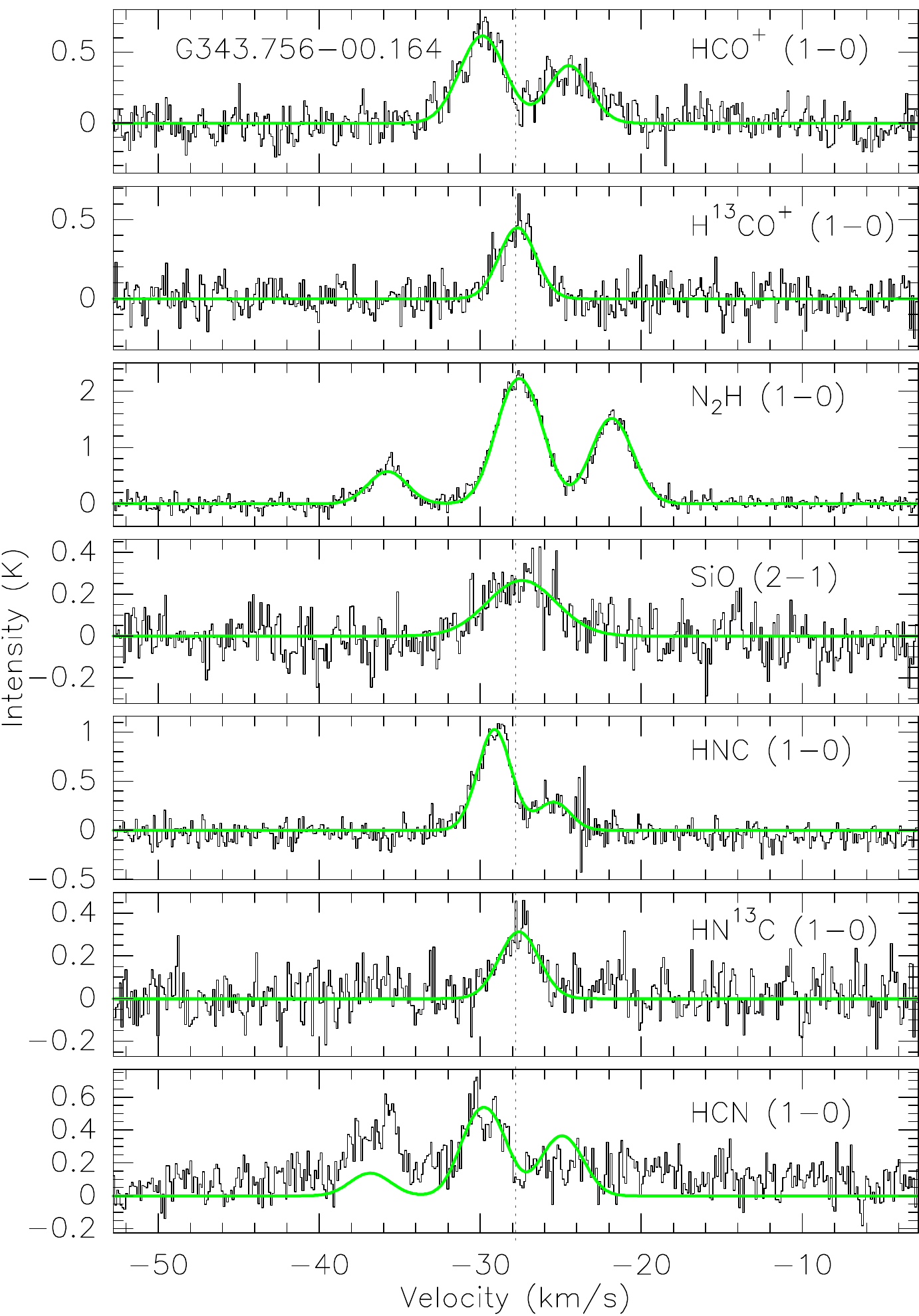}
\end{figure*}
\addtocounter{figure}{-1}
\begin{figure*}
\centering
\caption{Continued}
\includegraphics[width=0.80\textwidth]{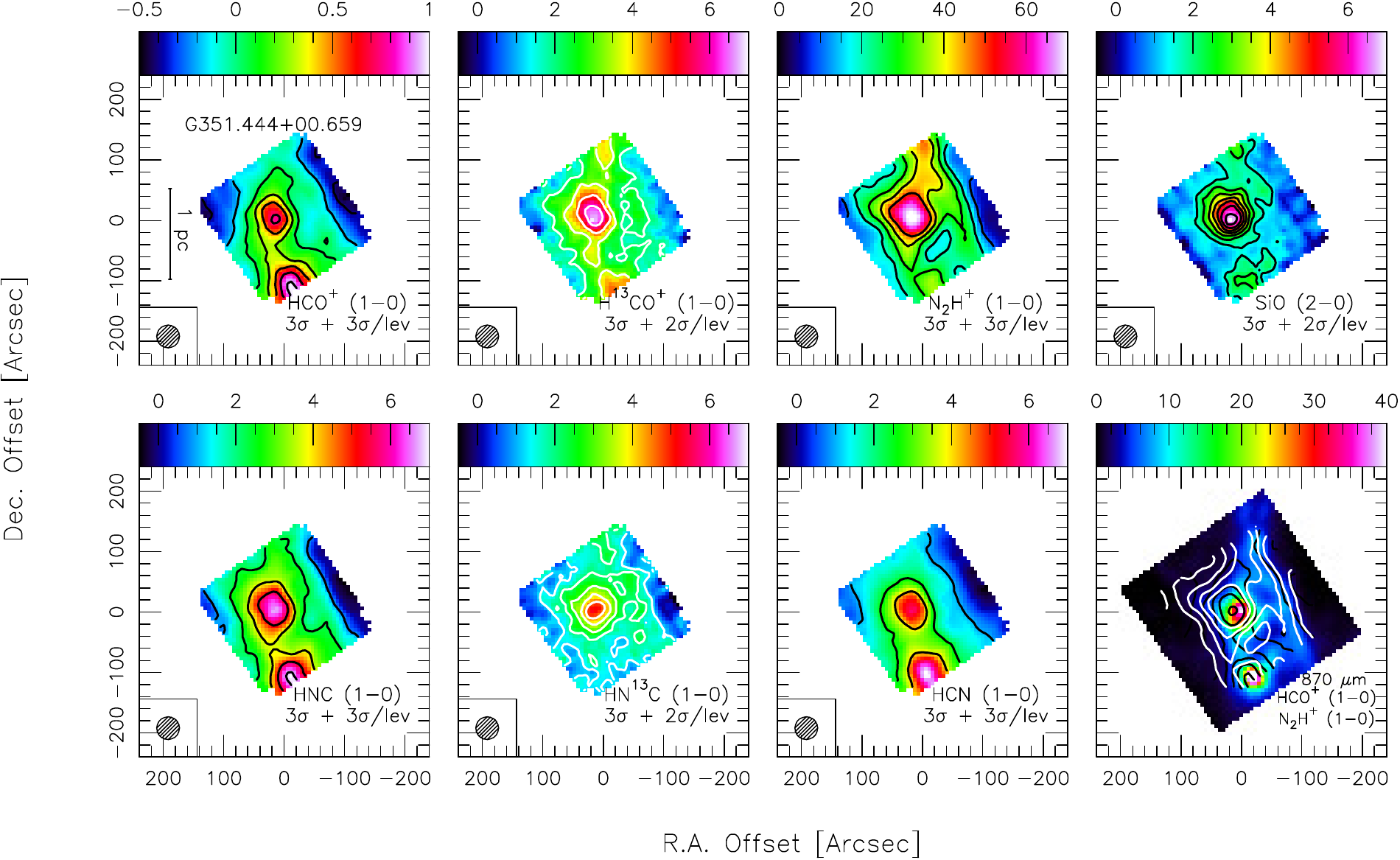}\vspace{0.5cm}
\includegraphics[width=0.50\textwidth]{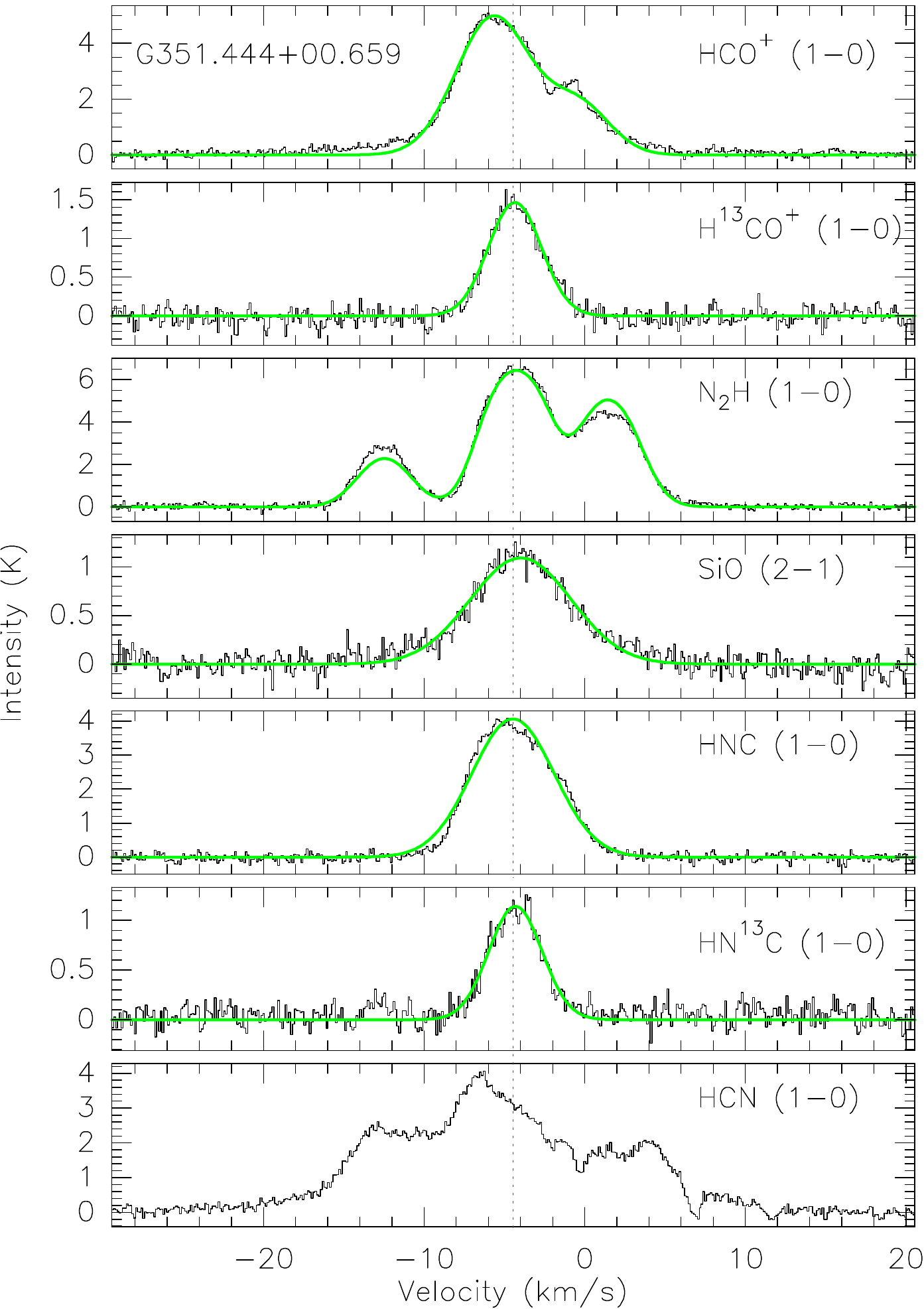}
\end{figure*}
\addtocounter{figure}{-1}
\begin{figure*}
\centering
\caption{Continued}
\includegraphics[width=0.80\textwidth]{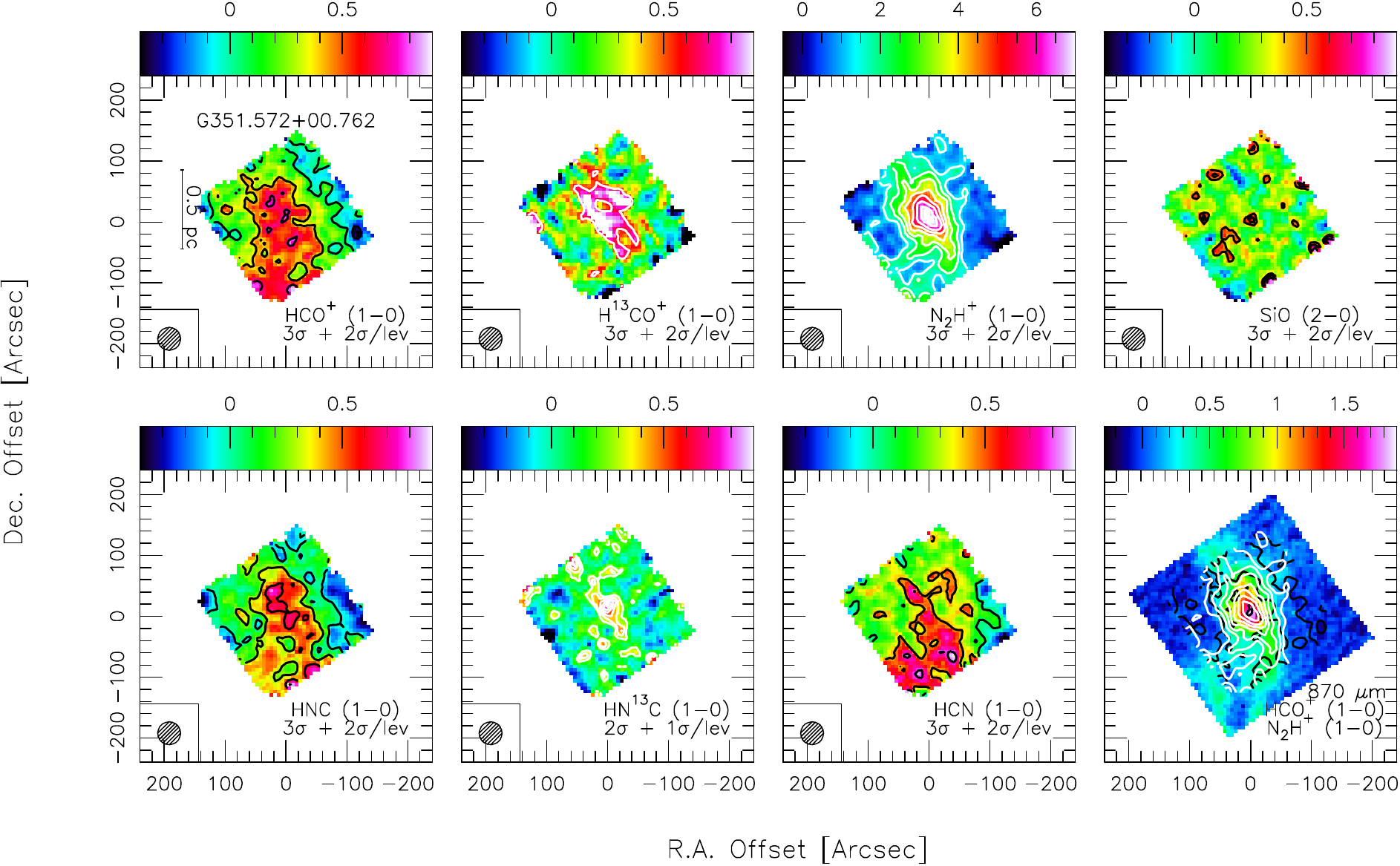}\vspace{0.5cm}
\includegraphics[width=0.50\textwidth]{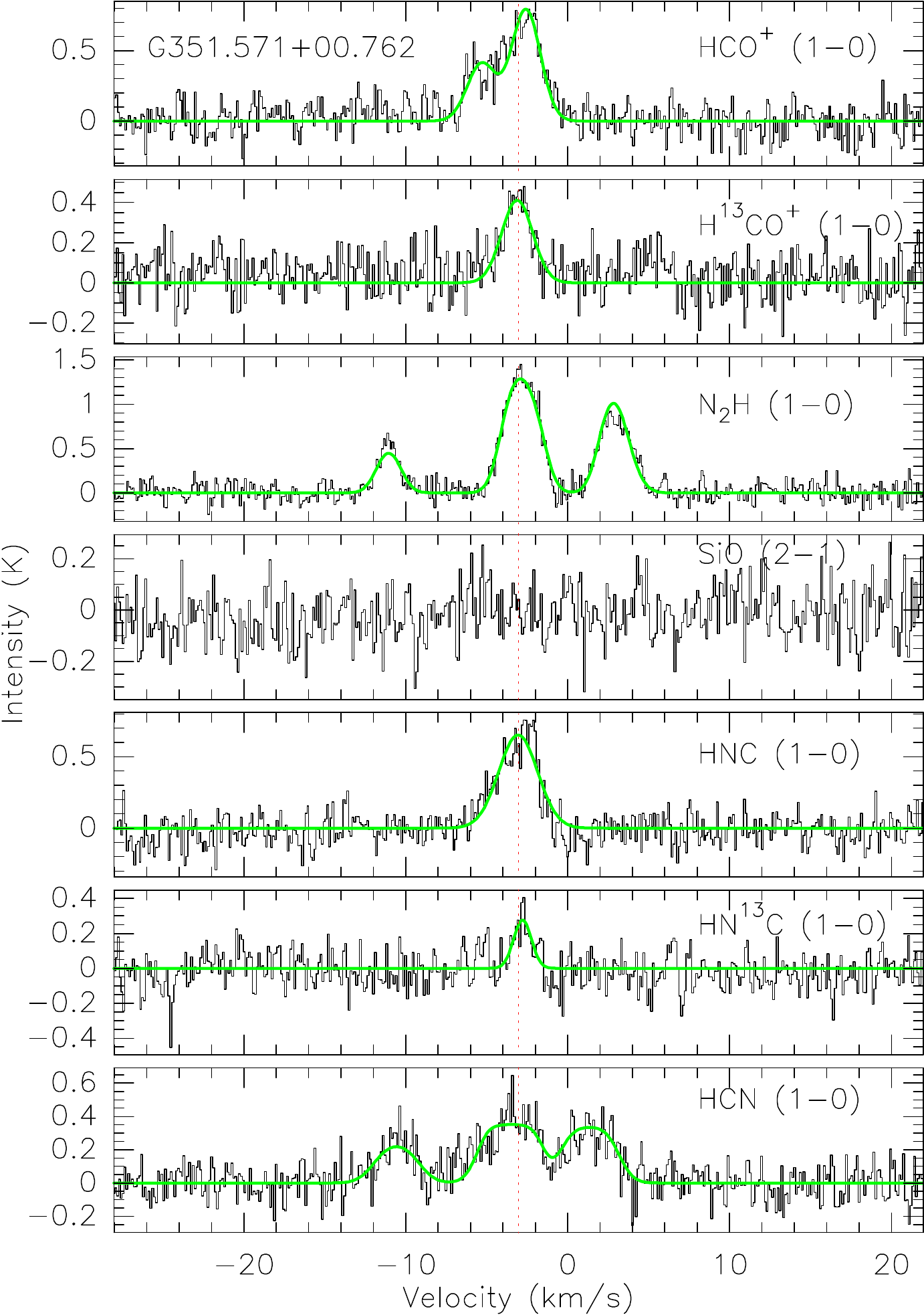}
\end{figure*}
\addtocounter{figure}{-1}
\begin{figure*}
\caption{Continued}
\centering
\includegraphics[width=0.80\textwidth]{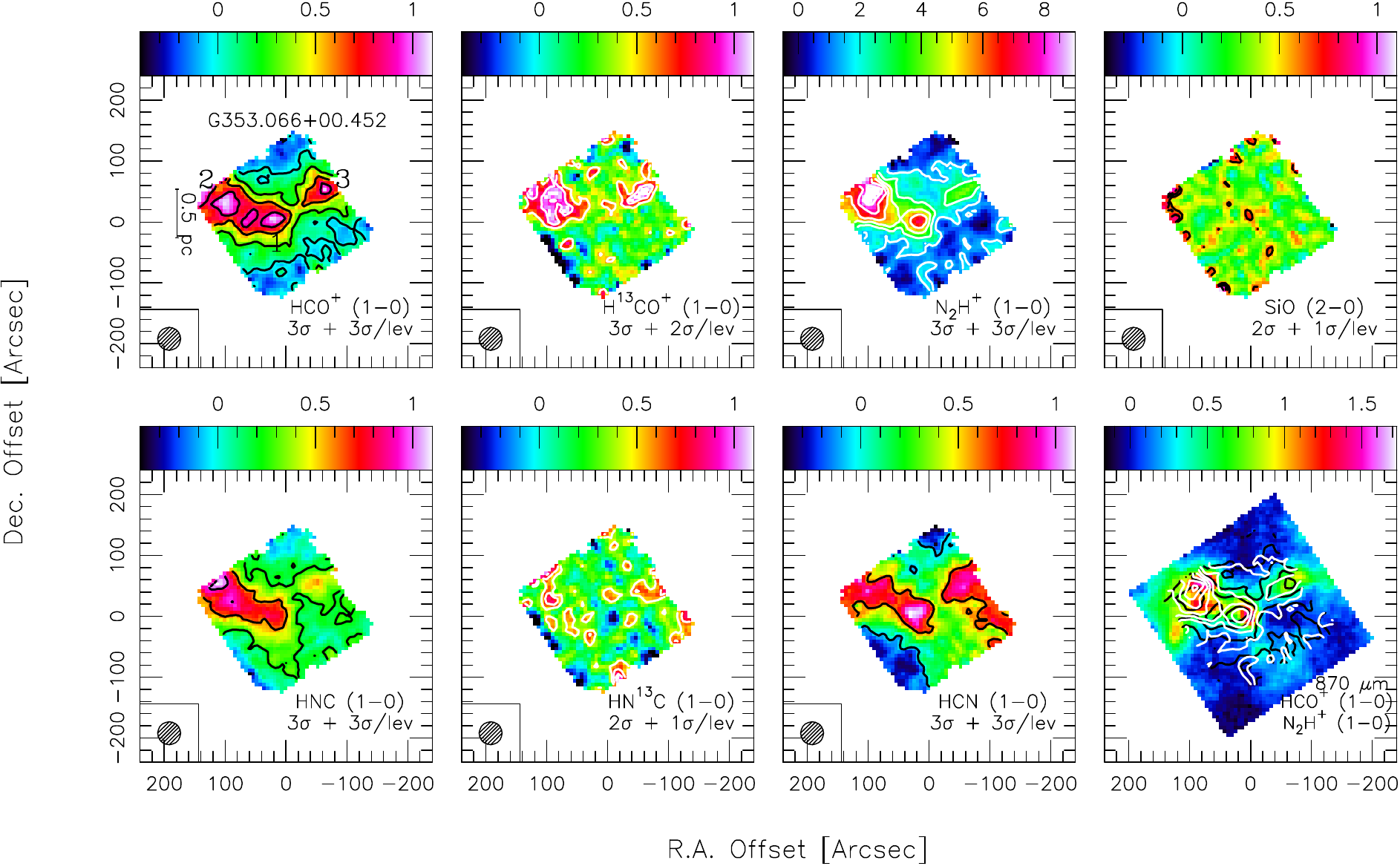}\\
\vspace{0.50cm}
%
\includegraphics[width=0.30\textwidth]{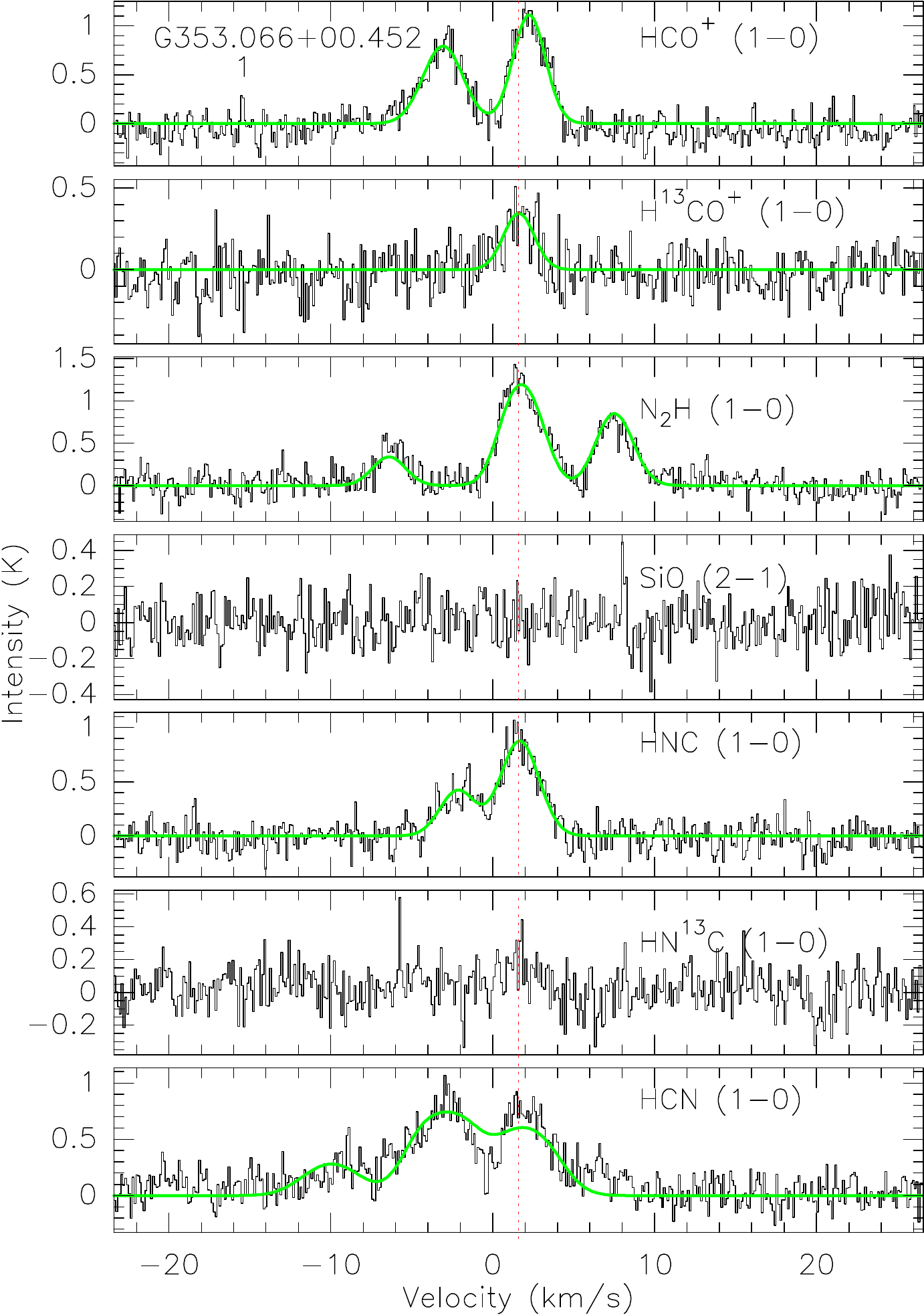}\hspace{0.3cm}
\includegraphics[width=0.30\textwidth]{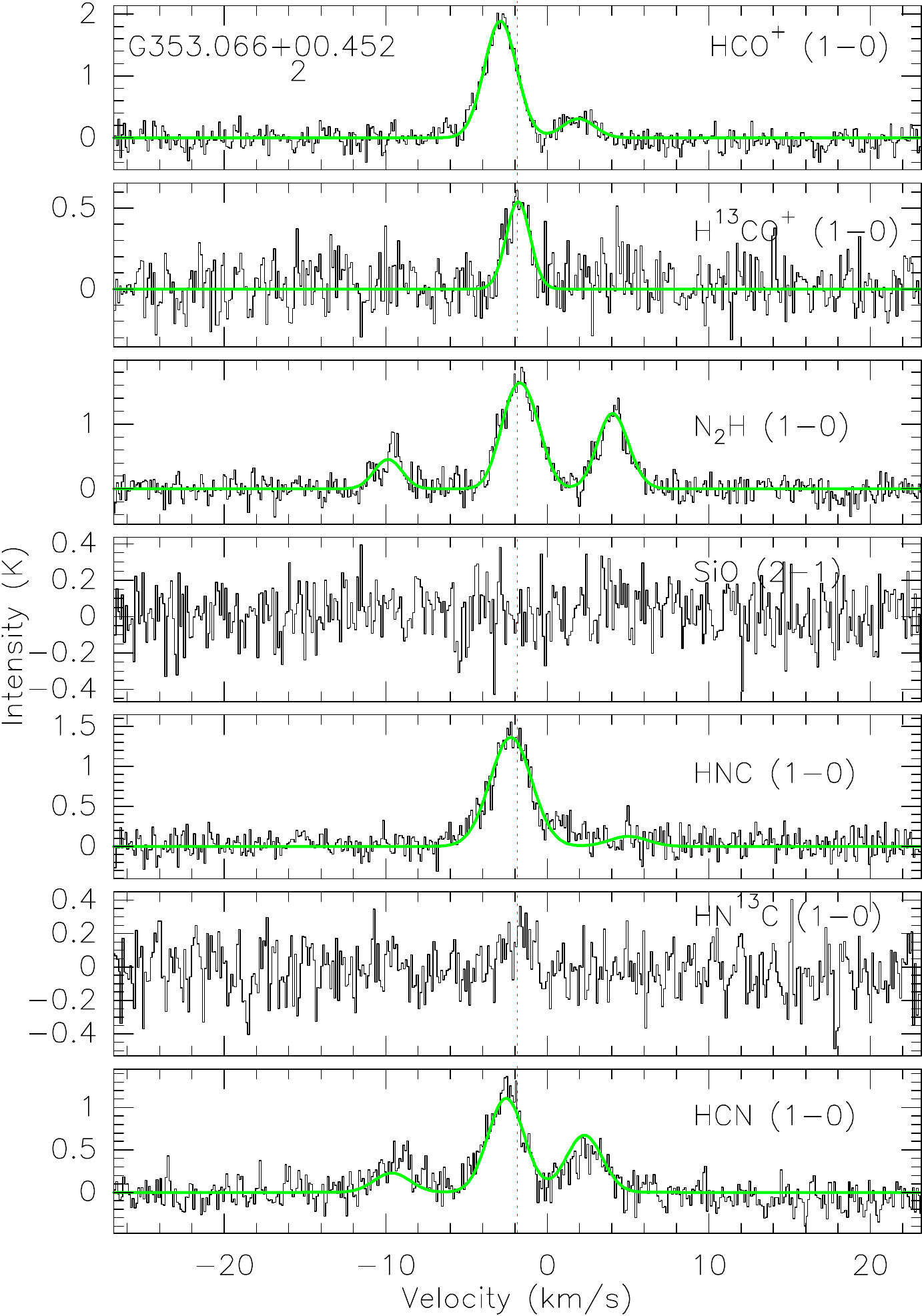}\hspace{0.3cm}
\includegraphics[width=0.30\textwidth]{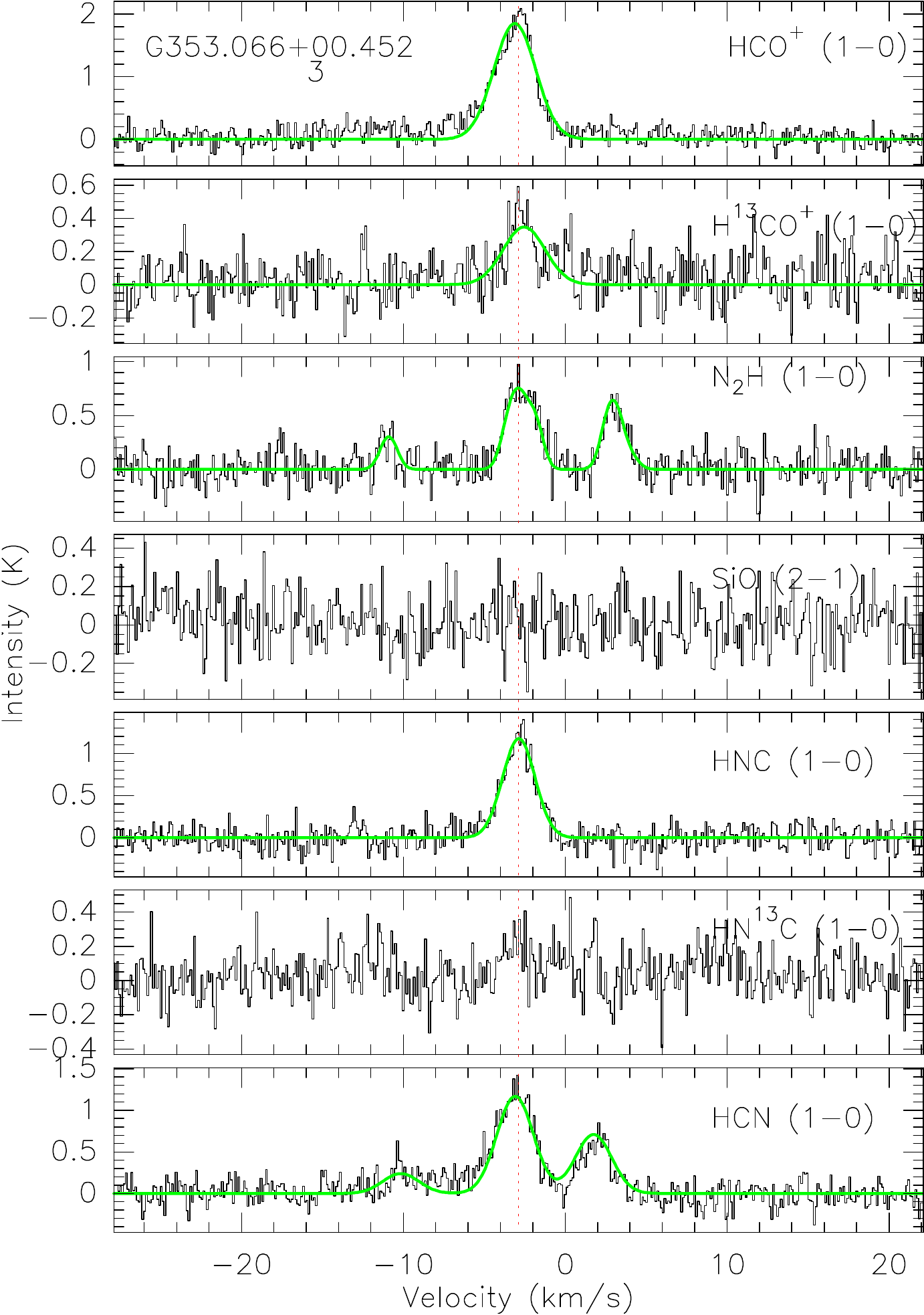}
\end{figure*}
\addtocounter{figure}{-1}
\begin{figure*}
\caption{Continued}
\centering
\includegraphics[width=0.80\textwidth]{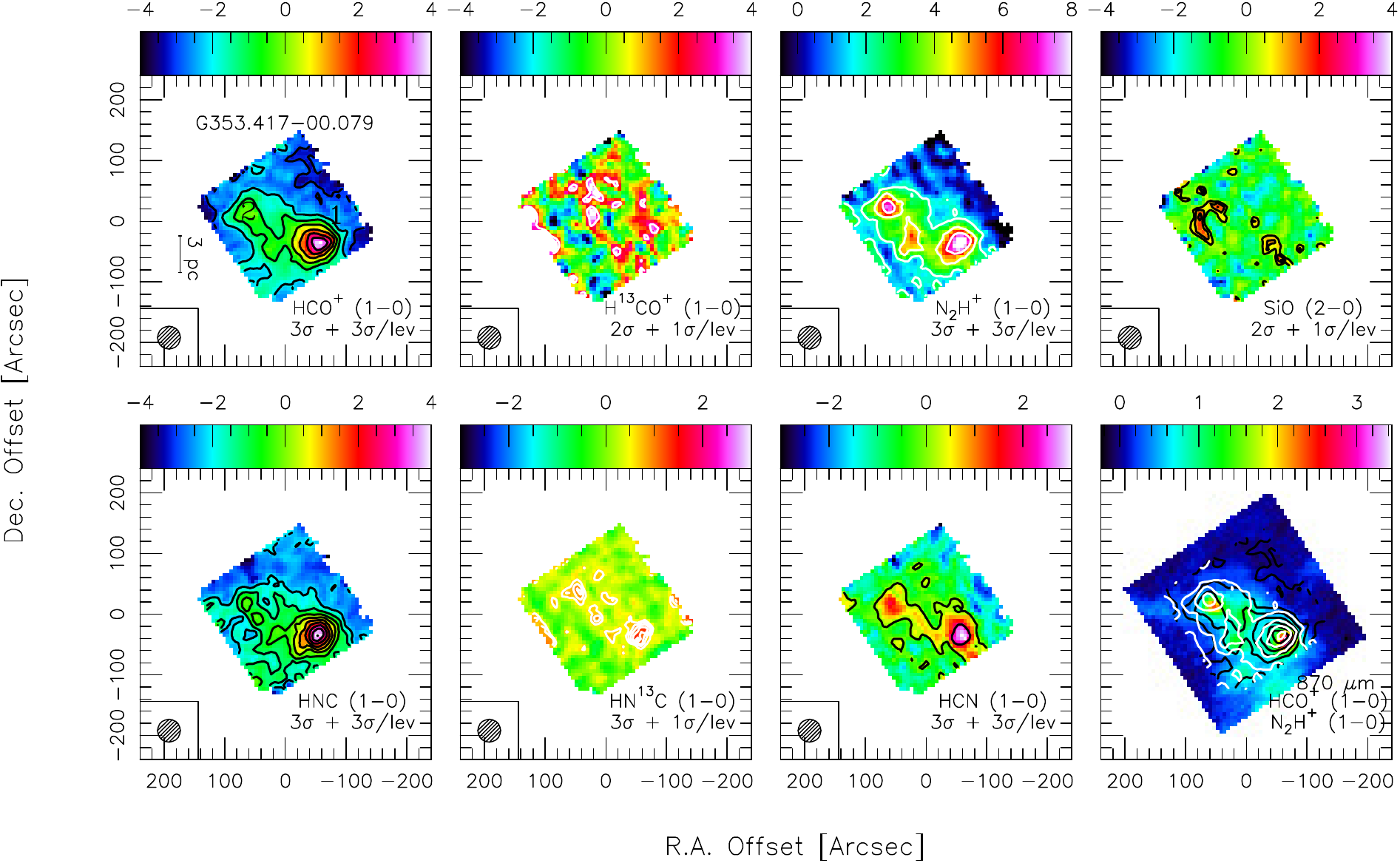}\vspace{0.50cm}
%
\includegraphics[width=0.45\textwidth]{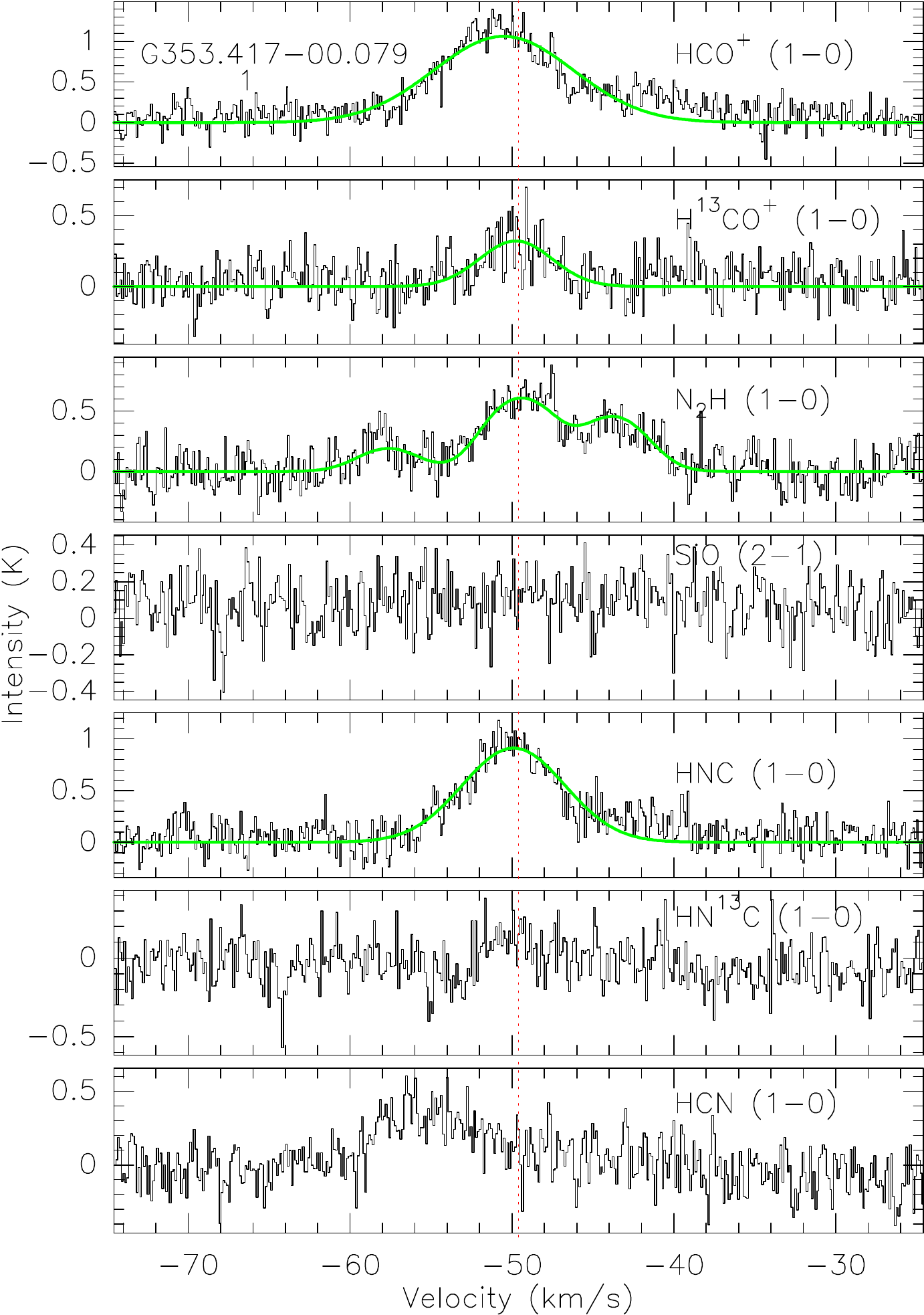}\hspace{0.50cm}
\includegraphics[width=0.45\textwidth]{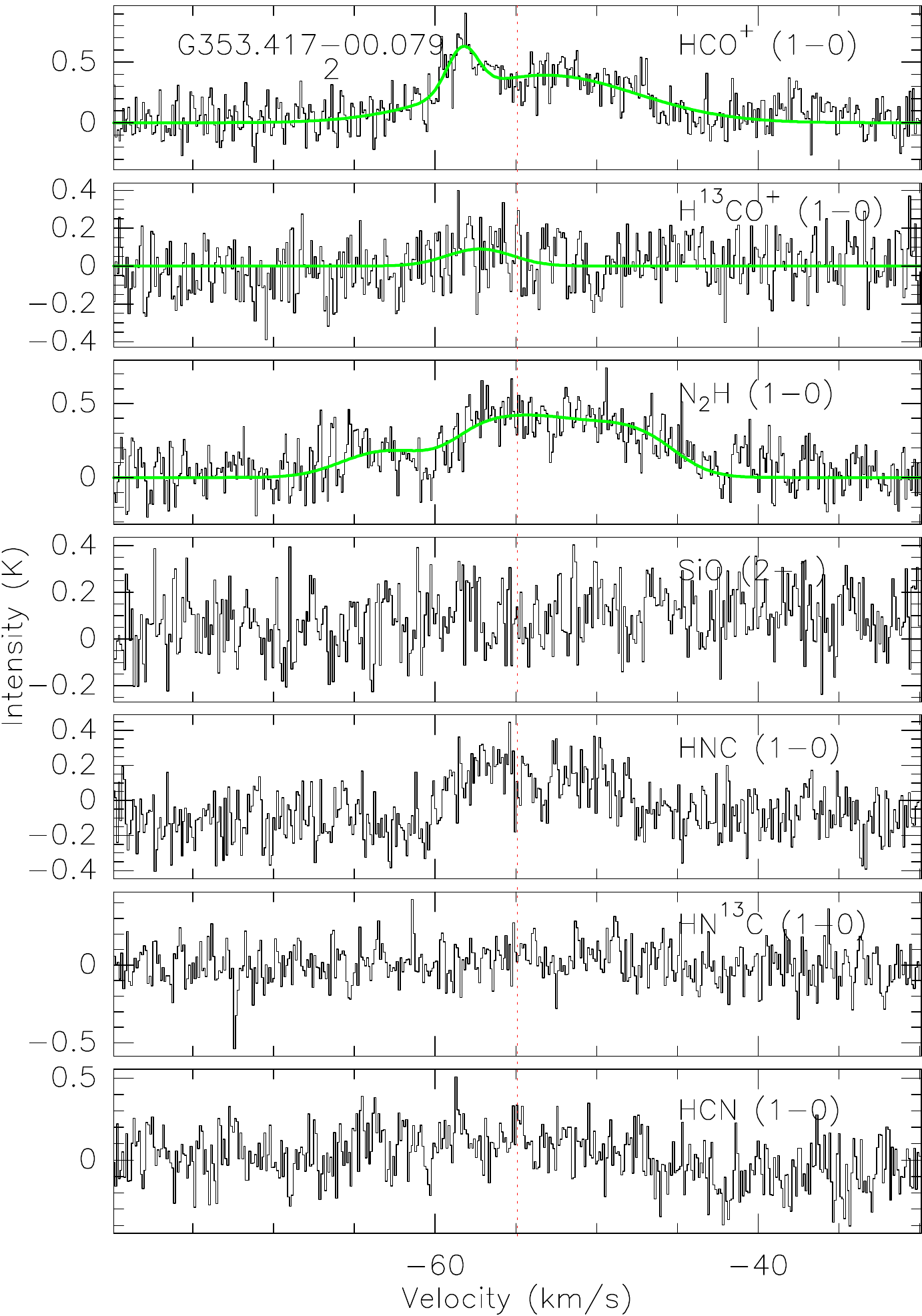}
\end{figure*}

\addtocounter{figure}{-1}

\begin{figure*}
\centering
\caption{Continued}
\includegraphics[width=0.80\textwidth]{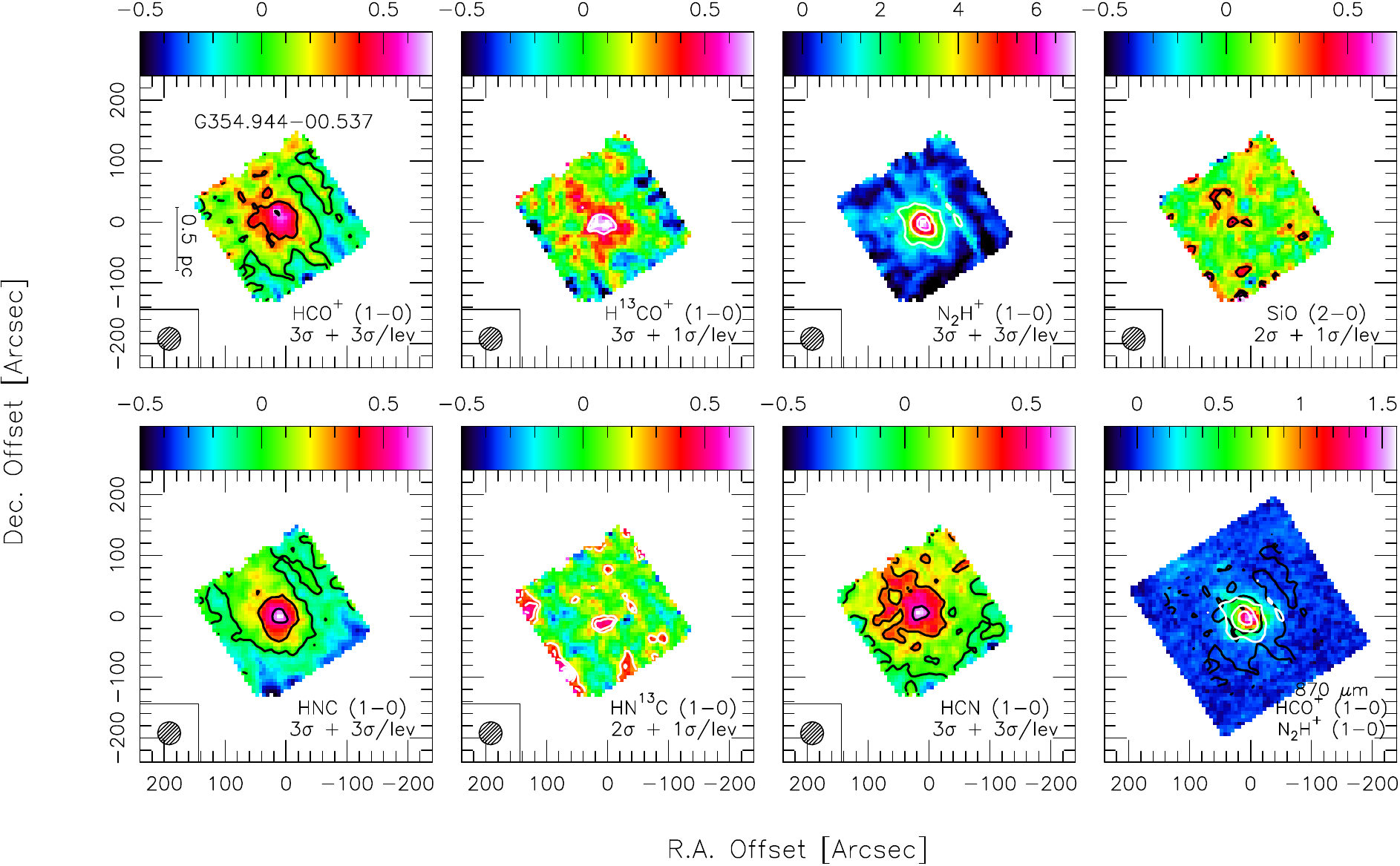}\vspace{0.5cm}
\includegraphics[width=0.50\textwidth]{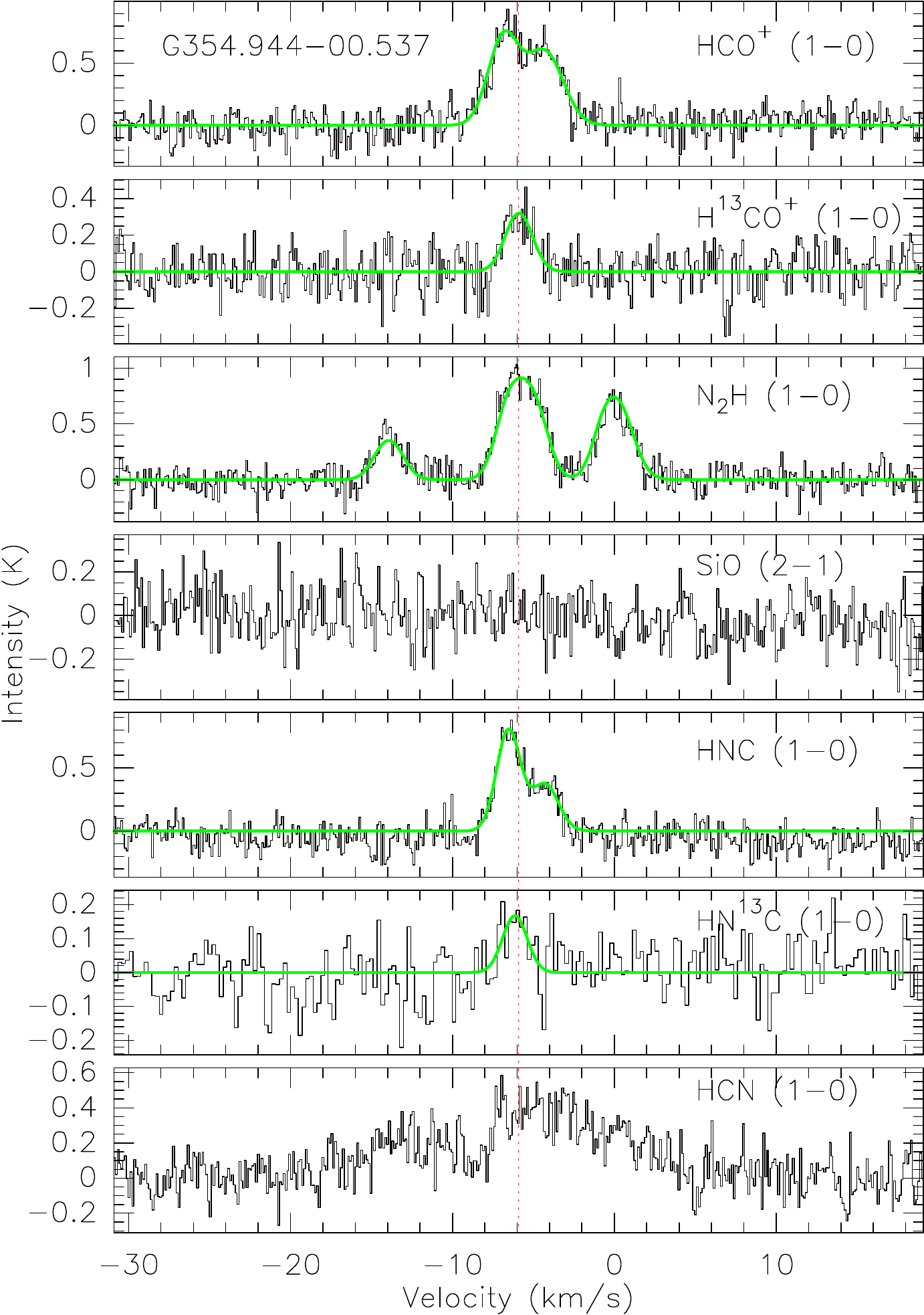}
\end{figure*}

\end{appendix}
\end{document}